\documentclass[12pt,a4paper]{elsarticle}
\usepackage{geometry}
\usepackage{newtxtext,newtxmath}
\usepackage{algorithmic}
\usepackage{amsfonts}
\usepackage{bm}
\usepackage{booktabs}
\usepackage{caption}
\usepackage{epstopdf}
\usepackage{extarrows}
\usepackage{fancyhdr}
\usepackage{float}
\usepackage{graphicx}
\usepackage{multirow}
\usepackage{nomencl}
\makenomenclature
\usepackage[numbers]{natbib}
\usepackage{pgfplots}
\pgfplotsset{compat=1.18}
\usepackage{subcaption}
\usepackage{siunitx}      
\usepackage{tikz}
\usepackage{xcolor}
\usepackage{color}
\usepackage[colorlinks,linkcolor=red,anchorcolor=blue,citecolor=green]{hyperref}

\usepackage{url}
\usepackage{appendix}
\usepackage[capitalise,nameinlink]{cleveref}
\crefname{equation}{Eq.}{Eqs.}
\biboptions{numbers,sort&compress}

\journal{Elsevier}

\begin{document}
	\begin{frontmatter}
		\title{\textbf{A GENERIC-guided active learning SPH method for viscoelastic fluids using Gaussian process regression}}
		
		\author[aff1,aff2]{Xuekai Dong}
		\author[aff3]{David Nieto Simavilla}
		\author[aff1]{Jie Ouyang}
		\author[aff1]{Xiaodong Wang\corref{cor1}}
		\ead{xiaodongwang@nwpu.edu.cn}
		\author[aff2,aff4,aff5]{Marco Ellero\corref{cor1}}
		\ead{mellero@bcamath.org}
		\cortext[cor1]{Corresponding author.}
		
		\address[aff1]{School of Mathematics and Statistics, Northwestern Polytechnical University, Xi'an, 710129, China}
		\address[aff2]{Basque Center for Applied Mathematics (BCAM), Alameda de Mazarredo 14, 48009, Bilbao, Spain}
		\address[aff3]{Dept. Energía y Combustibles, Escuela Técnica Superior de Ingenieros de Minas y Energía, Universidad Politécnica de Madrid, Calle Alenza 4, 28011, Madrid, Spain}
		\address[aff4]{IKERBASQUE, Basque Foundation for Science, Calle de Maria Diaz de Haro 3, 48013, Bilbao, Spain}
		\address[aff5]{Complex Fluids Research Group, Department of Chemical Engineering, Faculty of Science and Engineering, Swansea University, Swansea, SA1 8EN, United Kingdom\vspace{-1.1cm}}
		\date{June 5, 2025}
		\begin{abstract}
			When applying machine learning methods to learn viscoelastic constitutive relations, the polymer history dependence in viscoelastic fluids and the generalization ability of machine learning models are challenging. In this paper, guided by the general equation for nonequilibrium reversible-irreversible coupling (GENERIC) framework, a novel GENERIC-guided active learning smoothed particle hydrodynamics (${\rm{G^2ALSPH}}$) method is proposed to obtain effective constitutive relations for reliable simulations of viscoelastic flows. By utilizing the GENERIC framework, the target viscoelastic constitutive relation is reduced to a simple functional relation between the eigenvalues of the conformation tensor and the eigenvalues of its thermodynamically conjugated tensorial variable, which incorporates the flow-history-dependent memory effect. Based on data and Gaussian process regression (GPR), a new active learning strategy is developed to obtain the simplified constitutive relation, in which the generalization ability is ensured by actively acquiring more data points when needed. Moreover, a novel relative uncertainty is devised to establish an accuracy evaluation tool for the GPR prediction results, which reduces the number of required training data points while maintaining accuracy. Furthermore, the SPH method combined with the latest techniques serves as an effective macroscopic numerical method. Eventually, the Poiseuille flows and the flows around a periodic array of cylinders at different Weissenberg numbers are simulated to validate the effectiveness and accuracy of the ${\rm{G^2ALSPH}}$ method. The Oldroyd-B model is used as the ground truth constitutive relation to provide the required data for GPR, hence bringing analytical solutions for comparison. The excellent performance demonstrates that the ${\rm{G^2ALSPH}}$ method has a promising prospect of applications for data-driven simulations of viscoelastic fluids.
		\end{abstract}
		\begin{highlights}
			\item The GENERIC framework simplifies the constitutive relations of viscoelastic fluids.
			\item A GPR active learning strategy is proposed to gain simplified constitutive relations.
			\item A novel relative uncertainty reduces the required data while maintaining accuracy.
			\item The SPH method with the latest techniques serves as an effective macroscopic method.
			\item The ${\rm{G^2ALSPH}}$ method provides accurate data-driven simulations for viscoelastic flows.
		\end{highlights}
		\begin{keyword}
			Smoothed particle hydrodynamics; Viscoelastic fluids; GENERIC framework; Gaussian process regression; Active learning; Relative uncertainty; 
		\end{keyword}
	\end{frontmatter}
\section{Introduction}\label{sec1}
The research on viscoelastic fluids is of great significance~\cite{ma_adaptive_2024}. With the increasing computational capability of computers, numerical methods have become effective tools to study viscoelastic fluids~\cite{zohravi_Mesoscale_2025,oishi_Numerical_2012}. In order to accurately capture the flow characteristics of viscoelastic fluids, not only is the selection of macroscopic numerical simulation methods important, but the establishment of the constitutive relation is also essential.
\par
The smoothed particle hydrodynamics (SPH) method is a meshless Lagrangian method first proposed by Gingold and Monaghan and Lucy~\cite{gingold_Smoothed_1977,lucy_numerical_1977}. The SPH method can avoid some issues in mesh-based methods, including initial mesh generation and mesh distortion. Moreover, the SPH method does not require additional free surface tracking techniques. These advantages have led to its successful applications in a wide range of scientific problems, including free-surface flows~\cite{zhang_highorder_2025,sun_Inclusion_2023}, incompressible flows~\cite{meng_Highorder_2024,decourcy_Incompressible_2024}, fluid-structure interaction~\cite{guo_SmoothedInterface_2024}, surface tension~\cite{cen_singlephase_2024}, and geomechanics~\cite{hoang_Development_2024}. Ellero et al.~\cite{ellero_Viscoelastic_2002,ellero_SPH_2005} first applied the SPH method to numerical simulations of viscoelastic fluids. Since then, many researchers have utilized the SPH method to study various viscoelastic flow problems, such as droplet impact on a solid wall~\cite{xu_Numerical_2024}, cavity filling processes~\cite{xu_Multiscale_2023}, and non-isothermal filling processes~\cite{xu_2D_2025}. Examples of considered constitutive models include the Oldroyd-B model~\cite{ren_Simulation_2016}, the extended Pom–Pom (XPP) model~\cite{jiang_numerical_2014}, and the Phan–Thien–Tanner (PTT) model~\cite{evans_Numerical_2022}. In this paper, the SPH method serves as an effective macroscopic numerical method to provide reliable simulation results for viscoelastic fluids.
\par
The constitutive relation directly captures the complex time-dependent stress-strain relation inherent in the viscoelastic fluid, which is necessary to close the governing equations in the numerical simulation. The search for a valid constitutive relation for a specific viscoelastic fluid has become one of the most important aspects in the numerical simulation~\cite{dong_active_2024}. Two classical approaches to provide this relation include analytically-derived constitutive equations and multi-scale methods~\cite{jug_Learning_2024}. Constitutive equations can be obtained by either theoretical or phenomenological approaches. It is direct and computationally convenient to apply a constitutive relation in equation form to numerical simulations, for example, the Oldroyd-B model or the PTT model~\cite{xu_Numerical_2024,evans_Numerical_2022}. However, these approaches are limited, since the mesoscopic information can not be fully captured. The applicability of constitutive equations is generally restricted. Multi-scale methods can provide full microscopic information~\cite{xu_Multiscale_2023}, but mesoscopic numerical simulations are time-consuming. In recent years, machine learning methods have become effective tools to tackle this issue by reducing the number of required mesoscopic numerical simulations and extending the applicability of the learned constitutive relations.
\par
Probably the biggest challenge for applying machine learning methods to learn viscoelastic constitutive relations is the history dependence. To address this problem, some researchers have adopted various machine learning strategies to learn the time derivative of the stress or the time derivative of the microstructured polymer conformation tensor. Lei et al.~\cite{lei_Machinelearningbased_2020} have adopted a neural network method to learn the time derivative of the conformation tensor, named the deep non-Newtonian model (DeePN$^{2}$). Fang et al.~\cite{fang_DeePN2_2022} have further applied DeePN$^{2}$ to capture the viscoelastic responses of the chain- and star-shaped molecule suspensions. Lennon et al.~\cite{lennon_Scientific_2023} have built a neural network method to learn Rheological Universal Differential Equations (RUDEs). Some researchers have utilized neural networks to learn microstructure evolution. Young et al.~\cite{young_ScatteringInformed_2023a} have introduced a data-driven framework to predict the microstructure and stress evolution of complex flows by using scattering data. Mangal et al.~\cite{mangal_Learning_2025} have explored the application of neural operators to learn families of rheological constitutive models. Dabiri et al.~\cite{dabiri_Fractional_2023} have developed fractional rheology-informed neural networks (RhINNs) to identify fractional viscoelastic constitutive models from data. Besides, Vlachas et al.~\cite{vlachas_Multiscale_2022,vlachas_Accelerated_2022} have introduced the learning effective dynamics (LED) framework for molecular systems based on the neural network, and it has been further improved for modeling complex systems~\cite{kicic_Adaptive_2023,gao_Generative_2024}.
\par
The advantage of using neural network methods to learn constitutive relations is that it is general and straightforward~\cite{fang_DeePN2_2022}. As long as the neural network is built appropriately, good learning results can be obtained after training on a large number of data. However, two main shortcomings limit their applicability. The first is that the number of required data is large. The other is that the over-fitting phenomenon affects the generalization ability. When the learned constitutive relation is applied to a numerical case where the required range is beyond the training range, the results can be inaccurate. The accuracy of the entire simulation process can not be ensured at every time instant, and accumulated errors can further lead to inaccurate simulation results~\cite{mangal_Datadriven_2025}. These two shortcomings also appear in physics-informed neural networks (PINNs)~\cite{bonfanti_generalization_2024,cai_Physicsinformed_2021}.
\par
Compared to neural network methods, Gaussian process regression (GPR) provides the prediction results along with the corresponding direct uncertainties. Seryo et al.~\cite{seryo_Learning_2020} have employed GPR to learn the time derivative of the stress tensor for fluids with linear constitutive relations. Following that, Souta et al.~\cite{souta_Machinelearned_2023} have adopted the dual slip-link model to express non-linear stress response in well-entangled polymer fluids. These two works are straightforward and effective but come along with three limitations. The first is that the target function has a large number of variable inputs, which increases the complexity of GPR. The second is that the accuracy of the learned constitutive relation when applied to different numerical cases can not be ensured, because it is learned from a specific data set. The third is that the huge number of training data makes the application of the learning constitutive relation time-consuming. In this paper, the general equation for the nonequilibrium reversible-irreversible coupling (GENERIC) framework and the active learning strategy are adopted to address these challenges.
\par
To simplify the target constitutive relation and reduce the problem dimensionality, the GENERIC framework is introduced, which bridges the gap between dilute polymer solutions and macroscopic numerical simulations. The general equations of motion for a set of fluid particles carrying polymer molecules in suspension can be derived with the GENERIC framework~\cite{vazquez-quesada_Smoothed_2009,vazquez-quesada_Shear_2019}. The constitutive relation suggested by the GENERIC framework can be simplified to a functional relation between the eigenvalues of the conformation tensor $\bm{c}$ and the eigenvalues of the tensorial variable $\bm{\sigma}(\bm{c})$ for a viscoelastic fluid whose constitutive model is not pre-known. Here $\bm{\sigma}(\bm{c})$ is the tensorial variable thermodynamically conjugated to $\bm{c}$. Due to the physical symmetry, a simple function with multiple inputs and a single output becomes the final target function to be learned, which incorporates the flow-history-dependent memory effect.
\par
To reduce the number of required data while maintaining the accuracy of the entire simulation process, a new active learning strategy is proposed in this work to obtain the simplified target function. Zhao et al.~\cite{zhao_Active_2021} have already constructed a special constitutive equation based on GPR and mesoscopic data, which works well for a specific fluid but lacks generality since the learning is limited to parametrizing a constitutive relation with predefined equations. Following that work, Chang et al.~\cite{chang_multiscale_2023} have built a multi-scale framework to actively learn the constitutive relation of complex suspension flows. However, the direct uncertainty is chosen as an accuracy evaluation tool for GPR prediction results in their works, which is restrictive and requires more training data. In this paper, a novel relative uncertainty is devised, whose value at a given point is calculated by the ratio of the direct uncertainty to the absolute regularized value of the predicted mean. Furthermore, a new active learning strategy is established by employing the relative uncertainty as the accuracy evaluation tool for the GPR prediction results. The new active learning strategy reduces the required data while maintaining accuracy compared to the active learning strategy based on the direct uncertainty. Meanwhile, the generalization ability is well ensured by the new active learning strategy.
\par
Compared to the aforementioned methods, the GENERIC-guided active learning smoothed particle hydrodynamics (${\rm{G^2ALSPH}}$) method proposed here has three main advantages. Firstly, the adoption of the GENERIC framework brings universality for the ${\rm{G^2ALSPH}}$ method, allowing the parameterization of a family of constitutive models rather than a specific one. Moreover, the time-dependent memory effect in polymer simulations can be captured by learning only a simple function. Secondly, GPR can give prediction results along with the corresponding direct uncertainties. By a clever active learning strategy based on a novel relative uncertainty, the accuracy of the learned constitutive relation can be well guaranteed for the entire simulation process with a small data set. Thirdly, the ${\rm{G^2ALSPH}}$ method is computationally convenient and transferable, since it is easy to apply the learned constitutive relation to a different simulation process. In general, the required data for training should be obtained by experimental data or independent mesoscopic polymer simulations~\cite{moreno_Arbitrary_2021,simavilla_Mesoscopic_2022,simavilla_Nonaffine_2023,moreno_Generalized_2023}. For example, in the simplest case of the suspended Hookean dumbbells, the Oldroyd-B model would be recovered~\cite{vazquez-quesada_SPH_2017}. In this paper, in order to make comparisons and give validations, the Oldroyd-B model is adopted as an analytical constitutive relation to provide the required data. The GPR procedure reconstructs the constitutive relation from the generated data, discovering the underlying polymer physics, i.e., precisely the Oldroyd-B model in this work. Moreover, for the viscoelastic Poiseuille flow, the Oldroyd-B model brings analytical results~\cite{xue_Numerical_2004,ellero_SPH_2005}, which makes the validation process more direct and accurate.
\par
This article is structured as follows: In~\cref{sec2}, the governing equations and the SPH method are introduced. In~\cref{sec3}, GPR and the active learning strategy are elaborated. In~\cref{sec4}, the Poiseuille flows and the flows around a periodic array of cylinders at different Weissenberg numbers are simulated to validate the effectiveness and accuracy of the ${\rm{G^2ALSPH}}$ method. In~\cref{sec5}, the conclusions are presented. 
\section{The smoothed particle hydrodynamics method}\label{sec2}
The fundamental governing equations for viscoelastic fluid flow and their numerical discretization within the SPH framework are addressed in this section. The SPH method is established to effectively simulate viscoelastic flows by coupling the modified kernel gradient, the viscosity term discretization, the improved solid wall boundary conditions, and an optimized particle shifting technique.
\subsection{Governing equations}\label{sec2.1}
\subsubsection{The mass and momentum conservation equations}\label{sec2.1.1}
In the 2D Lagrangian frame, for an isothermal, transient, weakly compressible viscoelastic fluid, the basic governing equations can be written as~\cite{xu_Numerical_2024}
\begin{equation}
	\dot{\rho} = -\rho\nabla\cdot\bm{u},
	\label{eq:mass}
\end{equation}
\begin{equation}
	\bm{\dot{u}} = \frac{1}{\rho }\nabla\cdot\bm{\sigma}_{\rm{c}} + {\bm{F}},
	\label{eq:momentum}
\end{equation}
where the symbol \verb+"+$\cdot$\verb+"+ above the physical quantity indicates its material derivative. For example, the expression $\dot{\rho}$ represents $\partial\rho/\partial t+\bm{u}\cdot\nabla\rho$. The variables $\rho$ and $t$ denote the density and the time, respectively. The vectors $\bm{u}$ and $\bm{F}$ are the velocity and the acceleration from the external force, respectively. The tensor $\bm{\sigma}_{\rm{c}}$ is the Cauchy stress tensor, which is decomposed into the ordinary isotropic pressure $p$ and the extra stress tensor $\bm{\tau}$. For the viscoelastic fluid, the extra stress tensor $\bm{\tau}$ consists of the Newtonian solvent contribution $\bm{\tau}_{\rm{s}}$ and the polymer contribution $\bm{\tau}_{\rm{p}}$. The expression $\bm{\tau}=\bm{\tau}_{\rm{s}}+\bm{\tau}_{\rm{p}}$ is obtained. The Newtonian solvent stress tensor $\bm{\tau}_{\rm{s}}$ is computed by
\begin{equation}
	\bm{\tau}_{\rm{s}}=2 \eta_{\rm{s}} \bm{D},
	\label{eq:solvent_stress}
\end{equation}
\begin{equation}
	\bm{D}=\frac{1}{2}\left(\nabla \bm{u}+(\nabla \bm{u})^{\mathrm{T}}\right),
	\label{eq:deformation_rate}
\end{equation}
where $\bm{D}$ is the symmetric deformation rate tensor and $\eta_{\rm{s}}$ is the constant solvent viscosity. The symbol \verb+"+$\mathrm{T}$\verb+"+ denotes the transpose of a matrix. The total viscosity $\eta$ is defined by the expression $\eta=\eta_{\rm{s}}+\eta_{\rm{p}}$, where $\eta_{\rm{s}}$ and $\eta_{\rm{p}}$ are the solvent and polymer contributions to the total solution viscosity, respectively. Hence, the total stress tensor $\bm{\sigma}_{\rm{c}}$ is computed by~\cite{xu_2D_2025}
\begin{equation}
	\bm{\sigma}_{\rm{c}}=-p \bm{I}+2 \eta_{\rm{s}} \bm{D}+\bm{\tau}_{\rm{p}},
	\label{eq:total_stress}
\end{equation}
where $\bm{I}$ is the unit matrix.
\par
The weakly compressible assumption is used for the convenience of solving the pressure field. For the low-velocity flow, the compressibility of the fluid can be ignored, yielding $\nabla\cdot\bm{u}=0$. Therefore, by substituting the $\nabla\cdot\bm{u}=0$,~\cref{eq:deformation_rate}, and~\cref{eq:total_stress} into~\cref{eq:momentum}, the final momentum equation is obtained as~\cite{xu_Multiscale_2023}
\begin{equation}
	\bm{\dot{u}} = -\frac{1}{\rho }\nabla p + \frac{{{\eta_{\rm{s}}}}}{\rho }{\nabla ^2}{\bm{u}} + \frac{1}{\rho}\nabla\cdot\bm{\tau}_{\rm{p}} + {\bm{F}}.
	\label{eq:momentum2}
\end{equation}
The~\cref{eq:mass} and~\cref{eq:momentum2} form the final mass and momentum conservation equations. The polymer stress tensor $\bm{\tau}_{\rm{p}}$ is computed by the constitutive equation discussed in~\cref{sec2.1.3}.
\subsubsection{The equation of state}\label{sec2.1.2}
To close the governing equations, the equation computing the pressure is required. In the incompressible SPH method, the fluid pressure is implicitly computed either from a pressure Poisson equation or by constraining the dynamics of the mass density by means of Lagrangian multipliers~\cite{cummins_sph_1999,ellero_Incompressible_2007}. For the weakly compressible SPH method considered here, the equation of state is commonly adopted, which describes the relation between pressure and density. In this paper, a widely used form is adopted as~\cite{li_Extension_2024}
\begin{equation}
	p(\rho)=\frac{\rho_0 {c_{\rm{s}}}^2}{\gamma}\left(\left(\frac{\rho}{\rho_0}\right)^\gamma-1\right),
	\label{eq:state}
\end{equation}
where $\rho_0$ is the initial fluid density and $c_{\rm{s}}$ is the artificial sound speed. The parameter $\gamma$ is always set to $7$. To keep the variation of the fluid density within $1\%$, the constraint ${c_{\rm{s}}}\geq10 \max \left(|\bm{u}|\right)$ is required. This ensures that the flow behavior of the weakly compressible fluid is close enough to that of the truly incompressible fluid.
\subsubsection{The GENERIC framework for viscoelastic constitutive equations}\label{sec2.1.3}
In this paper, in order to maintain the full generality of the learned constitutive relation for viscoelastic fluids, dilute polymer solutions are considered. The state of the dilute polymer solution is characterized by the thermodynamic variables and the conformation tensor $\bm{c}$, which captures the average molecular state of the polymer molecules. Within the GENERIC framework, the equations of motion for a set of fluid particles carrying general polymer molecules can be derived~\cite{vazquez-quesada_Smoothed_2009,ellero_Thermodynamically_2003}. For a dilute polymer solution undergoing affine deformation, the evolution of the microstructured polymer conformation tensor $\bm{c}$ is obtained by~\cite{vazquez-quesada_Smoothed_2009}
\begin{equation}
	\dot{\bm{c}}=\bm{c} \cdot \bm{\kappa}+\bm{\kappa}^T \cdot \bm{c}+\frac{2}{\lambda_{\rm{p}} N_{\rm{p}} k_{\rm{B}} T} \bm{c} \cdot \bm{\sigma},
	\label{eq:conformation_tensor}
\end{equation}
where $\lambda_{\rm{p}}$, $N_{\rm{p}}$, $k_{\rm{B}}$, and $T$ are the polymeric relaxation time, the polymer number density, the Boltzmann constant, and the temperature, respectively. In~\cref{eq:conformation_tensor}, the velocity gradient is defined as $\bm{\kappa}=(\nabla \bm{u})^{\mathrm{T}}$, and $\bm{\sigma}(\bm{c})$ is the tensorial variable thermodynamically conjugated to $\bm{c}$, which is defined as 
\begin{equation}
	\frac{\bm{\sigma}}{T}=\frac{\partial \bm{s}_{\rm{p}}}{\partial \bm{c}},
	\label{eq:tensorial_variable}
\end{equation}
where the $\bm{s}_{\rm{p}}(\bm{c})$ is the polymeric entropy. Finally, the polymer stress tensor $\bm{\tau}_{\rm{p}}$ is computed by~\cite{vazquez-quesada_SPH_2017}
\begin{equation}
	\bm{\tau}_{\rm{p}}=-2\bm{c} \cdot \bm{\sigma}.
	\label{eq:polymer_stress}
\end{equation}
Since no specific force law is assumed for the polymer, the previous expressions have general validity for a dilute polymer solution~\cite{ottinger_equilibrium_2005}. Finally, the information of the entropy function $\bm{s}_{\rm{p}}(\bm{c})$ can directly close the thermodynamic-consistent constitutive equation for the dilute polymer solution. Thus, the tensorial variable tensor $\bm{\sigma}(\bm{c})$ in~\cref{eq:tensorial_variable} is chosen here as the target function. 
\par
In order to simplify the relevant equations and improve the efficiency of the regression, the evaluation equations of the above variables can be translated into the eigenvalue form due to rotational symmetry~\cite{vazquez-quesada_Smoothed_2009}. The conformation tensor $\bm{c}$ and the tensorial variable $\bm{\sigma}(\bm{c})$ diagonalize in the same basis, and the expressions $\bm{c}=\sum_\alpha c_\alpha \bm{v}_\alpha \bm{v}_\alpha^{\mathrm{T}}$ and $\bm{\sigma}(\bm{c})=\sum_\alpha \sigma_\alpha \bm{v}_\alpha \bm{v}_\alpha^{\mathrm{T}}$ can be used, where the $\{c_1, c_2\}$ and $\{\sigma_1, \sigma_2\}$ are the eigenvalues of $\bm{c}$ and $\bm{\sigma}$ in the two-dimensional situation $(\alpha=1, 2)$, respectively, and the condition $c_1 > c_2$ needs to be satisfied. The $\bm{v}_{\alpha} (\alpha=1 \, \rm{or} \, 2)$ is the normalized eigenvector.
\par
By taking the time derivative on both sides of the expression $\bm{c}=\sum_\alpha c_\alpha \bm{v}_\alpha \bm{v}_\alpha^{\mathrm{T}}$ and then multiplying the result with the eigenvectors left and right simultaneously, the following expression is obtained as
\begin{equation}
	\bm{v}_\alpha^{\mathrm{T}} \cdot \dot{\bm{c}} \cdot \bm{v}_\beta=\delta_{\alpha \beta} \dot{c}_\alpha+\left(c_\alpha-c_\beta\right) \dot{\bm{v}}_\alpha^{\mathrm{T}} \cdot \bm{v}_\beta,
	\label{eq:evaluation_c1}
\end{equation}
where the orthogonality of the eigenvectors is utilized, that is, the time derivative of $(\bm{v}_\alpha^{\mathrm{T}} \cdot \bm{v}_\beta)$ equals $0$. The $\delta_{\alpha \beta}$ is the Kronecker delta function. By multiplying~\cref{eq:conformation_tensor} with the eigenvectors left and right simultaneously, the following expression is obtained as
\begin{equation}
	\bm{v}_\alpha^{\mathrm{T}} \cdot \dot{\bm{c}} \cdot \bm{v}_\beta=c_\alpha \kappa_{\alpha \beta}+c_\beta \kappa_{\beta \alpha}+\frac{2}{\lambda_{\rm{p}} N_{\rm{p}} k_{\rm{B}} T} c_\alpha \sigma_\alpha \delta_{\alpha \beta},
	\label{eq:evaluation_c2}
\end{equation}
where the matrix element of the velocity gradient tensor in the eigenbasis of the conformation tensor is introduced as $\kappa_{\alpha\beta}=\bm{v}_\alpha^{\mathrm{T}} \cdot \bm{\kappa} \cdot \bm{v}_\beta$.
\par
By equating~\cref{eq:evaluation_c1} and~\cref{eq:evaluation_c2}, the evolution equations for the eigenvalues and eigenvectors are obtained as
\begin{equation}
	\dot{c}_\alpha=2 c_\alpha \kappa_{\alpha \alpha}+\frac{2}{\lambda_{\rm{p}} N_{\rm{p}} k_{\rm{B}} T} c_\alpha \sigma_\alpha,
	\label{eq:evaluation_eigenvalue}
\end{equation}
\begin{equation}
	\dot{\bm{v}}_\alpha = \sum\limits_\beta H_{\alpha \beta} \bm{v}_\beta,
	\label{eq:evaluation_eignvector}
\end{equation}
where the antisymmetric matrix $H_{\alpha \beta}$ is given by
\begin{equation}
	H_{\alpha \beta}= \begin{cases}\frac{1}{c_\alpha-c_\beta}\left[c_\alpha \kappa_{\alpha \beta}+c_\beta \kappa_{\beta \alpha}\right], & \text { if } c_\alpha \neq c_\beta, \\ 0, & \text { if } c_\alpha=c_\beta.\end{cases}
	\label{eq:evaluation_H}
\end{equation}
For more details, the reader is referred to the literature~\cite{vazquez-quesada_Smoothed_2009,vazquez-quesada_Shear_2019}.
\par
The Oldroyd-B constitutive equation can provide analytical results to make comparisons and hence provide validations. The Oldroyd-B constitutive model can be derived from the Hookean dumbbell model, for which the entropy reads~\cite{vazquez-quesada_SPH_2017}
\begin{equation}
	s_{\mathrm{p}}(\bm{c})=\frac{N_{\rm{p}} k_{\rm{B}}}{2}[\operatorname{tr}(\bm{I}-\bm{c})+\ln (\operatorname{det}(\bm{c}))].
	\label{eq:entropy}
\end{equation}
By coupling~\cref{eq:tensorial_variable}, the $\bm{\sigma}$ in this specific case is obtained as
\begin{equation}
	\bm{\sigma}=\frac{N_{\rm{p}} k_{\rm{B}} T}{2}\left(\bm{c}^{-1}-\bm{I}\right).
	\label{eq:tensorial_variable_OldB}
\end{equation}
The eigenvalues of $\bm{\sigma}$ are normalized as 
\begin{equation}
	\sigma_\alpha=\frac{N_{\rm{p}} k_{\rm{B}} T}{2}{\tilde{\sigma}}_\alpha,
	\label{eq:tensorial_variable_OldBeigen}
\end{equation}
where the ${\tilde{\sigma}}_\alpha$ is obtained as 
\begin{equation}
	{\tilde{\sigma}}_\alpha=\frac{1}{c_\alpha}-1.
	\label{eq:target_function}
\end{equation}
By substitution of~\cref{eq:tensorial_variable_OldB}, the~\cref{eq:conformation_tensor} becomes
\begin{equation}
	\dot{\bm{c}}=\bm{c} \cdot \bm{\kappa}+\bm{\kappa}^T \cdot \bm{c}+\frac{1}{\lambda_{\rm{p}}}\left(\bm{I}-\bm{c}\right),
	\label{eq:conformation_tensor_new}
\end{equation}
which corresponds to the standard Oldroyd-B model. By substitution of~\cref{eq:tensorial_variable_OldBeigen}, the~\cref{eq:evaluation_eigenvalue} becomes
\begin{equation}
	\dot{c}_\alpha=2 c_\alpha \kappa_{\alpha \alpha}+\frac{1}{\lambda_{\rm{p}}} c_\alpha {\tilde{\sigma}}_\alpha.
	\label{eq:evaluation_eigenvalue_new}
\end{equation}
Finally, by substituting~\cref{eq:tensorial_variable_OldB} into~\cref{eq:polymer_stress} and defining the polymeric viscosity as $\eta_{\rm{p}}=N_{\rm{p}} k_{\rm{B}} T \lambda_{\rm{p}}$, the polymer stress tensor $\bm{\tau}_{\rm{p}}$ for the Hookean dumbbell model is obtained by
\begin{equation}
	\bm{\tau}_{\rm{p}}=\frac{\eta_{\rm{p}}}{\lambda_{\rm{p}}}\left(\bm{c}-\bm{I}\right).
	\label{eq:polymer_stress_c}
\end{equation}
By combining the expression $\bm{\sigma}=\sum_\alpha \sigma_\alpha \bm{v}_\alpha \bm{v}_\alpha^{\mathrm{T}}$, the expression $\eta_{\rm{p}}=N_{\rm{p}} k_{\rm{B}} T \lambda_{\rm{p}}$,~\cref{eq:target_function}, and~\cref{eq:polymer_stress}, the polymer stress tensor $\bm{\tau}_{\rm{p}}$ can be computed by 
\begin{equation}
	\bm{\tau}_{\rm{p}}=-\frac{\eta_{\rm{p}}}{\lambda_{\rm{p}}}\sum_\alpha c_\alpha \tilde{\sigma}_\alpha \bm{v}_\alpha \bm{v}_\alpha^{\mathrm{T}}.
	\label{eq:target_function_eigen}
\end{equation}
\par
In~\cref{eq:target_function}, for the target function ${\tilde{\sigma}}_\alpha(c_1,c_2)$, the subscript $\alpha$ can be $1$ or $2$. Due to the symmetry of the eigenvalues of the conformation tensor, the analytical function for the Oldroyd-B model can be rewritten as
\begin{equation}
	f(x_1,x_2)=\frac{1}{x_1}-1,
	\label{eq:target_function1}
\end{equation}
where the point $(x_1,x_2)$ can be $(c_1,c_2)$ or $(c_2,c_1)$. For every fluid particle at each time instant, $(c_1,c_2)$ and $(c_2,c_1)$ need to be taken into consideration.
\par
Finally, for a prior \verb+"+unknown\verb+"+ Hookean dumbbell model, the~\cref{eq:target_function1} becomes the final target function. Once this function is learned, the time derivative of the eigenvalues and eigenvectors of the conformation tensor $\bm{c}$ can be computed by~\cref{eq:evaluation_eigenvalue_new} and ~\cref{eq:evaluation_eignvector}. After the update of the eigenvalues and eigenvectors of the conformation tensor, the polymer stress tensor $\bm{\tau}_{\rm{p}}$ can be computed by~\cref{eq:target_function_eigen}, and the closure of the governing equations is achieved.
\par
It is worth noting that the data of ${\tilde{\sigma}}_\alpha(c_1,c_2)$ can be obtained by mesoscopic polymer simulations~\cite{moreno_Arbitrary_2021,simavilla_Mesoscopic_2022,simavilla_Nonaffine_2023,moreno_Generalized_2023}. For the mesoscopic simulation of a specific dilute polymer solution, the parameters $\lambda_{\rm{p}}$, $N_{\rm{p}}$, $k_{\rm{B}}$, $T$, and $\eta_{\rm{p}}$ are pre-set constants. Thus, the target function ${\tilde{\sigma}}_\alpha(c_1,c_2)$ can directly determine a constitutive relation for this pre-defined dilute polymer solution, and the flow-history-dependent memory effect is consistently incorporated by the GENERIC framework. Moreover, as discussed above, the GENERIC framework is applicable to any dilute polymer solution. The energy-based kernel approach for viscoelastic fluids and some energy-compatible functions corresponding to different viscoelastic stress models are presented in the literature~\cite{otto_Machine_2024}, further demonstrating the theoretical underpinnings of the GENERIC-guided approach in our work.
\subsection{The smoothed particle hydrodynamics (SPH) method}\label{sec2.2}
\subsubsection{The SPH discretization}\label{sec2.2.1}
For the conventional SPH method, an arbitrary function $f(\bm{r})$ and its derivative at the position $\bm{r}_i$ are interpolated by the particles present in the support domain, which can be computed by the following expressions~\cite{dong_active_2024}:
\begin{equation}
	{f\left( {{\bm{r}_i}} \right)} = \sum\limits_j {\frac{{{m_j}}}{{{\rho _j}}}} f\left( {{\bm{r}_j}} \right){W_{ij}},
	\label{eq:SPH_f}
\end{equation}
\begin{equation}
	{\nabla \cdot f\left( {{{\bm{r}}_i}} \right)} = \frac{1}{{{\rho _i}}}\sum\limits_j {{m_j}\left[ {f\left( {{{\bm{r}}_j}} \right) - f\left( {{{\bm{r}}_i}} \right)} \right]} \cdot {\nabla _i}{W_{ij}},
	\label{eq:SPH_df1}
\end{equation}
\begin{equation}
	{\nabla \cdot f\left( {{{\bm{r}}_i}} \right)} = {\rho _i}\left[ {\sum\limits_j {{m_j}\left( {\frac{{f\left( {{{\bm{r}}_j}} \right)}}{{\rho _j^2}} + \frac{{f\left( {{{\bm{r}}_i}} \right)}}{{\rho _i^2}}} \right)} \cdot {\nabla _i}{W_{ij}}} \right],
	\label{eq:SPH_df2}
\end{equation}
where $m_j$ and $\rho _j$ are the mass and density of the $j$th particle in the support domain, respectively. The position $\bm{r}_i$ equals $(r_{x, i},r_{y, i})$ for two-dimensional problems. The kernel function $W$ is chosen as follows~\cite{morris_Modeling_1997}:
\begin{equation}
	W_{i j}=W(r_{ij}, h)=w_0 \begin{cases}(3-q)^5-6(2-q)^5+15(1-q)^5, & 0 \leqslant q<1, \\ (3-q)^5-6(2-q)^5, & 1 \leqslant q<2, \\ (3-q)^5, & 2 \leqslant q<3, \\ 0, & q \geqslant 3,\end{cases}
	\label{eq:kernel}
\end{equation}
where $r_{ij}=\left|\bm{r}_i-\bm{r}_j\right|$, $q=r_{ij}/h$, and the factor $w_0$ is chosen as $7/(478 \pi h^2)$ in two-dimensional problems. The parameter $h$ is the smoothing length defining the support domain of $W$. The $\nabla_i W_{i j}$ is given by
\begin{equation}
	\nabla_i W_{i j}=\frac{\bm{r}_{i j}}{r_{i j}} \frac{\partial W_{i j}}{\partial r_{i j}},
	\label{eq:dkernel}
\end{equation}
where the expression $\bm{r}_{ij}=\bm{r}_i-\bm{r}_j$ is considered.
The detailed derivation process can be found in our previous work~\cite{dong_active_2024}. In this paper, \cref{eq:SPH_df1,eq:SPH_df2} are applied to discretize the mass and momentum conservation equations, i.e., \cref{eq:mass,eq:momentum2}, respectively. 
\par
To get better accuracy in situations where the number of the particles in the support domain is insufficient or the particles are unevenly distributed, a modified kernel gradient~\cite{oger_improved_2007} is adopted as follows:
\begin{equation}
	\nabla_i^M W_{i j}=\bm{M}_i^{-1} \nabla_i W_{i j},
	\label{eq:mdkernel}
\end{equation}
\begin{equation}
	\bm{M}_i=\sum_j \frac{m_j}{\rho_j} \nabla_i W_{i j} \otimes \bm{r}_{j i},
	\label{eq:mdkernel_matrix}
\end{equation}
where $\bm{M}_i$ is the correction matrix. Sometimes, the insufficient number of particles in the support domain can lead to an ill-conditioned matrix, which produces wrong results. If the condition number is greater than $10^5$, the matrix $\bm{M}_i$ is considered to be ill-conditioned, and the original kernel gradient is used.
\par
The velocity gradient is discretized as~\cite{dong_active_2024}
\begin{equation}
	(\tilde{k}_{\alpha \beta})_{i} = {\left( {\frac{{\partial {u_{\alpha} }}}{{\partial {r_{\beta} }}}} \right)_i} = \sum\limits_j {\frac{{{m_j}}}{{{\rho _j}}}} \left(u_{\alpha, j} - u_{\alpha, i} \right)\frac{\partial^M W_{i j}}{\partial r_{\beta, i}},
	\label{eq:velocity_gradient}
\end{equation}
where $\alpha$ and $\beta$ indicate the $x$ direction or the $y$ direction in Cartesian coordinates. 
\par
The viscosity term discretization is employed to better capture physical viscosity~\cite{morris_Modeling_1997}. By the interpolation idea of finite difference, the second-order derivative of the kernel function is transformed into the discrete form of the first-order derivative of the kernel function, i.e.,
\begin{equation}
	(\frac{{{\eta _{\rm{s}}}}}{\rho }{\nabla ^2}{\bm{u}})_{i}=\sum_{j} \frac{m_{j}\left({\eta_{{\rm{s}},i}}+{\eta_{{\rm{s}},j}}\right) \bm{r}_{i j} \cdot \nabla_{i} W_{i j}}{\rho_{i}\rho_{j}\left|r_{i j}\right|^{2}} \bm{u}_{i j}.
	\label{eq:viscosity}
\end{equation}
\par
Finally, by substituting~\cref{eq:SPH_df1} into~\cref{eq:mass} and applying~\cref{eq:SPH_df2} into~\cref{eq:momentum2} and combining $\nabla_i^M W_{i j}$, the discretization results of the governing equations in the SPH method can be written as follows:
\begin{equation}
	\left\{\begin{array}{l}
		{\dot{\rho}_i} = \sum\limits_j {{m_j} ({{\bm{u}}_i} - {{\bm{u}}_j}) \cdot \nabla_i^M W_{i j}}, \\
		\dot{\bm{u}}_i=\sum\limits_j m_j\left[-\left(\frac{p_i}{\rho_i^2}+\frac{p_j}{\rho_j^2}\right) \nabla_i^M W_{i j}+ \frac{\left({\eta_{{\rm{s}},i}}+{\eta_{{\rm{s}},j}}\right) \bm{r}_{i j} \cdot \nabla_i^M W_{i j}}{\rho_{i}\rho_{j}\left|r_{i j}\right|^{2}} \bm{u}_{i j}+\left(\frac{\bm{\tau}_{{\rm{p}},i}}{\rho_i^2}+\frac{\bm{\tau}_{{\rm{p}},j}}{\rho_j^2}\right) \cdot \nabla_i^M W_{i j}\right] + \bm{F}_i, \\
		\dot{\bm{r}}_i=\bm{u}_i,\\
		p_i(\rho_i)=\frac{{\rho}_0 c_{\rm{s}}^2}{\gamma}\left(\left(\frac{\rho_i}{\rho_0}\right)^\gamma-1\right).
	\end{array}\right.
	\label{eq:final_discretization}
\end{equation}
Besides, the calculation of the polymer stress is given in~\cref{sec2.1.3} in detail. There is no special discretization treatment for those required equations, i.e.,~\cref{eq:evaluation_eigenvalue_new,eq:evaluation_eignvector,eq:evaluation_H,eq:target_function_eigen}.
\subsubsection{Time integration}\label{sec2.2.2}
The leapfrog scheme~\cite{dong_active_2024} is adopted to solve the system of~\cref{eq:final_discretization} for its second-order accuracy and high computational efficiency. To ensure the numerical stability, the time step $\Delta t$ should be restricted by the Courant–Friedrichs–Lewy condition, the viscous-diffusion condition, and the body force condition~\cite{fang_Improved_2009}, i.e.,
\begin{equation}
	\Delta t \leq \min \left(\frac{h}{c_{\rm{s}}}, 0.5 \frac{h^2}{\upsilon_0}, 
	\min_{i} \sqrt{\frac{h}{\left|\bm{F}_i\right|}}\right),
	\label{eq:time_step}
\end{equation}
where $\upsilon_0=\eta / \rho$ is kinematic viscosity and $\bm{F}_i$ is the external force acceleration.
\subsubsection{Boundary conditions}\label{sec2.2.3}
The periodic boundary condition and the solid wall boundary condition are employed in this paper. For the Poiseuille flow and the flow around a periodic array of cylinders, the periodic boundary condition is adopted in the direction of the fluid's transverse flow, which is the same as that applied in the literature~\cite{ren_Simulation_2016,vazquez-quesada_SPH_2012}. The solid wall boundary condition significantly affects the accuracy of the numerical simulation by serving two primary functions: (i) preventing particle penetration through the solid wall and (ii) patching up the support domain of the fluid particle near the solid wall. Several treatments for the solid wall boundary condition have been proposed, such as the repulsive force~\cite{monaghan_SPH_2009} and the fixed ghost particles~\cite{marrone_dSPH_2011}.
\par
In this paper, by combining the advantages of the dummy particle method proposed in the literature~\cite{ellero_SPH_2005,xu_Numerical_2024,adami_generalized_2012}, an improved dummy particle method is proposed. Four layers of dummy particles are fixed at solid walls to ensure that the support domains of fluid particles near the solid wall are fully replenished. The distance between dummy particles is the same as the distance between fluid particles in the initial placement. The initial setup and the update process for physical quantities at dummy particles are aligned. Throughout the numerical simulation, the velocities, density, and positions of dummy particles are kept constant. The velocities are set to zero in order to satisfy the no-slip boundary condition. The density is set to be the same as the initial density of fluid particles. Therefore, the volume and mass are also kept constant. Pressure and elastic stress are calculated by the following interpolation equation~\cite{ellero_SPH_2005}:
\begin{equation}
	A_{i}=\frac{\sum_{j}\left(m_{j} / \rho_{j}\right) A_{j} W_{i j}}{\sum_{j}\left(m_{j} / \rho_{j}\right) W_{i j}},
	\label{eq:interpolation}
\end{equation}
where $A$ denotes pressure or elastic stress. The $j$ is the index of fluid particles in the support domain of the dummy particle $i$. This configuration not only suppresses particle penetration but also ensures the computational accuracy of the evolution of the physical quantities on the fluid particles near the solid wall. The effectiveness of the proposed solid wall conditions is validated by the Poiseuille flow and the flow around a periodic array of cylinders in the next sections. The specific arrangement of the dummy particles can be found in the initial arrangements reported in~\cref{fig:Po_initial_b,fig:cy_initial_b}. 
\subsubsection{The optimized particle shifting technique}\label{sec2.2.4}
The particle shifting technique can enhance particle distribution and hence the simulation accuracy~\cite{sun_dplusSPH_2017}. Xu et al.~\cite{xu_Accuracy_2009} originally applied the particle shifting technique to the SPH method. In recent years, several particle shifting techniques have been proposed~\cite{sun_dplusSPH_2017,gao_new_2023}. In this paper, the optimized particle shifting technique presented in our previous work~\cite{dong_active_2024} is utilized, which is added after the position update of governing equations. For more details, the reader is referred to the paper~\cite{dong_active_2024}.
\section{The GENERIC-guided active learning smoothed particle hydrodynamics (${\rm{G^2ALSPH}}$) method}\label{sec3}
In this section, the theory of GPR is first outlined. Afterward, the definition of the newly proposed relative uncertainty is presented. Finally, the active learning strategy is established, and thus the ${\rm{G^2ALSPH}}$ method is accomplished. 
\subsection{Gaussian process regression (GPR)}\label{sec3.1}
GPR is a nonparametric model that uses a Gaussian process (GP) prior to regress and analyze data. A nonlinear relation between inputs and outputs can be constructed by GPR~\cite{duvenaud_Automatic_2014}. Based on Bayesian probability theory and a finite training data set, GPR provides a local nonlinear solution and the corresponding uncertainty assessment~\cite{williams_Gaussian_2006}. The advantages of GPR are that only a small number of data are required to provide the expected results and the direct uncertainties can be assessed at the corresponding locations simultaneously~\cite{deringer_Gaussian_2021}. These two features bring great application potential to GPR. The accuracy evaluation tool for the GPR prediction results can be established based on the uncertainty assessment, which brings an active learning idea.
\par
A GP stands for a collection of random variables. Any finite number of these variables satisfies a joint Gaussian distribution. GP is completely determined by its mean function and covariance function. For a process $f(\bm{x})$, its mean function $m(\bm{x})$ and covariance function $k(\bm{x},\bm{x}^{\prime})$ are defined as $\bm{E}[f(\bm{x})]$ and $\bm{E}[(f(\bm{x})-m(\bm{x}))(f(\bm{x}^{\prime})-m(\bm{x}^{\prime}))]$, respectively. Thus, GP is denoted as $f(\bm{x}) \sim {\rm{GP}}(m(\bm{x}),k(\bm{x},\bm{x}^{\prime}))$. For the sake of calculation simplicity, the data are generally preprocessed so that the expression $m(\bm{x})=0$ is satisfied. 
\par
In this paper, it is assumed that the data are preprocessed and noise-free. The case $y=f(\bm{x})$ is considered. The GPR with two inputs and one output is adopted in the proposed model. Specifically, the variable $\bm{x}=(x_1,x_2)$ indicates the input vector, whose source is the eigenvalues of the conformation tensor $\bm{c}$. The observation $y$ represents the final target function $f(x_1,x_2)$ in~\cref{eq:target_function1}. The GP prior of $f(\bm{x})$ is $f(\bm{x}) \sim {\rm{N}}(0,\bm{K})$. The composite kernel function is chosen as the covariance function $k(\bm{x},\bm{x}^{\prime})$ in the GPR~\cite{fuhg_physicsinformed_2022}, i.e.,
\begin{equation}
	k\left(\bm{x}, \bm{x}^{\prime}\right)=\sigma_f^2 \exp \left[\frac{-\left(x_1-x_1^{\prime}\right)^2}{2 l_1^2}+\frac{-\left(x_2-x_2^{\prime}\right)^2}{2 l_2^2}\right],
	\label{eq:covariance_function}
\end{equation}
where $l_1$ and $l_2$ are variance scales and $\sigma_f^2$ is signal variance. The variables $\bm{x}$ and $\bm{x^{\prime}}$ indicate $\bm{x}=(x_1,x_2)$ and $\bm{x^{\prime}}=(x_1^{\prime},x_2^{\prime})$, respectively. The joint prior distribution of the observations $\bm{y}$ and the observation $y_*$ in the test point $\bm{x_{*}}=(x_1^{*},x_2^{*})$ is obtained as
\begin{equation}
	\left[\begin{array}{l}
		\bm{y} \\
		y_*
	\end{array}\right] \sim \mathrm{N}\left(0,\left[\begin{array}{ll}
		\bm{K} & \bm{K}_*^{\mathrm{T}} \\
		\bm{K}_* & \bm{K}_{* *}
	\end{array}\right]\right),
	\label{eq:observation}
\end{equation}
where the matrices $\bm{K}$, $\bm{K}_*$ and $\bm{K}_{* *}$ are defined as follows:
\begin{equation}
	\bm{K}=\left[\begin{array}{cccc}
		k\left(\bm{x}_1, \bm{x}_1\right) & k\left(\bm{x}_1, \bm{x}_2\right) & \cdots & k\left(\bm{x}_1, \bm{x}_n\right) \\
		k\left(\bm{x}_2, \bm{x}_1\right) & k\left(\bm{x}_2, \bm{x}_2\right) & \cdots & k\left(\bm{x}_2, \bm{x}_n\right) \\
		\vdots & \vdots & \ddots & \vdots \\
		k\left(\bm{x}_n, \bm{x}_1\right) & k\left(\bm{x}_n, \bm{x}_2\right) & \cdots & k\left(\bm{x}_n, \bm{x}_n\right)
	\end{array}\right],
	\label{eq:matrix_K}
\end{equation}
\begin{equation}
	\bm{K}_*=\left[\begin{array}{llll}
		k\left(\bm{x}_{*}, \bm{x}_1\right) & k\left(\bm{x}_{*}, \bm{x}_2\right) & \cdots & k\left(\bm{x}_{*}, \bm{x}_n\right)
	\end{array}\right],
	\label{eq:matrix_K*}
\end{equation}
\begin{equation}
	\bm{K}_{**}=k\left(\bm{x}_{*}, \bm{x}_{*}\right).
	\label{eq:matrix_K**}
\end{equation}
\par
According to Bayesian theory~\cite{williams_Gaussian_2006}, the posterior distribution of the predicted value $y_*$ is
\begin{equation}
	y_* \mid \bm{y} \sim \mathrm{N}\left(\bm{K}_* \bm{K}^{-1} \bm{y}, \bm{K}_{* *}-\bm{K}_* \bm{K}^{-1} \bm{K}_*^{\mathrm{T}}\right).
	\label{eq:predicted_y*}
\end{equation}
The mean and variance of the predicted value $y_*$ are obtained as follows:
\begin{equation}
	\bar{y}_*=\bm{K}_* \cdot \bm{K}^{-1} \bm{y},
	\label{eq:y*}
\end{equation}
\begin{equation}
	\sigma_{y_*}^2=\bm{K}_{* *}-\bm{K}_* \bm{K}^{-1} \bm{K}_*^{\mathrm{T}}.
	\label{eq:var_y*}
\end{equation}
\par
The credibility of GPR depends on the covariance function $k(\bm{x},\bm{x}^{\prime})$. In~\cref{eq:covariance_function}, the function $k(\bm{x},\bm{x}^{\prime})$ depends on the hyperparameters $l_1$, $l_2$, and $\sigma_f$. Thus, the hyperparameter $\bm{\theta}=\left\{l_1, l_2, \sigma_f\right\}$ directly determines the quality of the predicted result, which can be obtained by the great likelihood estimation. The negative logarithm of the conditional probability of the training samples is taken as the likelihood function $L(\bm{\theta})=-{\rm{log}} \; p(\bm{y}|\bm{x},\bm{\theta})$. According to the expression $\bm{y} \sim {\rm{N}}(0,\bm{K})$, the function $L(\bm{\theta})$ is obtained as
\begin{equation}
	L(\bm{\theta})=\frac{1}{2} \bm{y}^T \bm{K}^{-1} \bm{y}+\frac{1}{2} \log |\bm{K}|+\frac{n}{2} \log 2 \pi.
	\label{eq:function_L}
\end{equation}
By taking the partial derivative of the above equation with respect to $\bm{\theta}$, the result is obtained as
\begin{equation}
	\frac{\partial L(\bm{\theta})}{\partial \theta_i}=\frac{1}{2} \operatorname{tr}\left(\left(\left(\bm{K}^{-1} \bm{y}\right)\left(\bm{K}^{-1} \bm{y}\right)^T-\bm{K}^{-1}\right) \frac{\partial \bm{K}}{\partial \theta_i}\right),\
	\label{eq:function_dL}
\end{equation}
where \verb+"+$\operatorname{tr}$\verb+"+ denotes the trace of a matrix. The optimal solution of the hyperparameter $\bm{\theta}$ can be obtained by minimizing and solving~\cref{eq:function_dL} using the limited-memory Broyden-Fletcher-Goldfarb-Shanno (L-BFGS) method. After the completion of the training process for the hyperparameter $\bm{\theta}$, the mean $\bar{y}_*$ and variance $\operatorname{var}\left(y_*\right)$ of the predicted value $y_*$ can be obtained by substituting the optimized hyperparameter $\bm{\theta}$ in~\cref{eq:covariance_function,eq:y*,eq:var_y*}.
\par
For the GPR with $N$ training data points, the overall time complexity is $\mathcal{O}(N^3)$~\cite{dong_active_2024}. If the proposed ${\rm{G^2ALSPH}}$ method is applied for multi-scale modeling, the required data are obtained by mesoscopic simulations, which are typically very time-consuming. Thus, for the sake of generality, it is important to propose a clever accuracy evaluation tool for GPR prediction results to enable the training data set to be as small as possible while maintaining accuracy.
\par
It is worth noting that the learned constitutive relation is determined by the training data set, the optimized hyperparameters, and the selected covariance function, which means the learned constitutive relation can be stored by these three components. When applied to a new numerical case as an initial local constitutive relation, the learned constitutive relation can be recovered by inputting the training data set, the optimized hyperparameters, and the selected covariance function to GPR.
\subsection{The relative uncertainty}\label{sec3.2}
The accuracy evaluation tool for GPR prediction results determines the active learning strategy. This is because the central component of the active learning strategy is initiating the active learning procedure when the results obtained by the accuracy evaluation tool exceed the pre-set tolerance. Commonly, the $95\%$ confidence interval is chosen to set the direct uncertainty to serve as an accuracy evaluation tool for GPR prediction results~\cite{chang_multiscale_2023,zhao_Active_2021}. For the case described in~\cref{sec3.1}, the mean and variance of the predicted value $y_*$ are obtained by~\cref{eq:y*,eq:var_y*}. On this basis, the standard deviation $E_{\rm{s}}$ of the predicted value $y_*$ can be obtained by calculating the square root of the diagonal element of the variance $\sigma_{y_*}^2$. The $95\%$ confidence interval $[\bar{y}_*-1.96 E_{\rm{s}}, \bar{y}_*+1.96 E_{\rm{s}}]$ is determined by a specific scale, i.e., $1.96$ times the standard deviation $E_{\rm{s}}$.
\par
In this paper, the direct uncertainty is also set to be $1.96 E_{\rm{s}}$. The newly proposed relative uncertainty $U_{\rm{r}}$ at a given point is defined as the ratio of the direct uncertainty to the absolute value of the predicted mean plus a small number $\varepsilon$ and presented in the following equation:
\begin{equation}
	U_{\rm{r}}=\frac{1.96 E_{\rm{s}}}{\left|\bar{y}_*\right|+\varepsilon},
	\label{eq:relative_uncertainty}
\end{equation}
where the $\varepsilon$ is a small parameter. The introduction of $\varepsilon$ prevents the denominator from vanishing, which ensures the stability of the active learning process. It is important to note that the value of $\varepsilon$ can be neither too large nor too small. If the value of $\varepsilon$ is too large, the accuracy of the learned constitutive relation can not be ensured. If it is too small, the active learning process can not perform well in the area where the value of $\left|\bar{y}_*\right|$ is very close to $0$. In this work, the $\varepsilon=0.2$ is chosen.
\par
The superiority of the relative uncertainty over the direct uncertainty is clear. For a given tolerance to the relative uncertainty, when the value of $\bar{y}_*$ at a certain location becomes larger during simulation, the restriction on $1.96 E_{\rm{s}}$, that is, the corresponding direct uncertainty, can become somewhat looser, while the percentage form of $U_{\rm{r}}$ can still maintain accuracy. The biggest advantage of the relative uncertainty is that it can reduce the number of required training data points in the active learning process while maintaining the accuracy of the entire simulation process. The establishment of the relative uncertainty makes the corresponding accuracy evaluation tool and active learning strategy more general and robust. 
\subsection{The active learning strategy}\label{sec3.3}
For any dilute polymer solution, the target constitutive relation can be simplified by the GENERIC framework, which incorporates the flow-history-dependent memory effect. For example, in~\cref{eq:target_function1}, the final target function $f(x_1,x_2)$ only has two inputs and one output. In this paper, to validate the ${\rm{G^2ALSPH}}$ method, the Hookean dumbbell model is employed, and the necessary data are obtained from~\cref{eq:target_function1} directly. Moreover, the GPR with two inputs and one output is adopted here. The active learning procedure is activated on demand to update the target constitutive relation. A novel accuracy evaluation tool is devised by imposing the newly proposed relative uncertainty $U_{\rm{r}}$. The location of the new sampling point is decided by this novel accuracy evaluation tool. The detailed active learning strategy of the ${\rm{G^2ALSPH}}$ method is presented as follows:
\begin{list}{\textbullet}{\leftmargin=3em \rightmargin=0em \itemsep=0pt \parsep=0pt}
	\item[(1)]\textbf{Set the value of the tolerance $\bm{\delta_{\rm{tol}}}$.} The accuracy evaluation tool consisting of the proposed relative uncertainty $U_{\rm{r}}$ is adopted. At the beginning, the tolerance $\delta_{\rm{tol}}$ is set to be a certain value, e.g., $0.01$. The magnitude of this value determines the accuracy of the entire numerical simulation process and the number of required training data points.
	\item[(2)]\textbf{Learn an initial local constitutive relation.} Some training points are selected at the beginning. An initial local constitutive relation is learned by regressing these points. There is no specific constitutive relation in equation form included in the governing equations. This local learned constitutive relation is necessary for the initiation of the macroscopic SPH simulation.
	\item[(3)]\textbf{Start the macroscopic SPH simulation.} At each time instant, calculate the relative uncertainty $U_{\rm{r}}$ for all fluid particles. Record the maximum value of the relative uncertainty $U_{\rm{r}}$ for all fluid particles. Determine whether this maximum value at the current time instant is less than the tolerance $\delta_{\rm{tol}}$ or not. If so, the learned constitutive relation can be directly used, and the macroscopic SPH simulation can update the time instant. On the contrary, the macroscopic SPH simulation is paused to wait for a new learned constitutive relation. The active learning procedure is then initiated.
	\item[(4)]\textbf{Repeat the active learning procedure until the maximum value of the relative uncertainty $\bm{U_{\rm{r}}}$ for all fluid particles at the current time instant is below the pre-set tolerance $\bm{\delta_{\rm{tol}}}$.} At this time instant, the maximum value of the relative uncertainty $U_{\rm{r}}$ for all fluid particles is above the tolerance $\delta_{\rm{tol}}$. This corresponds to having one or more points where the relative uncertainty results do not meet the accuracy requirement. The point that implies the maximum relative uncertainty is chosen as the location of the next sampling point. This new sampling point is included to form a new training data set. Then, the training process of GPR is performed by using this new training data set to obtain a new learned constitutive relation. Determine whether the predicted results for all fluid particles satisfy the tolerance limit for this new learned constitutive relation. If not, continue to incorporate the new location with the largest relative uncertainty into the training set. The above steps are repeated until the predicted results for all fluid particles obtained by the current learned constitutive relation satisfy the tolerance limit. The learned constitutive relation obtained at the end of the iteration can thus be applied to the subsequent macroscopic SPH simulation.
	\item[(5)]\textbf{Continue the macroscopic SPH simulation until the tolerance limit is broken.} When the maximum value of the relative uncertainty $U_{\rm{r}}$ for all fluid particles exceeds the tolerance $\delta_{\rm{tol}}$ at a certain time instant, the active learning procedure in step (4) is initiated again. 
	\item[(6)]\textbf{Repeat steps (4) and (5) until the end of the macroscopic SPH simulation.}
\end{list}
\par
\begin{figure}[H]
	\centering		
	\includegraphics[width=15cm]{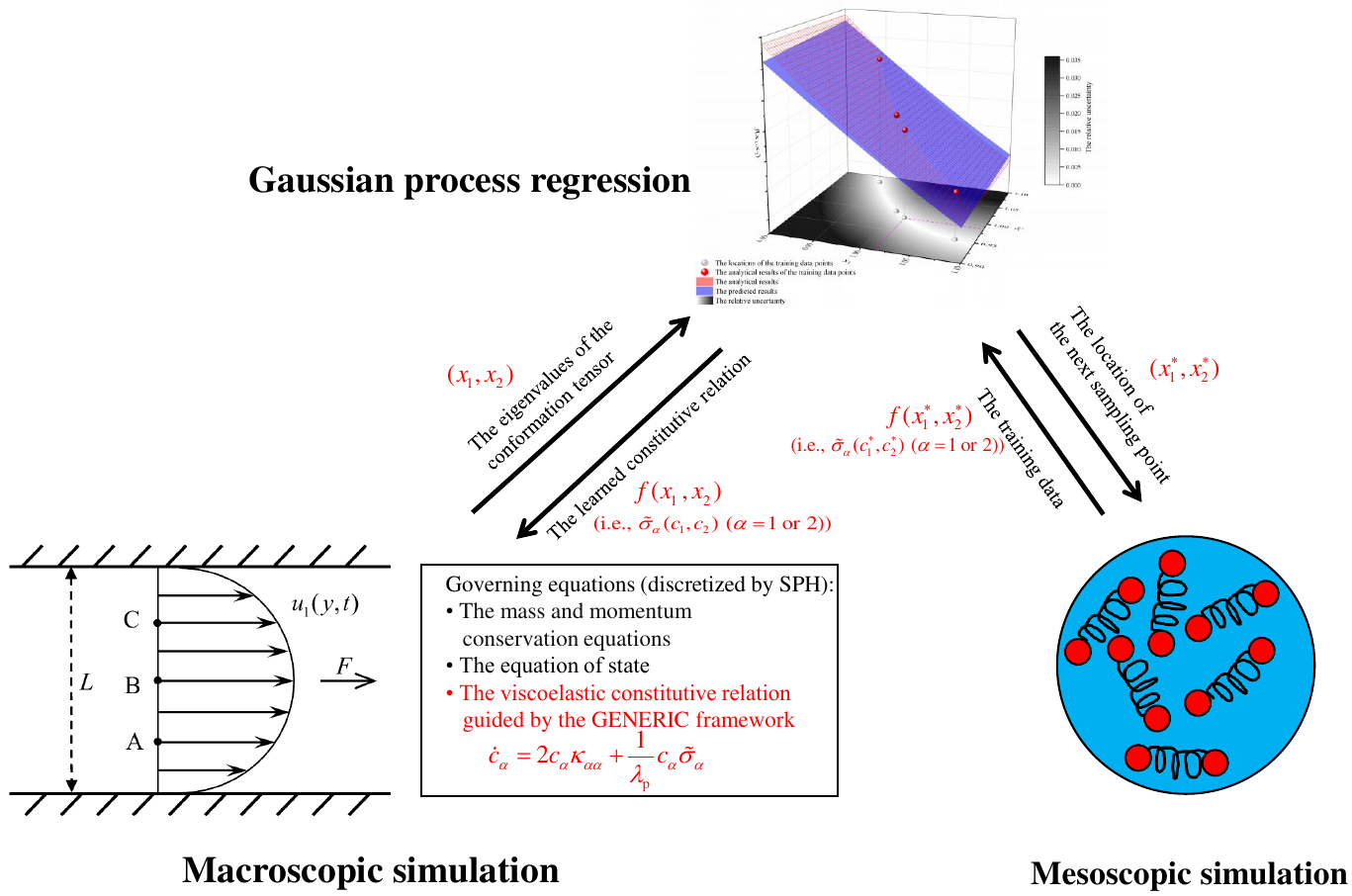}
	\caption{The schematic diagram of the ${\rm{G^2ALSPH}}$ method.}
	\label{fig:idea}
\end{figure}
\begin{figure}[H]
	\centering		
	\includegraphics[width=14cm]{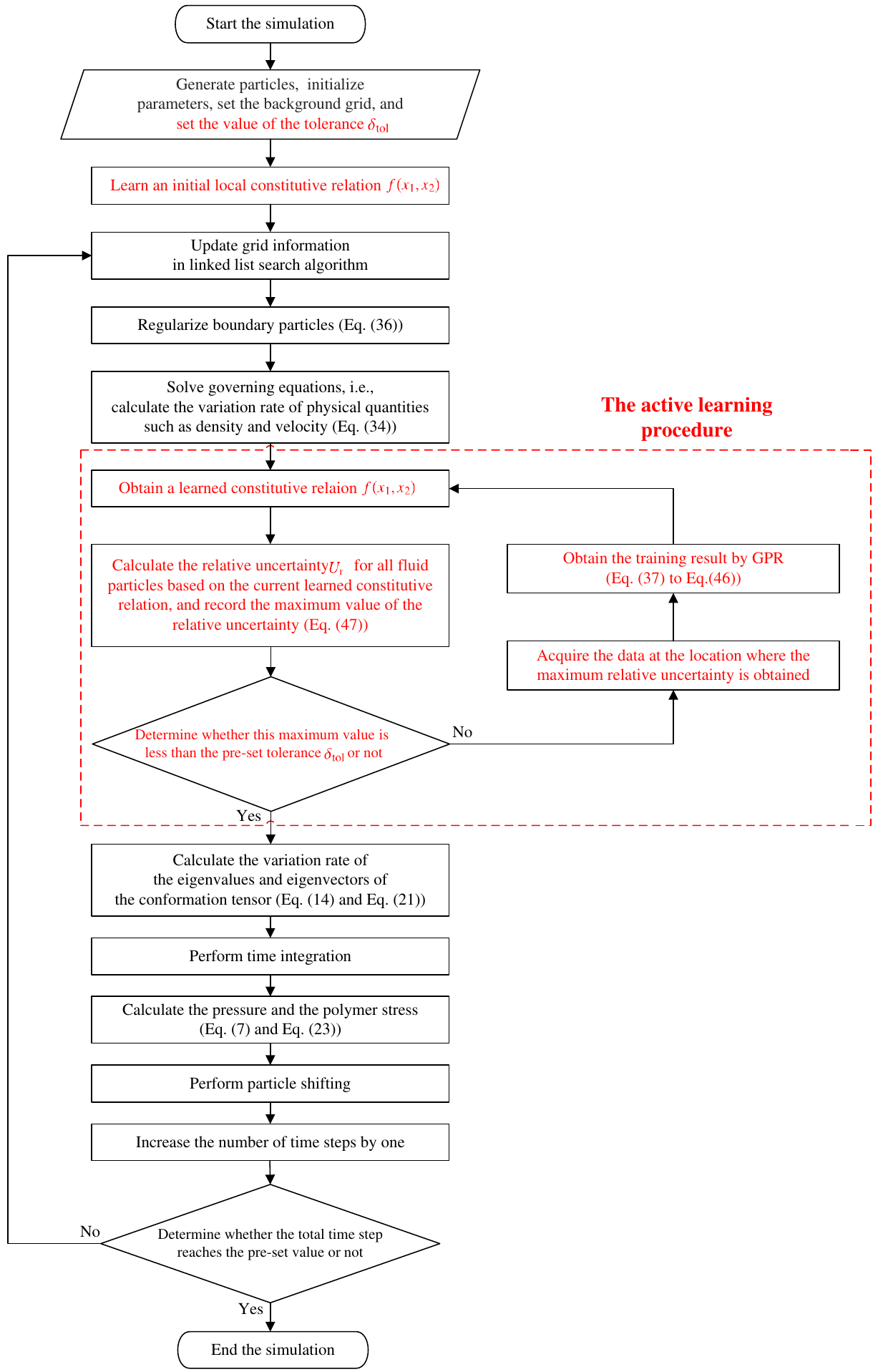}
	\caption{The flowchart of the ${\rm{G^2ALSPH}}$ method.}
	\label{fig:flow_chart}
\end{figure}

The schematic diagram and the flowchart of the ${\rm{G^2ALSPH}}$ method are given by~\cref{fig:idea,fig:flow_chart}, respectively. The active learning strategies based on the relative and direct uncertainties are denoted as the relative and direct uncertainty strategies, respectively. In the following sections, the superiority of the relative uncertainty strategy over the direct uncertainty strategy will be studied.
\section{Numerical examples}\label{sec4}
In this section, the Poiseuille flows and the flows around a periodic array of cylinders at different Weissenberg numbers are numerically simulated to validate the proposed ${\rm{G^2ALSPH}}$ method. The macroscopic simulations are implemented in C++ programs, while the GPR is achieved in Python programs by the scikit-learn library~\cite{bisong_Introduction_2019}. The information transfer of the learned constitutive relation between C++ and Python programs is in the form of files, where the training data set and the optimized hyperparameters are stored. For a better description of the flow, the non-dimensional physical quantities are employed for all numerical cases in this paper. The non-dimensionalization of the governing equations is given in the~\ref{app:non-dimensionalization}.
\subsection{The Poiseuille flow}\label{sec4.1}
The Oldroyd-B model is used as an unknown target constitutive relation in this work, and it is utilized to generate the required data for GPR. Hence, the results obtained by the SPH method with the explicit Oldroyd-B model become the reference analytical results for comparison. For the macroscopic simulation case, the viscoelastic Poiseuille flow is chosen first, which is a classical benchmark to validate the transient accuracy of a numerical method. The availability of the analytical solutions of the Poiseuille flow with the Oldroyd-B model allows for quantitative comparisons~\cite{xue_Numerical_2004,ellero_SPH_2005}. The non-dimensional Reynolds number is defined as $Re=\rho_0 V L/\eta$, where the variables $V$ and $L$ are characteristic velocity and characteristic length, respectively. The non-dimensional Weissenberg number is defined as $Wi=\lambda_{\rm{p}} V/L$, where $\lambda_{\rm{p}}$ is the polymeric relaxation time. The solvent viscosity ratio $\beta$ is defined as $\beta=\eta_{\rm{s}}/\eta$.
\par
\begin{figure}[H]
	\centering
	\begin{minipage}{0.513\linewidth}
		\begin{subfigure}[b]{\linewidth}
			\centering
			\includegraphics[width=\linewidth]{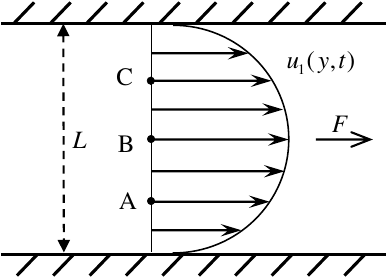}
			\caption{}
			\label{fig:Po_initial_a}
		\end{subfigure}%
	\end{minipage}%
	\hspace{10mm}
	\begin{minipage}{0.2\linewidth}
		\vspace{-1mm}
		\begin{subfigure}[b]{\linewidth}
			\centering
			\includegraphics[width=\linewidth]{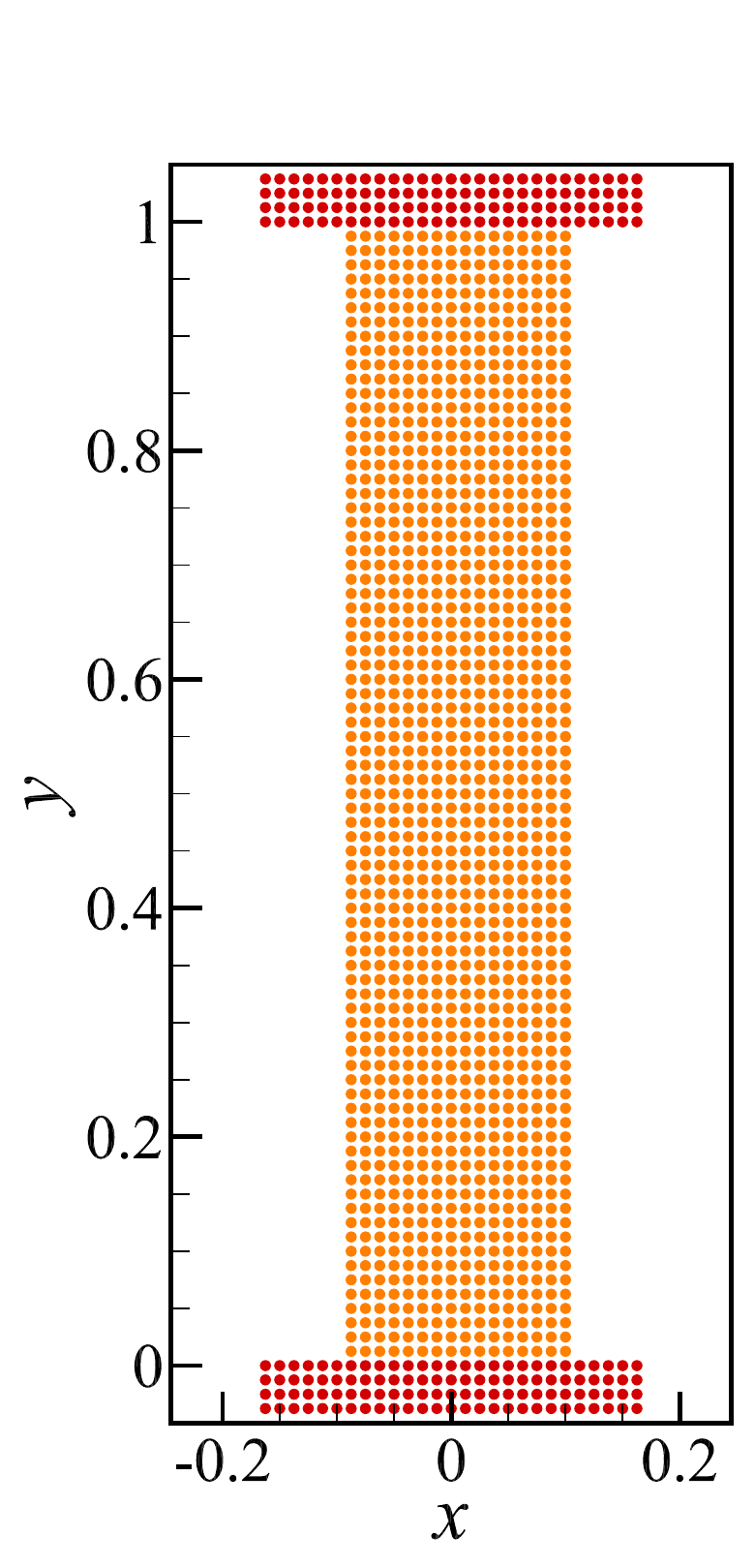}
			\caption{}
			\label{fig:Po_initial_b}
		\end{subfigure}%
	\end{minipage}
	\caption{The initial setup of the Poiseuille flow: (a) the computational model and (b) the particle distribution at the initial instant.}
	\label{fig:Po_initial}
\end{figure}

The initial setup of the Poiseuille flow is illustrated by~\cref{fig:Po_initial}. The height of the channel is set as $L=1.0$, which is considered the characteristic length. The width of the domain to be studied is set to be $0.2$. The fluid density is set as $\rho_0=1.0$. The solvent viscosity ratio is set as $\beta=0.1$. The artificial sound speed is set as $c_{\rm{s}}=2.0$. A constant external force acceleration $F_0=1.0$ is employed in the transverse flow direction to drive the flow of the fluid. The Weissenberg and Reynolds numbers are set as $Wi=0.1$ and $Re=1.0$, respectively. As shown in~\cref{fig:Po_initial_b}, the initial particle spacing is set as $d_0=0.0125$, corresponding to $79$ fluid particles along the y-direction. The total numbers of the fluid and wall particles are $1264$ and $216$, respectively. The smoothing length is set as $h=0.9d_0$. The time step is set as $\Delta t=5\times {10}^{-5}$. The total simulation time is set to be $2.0$, which allows the Poiseuille flow to reach a steady state. Three probe points, A, B, and C, are set in the channel, whose distances to the lower boundary are $L/4$, $L/2$, and $3L/4$, respectively. The analytical solutions under start-up flow are given in the~\ref{app:Poiseuille_analytical}. More details about the analytical solutions can be found in the literature~\cite{xue_Numerical_2004,ellero_SPH_2005}.
\subsubsection{The accuracy and convergence of the proposed SPH method}\label{sec4.1.1}
\begin{figure}[H]
	\centering
	\begin{subfigure}[b]{0.5\linewidth}
		\includegraphics[width=\linewidth]{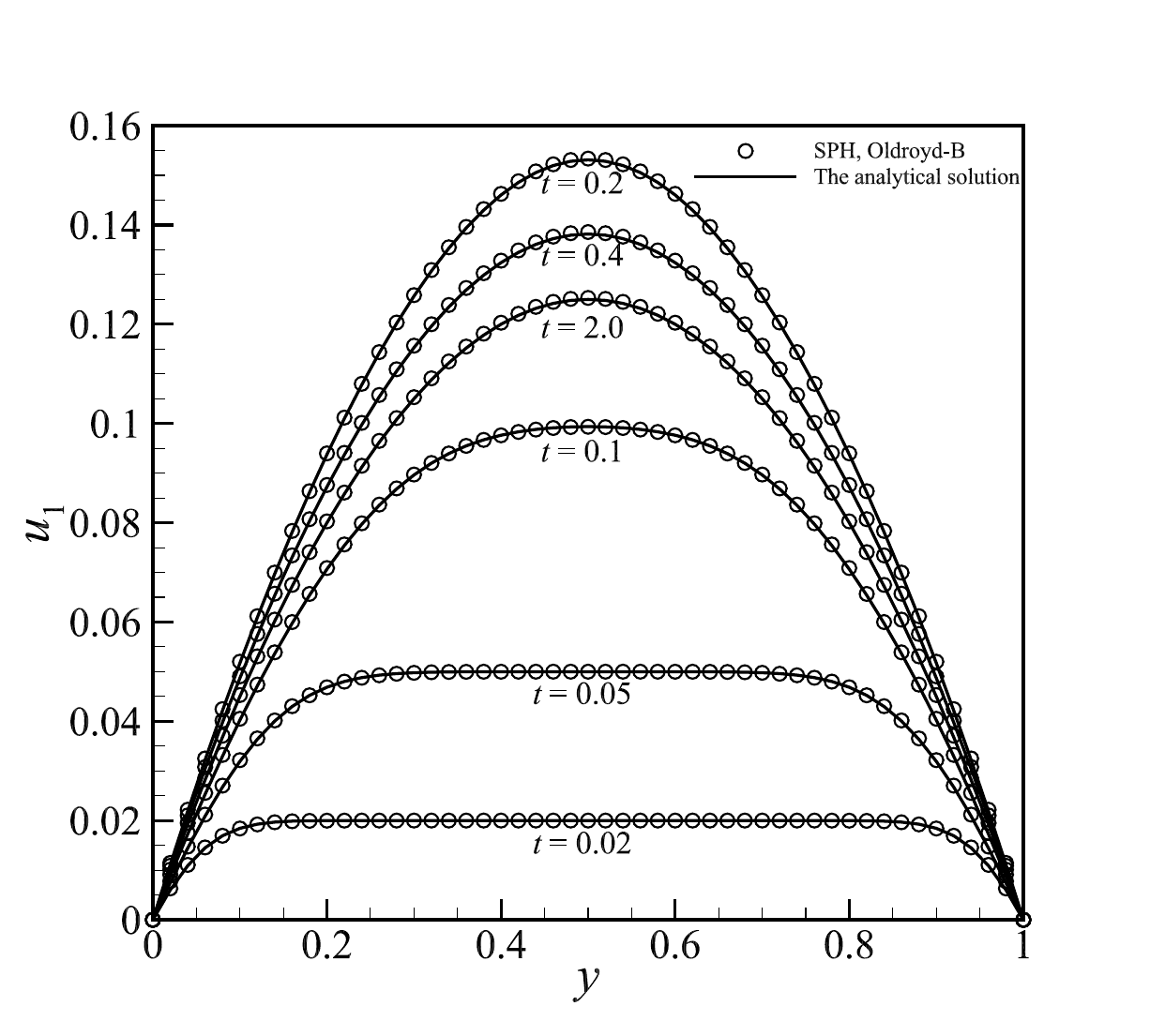}
		\caption{}
		\label{fig:Po_ve_a}
	\end{subfigure}%
	\begin{subfigure}[b]{0.5\linewidth}
		\includegraphics[width=\linewidth]{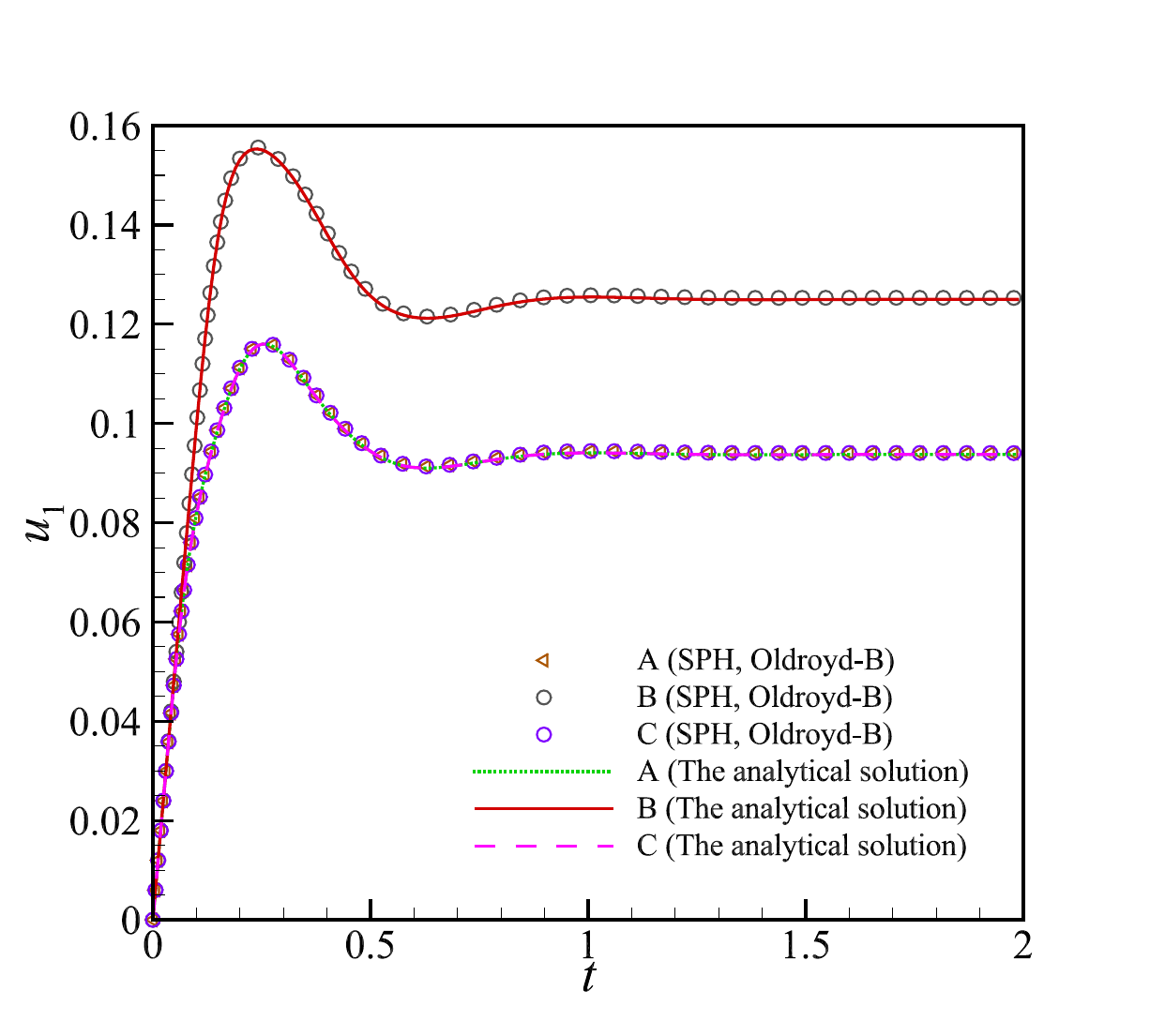}
		\caption{}
		\label{fig:Po_ve_b}
	\end{subfigure}%
	\caption{Comparison among the results of the Poiseuille flow ($Wi=0.1$, $Re=1.0$, $\beta=0.1$) at the steady state obtained by the SPH method with the Oldroyd-B model ($d_0=0.0125$) and the analytical solution: (a) the distributions of the velocity $u_1$ at some specific time instants and (b) the variations of the velocity $u_1$ over time at the probe points A, B, and C.}
	\label{fig:Po_ve}
\end{figure}

The accuracy and convergence of the SPH method in simulating viscoelastic fluids is first analyzed here. The explicit Oldroyd-B model serves as the constitutive relation here. In~\cref{fig:Po_ve_a}, the distributions of the velocity $u_1$ at some specific time instants, $t=0.02$, $t=0.05$, $t=0.1$, $t=0.2$, $t=0.4$, and $t=2.0$, are presented. The results obtained by the SPH method are in excellent agreement with the corresponding analytical solutions. In~\cref{fig:Po_ve_b}, the variations of the velocity $u_1$ over time at the probe points A, B, and C are presented. Due to the symmetry of the channel flow, the results of the velocity $u_1$ at probe point A are in agreement with those at probe point C. Due to elasticity, the fluid overshoots and then bounces back. After a slight oscillation, the velocity reaches a steady state at $t=2.0$. As shown in~\cref{fig:Po_ve_b}, the results obtained by the SPH method are in good agreement with the analytical solutions~\cite{xue_Numerical_2004,ellero_SPH_2005}.
\par
\begin{figure}[H]
	\centering
	\begin{subfigure}[b]{0.333\linewidth}
		\includegraphics[width=\linewidth]{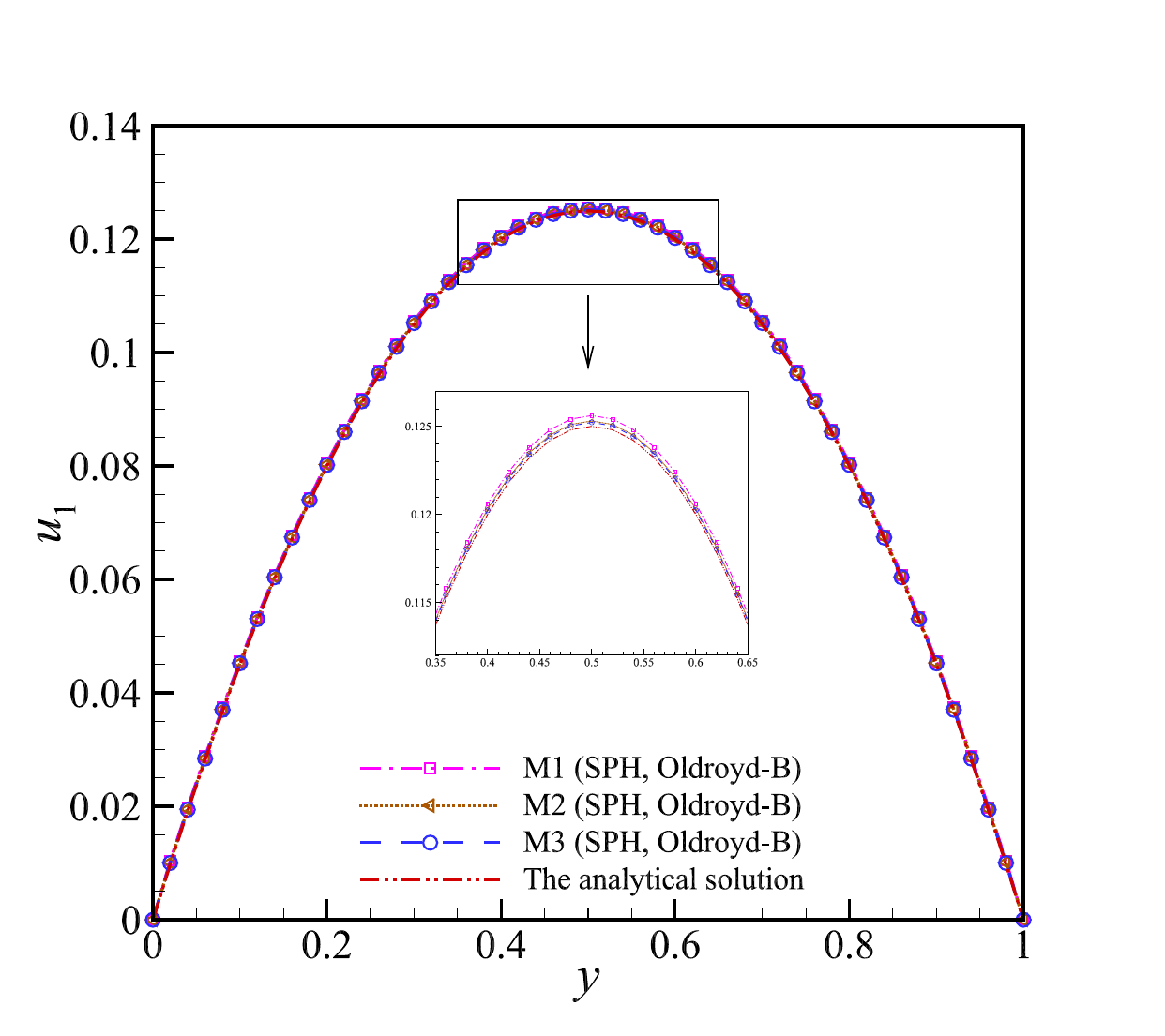}
		\caption{}
		\label{fig:Po_convergence_a}
	\end{subfigure}%
	\begin{subfigure}[b]{0.333\linewidth}
		\includegraphics[width=\linewidth]{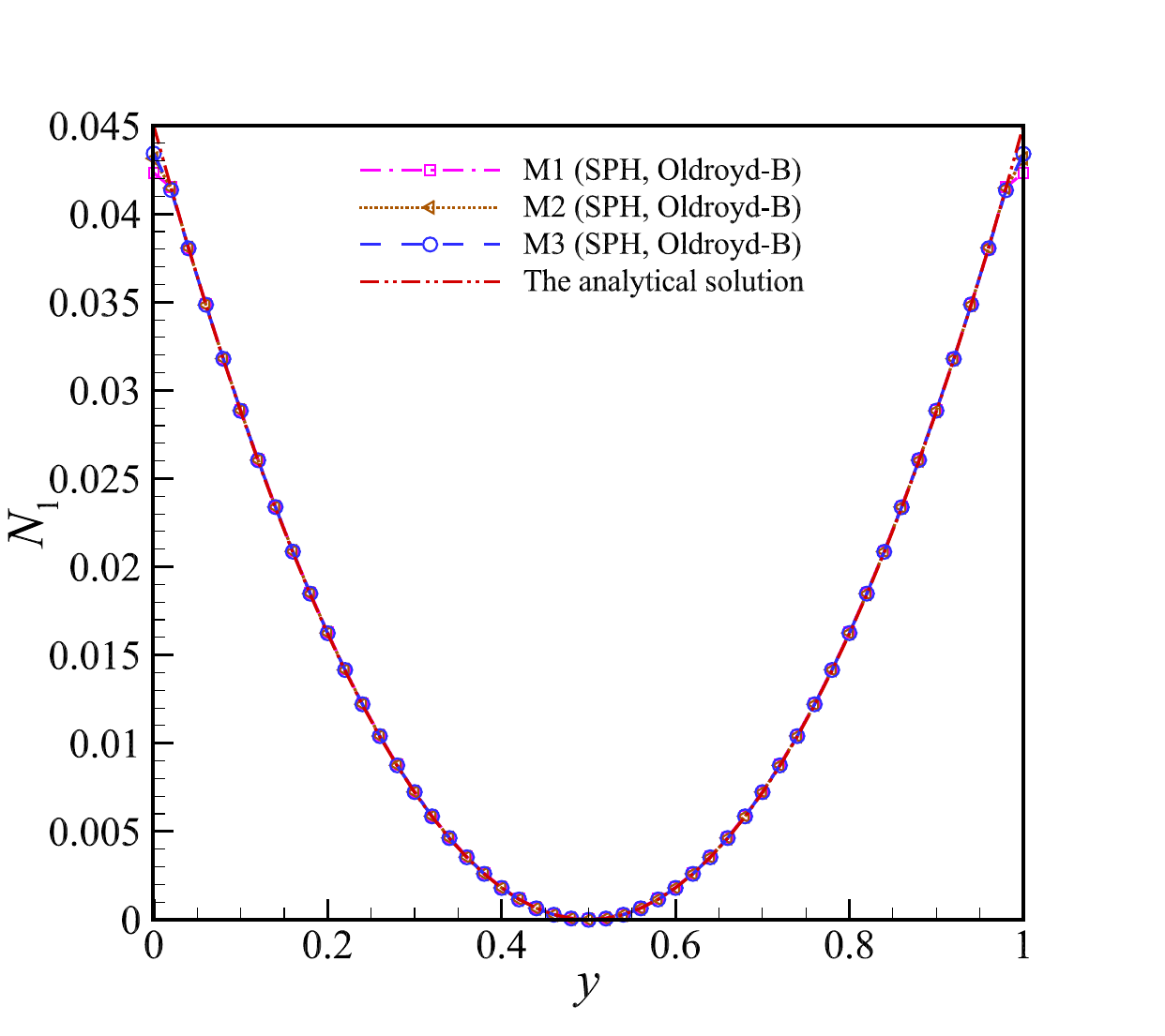}
		\caption{}
		\label{fig:Po_convergence_b}
	\end{subfigure}%
	\begin{subfigure}[b]{0.333\linewidth}
		\includegraphics[width=\linewidth]{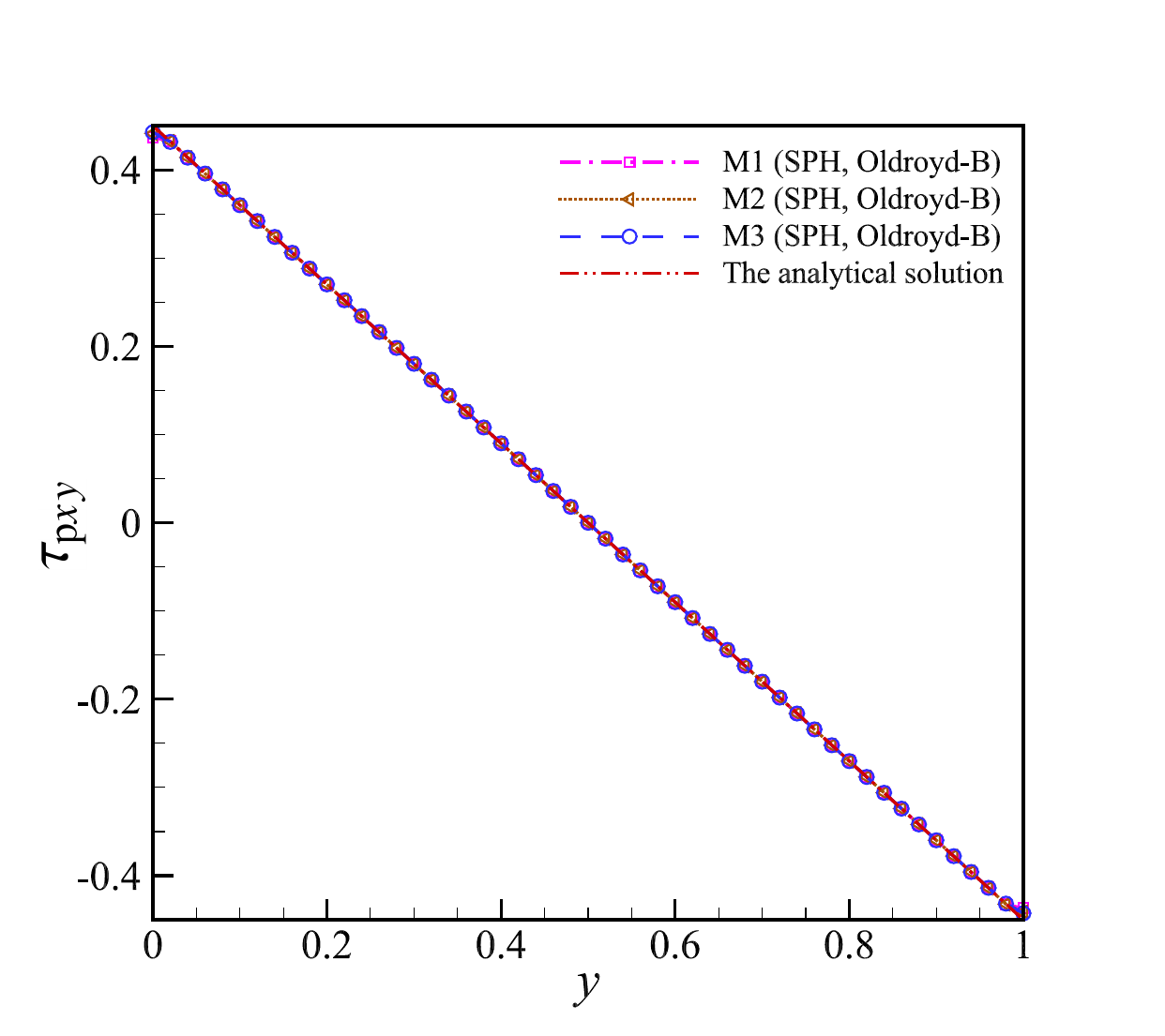}
		\caption{}
		\label{fig:Po_convergence_c}
	\end{subfigure}%
	\caption{Comparison among the results of the Poiseuille flows ($Wi=0.1$, $Re=1.0$, $\beta=0.1$) at the steady state obtained by the SPH method with the Oldroyd-B model at three different resolutions ($d_0=0.02$, $d_0=0.0125$, and $d_0=0.01$) and the analytical solutions: (a) the velocity $u_1$, (b) the first normal stress difference $N_1$, and (c) the shear stress $\tau_{{\rm{p}}xy}$.}
	\label{fig:Po_convergence}
\end{figure}

To further assess the convergence of the SPH method, the Poiseuille flow is simulated for two more resolutions. Three resolutions are denoted as M1, M2, and M3, which correspond to initial particle spacing settings of $0.02$, $0.0125$, and $0.01$, respectively. Except for the resolution, the other physical parameters and computational model settings are kept unchanged. In~\cref{fig:Po_convergence}, the results of the velocity $u_1$, the first normal stress difference $N_1=\tau_{{\rm{p}}xx}-\tau_{{\rm{p}}yy}$, the shear stress $\tau_{{\rm{p}}xy}$ at the steady state at three resolutions are given. As the initial particle spacing $d_0$ decreases, the results obtained by the SPH method converge to the analytical solutions. Quantitative convergence comparisons with respect to the velocity $u_1$ at the steady state are given in~\cref{tableerror}. The relative error is computed by the $l_2$-norm equation, i.e.,
\begin{equation}
	\|E\|_2=\sqrt{\frac{\sum\left(R_{\rm{A}}-R_{\rm{M}}\right)^2}{\sum\left(R_{\rm{A}}\right)^2}},\label{eq41}
\end{equation}
where $R_{\rm{M}}$ stands for the results obtained at the resolutions M1, M2, and M3, and $R_{\rm{A}}$ stands for the analytical solutions. As shown in~\cref{tableerror}, the $l_2$-norm error becomes smaller as the initial particle spacing $d_0$ decreases, which proves the convergence and accuracy of the SPH method in simulating viscoelastic fluids.
\begin{table}[H]
	\centering
	\caption{The $l_2$-norm errors at three resolutions M1, M2, and M3.}
	\resizebox{1.0\textwidth}{!}{%
		\begin{tabular}{c c c c c}
			\hline
			\multirow{1}{*}{Resolution}&\multirow{1}{*}{$d_0$}&\multirow{1}{*}{The number of the fluid particles along y-direction}& \multirow{1}{*}{The overall number of particles}& \multirow{1}{*}{The $l_2$-norm error} \\
			\hline
			M1 &0.02   &49    & 658  & 0.6539\% \\
			M2 &0.0125  &79    & 1480 & 0.3213\% \\
			M3 &0.01    &99    & 2228 & 0.2285\% \\
			\hline
		\end{tabular}\label{tableerror}
	}
\end{table}
\subsubsection{The active learning process for the Poiseuille flow}\label{sec4.1.2}
In this subsection, the ${\rm{G^2ALSPH}}$ method is applied to simulate the viscoelastic Poiseuille flow. A prior \verb+"+unknown\verb+"+ constitutive relation is actively learned by GPR, where the required data are provided by the underlying Oldroyd-B model. Except for that, the other physical parameters and computational model settings are kept unchanged. The resolution M2 is chosen for the macroscopic simulation, i.e., $d_0=0.0125$. The active learning strategy described in~\cref{sec3.3} is specifically applied here. The specific active learning process of the ${\rm{G^2ALSPH}}$ method with the tolerance setting $\delta_{\rm{tol}}=0.005$ is stated in detail in the following paragraphs.
\par
\begin{figure}[H]
	\centering
	\raisebox{6.0\height}{\rotatebox{90}{\scriptsize{$t=0$}}}
	\begin{minipage}[t]{0.9\textwidth}
		\centering
		\begin{subfigure}[b]{0.49\linewidth}
			\raisebox{0.4\height}{\rotatebox{0}{\scriptsize{The distribution of the velocity $u_1$}}}
			\centering
			\includegraphics[width=\linewidth]{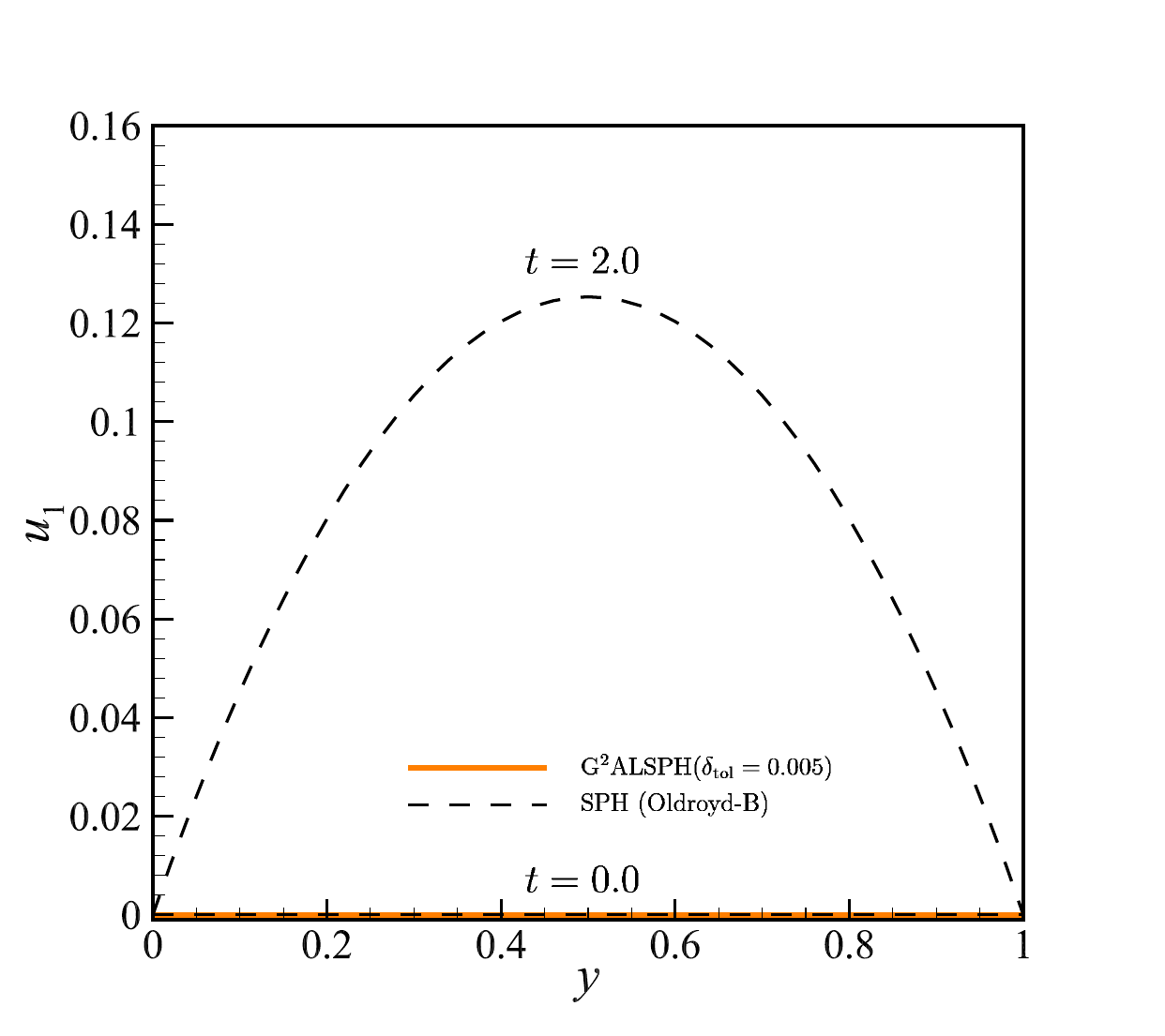}
			\caption{}
			\label{fig:al_procedure_a1}
		\end{subfigure}%
		\begin{subfigure}[b]{0.49\linewidth}
			\raisebox{4.0\height}{\rotatebox{0}{\scriptsize{The learned constitutive relation}}}
			\centering
			\includegraphics[width=\linewidth]{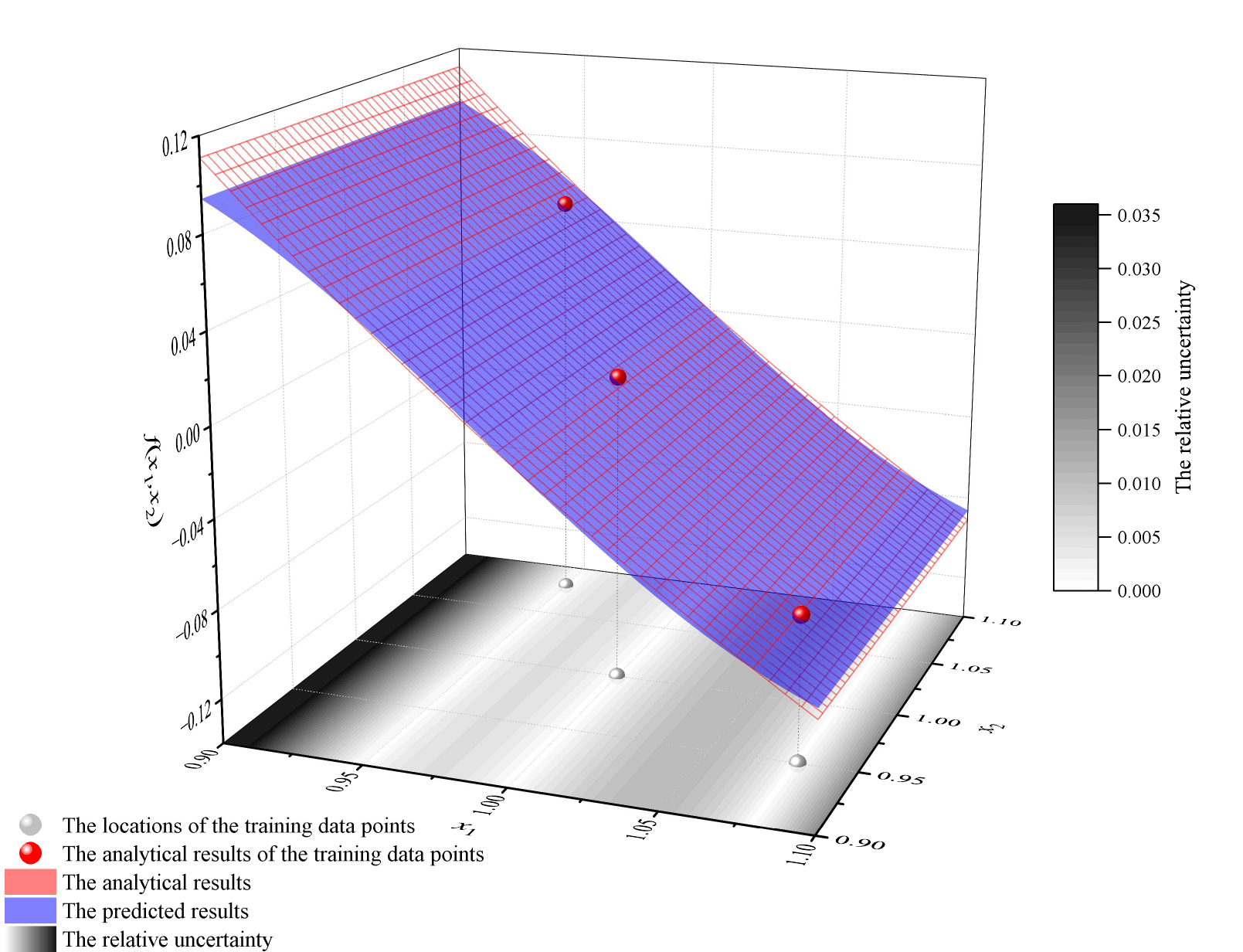}
			\caption{}
			\label{fig:al_procedure_a2}
		\end{subfigure}%
	\end{minipage}
	\vfill
	\centering
	\raisebox{2.5\height}{\rotatebox{90}{\scriptsize{$t=0.0608$}}}
	\begin{minipage}[t]{0.9\textwidth}
		\centering
		\begin{subfigure}[b]{0.49\linewidth}
			\centering
			\includegraphics[width=\linewidth]{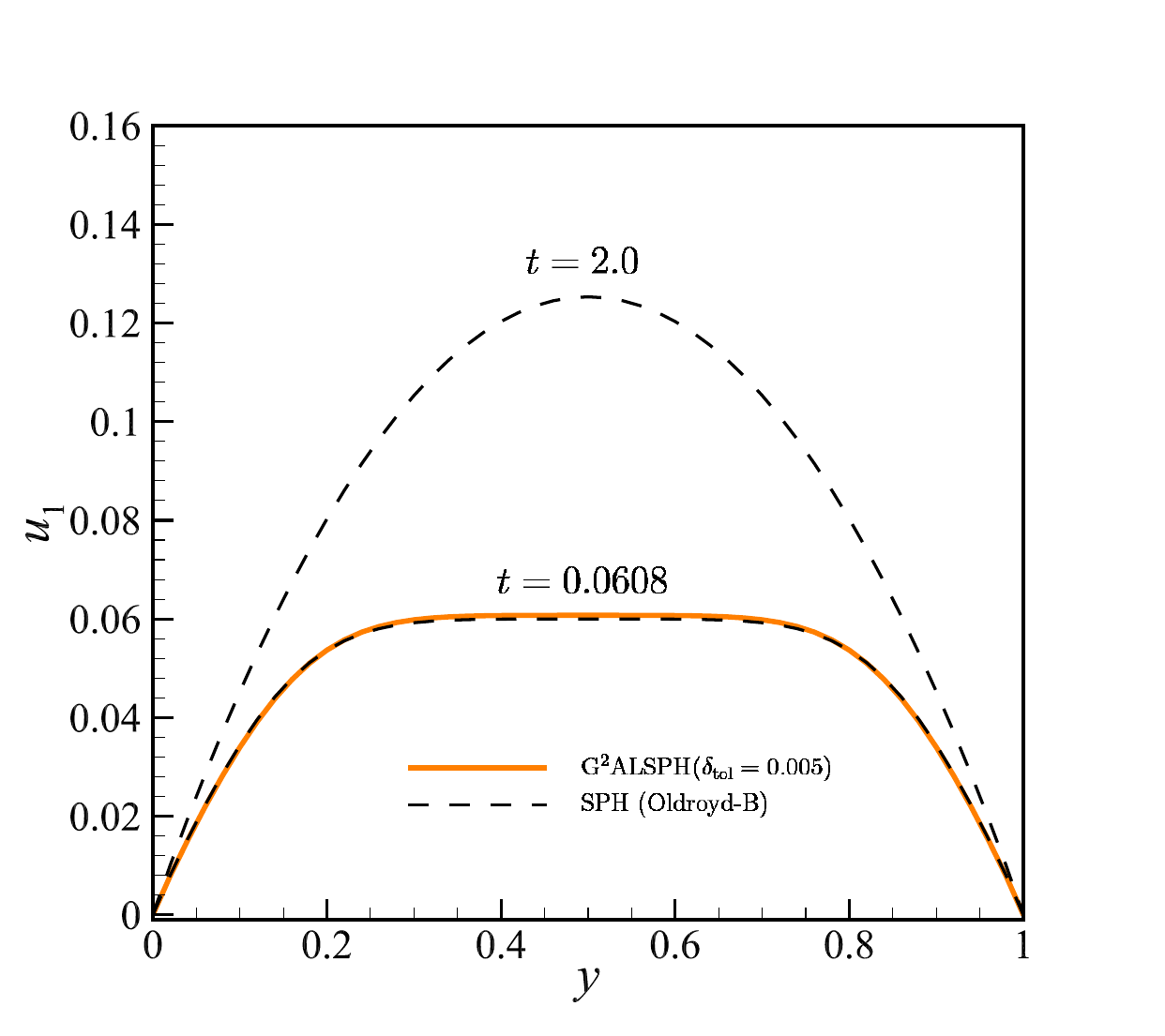}
			\caption{}
			\label{fig:al_procedure_b1}
		\end{subfigure}%
		\begin{subfigure}[b]{0.49\linewidth}
			\centering
			\includegraphics[width=\linewidth]{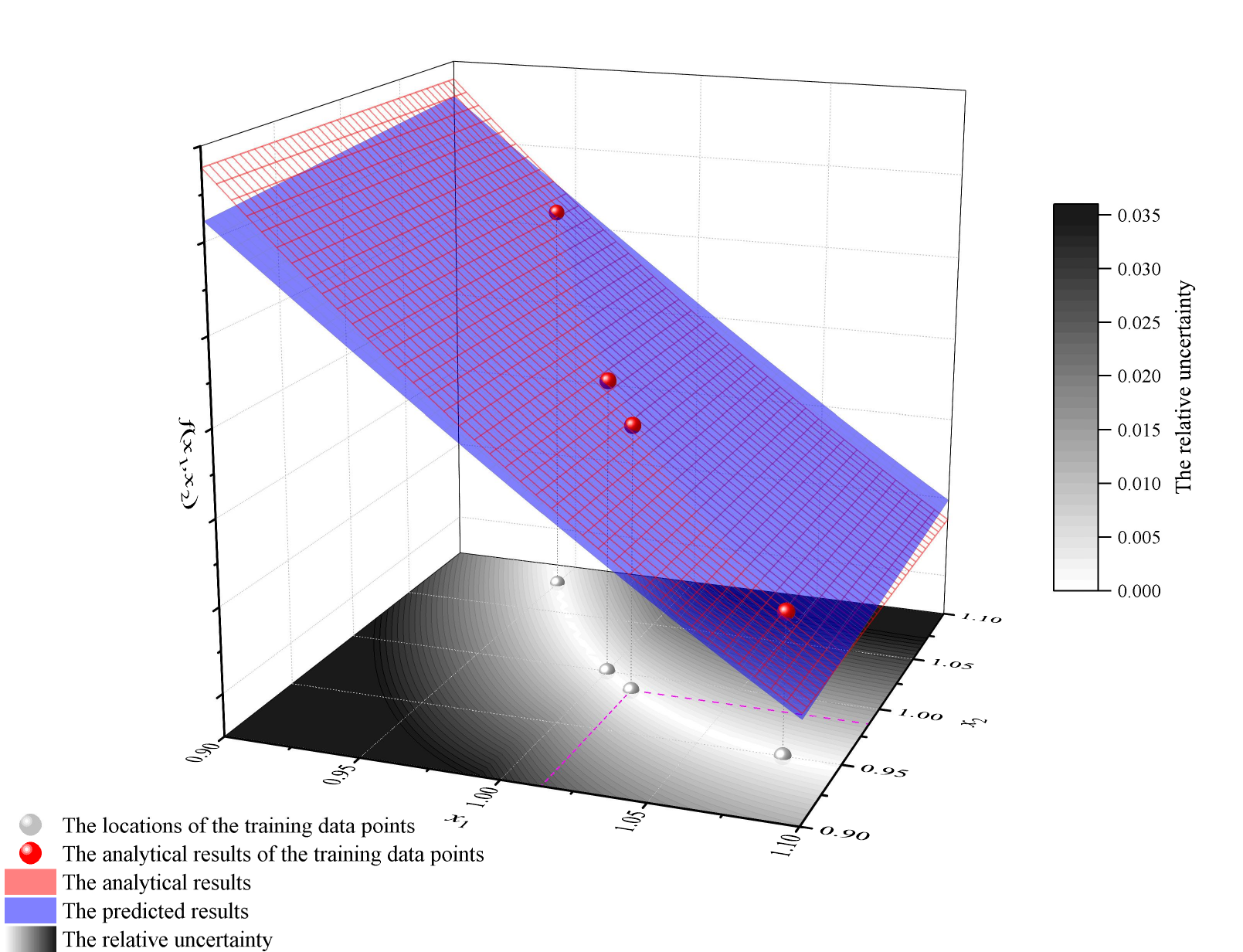}
			\caption{}
			\label{fig:al_procedure_b2}
		\end{subfigure}%
	\end{minipage}
	\vfill
	\centering
	\raisebox{2.5\height}{\rotatebox{90}{\scriptsize{$t=0.2203$}}}
	\begin{minipage}[t]{0.9\textwidth}
		\centering
		\begin{subfigure}[b]{0.49\linewidth}
			\centering
			\includegraphics[width=\linewidth]{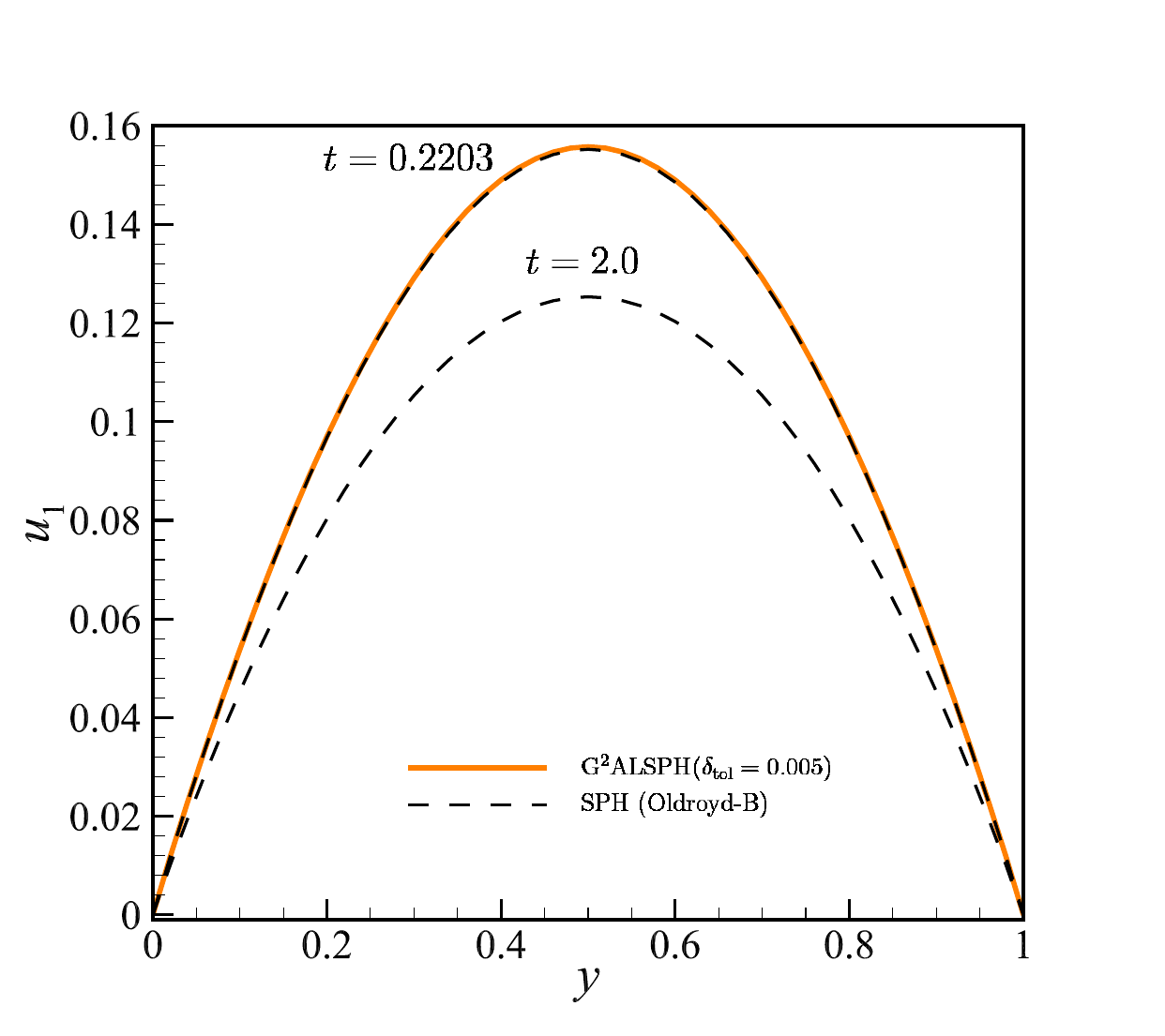}
			\caption{}
			\label{fig:al_procedure_c1}
		\end{subfigure}%
		\begin{subfigure}[b]{0.49\linewidth}
			\centering
			\includegraphics[width=\linewidth]{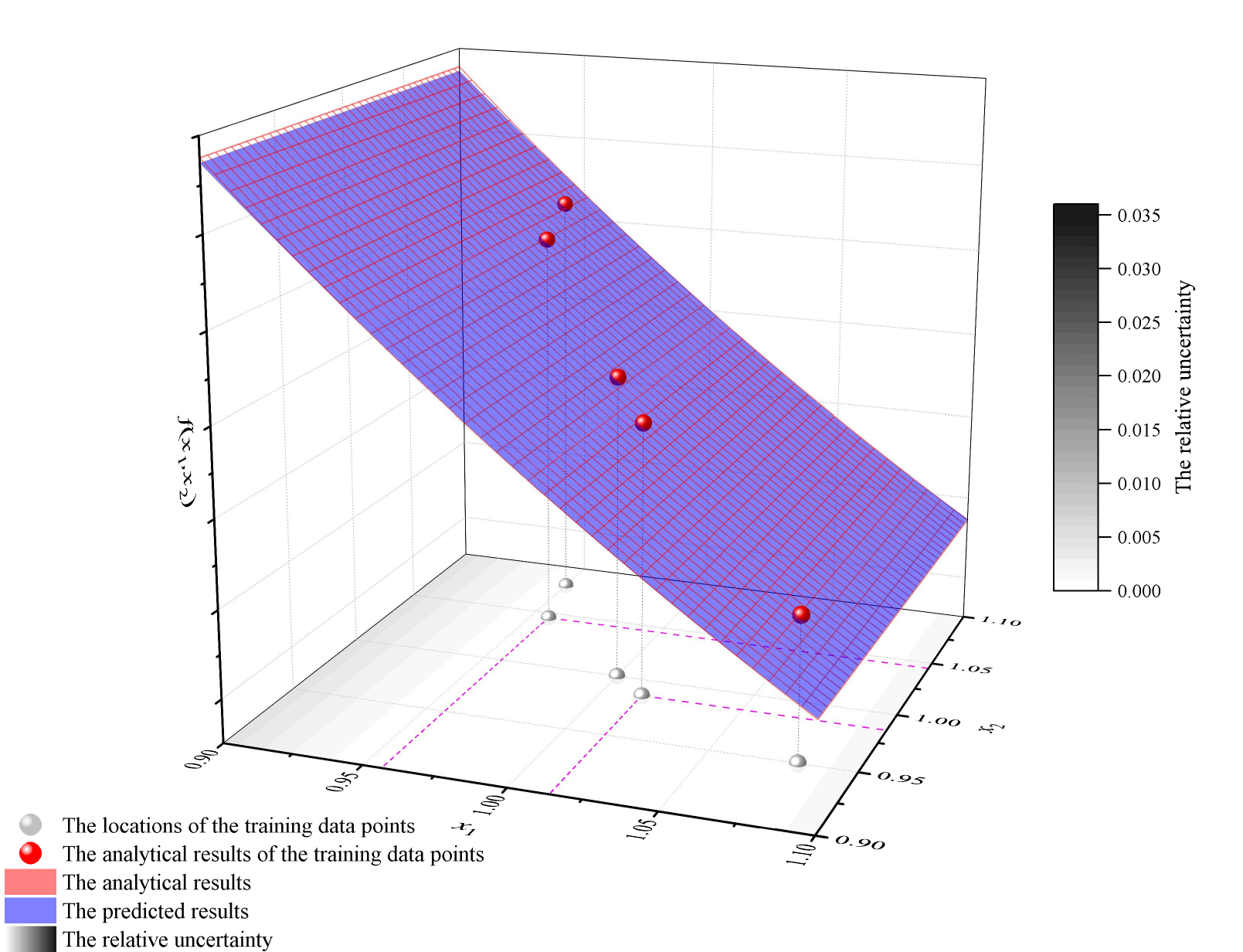}
			\caption{}
			\label{fig:al_procedure_c2}
		\end{subfigure}%
	\end{minipage}
	\vspace{-1mm}
	\caption{The active learning process for the Poiseuille flow ($Wi=0.1, \, \delta_{\rm{tol}}=0.005$): the distributions of the velocity $u_1$ and the learned constitutive relations at three time instants when $t=0$ is the initial instant and the active learning procedures are initiated at instants $t=0.0608$ and $t=0.2203$.}
	\label{fig:al_procedure}
\end{figure}

In~\cref{fig:al_procedure}, the distributions of the velocity $u_1$ and the learned constitutive relations at three time instants $t=0$, $t=0.0608$, and $t=0.2203$ are depicted. At $t=0$, the tolerance is set as $\delta_{\rm{tol}}=0.005$. Three training points are randomly selected to form an initial training data set to be inputted into GPR, and an initial local constitutive relation is obtained through GPR training, as shown in~\cref{fig:al_procedure_a2}. Then, the macroscopic SPH simulation is started by continuously monitoring whether the maximum value of the relative uncertainty $U_{\rm{r}}$ for all fluid particles at the current instant exceeds the tolerance $\delta_{\rm{tol}}$ or not. At $t=0.0608$, that maximum value first exceeds the tolerance $\delta_{\rm{tol}}$, and hence the active learning procedure is initiated. The point that implies the maximum relative uncertainty is chosen as the location of the next sampling point, which is included to form a new training data set. The GPR is performed on the newly formed training data set. A new constitutive relation is thus learned, which happens to meet the tolerance limit, that is, no more data points need to be acquired, as shown in~\cref{fig:al_procedure_b2}. The macroscopic SPH simulation is initiated again by continuously monitoring whether the maximum value of the relative uncertainty $U_{\rm{r}}$ for all fluid particles at any subsequent instant exceeds the tolerance $\delta_{\rm{tol}}$. At $t=0.2203$, that maximum value exceeds the tolerance $\delta_{\rm{tol}}$ for the second time, and hence the active learning procedure is re-initiated. One more training data point, which implies the maximum value of the relative uncertainty $U_{\rm{r}}$ for all fluid particles, is acquired, and the GPR is performed on the new training data set. The new learned constitutive relation meets the tolerance limit again, as shown in~\cref{fig:al_procedure_c2}. And the macroscopic SPH simulation is initiated. In subsequent time instants until the steady state, for the current learned constitutive relation, the maximum value of the relative uncertainty $U_{\rm{r}}$ for all fluid particles is always below the tolerance $\delta_{\rm{tol}}$. The active learning procedure does not need to be initiated again until the end of the simulation. Thus, \cref{fig:al_procedure_c2} depicts the final learned constitutive relation.
\par
The dashed lines in~\cref{fig:al_procedure_a1,fig:al_procedure_b1,fig:al_procedure_c1} and the red meshes in~\cref{fig:al_procedure_a2,fig:al_procedure_b2,fig:al_procedure_c2} represent the results obtained by the explicit Oldroyd-B model, which are deemed to be the reference analytical results for the ${\rm{G^2ALSPH}}$ method. As shown in~\cref{fig:al_procedure_b1,fig:al_procedure_c1}, the distributions of the velocity $u_1$ obtained by the ${\rm{G^2ALSPH}}$ method match well the reference analytical results, which indicates the effectiveness of the ${\rm{G^2ALSPH}}$ method.
\par
From the learned constitutive relation in~\cref{fig:al_procedure_a2} to that in~\cref{fig:al_procedure_b2}, an additional point is sampled to obtain the new learned constitutive relation, and some changes appear in the relative uncertainty results. The results improve in the region where the particles are concentrated, while they deteriorate in the region far from where the particles are concentrated. This shows one of the greatest characteristics of GPR, that is, the closer the region to training points, the higher its predictive accuracy. Adding a new training point alters the concentration of data points, thereby impacting the learning outcomes across the entire region. As depicted in~\cref{fig:al_procedure_c2}, with the inclusion of an additional training data point compared to~\cref{fig:al_procedure_b2}, the relative uncertainty results and prediction accuracy in the region $[0.9,1.1] \times [0.9,1.1]$ are improved.
\subsubsection{The Poiseuille flows with three different tolerances ($\mathit{Wi=0.1}$)}\label{sec4.1.3}
Two additional values, $\delta_{\rm{tol}}=0.05$ and $\delta_{\rm{tol}}=0.01$, are chosen to investigate the effects of the setting of the pre-set tolerance on the active learning and numerical simulation results. Their initial local learned constitutive relations are kept the same as that in~\cref{sec4.1.2}. The other physical parameters and computational model settings are also kept unchanged.
\par
\begin{figure}[H]
	\centering
	\begin{subfigure}[b]{0.333\linewidth}
		\includegraphics[width=\linewidth]{figures/GPRTWOIN530_238.pdf}
		\caption{$\delta_{\rm{tol}}=0.05$}
		\label{fig:Po_al-results1_a}
	\end{subfigure}%
	\begin{subfigure}[b]{0.333\linewidth}
		\includegraphics[width=\linewidth]{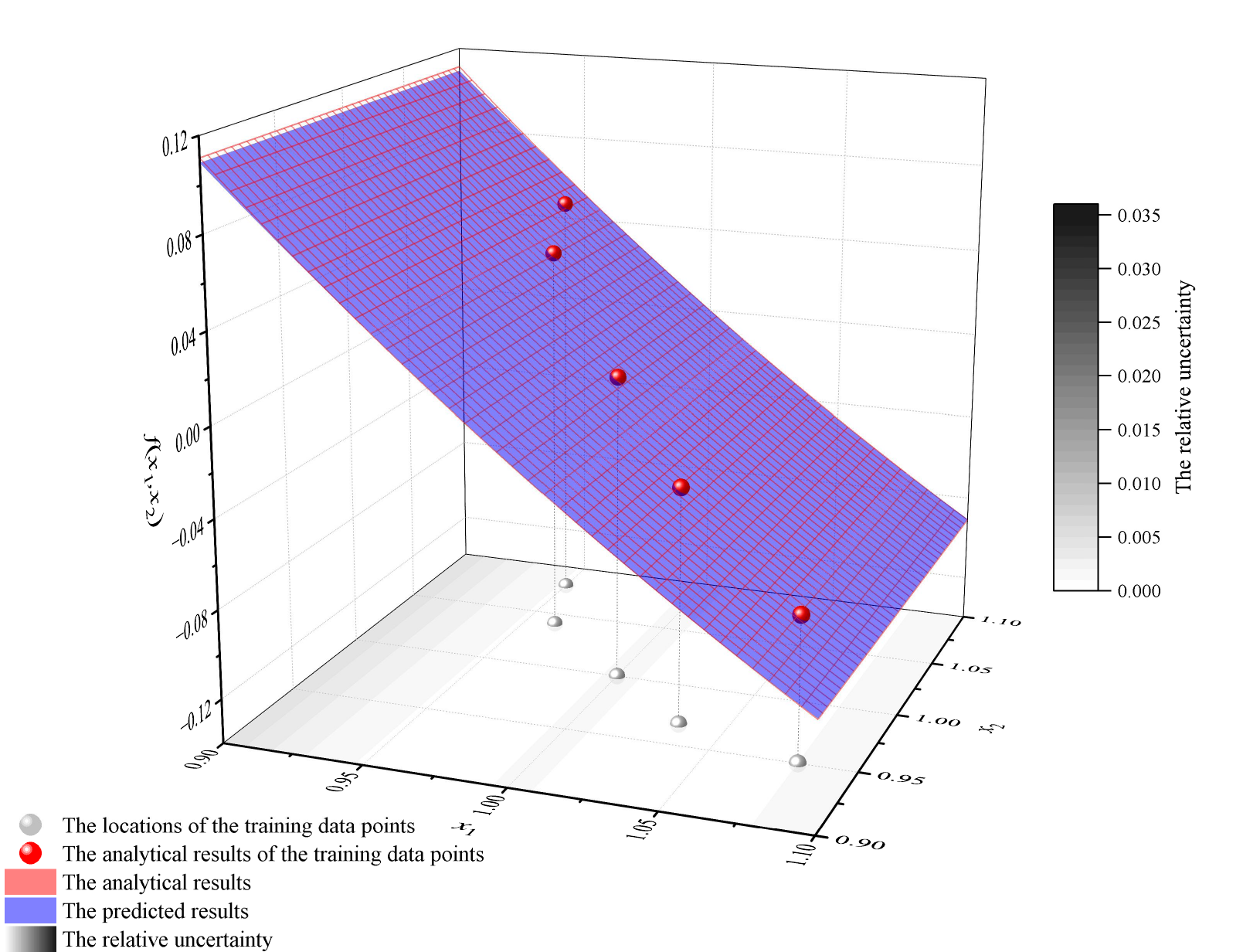}
		\caption{$\delta_{\rm{tol}}=0.01$}
		\label{fig:Po_al-results1_b}
	\end{subfigure}%
	\begin{subfigure}[b]{0.333\linewidth}
		\includegraphics[width=\linewidth]{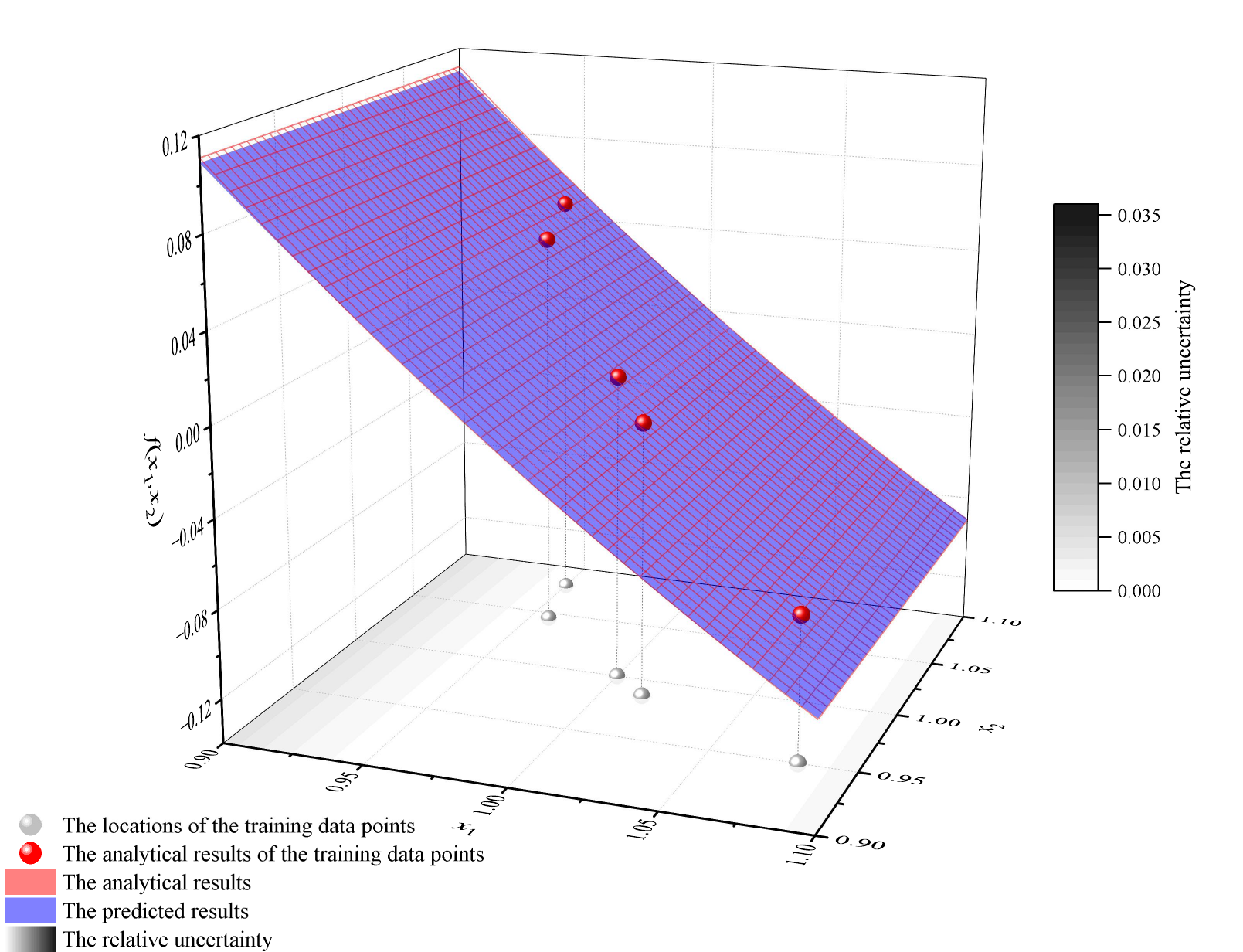}
		\caption{$\delta_{\rm{tol}}=0.005$}
		\label{fig:Po_al-results1_c}
	\end{subfigure}%
	\caption{The final learned constitutive relations obtained at three different tolerance settings for the Poiseuille flows ($Wi=0.1$): (a) $\delta_{\rm{tol}}=0.05$ (denoted as R1), (b) $\delta_{\rm{tol}}=0.01$ (denoted as R2), and (c) $\delta_{\rm{tol}}=0.005$ (denoted as R3).}
	\label{fig:Po_al-results1}
\end{figure}

Three final learned constitutive relations are presented in~\cref{fig:Po_al-results1}. It can be observed that as the tolerance decreases, the number of the required training data points increases. As more training data points are added, the results of the relative uncertainty in some regions improve. Eventually, for a tolerance setting of $\delta_{\rm{tol}}=0.05$, the initial local constitutive relation learned with three training data points satisfies the tolerance limit throughout the entire simulation process, thus eliminating the need for an active learning procedure. In contrast, for tolerance settings $\delta_{\rm{tol}}=0.01$ and $\delta_{\rm{tol}}=0.005$, two additional training data points are respectively acquired, resulting in a total of five training data points. For a convenient presentation, these three learned constitutive relations are denoted as R1, R2, and R3, whose pre-set tolerances are $\delta_{\rm{tol}}=0.05$, $\delta_{\rm{tol}}=0.01$, and $\delta_{\rm{tol}}=0.005$, respectively.
\par
\begin{figure}[H]
	\centering
	\begin{subfigure}[b]{0.5\linewidth}
		\includegraphics[width=\linewidth]{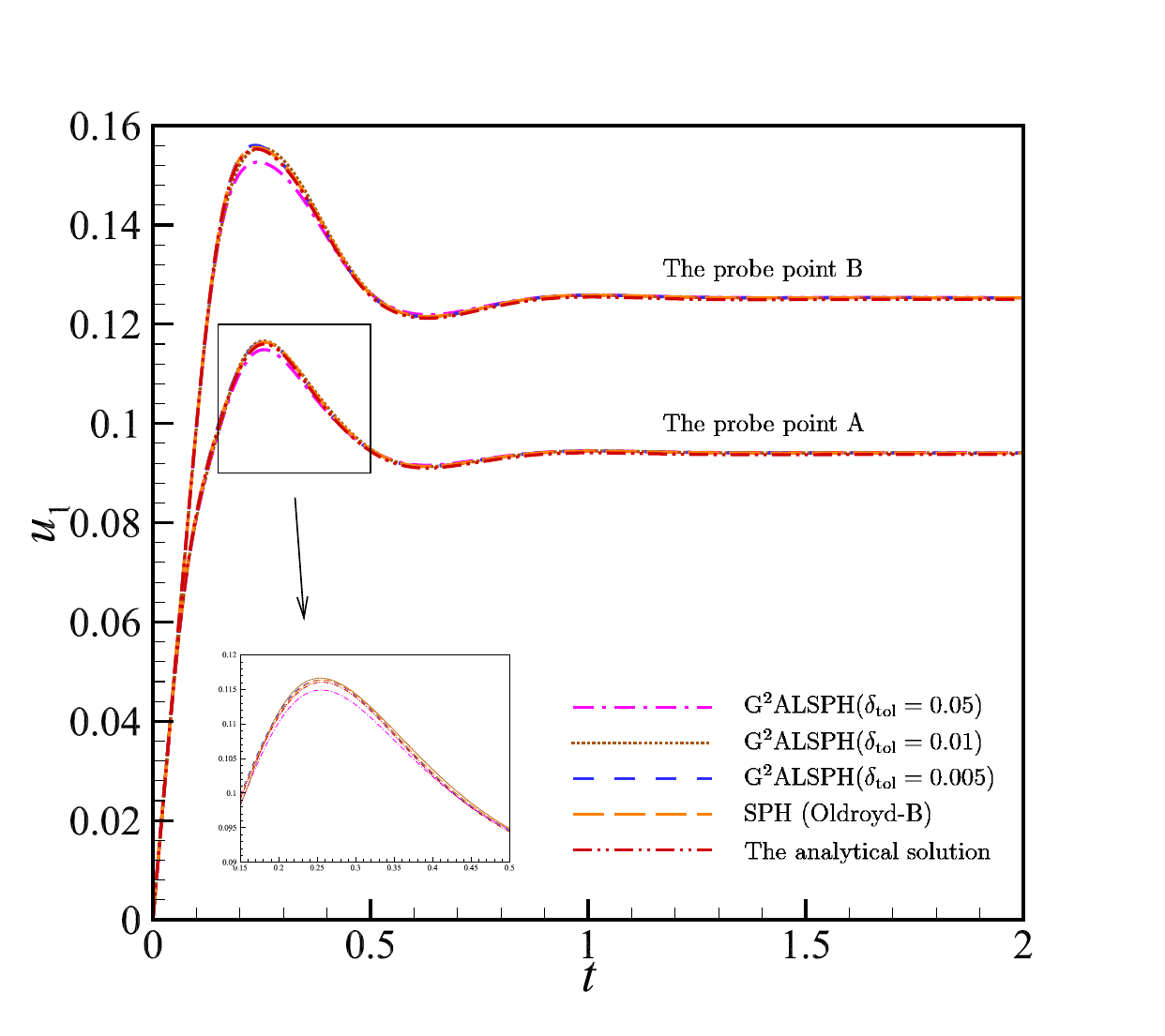}
		\caption{}
		\label{fig:Po1-results1_a}
	\end{subfigure}%
	\begin{subfigure}[b]{0.5\linewidth}
		\includegraphics[width=\linewidth]{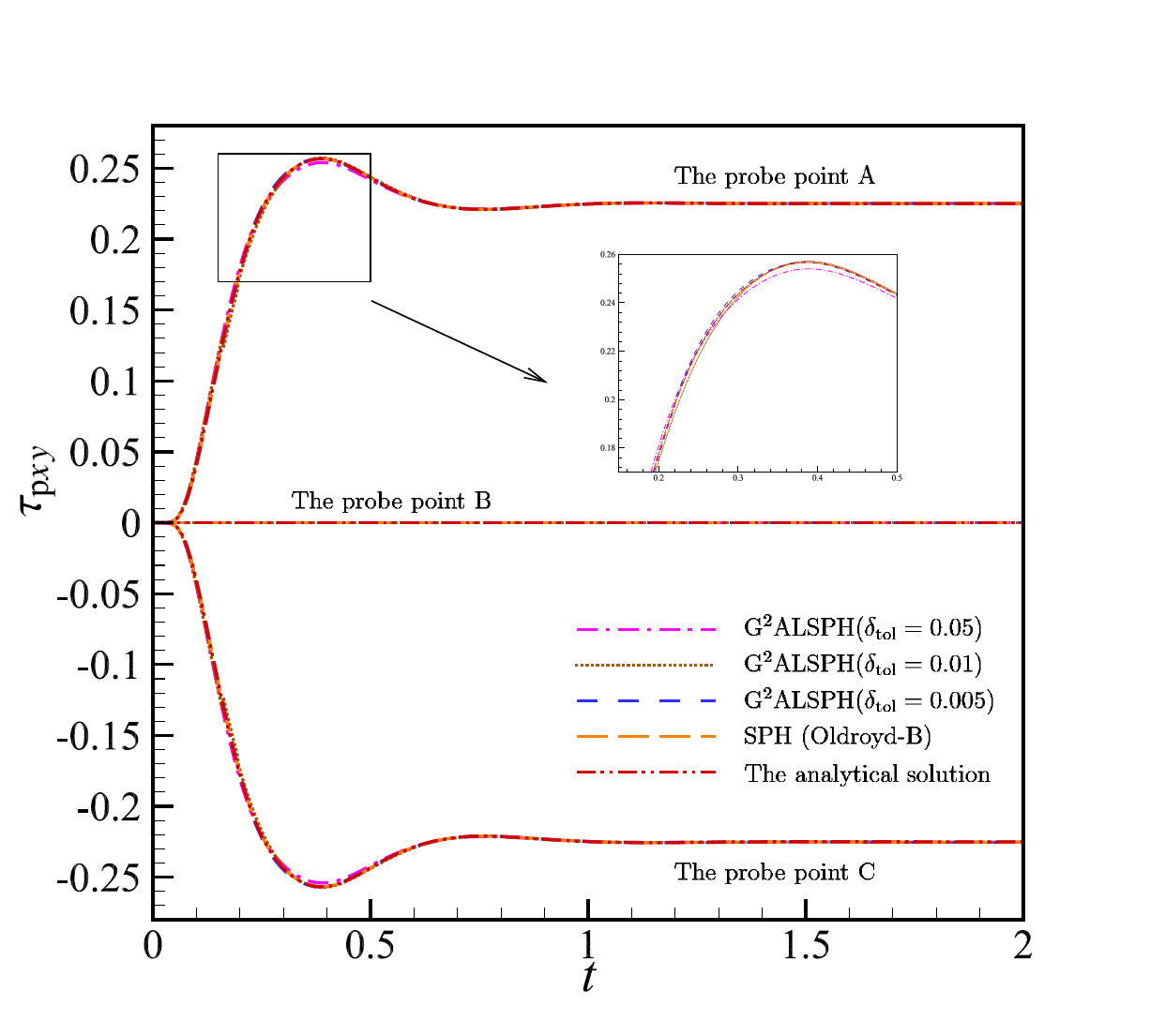}
		\caption{}
		\label{fig:Po1-results1_b}
	\end{subfigure}%
	\caption{Comparison among the results in the probe points A, B, and C in the Poiseuille flows ($Wi=0.1$), which are obtained by the ${\rm{G^2ALSPH}}$ method with three different tolerance settings ($\delta_{\rm{tol}}=0.05$, $\delta_{\rm{tol}}=0.01$, and $\delta_{\rm{tol}}=0.005$) and the SPH method with the Oldroyd-B model and the analytical solutions: (a) the variations of the velocity $u_1$ over time and (b) the variations of the shear stress $\tau_{{\rm{p}}xy}$ over time.}
	\label{fig:Po1-results1}
\end{figure}

Regarding the simulation results, in~\cref{fig:Po1-results1}, the variations of the velocity $u_1$ and the shear stress $\tau_{{\rm{p}}xy}$ over time at probe points A, B, and C obtained by the ${\rm{G^2ALSPH}}$ method at three different tolerance settings $\delta_{\rm{tol}}=0.05$, $\delta_{\rm{tol}}=0.01$, and $\delta_{\rm{tol}}=0.005$, and the SPH method with the Oldroyd-B model and the analytical solutions are presented all together. The red lines represent the analytical solutions obtained through the formulas presented in~\ref{app:Poiseuille_analytical}. The results obtained by the SPH method with the Oldroyd-B model match very well the analytical solutions. As the tolerance decreases, the results of the variations of the velocity $u_1$ and the shear stress $\tau_{{\rm{p}}xy}$ over time obtained by the ${\rm{G^2ALSPH}}$ method converge to the reference analytical results obtained by the SPH method with the Oldroyd-B model. When the tolerance is less than $0.01$, the results obtained by the ${\rm{G^2ALSPH}}$ method are close enough to the reference analytical results, which proves that the accuracy of the entire simulation process can be ensured at every time instant by utilizing the ${\rm{G^2ALSPH}}$ method with an appropriate tolerance.
\par
\begin{figure}[htbp]
	\centering
	\raisebox{1.3\height}{\rotatebox{90}{\scriptsize{The eigenvalue $c_1$}}}
	\begin{minipage}[t]{0.58\linewidth}
		\centering
		\begin{subfigure}[b]{0.24\linewidth}
			\raisebox{1.5\height}{\rotatebox{0}{\scriptsize{${\rm{G^2ALSPH}}$}}}
			\centering
			\includegraphics[width=\linewidth]{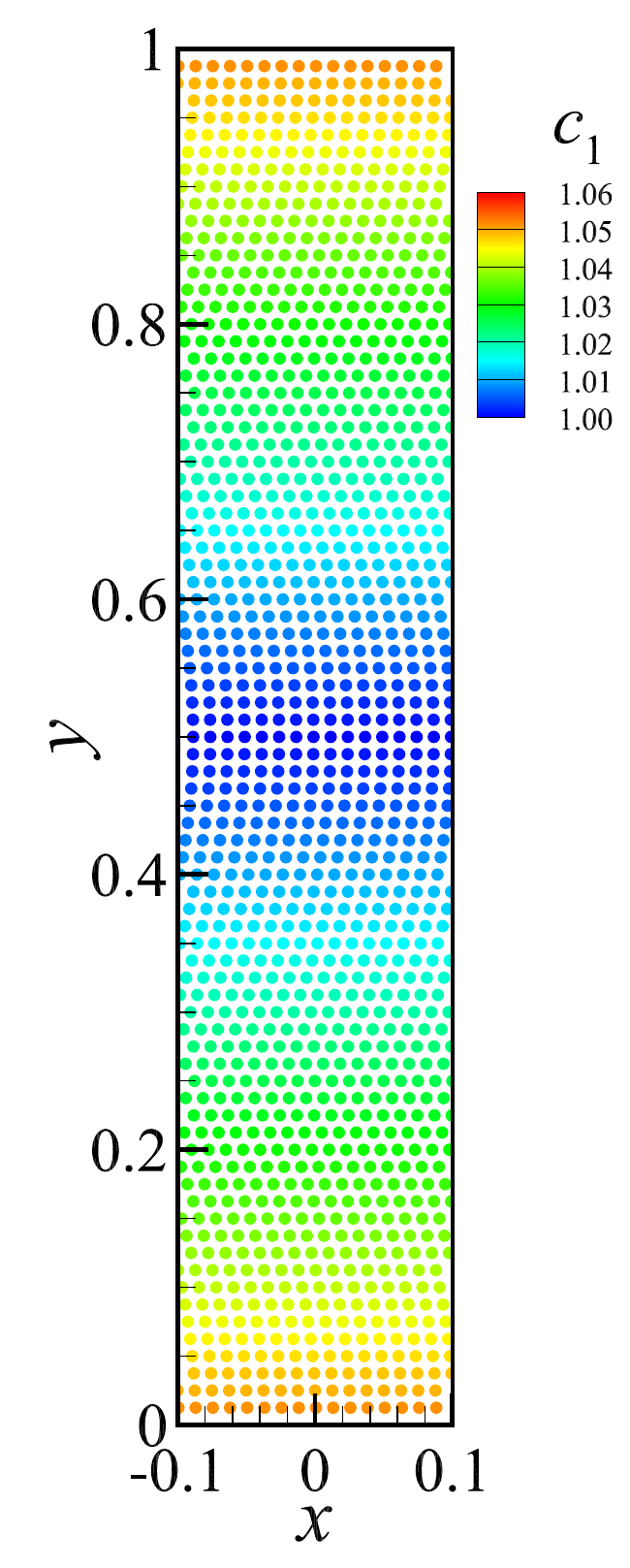}
			\label{fig:Po1_distribution_a1}
		\end{subfigure}%
		\hfill
		\begin{subfigure}[b]{0.24\linewidth}
			\raisebox{1.5\height}{\rotatebox{0}{\scriptsize{SPH (Oldroyd-B)}}}
			\centering
			\includegraphics[width=\linewidth]{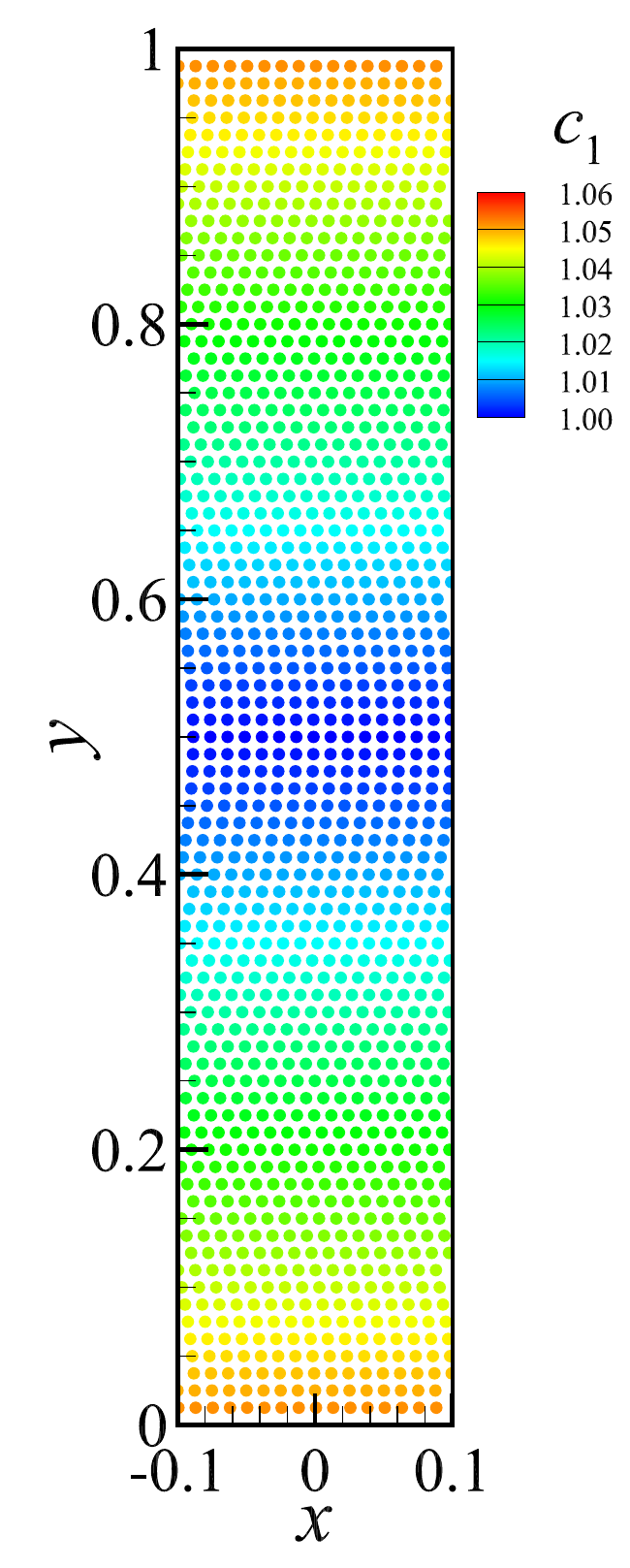}
			\label{fig:Po1_distribution_b1}
		\end{subfigure}%
		\hfill
		\begin{subfigure}[b]{0.24\linewidth}
			\raisebox{1.5\height}{\rotatebox{0}{\scriptsize{The relative errors}}}
			\centering
			\includegraphics[width=\linewidth]{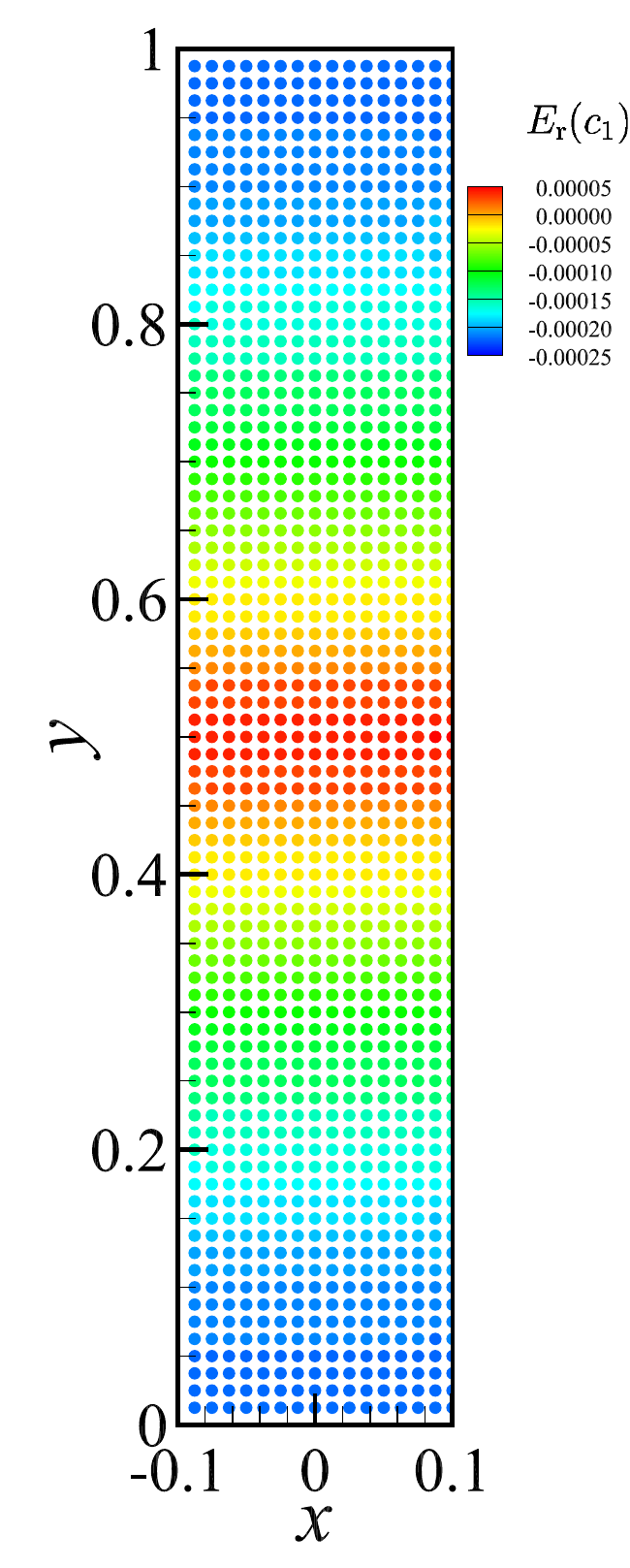}
			\label{fig:Po1_distribution_c1}
		\end{subfigure}%
	\end{minipage}
	\vspace{-8mm}
	\vfill
	\centering
	\raisebox{1.3\height}{\rotatebox{90}{\scriptsize{The eigenvalue $c_2$}}}
	\begin{minipage}[t]{0.58\linewidth}
		\centering
		\begin{subfigure}[b]{0.24\linewidth}
			\centering
			\includegraphics[width=\linewidth]{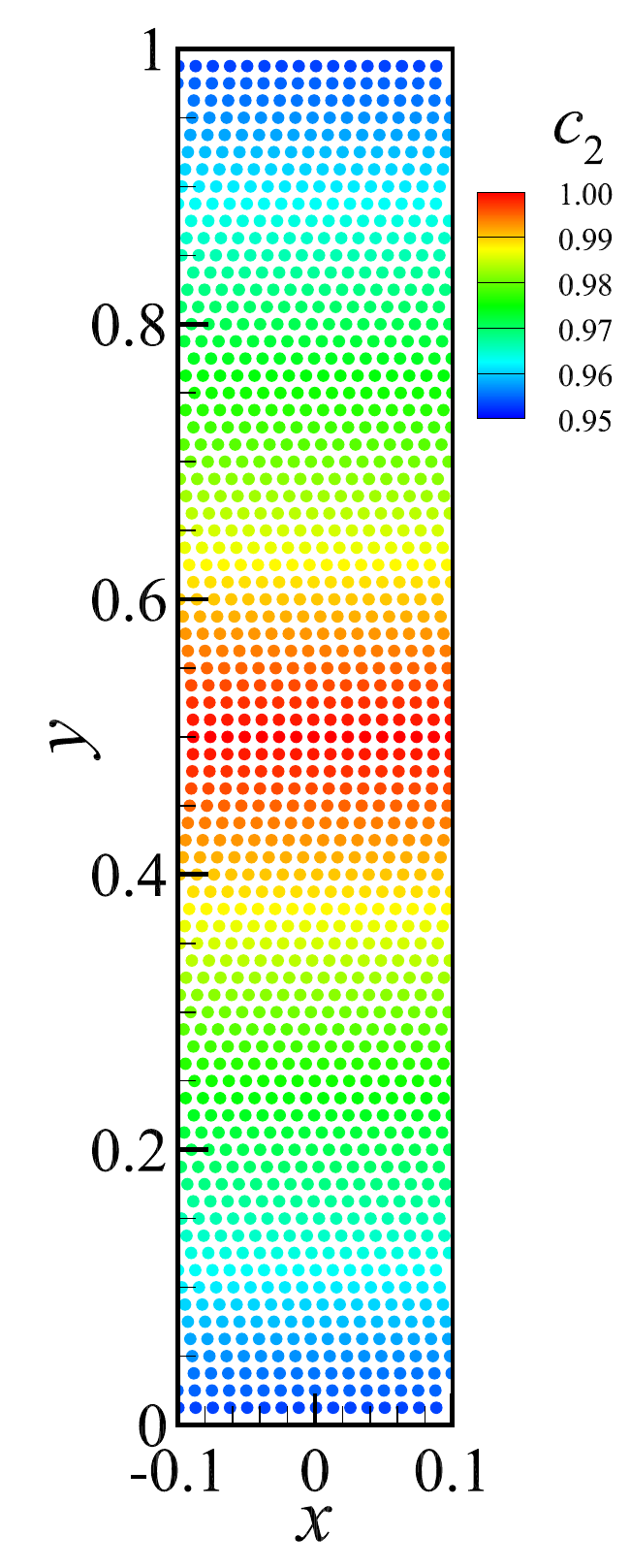}
			\label{fig:Po1_distribution_a2}
		\end{subfigure}%
		\hfill
		\begin{subfigure}[b]{0.24\linewidth}
			\centering
			\includegraphics[width=\linewidth]{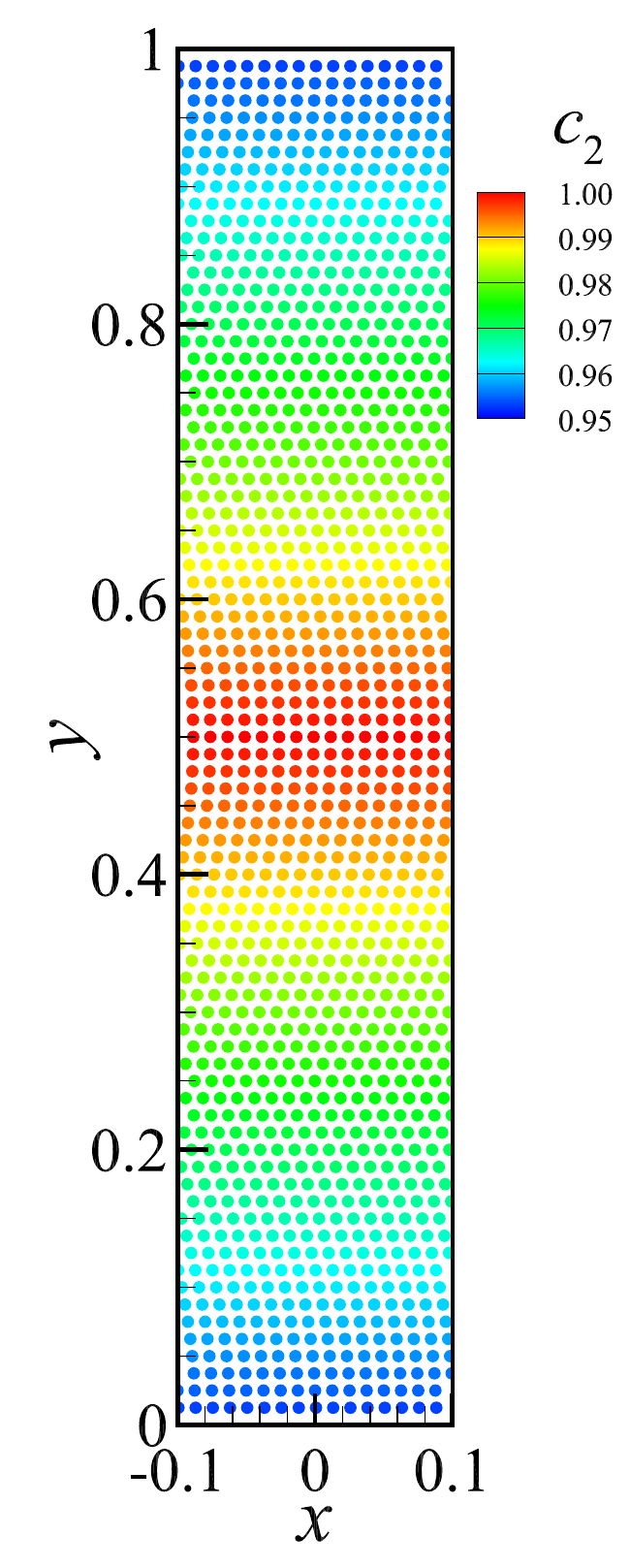}
			\label{fig:Po1_distribution_b2}
		\end{subfigure}%
		\hfill
		\begin{subfigure}[b]{0.24\linewidth}
			\centering
			\includegraphics[width=\linewidth]{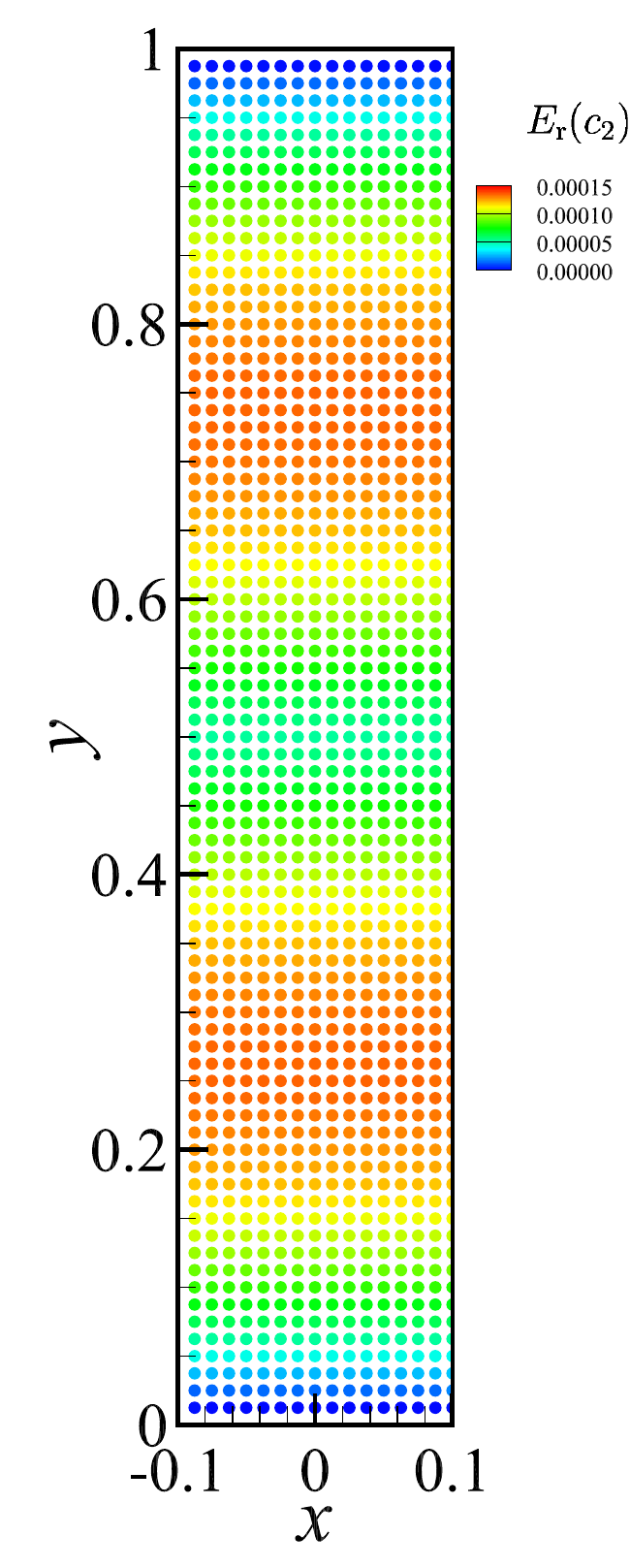}
			\label{fig:Po1_distribution_c2}
		\end{subfigure}%
	\end{minipage}
	\vspace{-8mm}
	\vfill
	\centering
	\raisebox{1.6\height}{\rotatebox{90}{\scriptsize{The velocity $u_1$}}}
	\begin{minipage}[t]{0.58\linewidth}
		\centering
		\begin{subfigure}[b]{0.24\linewidth}
			\centering
			\includegraphics[width=\linewidth]{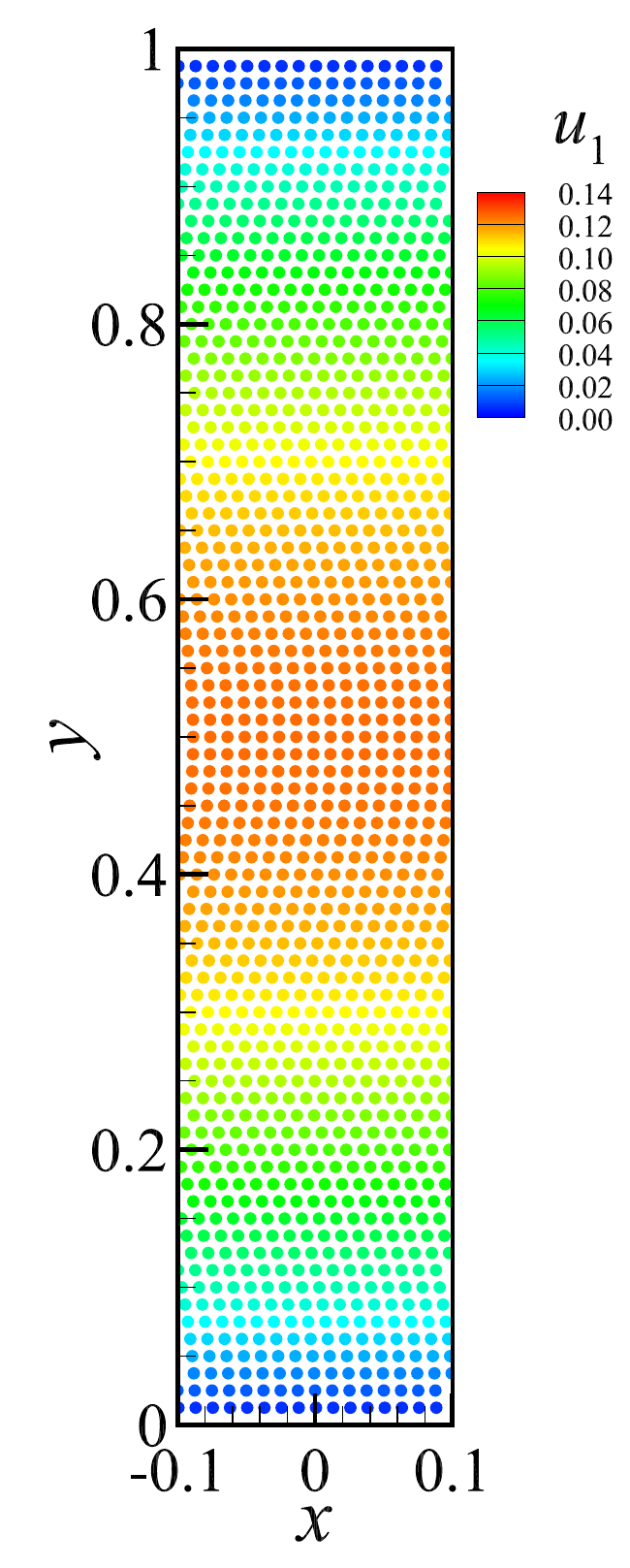}
			\label{fig:Po1_distribution_a3}
		\end{subfigure}%
		\hfill
		\begin{subfigure}[b]{0.24\linewidth}
			\centering
			\includegraphics[width=\linewidth]{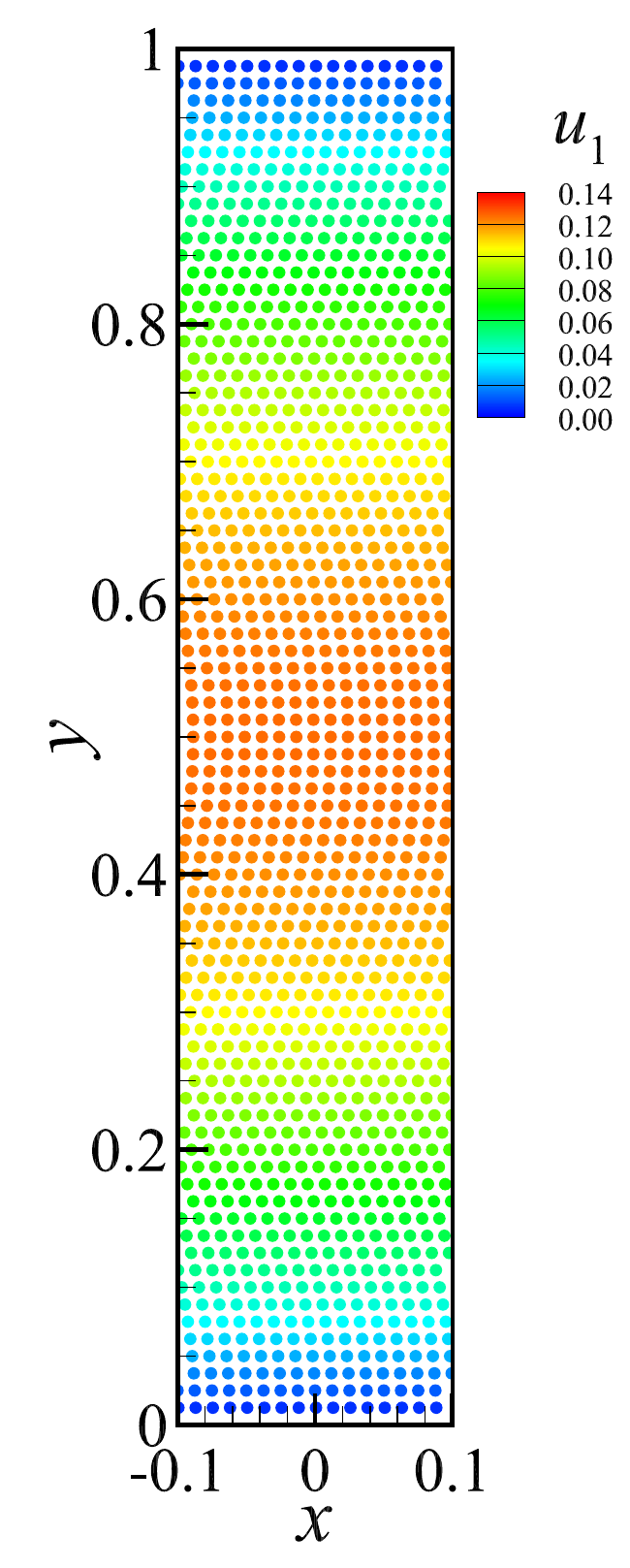}
			\label{fig:Po1_distribution_b3}
		\end{subfigure}%
		\hfill
		\begin{subfigure}[b]{0.24\linewidth}
			\centering
			\includegraphics[width=\linewidth]{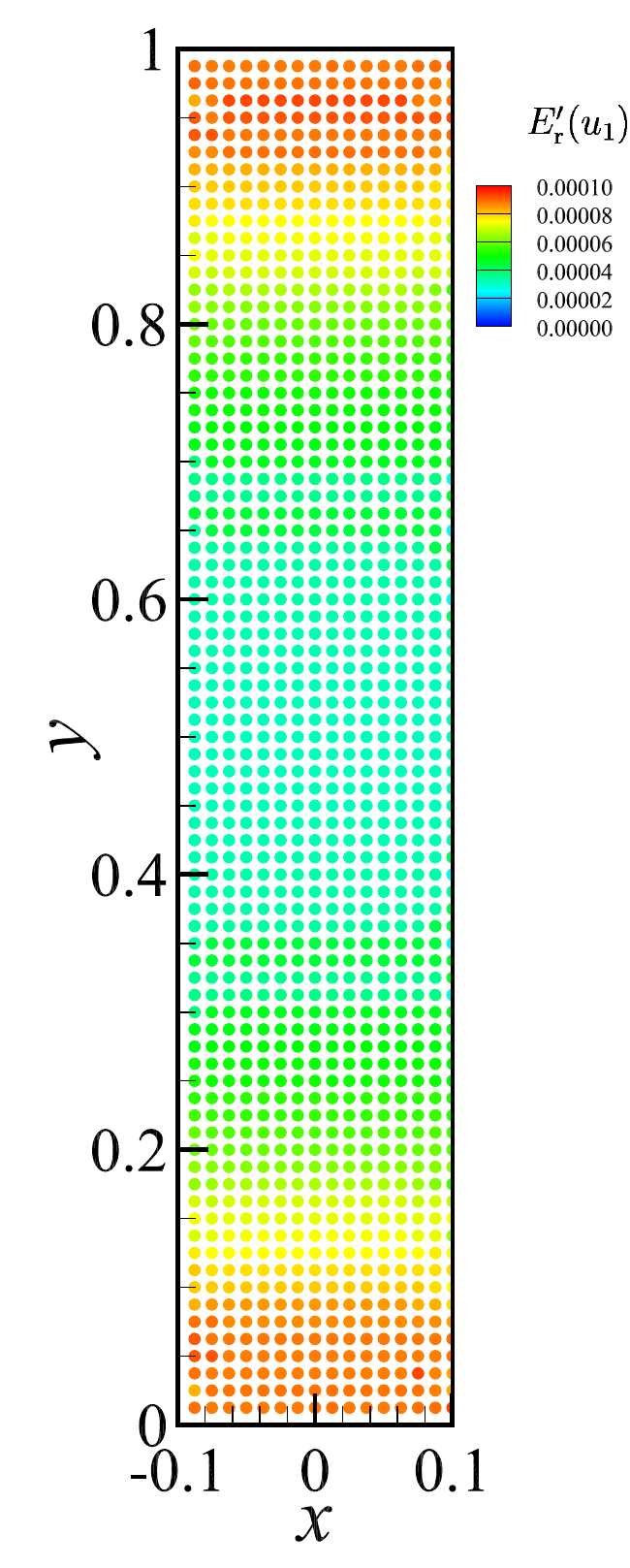}
			\label{fig:Po1_distribution_c3}
		\end{subfigure}%
	\end{minipage}
	\vspace{-8mm}
	\vfill
	\centering
	\raisebox{0.4\height}{\rotatebox{90}{\scriptsize{The first normal stress difference $N_1$}}}
	\begin{minipage}[t]{0.58\linewidth}
		\centering
		\begin{subfigure}[b]{0.24\linewidth}
			\centering
			\includegraphics[width=\linewidth]{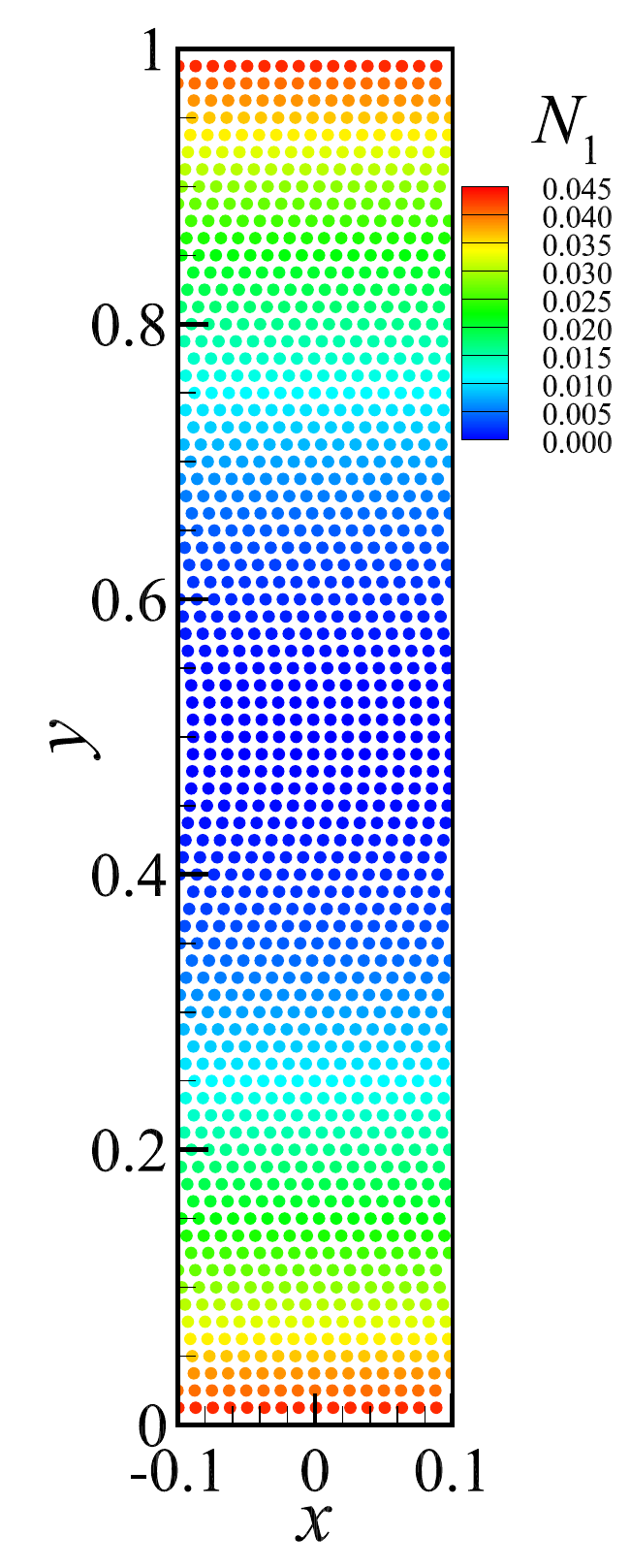}
			\caption{}
			\label{fig:Po1_distribution_a4}
		\end{subfigure}%
		\hfill
		\begin{subfigure}[b]{0.24\linewidth}
			\centering
			\includegraphics[width=\linewidth]{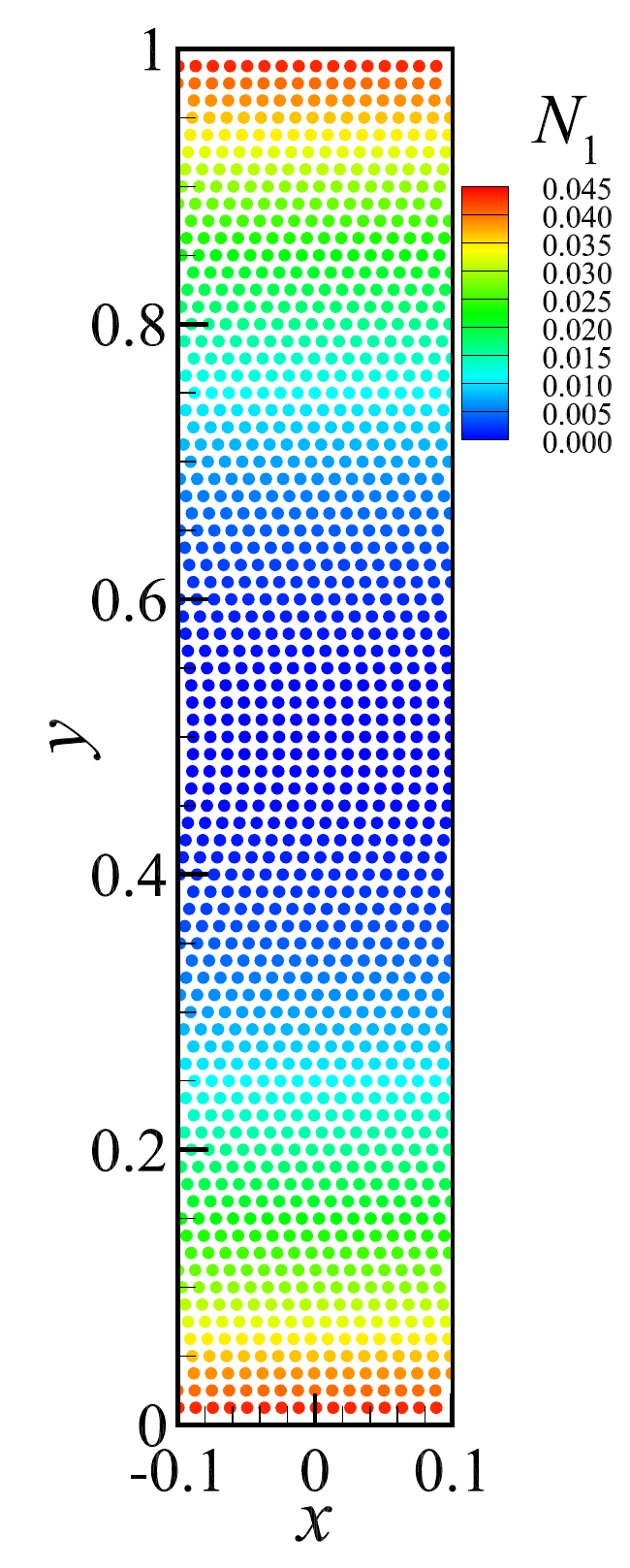}
			\caption{}
			\label{fig:Po1_distribution_b4}
		\end{subfigure}%
		\hfill
		\begin{subfigure}[b]{0.24\linewidth}
			\centering
			\includegraphics[width=\linewidth]{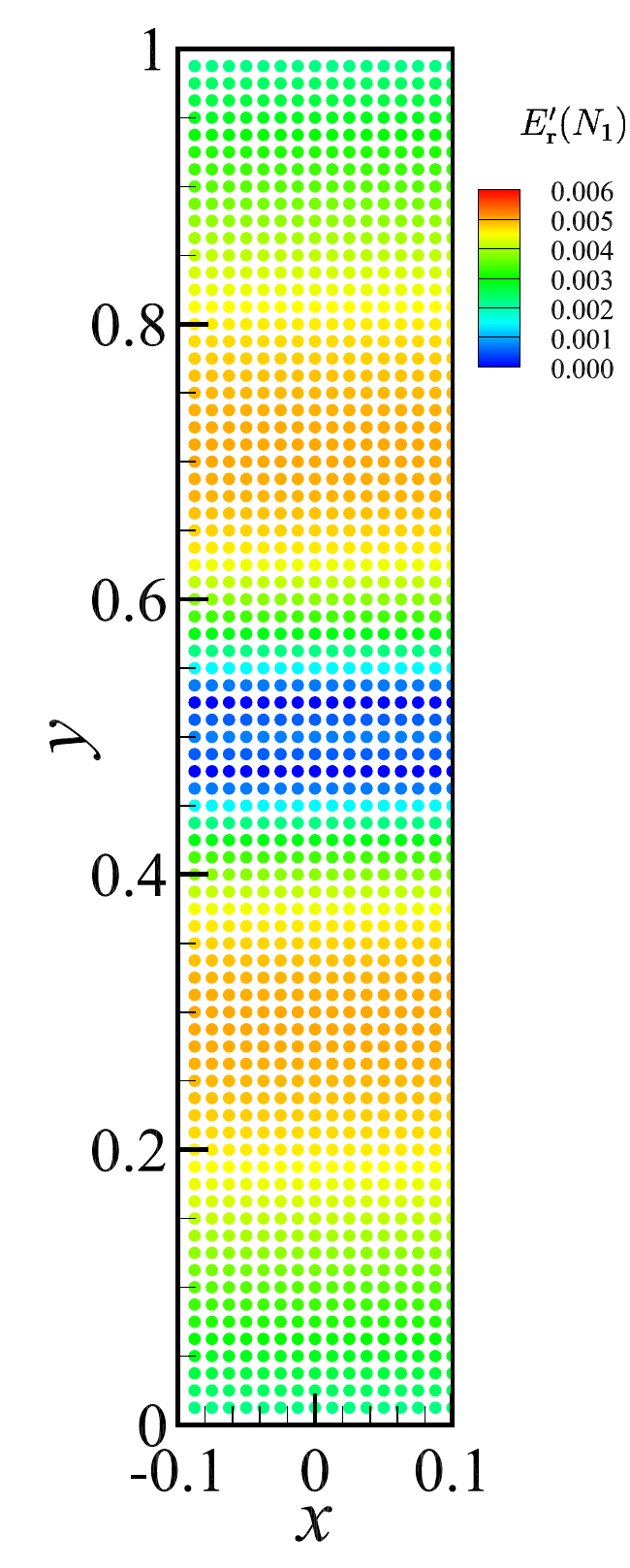}
			\caption{}
			\label{fig:Po1_distribution_c4}
		\end{subfigure}%
	\end{minipage}
	\vspace{-1mm}
	\captionsetup{justification=justified}
	\caption{The distributions of the eigenvalues $c_1$ and $c_2$, the velocity $u_1$, and the first normal stress difference $N_1$ at the steady state for the Poiseuille flows ($Wi=0.1$): (a) the results obtained by the ${\rm{G^2ALSPH}}$ method ($\delta_{\rm{tol}}=0.005$), (b) the results obtained by the SPH method with the Oldroyd-B model, and (c) the relative errors.}
	\label{fig:Po1_distribution}
\end{figure}

In~\cref{fig:Po1_distribution}, the distributions of the eigenvalues $c_1$ and $c_2$, the velocity $u_1$, and the first normal stress difference $N_1$ at the steady state for the Poiseuille flows at $Wi=0.1$ are given. \cref{fig:Po1_distribution_a4} depicts the results obtained by the ${\rm{G^2ALSPH}}$ method with the tolerance setting $\delta_{\rm{tol}}=0.005$. \cref{fig:Po1_distribution_b4} depicts the results obtained by the SPH method with the Oldroyd-B model. \cref{fig:Po1_distribution_c4} depicts the distributions of the relative errors between the results obtained by the ${\rm{G^2ALSPH}}$ method and the SPH method with the Oldroyd-B model. There are two types of equations for calculating the relative error. For the eigenvalues $c_1$ and $c_2$, the commonly used equation for calculating the relative error is as follows:
\begin{equation}
	E_{\rm{r}}(A_i)=\frac{A_i-A_{{\rm{analytical}},i}}{A_{{\rm{analytical}},i}},\label{eq:relative_error0}
\end{equation}
where $A_i$ indicates the result of a physical variable on the $i{\rm{th}}$ fluid particle obtained by the ${\rm{G^2ALSPH}}$ method and $A_{{\rm{analytical}},i}$ indicates the corresponding reference analytical result obtained by the SPH method with the Oldroyd-B model. For the velocity $u_1$ and the first normal stress difference $N_1$, where the vanishing values may appear, the following relative error equation is employed as
\begin{equation}
	E^{\prime}_{\rm{r}}(A_i)=\frac{\left|A_i-A_{{\rm{analytical}},i}\right|}{\left|A_{{\rm{analytical}},i}\right|+0.001},\label{eq:relative_error1}
\end{equation}
where the physical meanings of $A_i$ and $A_{{\rm{analytical}},i}$ are the same as those in~\cref{eq:relative_error0}. Since the simulation results in~\cref{fig:Po1_distribution_a4,fig:Po1_distribution_b4} are provided by two different methods at the same instant $t=2.0$, the positions of the fluid particles are not correspondingly the same. Thus, the arrangement of the fluid particles at the initial instant in~\cref{fig:Po_initial_b} is used, and the physical quantities on the fluid particles are interpolated from the results in~\cref{fig:Po1_distribution_a4,fig:Po1_distribution_b4} by the interpolation equation~\cref{eq:interpolation}. Afterward, the relative errors can be accurately calculated.
\par
As shown in~\cref{fig:Po1_distribution_c4}, for the eigenvalues $c_1$ and $c_2$, all the absolute values of the relative errors on fluid particles are below $0.025\%$. For the velocity $u_1$, all the values of the relative errors on fluid particles are below $0.01\%$, whereas, for the first normal stress difference $N_1$, they are below $0.6\%$. The results in~\cref{fig:Po1-results1,fig:Po1_distribution} fully validate the effectiveness of the ${\rm{G^2ALSPH}}$ method in simulating viscoelastic flows.
\subsection{The Poiseuille flows at larger Weissenberg numbers}\label{sec4.2.}
In order to study the effects of the setting of the pre-set tolerance and the choice of the initial local constitutive relation and study the superiority of the new relative uncertainty strategy over the direct uncertainty strategy, the Poiseuille flows at $Wi=0.5$ and $Wi=1.0$ are further considered. The artificial sound speed $c_{\rm{s}}$ and smoothing length $h$ are set to $c_{\rm{s}}=3.0$ and $h=1.0d_0$ for $Wi=0.5$, and to $c_{\rm{s}}=5.0$ and $h=1.0d_0$ for $Wi=1.0$. All other physical parameters and computational model settings remain the same as those in~\cref{sec4.1}. An increase in the Weissenberg number $Wi$ from $0.1$ to $0.5$ or $1.0$ leads to an increase in the elasticity of the fluid and hence an increase in the maximum velocity of the fluid due to overshooting. Thus, a longer simulation time is needed to reach the steady state for the Poiseuille flow, and a larger artificial sound speed is necessary to enforce fluid incompressibility.
\subsubsection{The Poiseuille flows with three different tolerances ($\mathit{Wi=0.5}$)}\label{sec4.2.1}
\begin{figure}[H]
	\centering
	\begin{subfigure}[b]{0.333\linewidth}
		\includegraphics[width=\linewidth]{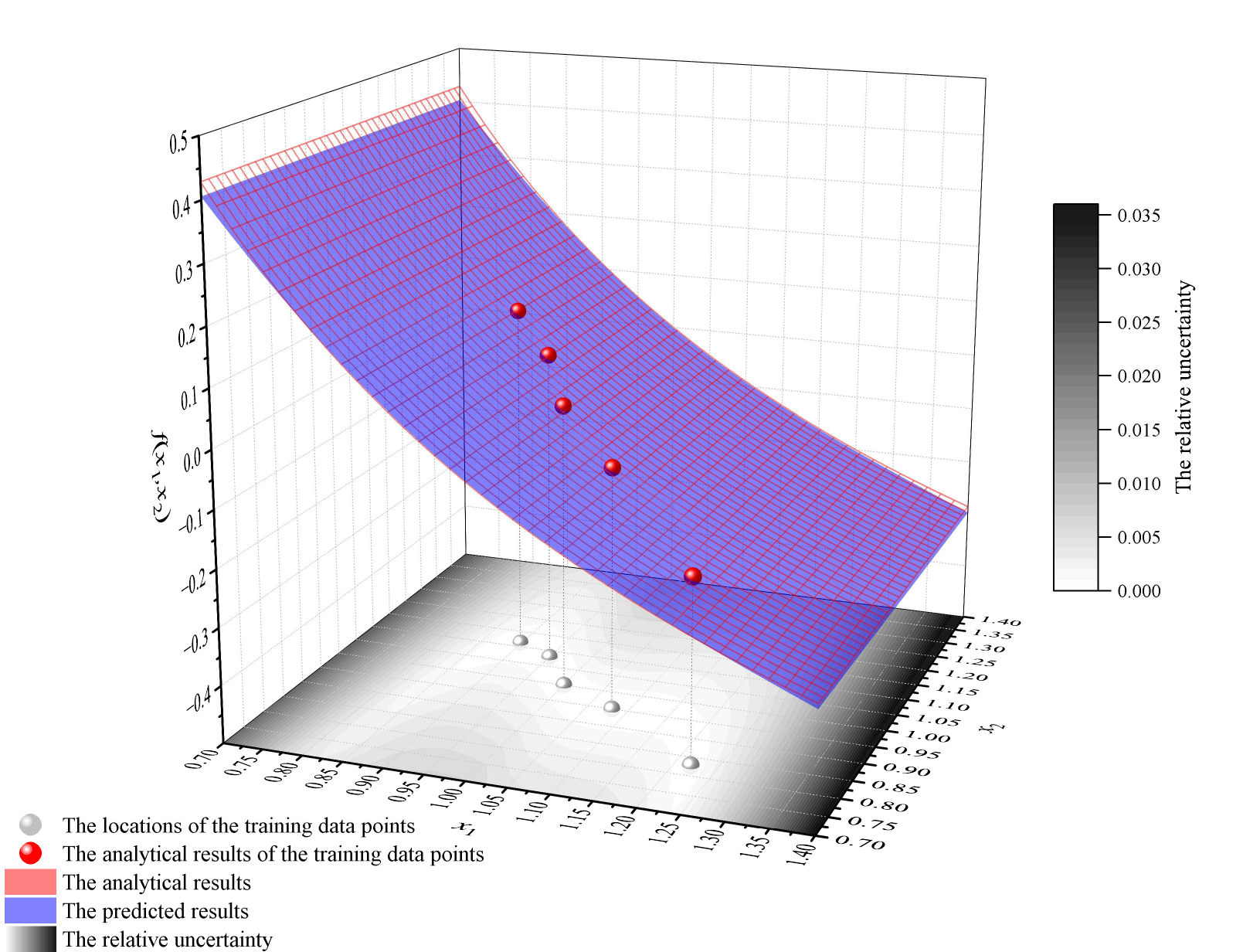}
		\caption{$\delta_{\rm{tol}}=0.05$ with R1}
		\label{fig:Po_al-results2_a}
	\end{subfigure}%
	\begin{subfigure}[b]{0.333\linewidth}
		\includegraphics[width=\linewidth]{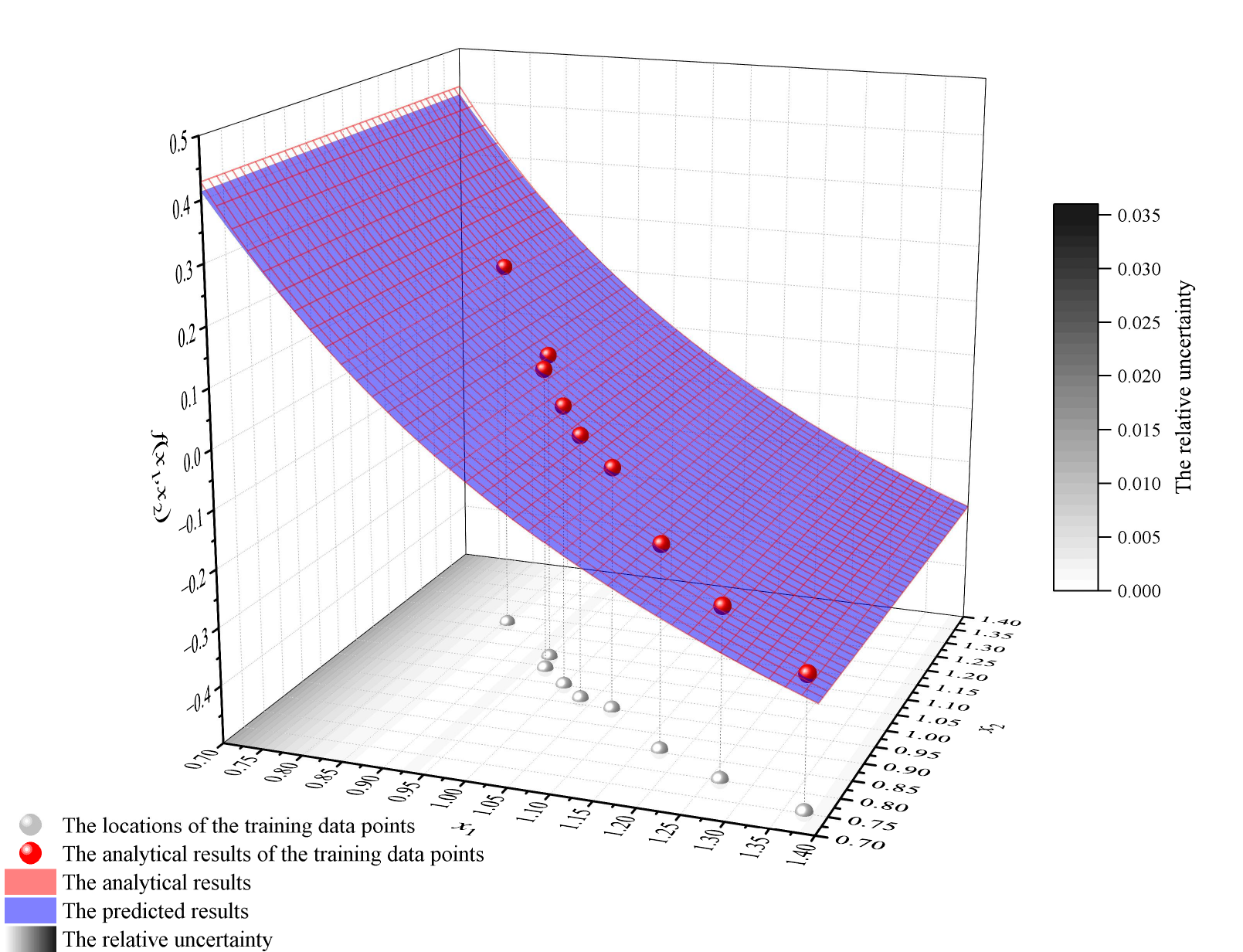}
		\caption{$\delta_{\rm{tol}}=0.01$ with R1}
		\label{fig:Po_al-results2_b}
	\end{subfigure}%
	\begin{subfigure}[b]{0.333\linewidth}
		\includegraphics[width=\linewidth]{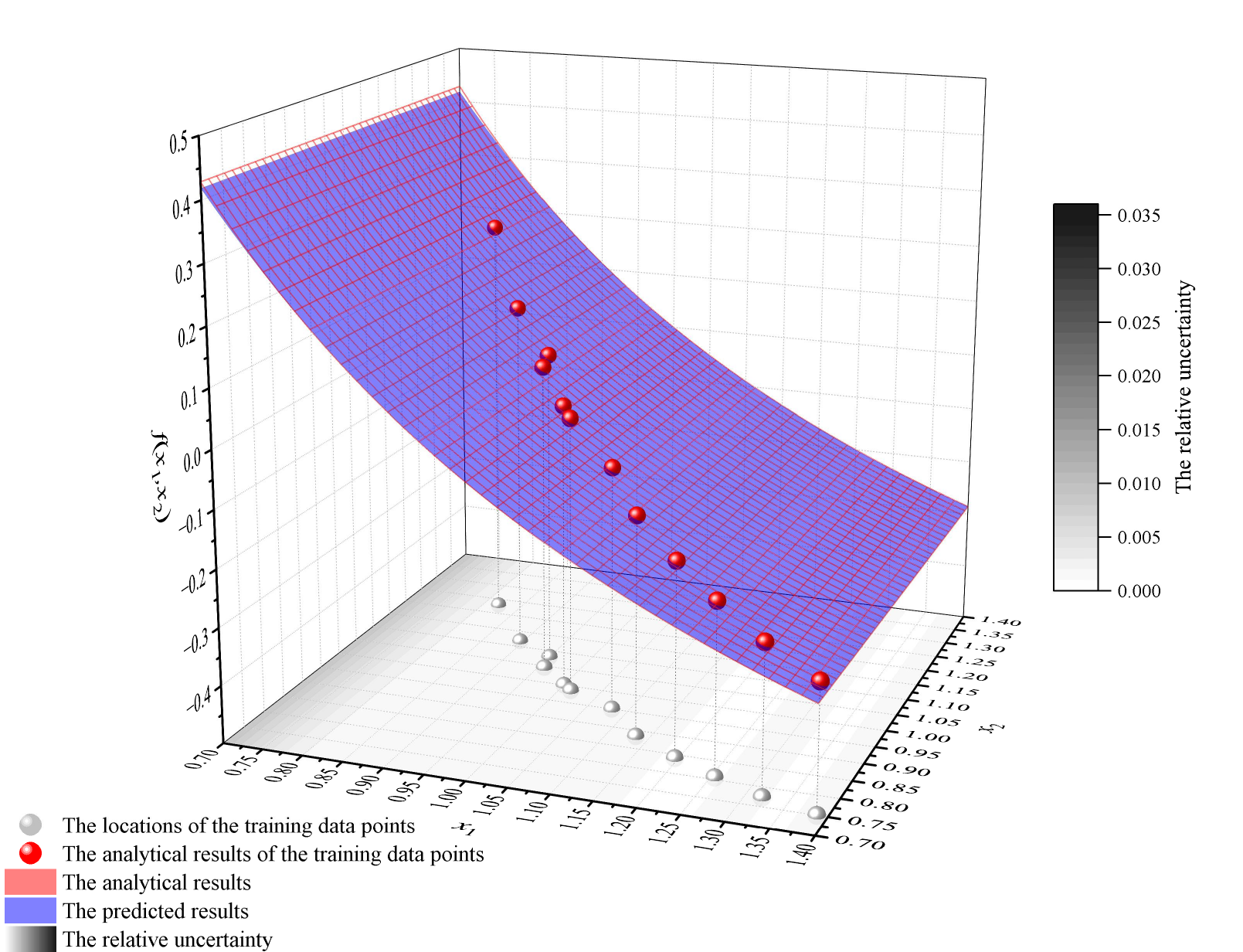}
		\caption{$\delta_{\rm{tol}}=0.005$ with R1}
		\label{fig:Po_al-results2_c}
	\end{subfigure}%
	\caption{The final learned constitutive relations obtained by the ${\rm{G^2ALSPH}}$ method with three different tolerance settings for the Poiseuille flows ($Wi=0.5$): (a) $\delta_{\rm{tol}}=0.05$, (b) $\delta_{\rm{tol}}=0.01$, and (c) $\delta_{\rm{tol}}=0.005$, where three initial local constitutive relations are the same as R1.}
	\label{fig:Po_al-results2}
\end{figure}

Compared to the cases in~\cref{sec4.1.3}, a larger Weissenberg number $Wi=0.5$ is adopted here while three tolerance settings are kept unchanged. The initial local constitutive relations are chosen the same as R1. The similar results as in~\cref{sec4.1.3} can be obtained in~\cref{fig:Po_al-results2}. As the pre-set tolerance decreases, more data points are needed to complete the entire numerical simulation. Compared to the relative uncertainty results in~\cref{fig:Po_al-results2_a}, those in~\cref{fig:Po_al-results2_b} get improved in the region $[0.7,1.4] \times [0.7,1.4]$. Compared to the relative uncertainty results in~\cref{fig:Po_al-results2_b}, those in~\cref{fig:Po_al-results2_c} get better in some specific regions but a little worse in other regions. This is because the different concentration distributions of training data points slightly affect the GPR prediction results. However, throughout the simulations, the learned constitutive relations satisfy the corresponding tolerance limits at every time instant.
\par
\begin{figure}[H]
	\centering
	\begin{subfigure}[b]{0.5\linewidth}
		\includegraphics[width=\linewidth]{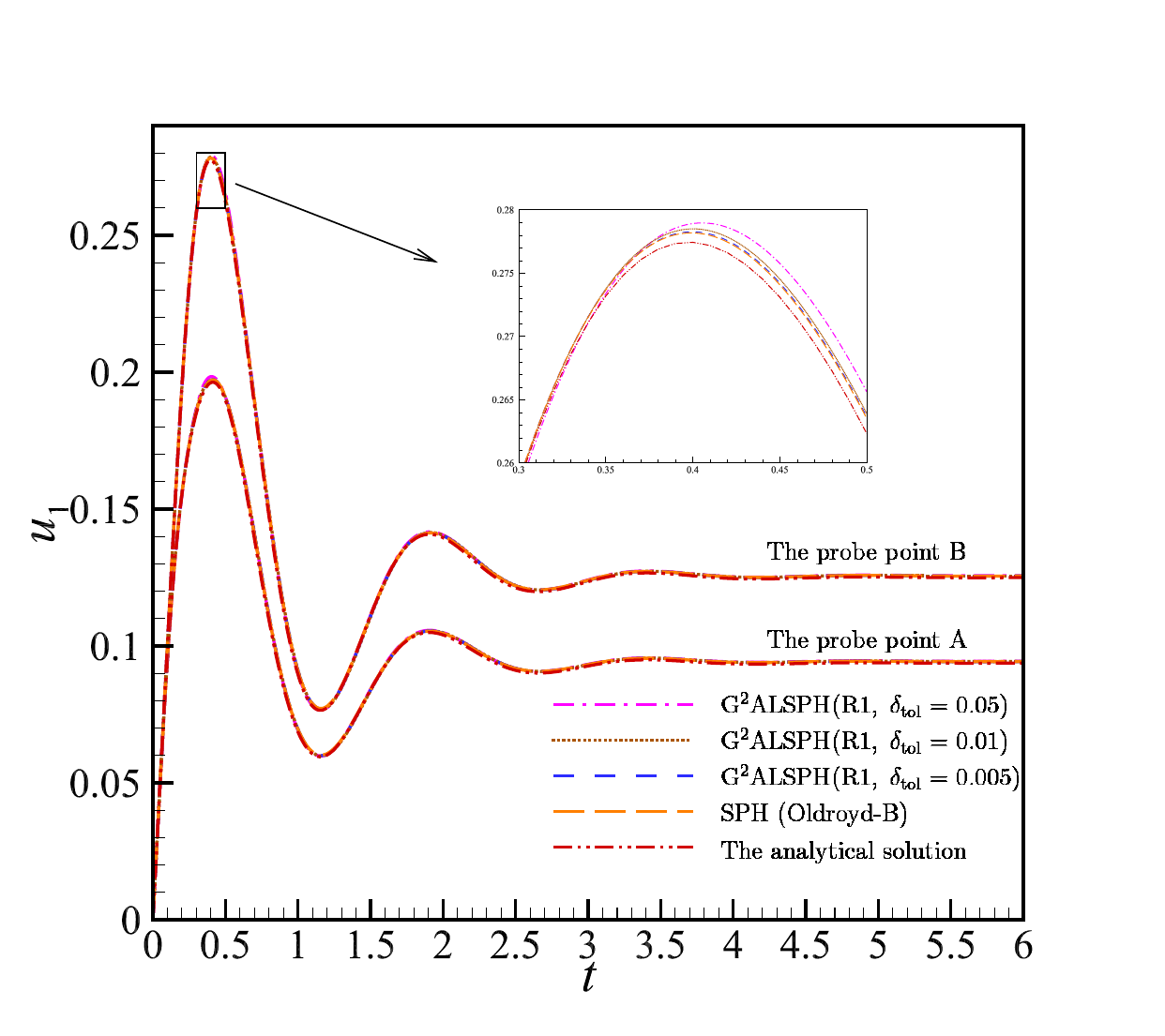}
		\caption{}
		\label{fig:Po2-tol-results_a}
	\end{subfigure}%
	\begin{subfigure}[b]{0.5\linewidth}
		\includegraphics[width=\linewidth]{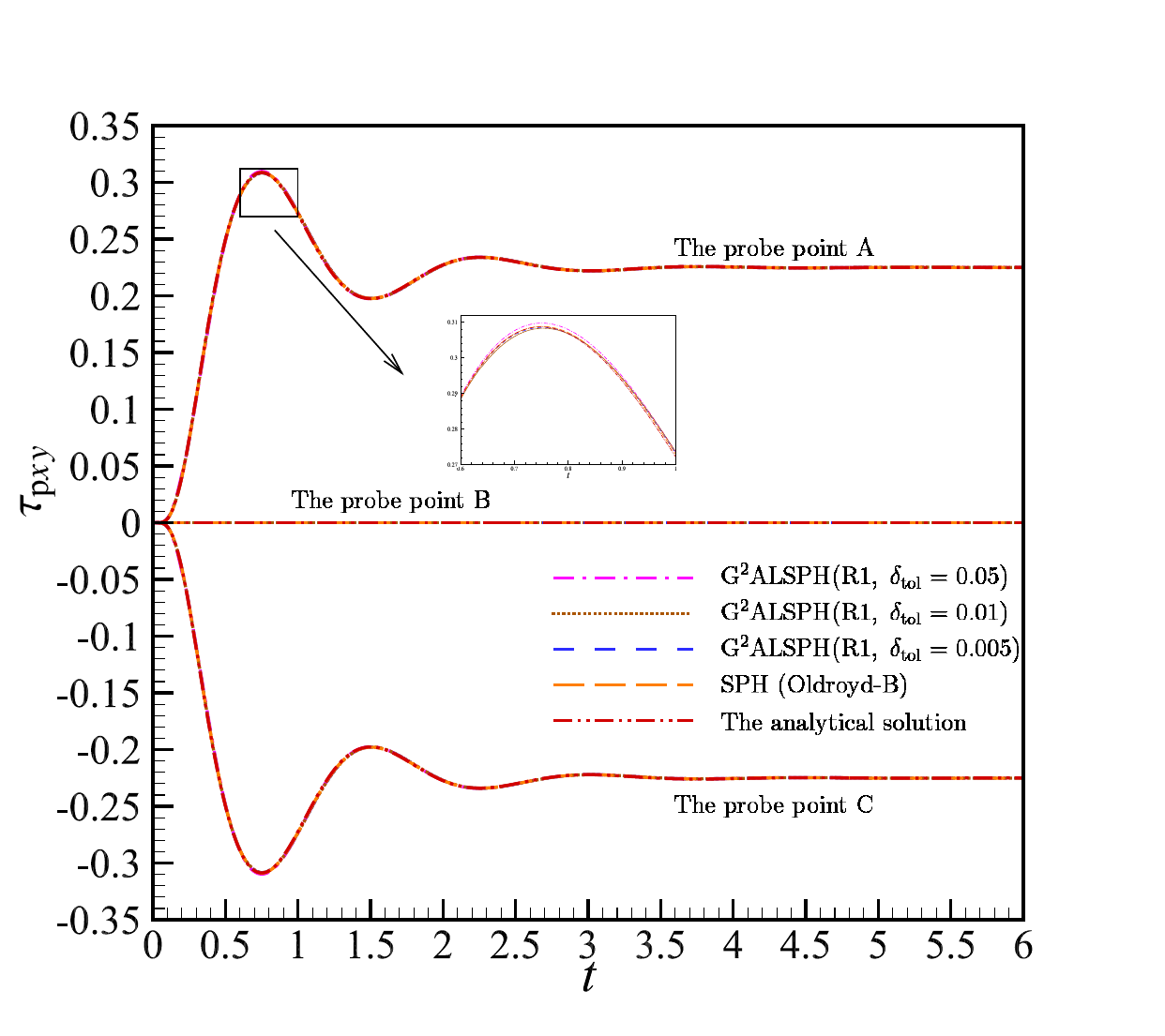}
		\caption{}
		\label{fig:Po2-tol-results_b}
	\end{subfigure}%
	\caption{Comparison among the results in the probe points A, B, and C in the Poiseuille flows ($Wi=0.5$), which are obtained by the ${\rm{G^2ALSPH}}$ method with three different tolerance settings ($\delta_{\rm{tol}}=0.05$, $\delta_{\rm{tol}}=0.01$, and $\delta_{\rm{tol}}=0.005$) and the SPH method with the Oldroyd-B model and the analytical solutions: (a) the variations of the velocity $u_1$ over time and (b) the variations of the shear stress $\tau_{{\rm{p}}xy}$ over time.}
	\label{fig:Po2-tol-results}
\end{figure}

In~\cref{fig:Po2-tol-results}, the variations of the velocity $u_1$ and the shear stress $\tau_{{\rm{p}}xy}$ over time in the probe points A, B, and C in the Poiseuille flows at $Wi=0.5$ are depicted. As shown in~\cref{fig:Po2-tol-results}, as the tolerance decreases, the results of the variations of the velocity $u_1$ and the shear stress $\tau_{{\rm{p}}xy}$ over time obtained by the ${\rm{G^2ALSPH}}$ method converge to the reference analytical results. This demonstrates that as the tolerance decreases, the number of training data points required to complete the entire numerical simulation increases, and the accuracy of the simulation results becomes higher throughout the simulation. 
\subsubsection{The Poiseuille flows with three different initial local constitutive relations ($\mathit{Wi=0.5}$)}\label{sec4.2.2}
In~\cref{sec4.1.2}, three final learned constitutive relations are obtained by the ${\rm{G^2ALSPH}}$ method and denoted as R1, R2, and R3, which are chosen as three different initial local constitutive relations here to study the effects of the choice of the initial local constitutive relation and the applicability of the learned constitutive relation. The pre-set tolerance settings are kept the same as $\delta_{\rm{tol}}=0.01$. 
\par
\begin{figure}[H]
	\centering
	\begin{subfigure}[b]{0.333\linewidth}
		\includegraphics[width=\linewidth]{figures/GPRTWOIN504_372.pdf}
		\caption{$\delta_{\rm{tol}}=0.01$ with R1}
		\label{fig:Po2_al-results2_a}
	\end{subfigure}%
	\begin{subfigure}[b]{0.333\linewidth}
		\includegraphics[width=\linewidth]{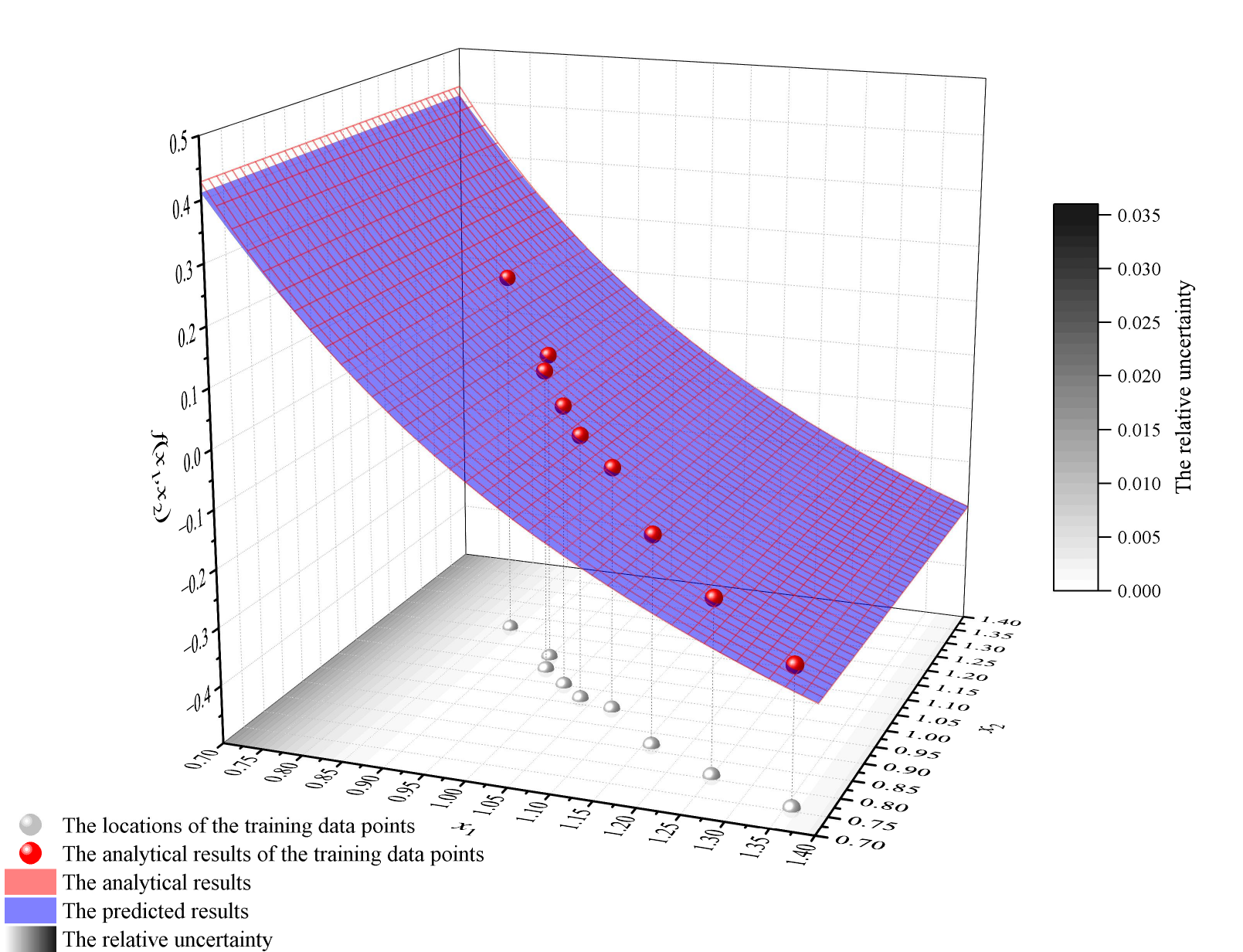}
		\caption{$\delta_{\rm{tol}}=0.01$ with R2}
		\label{fig:Po2_al-results2_b}
	\end{subfigure}%
	\begin{subfigure}[b]{0.333\linewidth}
		\includegraphics[width=\linewidth]{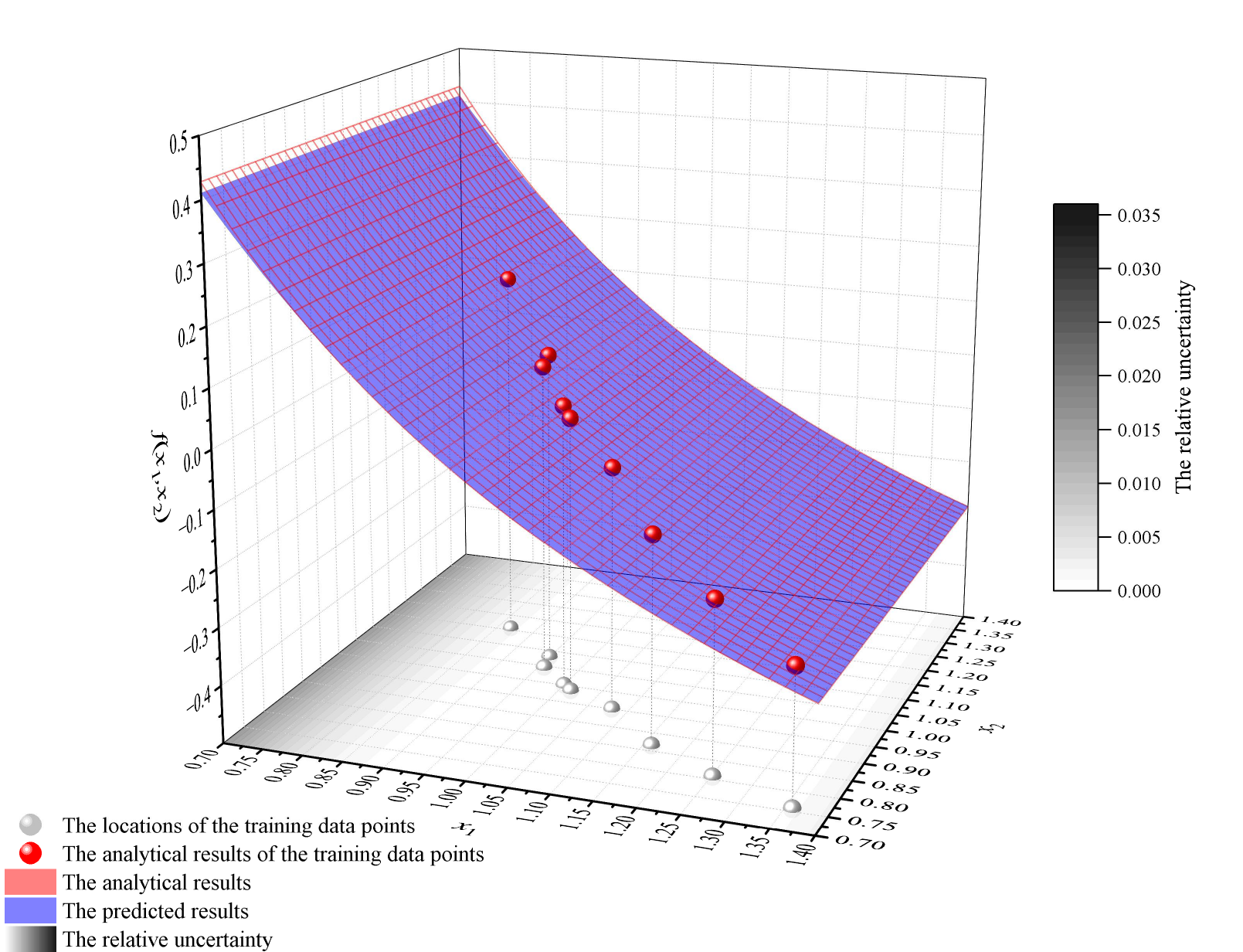}
		\caption{$\delta_{\rm{tol}}=0.01$ with R3}
		\label{fig:Po2_al-results2_c}
	\end{subfigure}%
	\caption{The final learned constitutive relations obtained by the ${\rm{G^2ALSPH}}$ method with three different initial local constitutive relation settings for the Poiseuille flows ($Wi=0.5$): (a) R1 serving as the initial local constitutive relation, (b) R2 serving as the initial local constitutive relation, and (c) R3 serving as the initial local constitutive relation, where three tolerance settings are the same as $\delta_{\rm{tol}}=0.01$.}
	\label{fig:Po2_al-results2}
\end{figure}

Three final learned constitutive relations are presented in~\cref{fig:Po2_al-results2}. Although the initial local constitutive relations are different, the same tolerance $\delta_{\rm{tol}}=0.01$ leads to similar final learned constitutive relations, the accuracy of which can be guaranteed in the region $[0.7,1.4] \times [0.7,1.4]$. This proves that the initial learned constitutive relation has only small effects on the final learned constitutive relation, and it can be safely applied to new numerical simulations. The application range of the learned constitutive relation is not a limitation, since the active learning strategy automatically ensures that the learned constitutive relation meets the required accuracy everywhere and at every time instant.
\par
\begin{figure}[H]
	\centering
	\begin{subfigure}[b]{0.5\linewidth}
		\includegraphics[width=\linewidth]{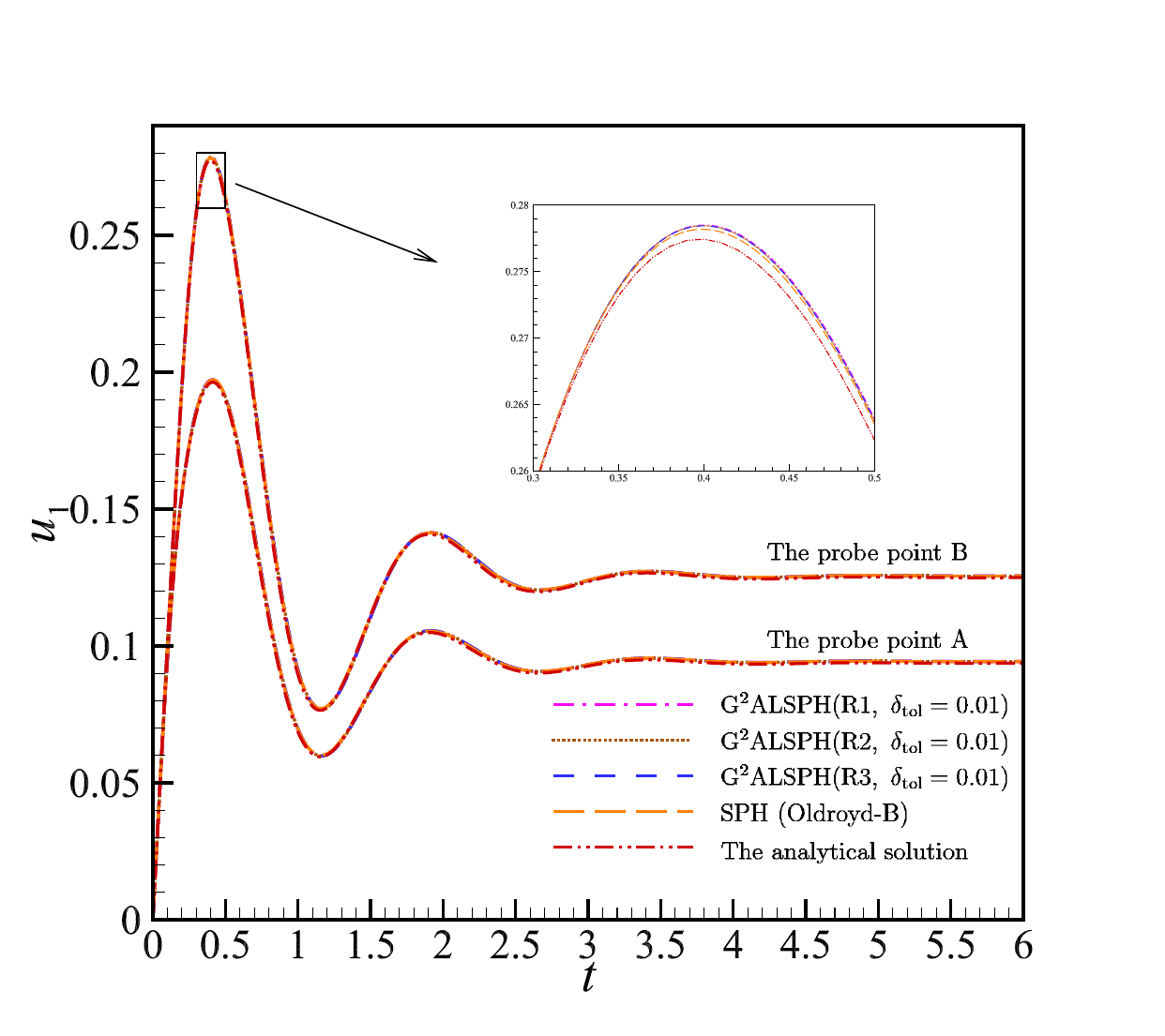}
		\caption{}
		\label{fig:Po2-results1_a}
	\end{subfigure}%
	\begin{subfigure}[b]{0.5\linewidth}
		\includegraphics[width=\linewidth]{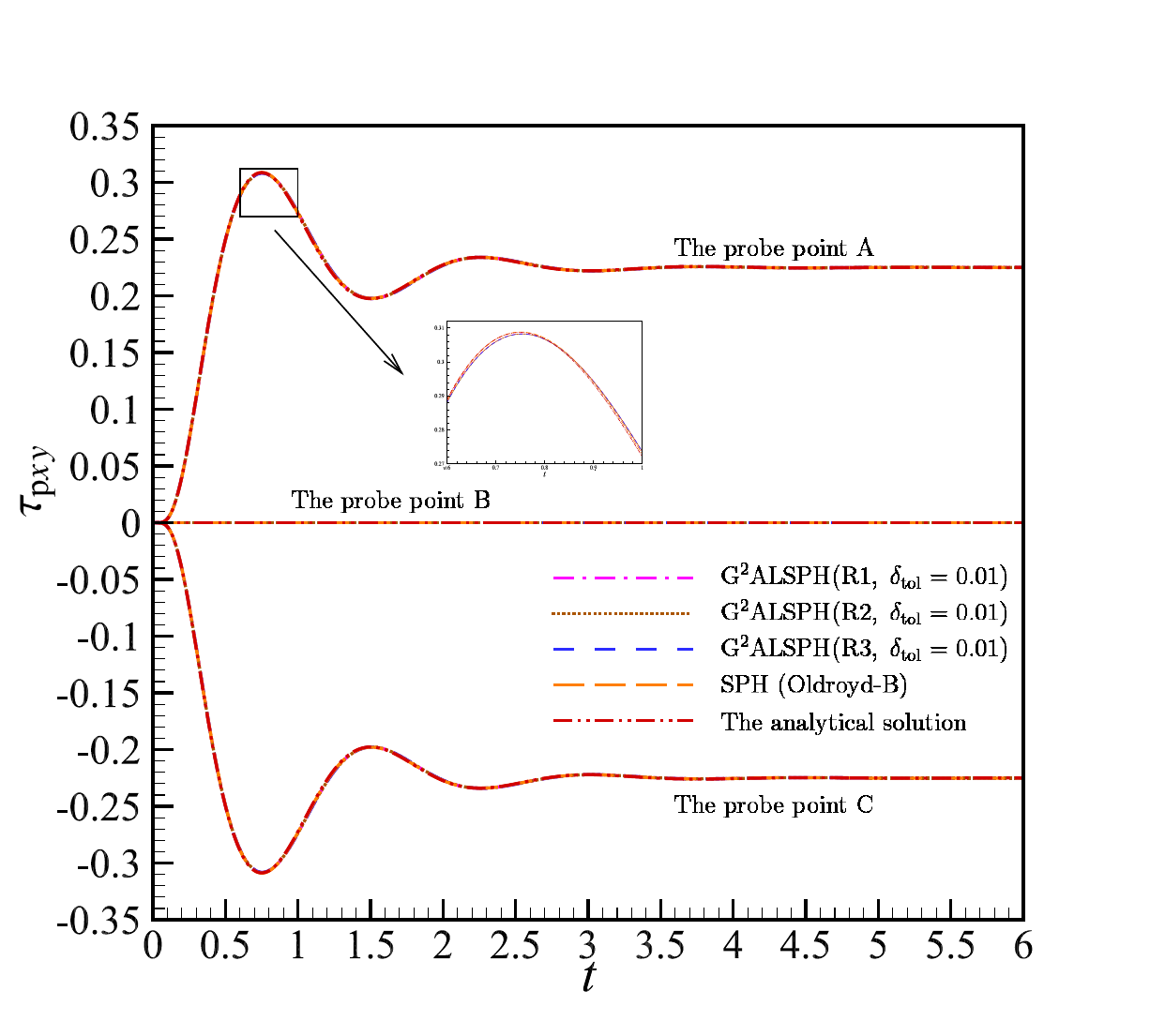}
		\caption{}
		\label{fig:Po2-results1_b}
	\end{subfigure}%
	\caption{Comparison among the results in the probe points A, B, and C in the Poiseuille flows ($Wi=0.5$), which are obtained by the ${\rm{G^2ALSPH}}$ method with three different initial local constitutive relation settings (R1, R2, and R3) and the SPH method with the Oldroyd-B model and the analytical solutions: (a) the variations of the velocity $u_1$ over time and (b) the variations of the shear stress $\tau_{{\rm{p}}xy}$ over time.}
	\label{fig:Po2-results1}
\end{figure}

Then we study the simulation results. The variations of the velocity $u_1$ and the shear stress $\tau_{{\rm{p}}xy}$ over time in the probe points A, B, and C are presented in~\cref{fig:Po2-results1}. The ${\rm{G^2ALSPH}}$ method, when employed with three different initial local constitutive relations, yields results that are in excellent agreement with the reference analytical results and exhibit a consistently high level of accuracy. \cref{fig:Po2-results1} thus shows that the choice of the initial constitutive relation hardly affects the accuracy of the ${\rm{G^2ALSPH}}$ simulation results, which is uniquely determined by the pre-set tolerance.
\par
From the results in~\cref{sec4.2.1,sec4.2.2}, it can be drawn that continuing to decrease the tolerance does not significantly increase the accuracy further, but it increases the number of required training data points. For the optimal setting of the pre-set tolerance, since the analytical solutions of a numerical simulation are not pre-known in real applications, it is recommended to conduct two or more numerical experiments at different pre-set tolerance settings and compare their results to determine an appropriate setting of the pre-set tolerance. Thanks to the convergence of the ${\rm{G^2ALSPH}}$ method, it is also possible to compare the final simulation results obtained at different tolerance settings to determine whether the final simulation results are acceptable or not.
\subsubsection{The superiority of the relative uncertainty strategy over the direct uncertainty strategy}\label{sec4.2.3}
The active learning strategies based on the relative and direct uncertainties are denoted as the relative and direct uncertainty strategies, respectively, which are compared in this subsection. The Poiseuille flows at $Wi=0.1$, $Wi=0.5$, and $Wi=1.0$ are simulated by the ${\rm{G^2ALSPH}}$ method with the relative and direct uncertainty strategies, separately. The initial local constitutive relations are kept the same as R1. Except for the accuracy evaluation tools, i.e., the relative uncertainty and the direct uncertainty, in the active learning strategies and the corresponding tolerance settings, the other physical parameters and model settings are the same for the cases with the same Weissenberg number.
\par
In the cases with the direct uncertainty strategy, the $1.96E_{\rm{s}}$ serves as the accuracy evaluation tool, and the tolerance is set as $\delta^{\prime}_{\rm{tol}}=0.002$. For the cases with relative uncertainty strategy, the variable $U_{\rm{r}}$ defined in~\cref{eq:relative_uncertainty} serves as the accuracy evaluation tool, and the tolerance is set as $\delta_{\rm{tol}}=0.01$. The value of $\varepsilon$ in~\cref{eq:relative_uncertainty} is set to be $0.2$ in this work, and actually $\delta^{\prime}_{\rm{tol}}=0.002$ is a restriction on the numerator of the right part in~\cref{eq:relative_uncertainty}. The value of the predicted mean $\bar{y}_*$ in the corresponding denominator can be $0$. The value $0.002$ divided by $0.2$ equals $0.01$. Thus, the tolerance $\delta_{\rm{tol}}=0.01$ is derived, which plays the equivalent role in the relative uncertainty strategy as the tolerance $\delta^{\prime}_{\rm{tol}}=0.002$ does in the direct uncertainty strategy. The values of $1.96E_{\rm{s}}$ need to be kept strictly less than $0.002$ throughout the numerical simulations with the direct uncertainty strategy. On the contrary, for the numerical simulations with the relative uncertainty strategy, the limits on the values of $1.96E_{\rm{s}}$ can be more lenient where the values predicted mean $\bar{y}_*$ are not equal to $0$. Therefore, the tolerance $\delta^{\prime}_{\rm{tol}}=0.002$ in the direct uncertainty strategy becomes the stricter one in real applications.
\par
\begin{figure}[H]
	\centering
	\begin{minipage}[t]{0.9\linewidth}
		\centering
		\raisebox{2.5\height}{\rotatebox{90}{\scriptsize{$Wi=0.1$}}}
		\begin{subfigure}[b]{0.49\linewidth}
			\raisebox{1.5\height}{\rotatebox{0}{\scriptsize{The relative uncertainty strategy}}}
			\centering
			\includegraphics[width=\linewidth]{figures/GPRTWOIN502_366.pdf}
			\label{fig:Po_constitutive_a1}
		\end{subfigure}%
		\hfill
		\begin{subfigure}[b]{0.49\linewidth}
			\raisebox{1.5\height}{\rotatebox{0}{\scriptsize{The direct uncertainty strategy}}}
			\centering
			\includegraphics[width=\linewidth]{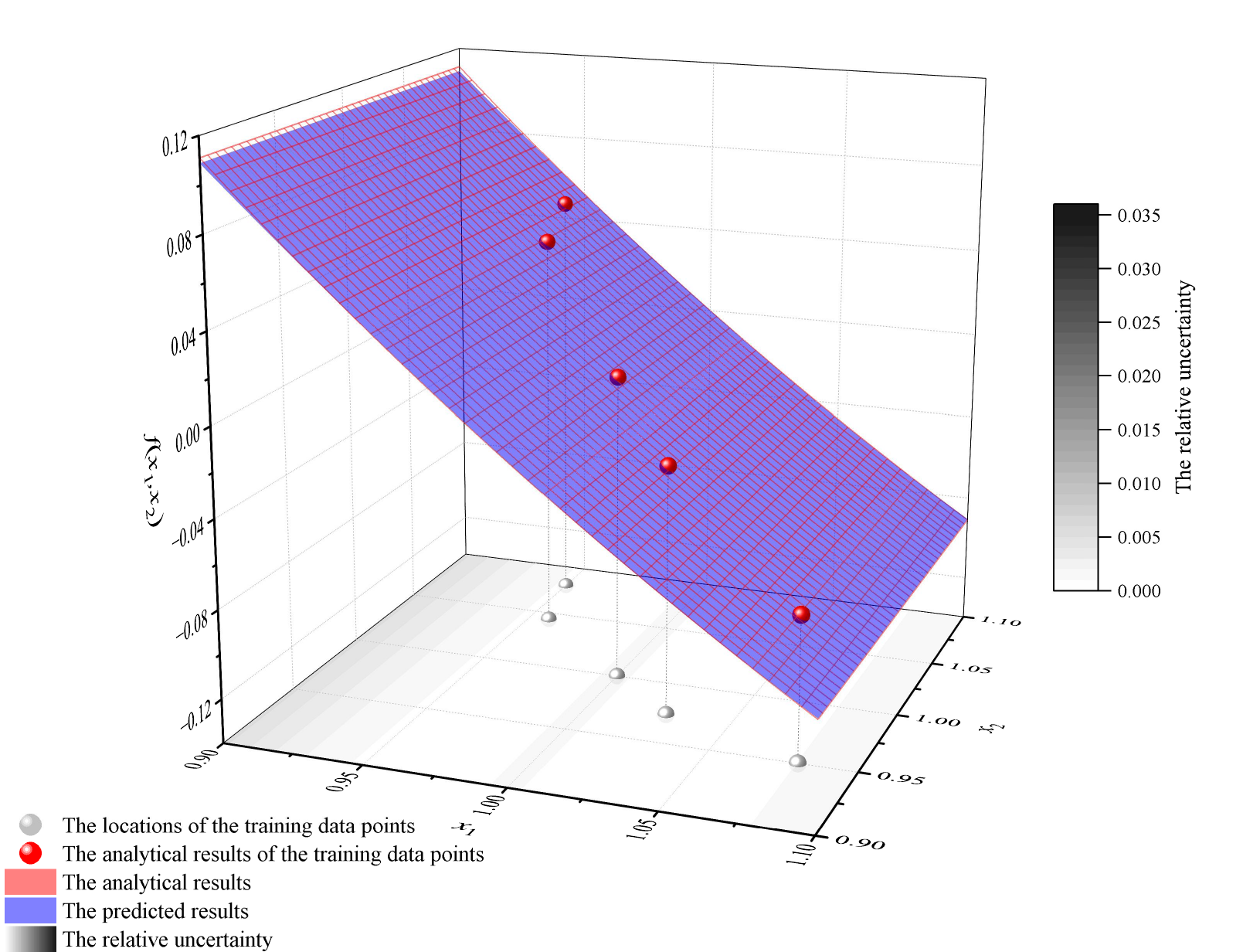}
			\label{fig:Po_constitutive_b1}
		\end{subfigure}%
	\end{minipage}
	\vspace{-6mm}
	\vfill
	\centering
	\raisebox{2.5\height}{\rotatebox{90}{\scriptsize{$Wi=0.5$}}}
	\begin{minipage}[t]{0.9\linewidth}
		\centering
		\begin{subfigure}[b]{0.49\linewidth}
			\centering
			\includegraphics[width=\linewidth]{figures/GPRTWOIN504_372.pdf}
			\label{fig:Po_constitutive_a2}
		\end{subfigure}%
		\hfill
		\begin{subfigure}[b]{0.49\linewidth}
			\centering
			\includegraphics[width=\linewidth]{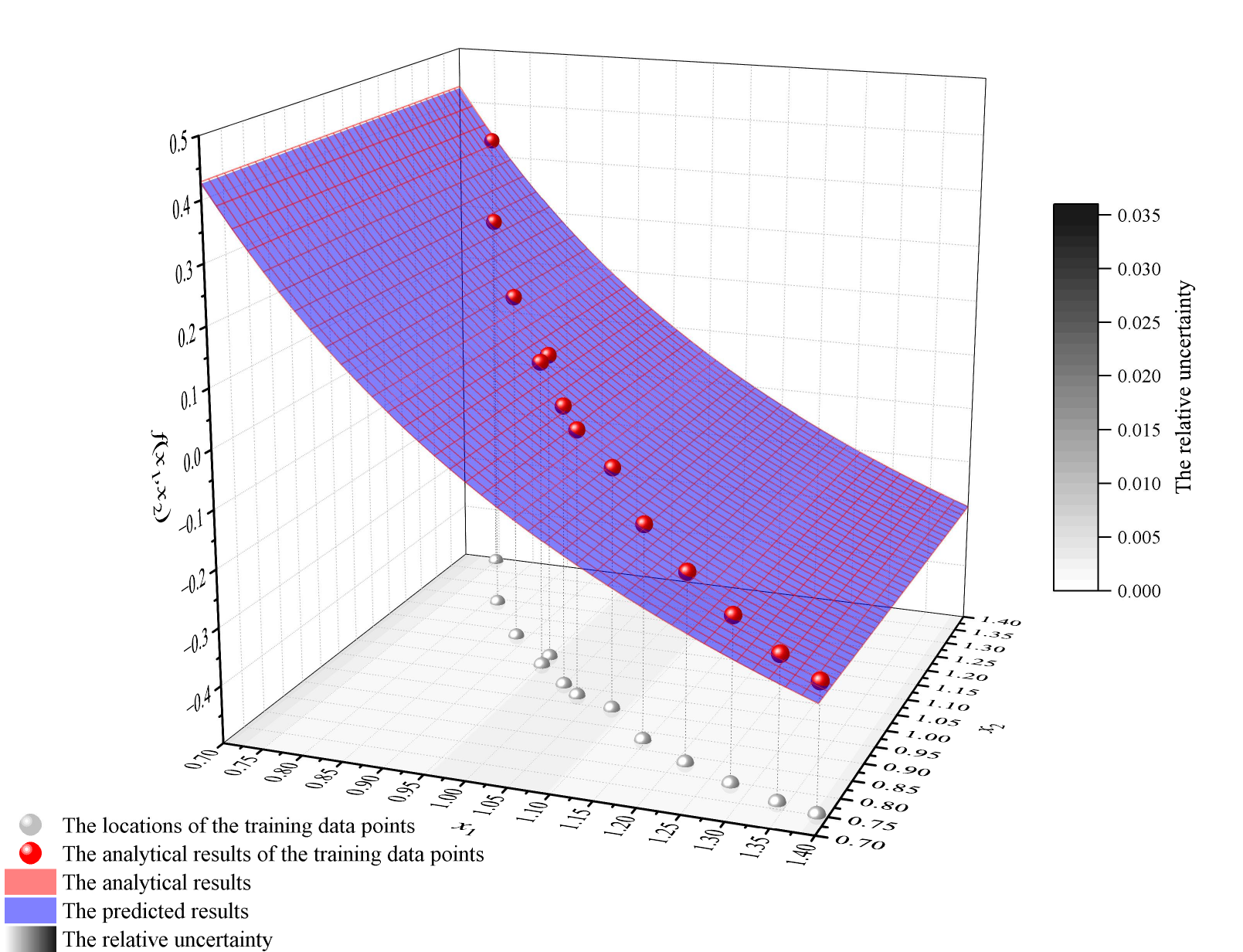}
			\label{fig:Po_constitutive_b2}
		\end{subfigure}%
	\end{minipage}
	\vspace{-6mm}
	\vfill
	\centering
	\raisebox{2.5\height}{\rotatebox{90}{\scriptsize{$Wi=1.0$}}}
	\begin{minipage}[t]{0.9\linewidth}
		\centering
		\begin{subfigure}[b]{0.49\linewidth}
			\centering
			\includegraphics[width=\linewidth]{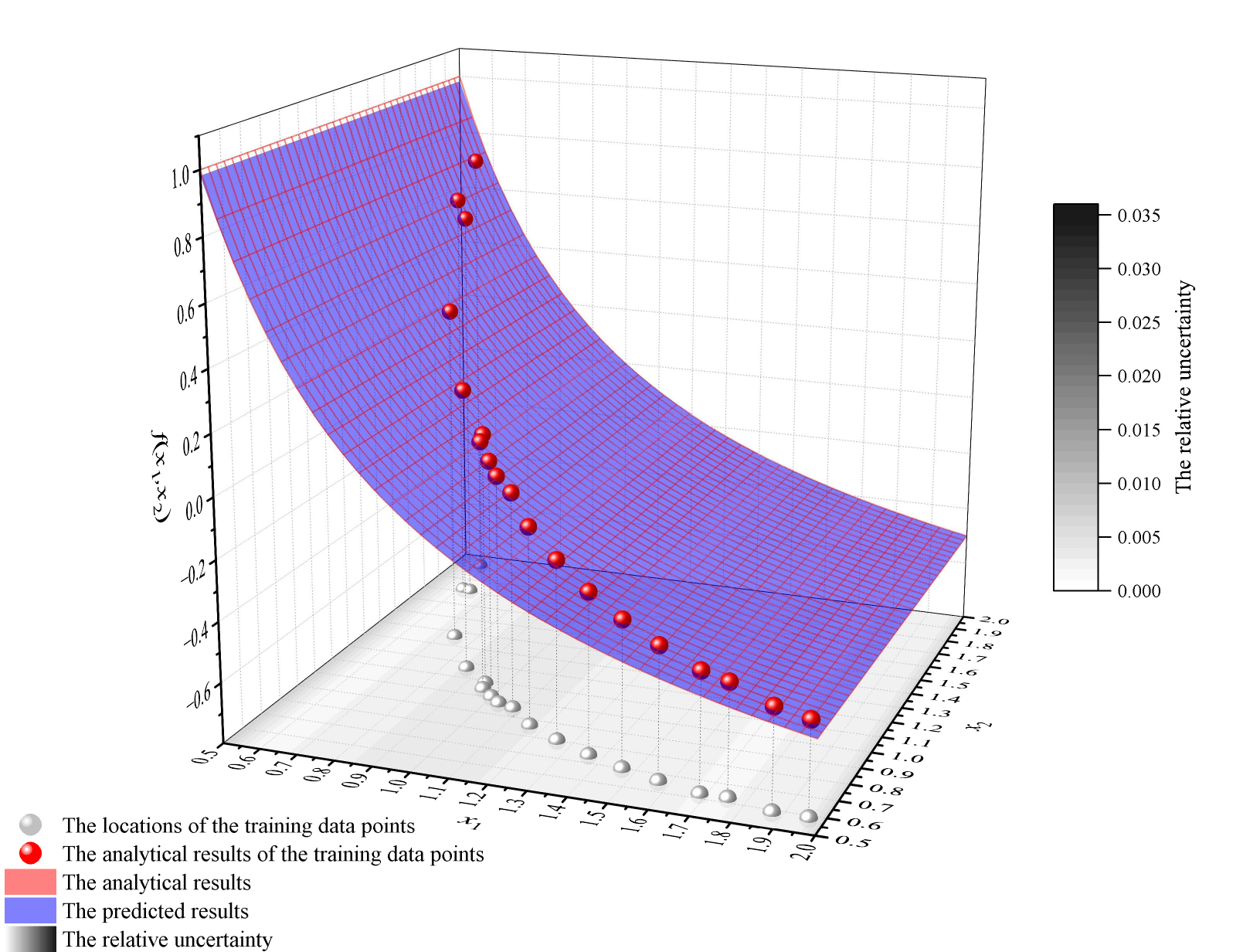}
			\caption{$\delta_{\rm{tol}}=0.01$ with R1}
			\label{fig:Po_constitutive_a3}
		\end{subfigure}%
		\hfill
		\begin{subfigure}[b]{0.49\linewidth}
			\centering
			\includegraphics[width=\linewidth]{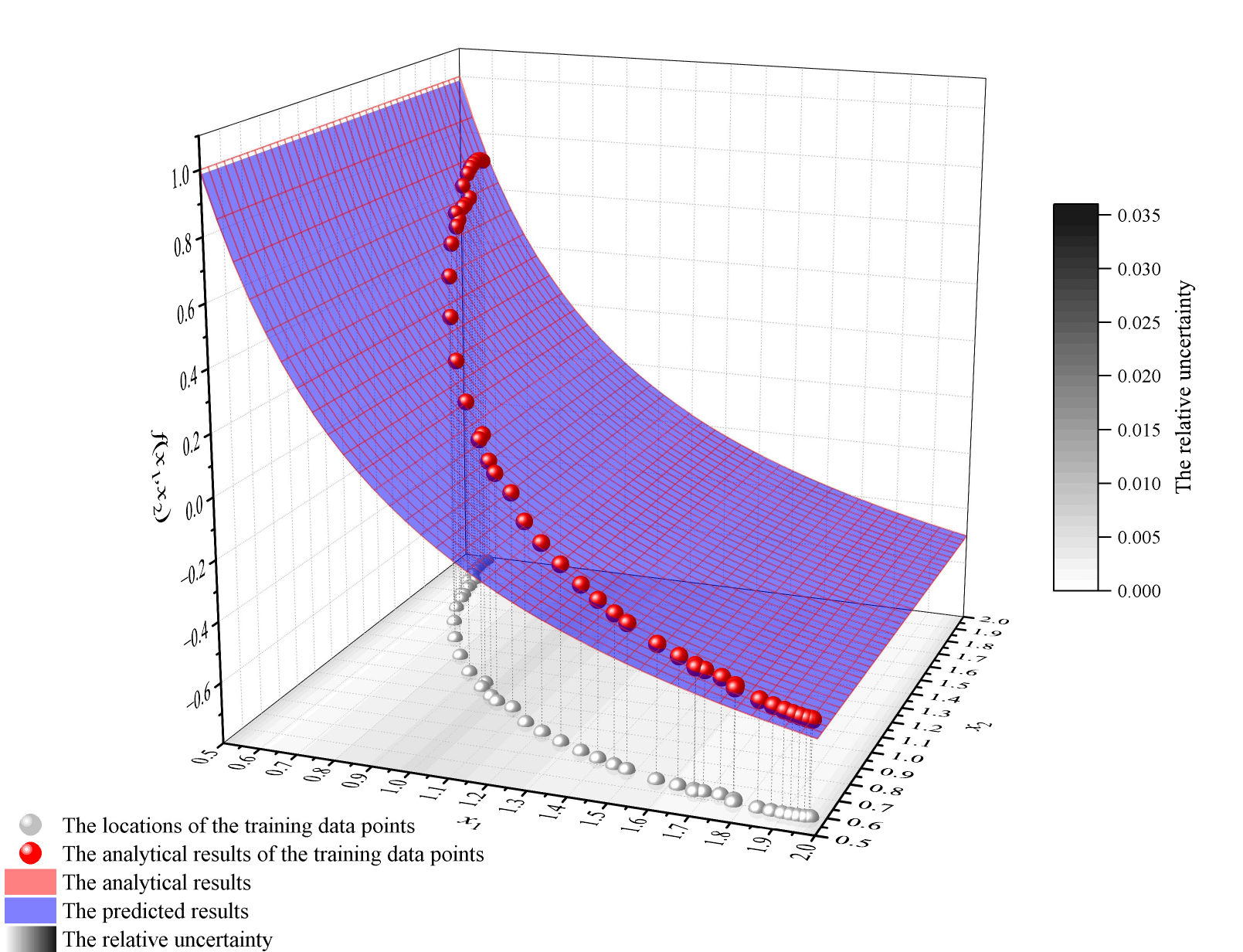}
			\caption{$\delta^{\prime}_{\rm{tol}}=0.002$ with R1}
			\label{fig:Po_constitutiven_b3}
		\end{subfigure}%
	\end{minipage}
	\vspace{-1mm}
	\captionsetup{justification=justified}
	\caption{The final learned constitutive relations obtained by the ${\rm{G^2ALSPH}}$ method for the Poiseuille flows ($Wi=0.1$, $Wi=0.5$, and $Wi=1.0$): (a) the relative uncertainty strategy ($\delta_{\rm{tol}}=0.01$ with R1 serving as the initial local constitutive relation) and (b) the direct uncertainty strategy ($\delta^{\prime}_{\rm{tol}}=0.002$ with R1 serving as the initial local constitutive relation), where $\delta_{\rm{tol}}=0.01$ plays the equivalent role in the relative uncertainty strategy as $\delta^{\prime}_{\rm{tol}}=0.002$ does in the direct uncertainty strategy.}
	\label{fig:Po_constitutive}
\end{figure}

The final learned constitutive relations are first studied, which are presented in~\cref{fig:Po_constitutive}. In order to be able to intuitively compare the final learned constitutive relations, the uncertainty assessments in~\cref{fig:Po_constitutive} are unified to the relative uncertainty results. As shown in~\cref{fig:Po_constitutive}, the direct uncertainty strategy yields better results in some regions but slightly worse in others compared to the relative uncertainty strategy. This is due to the varying concentration distributions of training data points affecting GPR prediction results. 
\par
\begin{figure}[H]
	\centering
	\begin{minipage}[t]{0.9\linewidth}
		\centering
		\raisebox{2.9\height}{\rotatebox{90}{\scriptsize{$Wi=0.1$}}}
		\begin{subfigure}[b]{0.49\linewidth}
			\raisebox{0.1\height}{\rotatebox{0}{\scriptsize{The velocity variations}}}
			\centering
			\includegraphics[width=\linewidth]{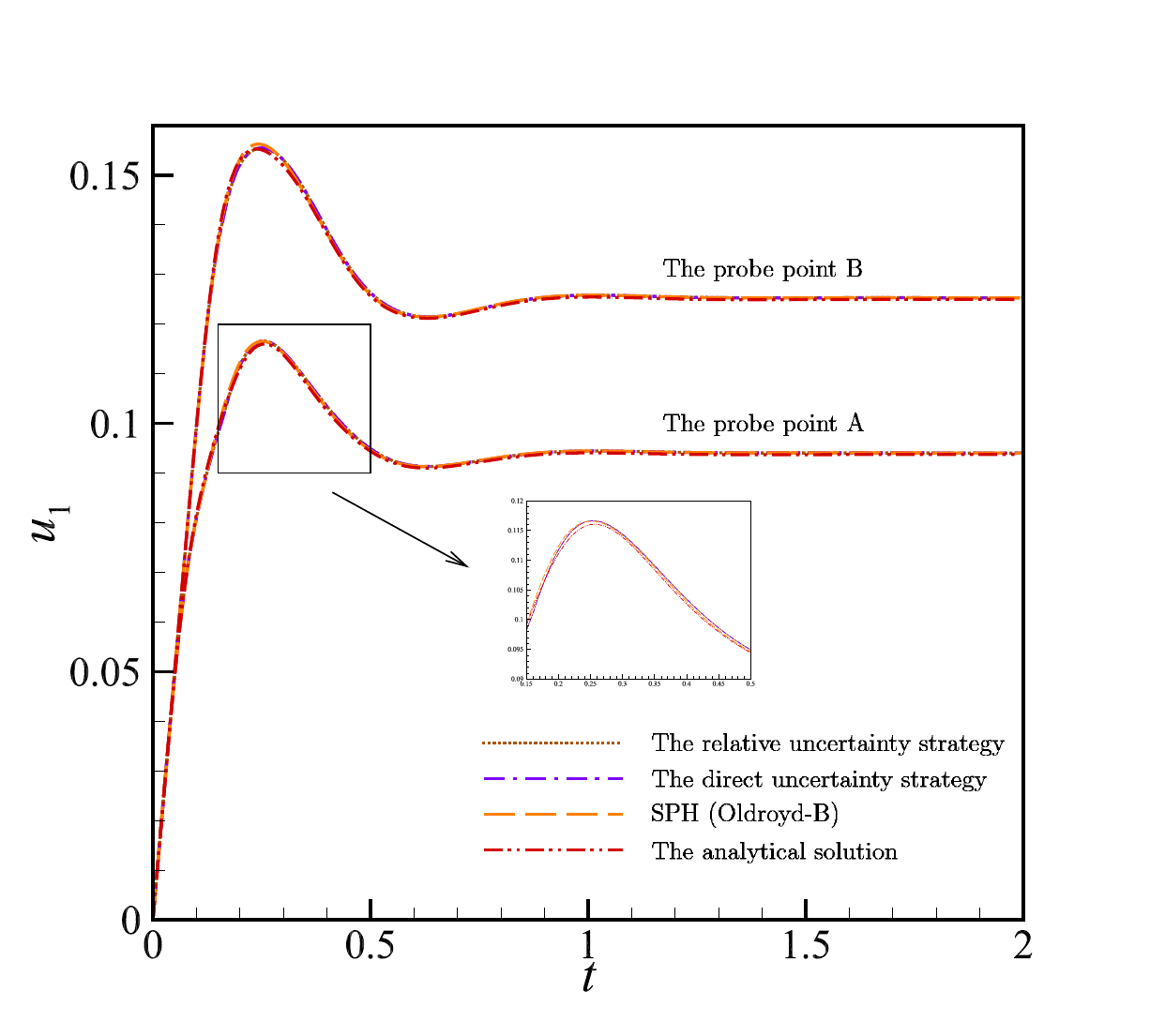}
			\label{fig:Po_results_a1}
		\end{subfigure}%
		\hfill
		\begin{subfigure}[b]{0.49\linewidth}
			\raisebox{0.1\height}{\rotatebox{0}{\scriptsize{The shear stress variations}}}
			\centering
			\includegraphics[width=\linewidth]{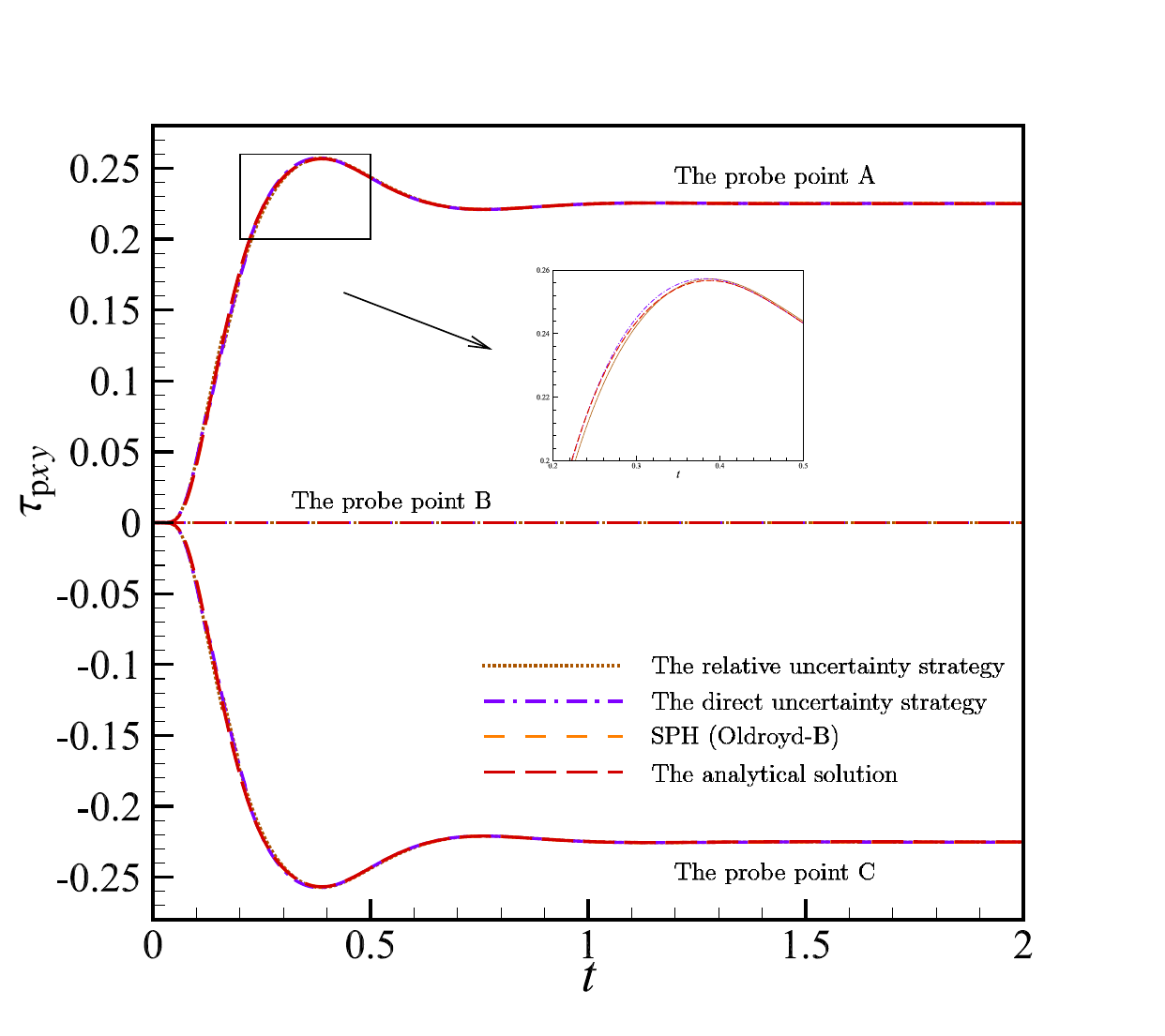}
			\label{fig:Po_results_b1}
		\end{subfigure}%
	\end{minipage}
	\vspace{-6mm}
	\vfill
	\centering
	\raisebox{2.9\height}{\rotatebox{90}{\scriptsize{$Wi=0.5$}}}
	\begin{minipage}[t]{0.9\linewidth}
		\centering
		\begin{subfigure}[b]{0.49\linewidth}
			\centering
			\includegraphics[width=\linewidth]{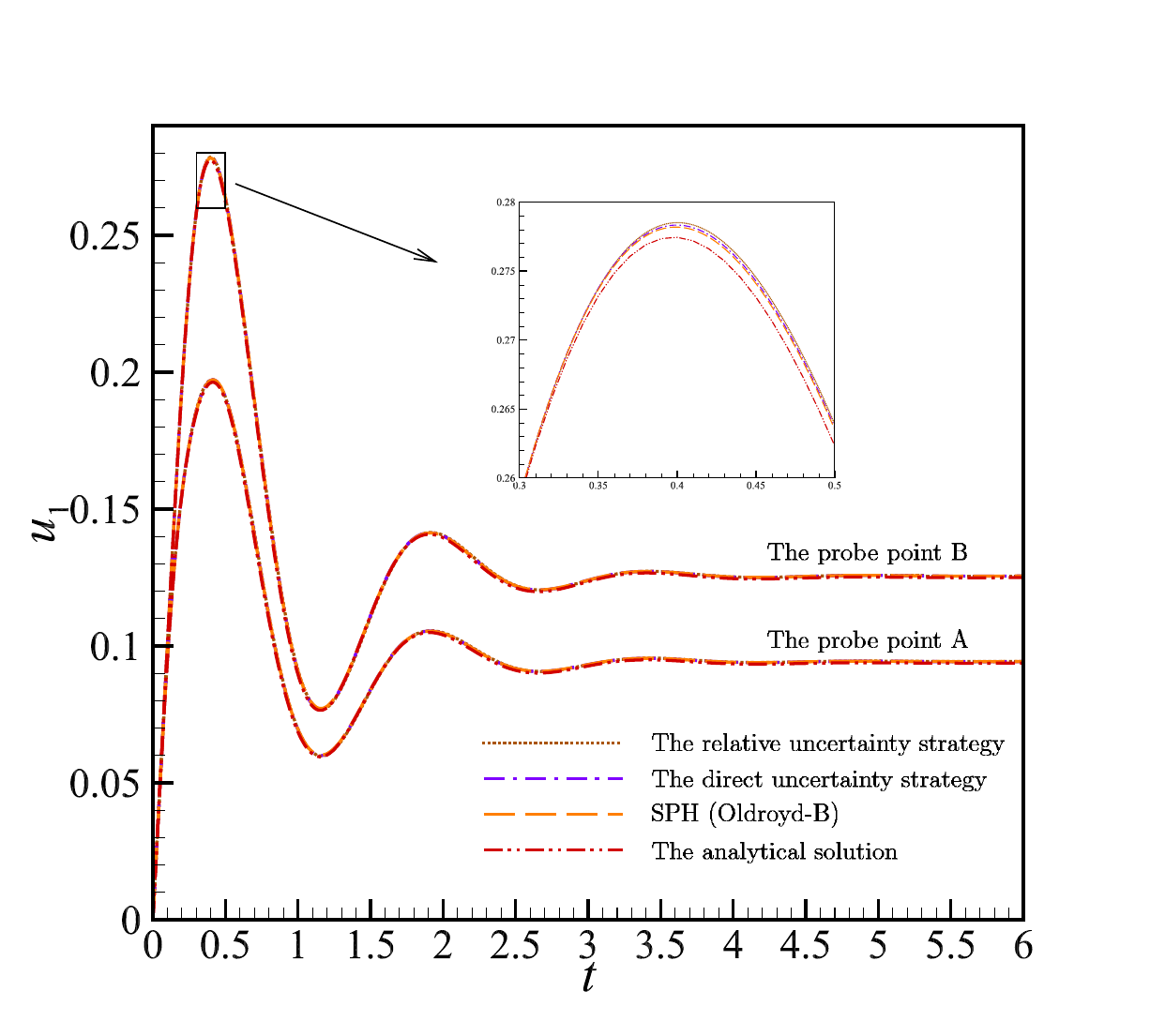}
			\label{fig:Po_results_a2}
		\end{subfigure}%
		\hfill
		\begin{subfigure}[b]{0.49\linewidth}
			\centering
			\includegraphics[width=\linewidth]{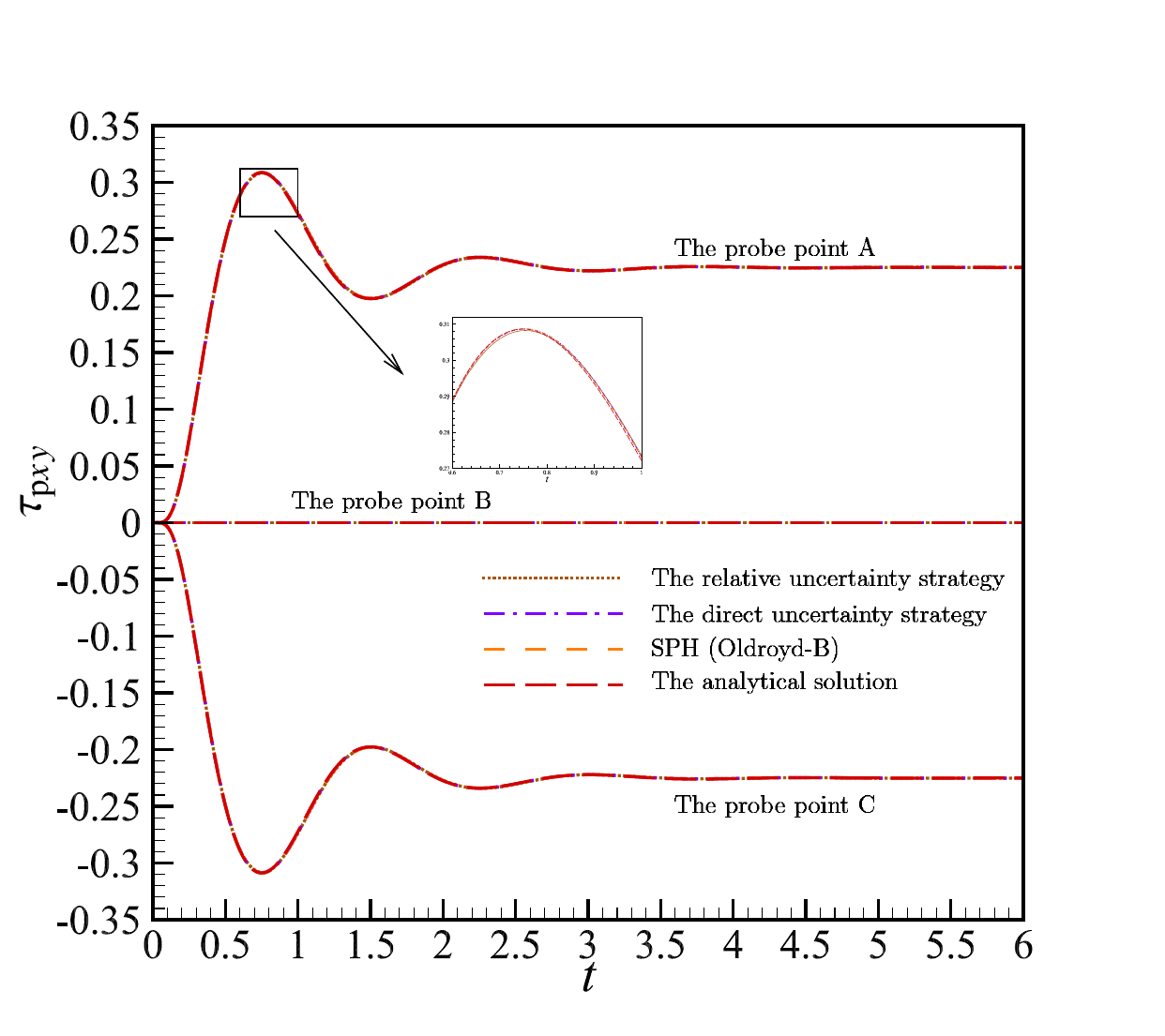}
			\label{fig:Po_results_b2}
		\end{subfigure}%
	\end{minipage}
	\vspace{-6mm}
	\vfill
	\centering
	\raisebox{2.9\height}{\rotatebox{90}{\scriptsize{$Wi=1.0$}}}
	\begin{minipage}[t]{0.9\linewidth}
		\centering
		\begin{subfigure}[b]{0.49\linewidth}
			\centering
			\includegraphics[width=\linewidth]{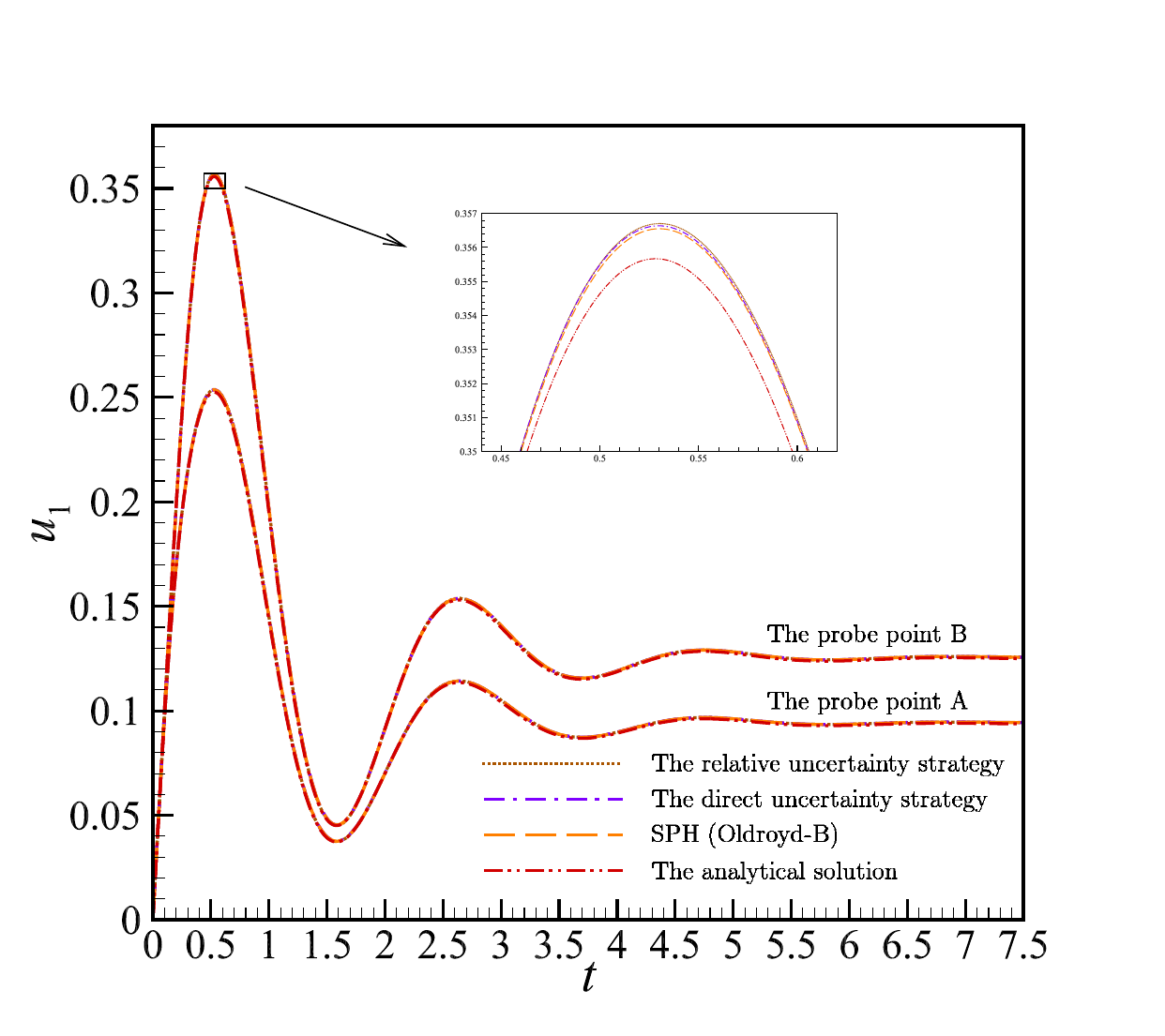}
			\caption{}
			\label{fig:Po_results_a3}
		\end{subfigure}%
		\hfill
		\begin{subfigure}[b]{0.49\linewidth}
			\centering
			\includegraphics[width=\linewidth]{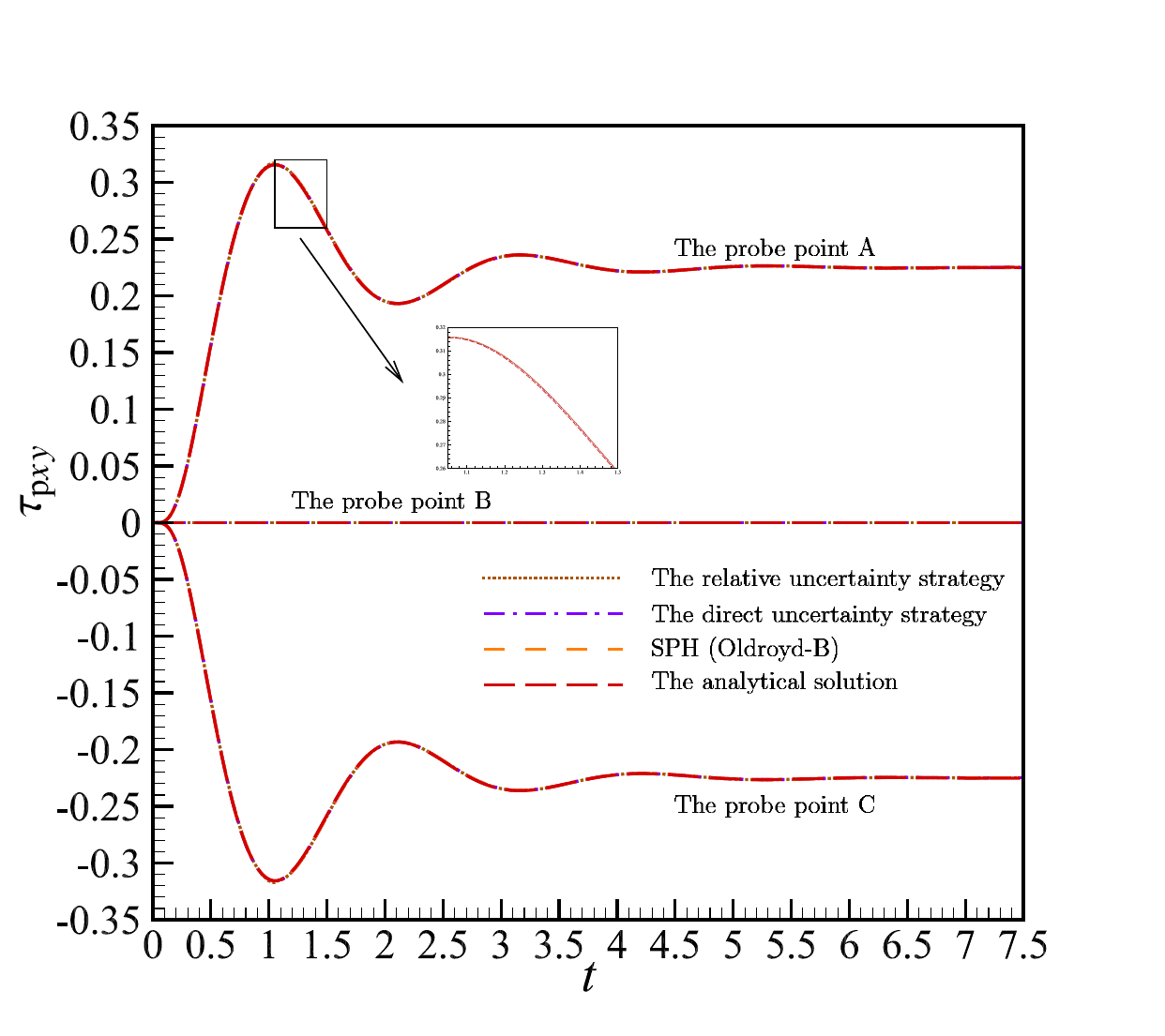}
			\caption{}
			\label{fig:Po_results_b3}
		\end{subfigure}%
	\end{minipage}
	\vspace{-1mm}
	\captionsetup{justification=justified}
	\caption{Comparison among the results in the probe points A, B, and C in the Poiseuille flows ($Wi=0.1$, $Wi=0.5$, and $Wi=1.0$), which are obtained by the ${\rm{G^2ALSPH}}$ method with relative and direct uncertainty strategies and the SPH method with the Oldroyd-B model and the analytical solutions: (a) the variations of the velocity $u_1$ over time and (b) the variations of the shear stress $\tau_{{\rm{p}}xy}$ over time.}
	\label{fig:Po_results}
\end{figure}

The simulation results are then studied. In~\cref{fig:Po_results}, the results of the variations of the velocity $u_1$ and the shear stress $\tau_{{\rm{p}}xy}$ over time in the probe points A, B, and C are shown. Compared with the results obtained under the relative uncertainty strategy with $\delta_{\rm{tol}}=0.01$, the results obtained under the direct uncertainty strategy with $\delta^{\prime}_{\rm{tol}}=0.002$ are closer to the reference analytical results. This is because of the stricter tolerance limit. However, the results obtained under the relative and direct uncertainty strategies are very accurate and agree quite well with each other, which demonstrates that the relative uncertainty strategy can achieve the same accuracy as the direct uncertainty strategy at the equivalent level of tolerance.
\par
\begin{figure}[H]
	\centering		
	\includegraphics[width=9cm]{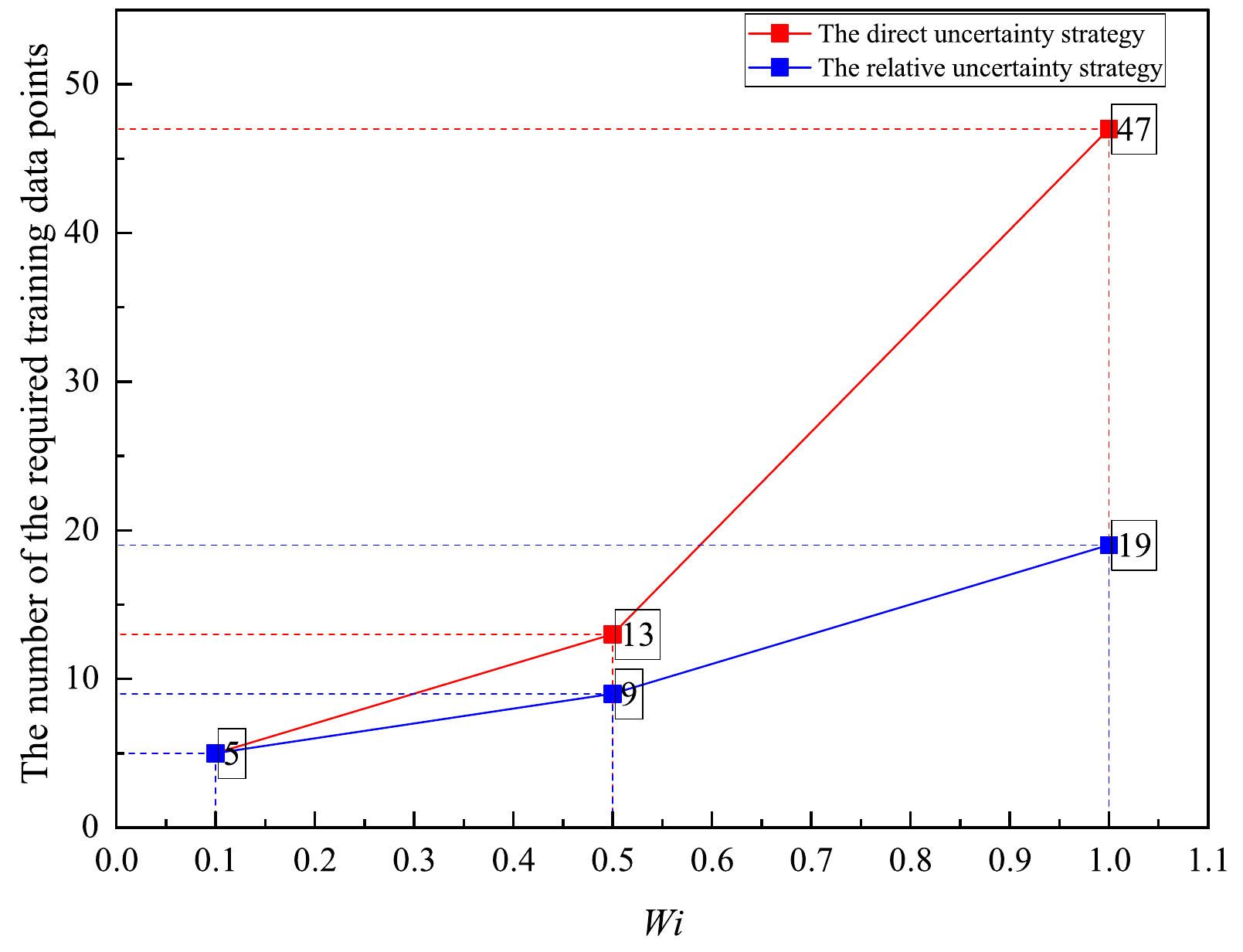}
	\caption{Comparison of the numbers of the training data points in the final learned constitutive relations obtained by the relative and direct uncertainty strategies in the Poiseuille flows ($Wi=0.1$, $Wi=0.5$, and $Wi=1.0$).}
	\label{fig:number1}
\end{figure}

In~\cref{fig:number1}, the comparison of the numbers of the training data points in the final learned constitutive relations is depicted. As the Weissenberg number increases, the region covered by the eigenvalues expands, and the accuracy of the learned constitutive relations needs to be guaranteed gradually in larger regions. As a consequence, the number of training data points in the final learned constitutive relation increases sharply for the direct uncertainty strategy but only gently for the relative uncertainty strategy.
\par
To sum up, the above results demonstrate that the relative uncertainty strategy can reduce the number of required training data points while maintaining the same accuracy during the simulation compared to the direct uncertainty strategy at the equivalent level of tolerance.
\subsubsection{Analysis of the space explored by the eigenvalues $c_1$ and $c_2$ in Poiseuille flows}\label{sec4.2.4}
\begin{figure}[H]
	\centering		
	\includegraphics[width=9.0cm]{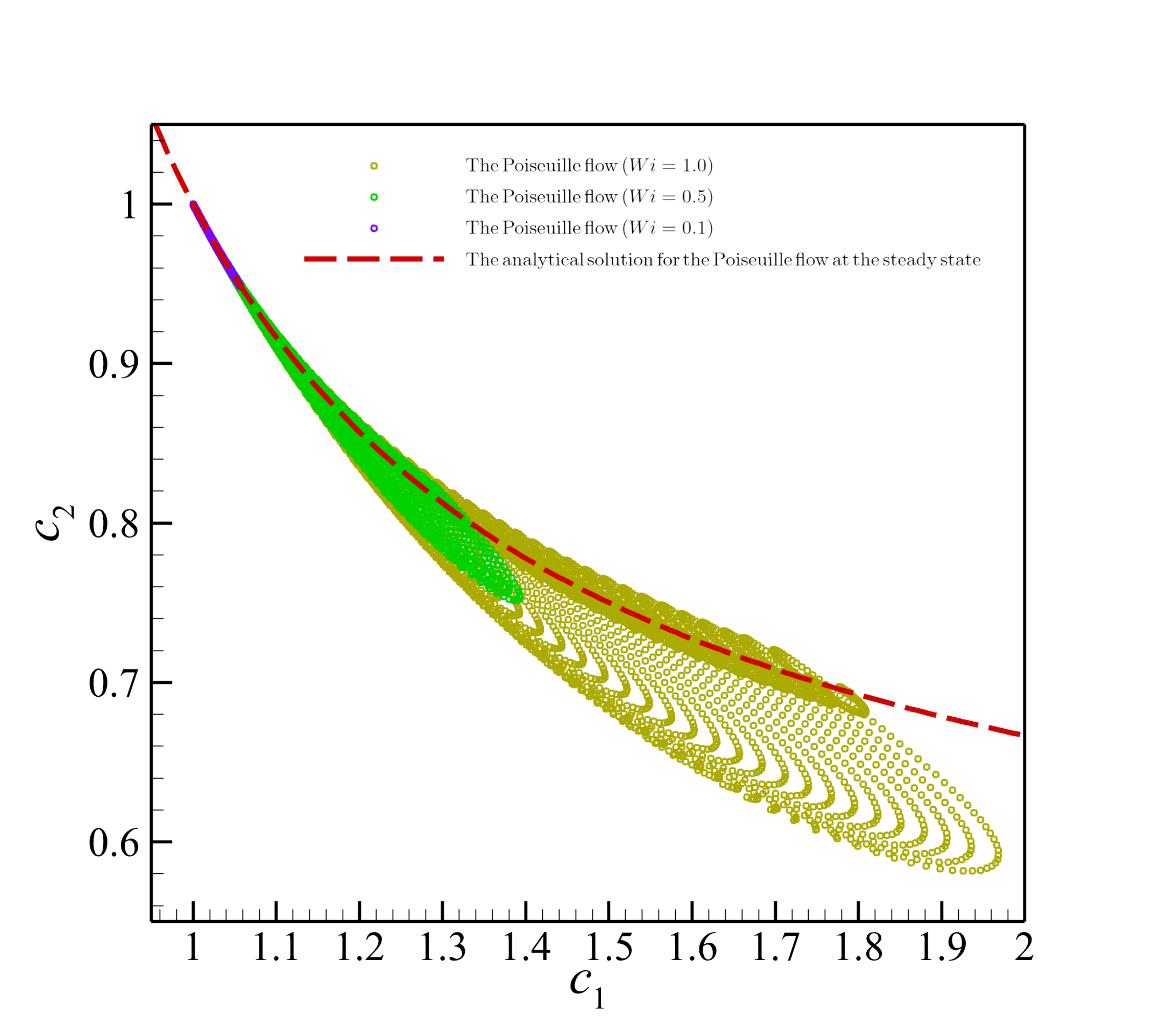}
	\caption{The locations of the eigenvalues $c_1$ and $c_2$ outputted every $500$ time iterations in the Poiseuille flows ($Wi=0.1$, $Wi=0.5$, and $Wi=1.0$) and the analytical solution for the Poiseuille flow at the steady state.}
	\label{fig:Po2_c1c2}
\end{figure}

The active learning processes during the numerical simulations are affected by the variation of the locations of the eigenvalues $c_1$ and $c_2$. Thus, the results of the transient Poiseuille flow at $Wi=0.1$, $Wi=0.5$, and $Wi=1.0$ are outputted every $500$ time iterations, which correspond to the purple, green, and yellow circle symbols in~\cref{fig:Po2_c1c2}, respectively. The analytical solution for the Poiseuille flow with the Oldroyd-B model at the steady state is also presented by the red dashed line in~\cref{fig:Po2_c1c2}, whose expression for $c_2$ is as follows~\cite{simavilla_Nonaffine_2023}:
\begin{equation}
	c_2=\frac{c_1}{2 c_1-1}.\label{eq:c_2}
\end{equation} 
The~\cref{eq:c_2} is derived from introducing the Oldroyd-B model into the steady-state version of the evolution equation of $c_2$ in~\cref{eq:evaluation_eigenvalue_new} with a velocity gradient.
\par
As shown in~\cref{fig:Po2_c1c2}, as the Weissenberg number increases, the corresponding regions covered by the eigenvalues $c_1$ and $c_2$ during the numerical simulation expand. The learned constitutive relations R1 in~\cref{sec4.1.3} can only guarantee the predictive accuracy in the region covered by the purple circle symbols and its neighborhood at the tolerance setting $\delta_{\rm{tol}}=0.05$. When it is applied as the initial local constitutive relation to the Poiseuille flows at higher Weissenberg numbers or stricter tolerance settings, more training data points need to be sampled to provide accurate constitutive relations in larger regions. In the ${\rm{G^2ALSPH}}$ method, the accuracy of the constitutive relation can be monitored at every time instant by the relative uncertainty and can be guaranteed by dynamically sampling more training data to update the learned constitutive relation, which ensures the generalization ability. 
\subsection{The flow around a periodic array of cylinders}\label{sec4.3}
\begin{figure}[H]
	\centering
	\begin{minipage}{0.6\linewidth}
		\begin{subfigure}[b]{\linewidth}
			\centering
			\includegraphics[width=\linewidth]{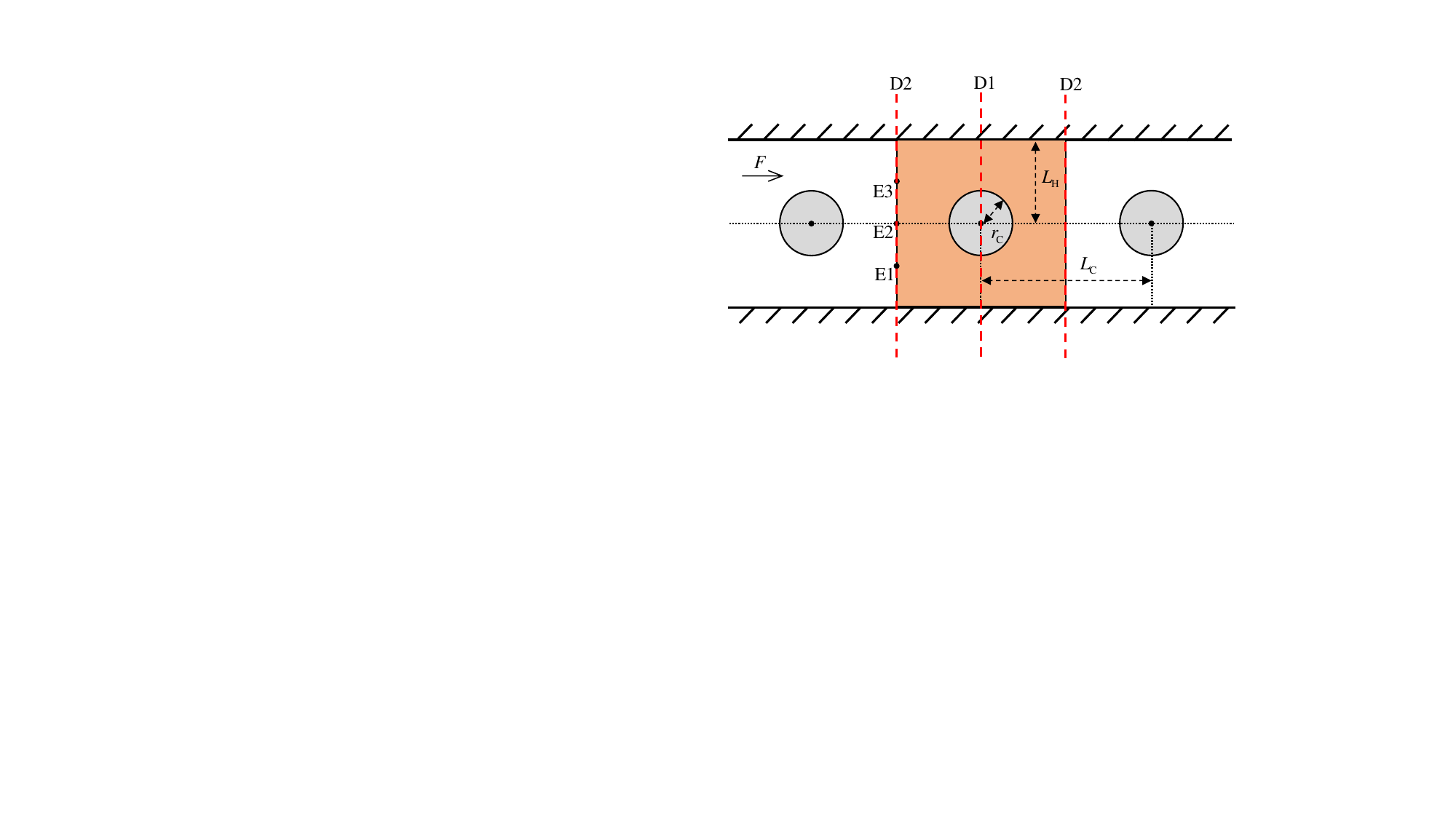}
			\caption{}
			\label{fig:cy_initial_a}
		\end{subfigure}%
	\end{minipage}%
	\hspace{5mm}
	\begin{minipage}{0.32\linewidth}
		\vspace{2.5mm}
		\begin{subfigure}[b]{\linewidth}
			\centering
			\includegraphics[width=\linewidth]{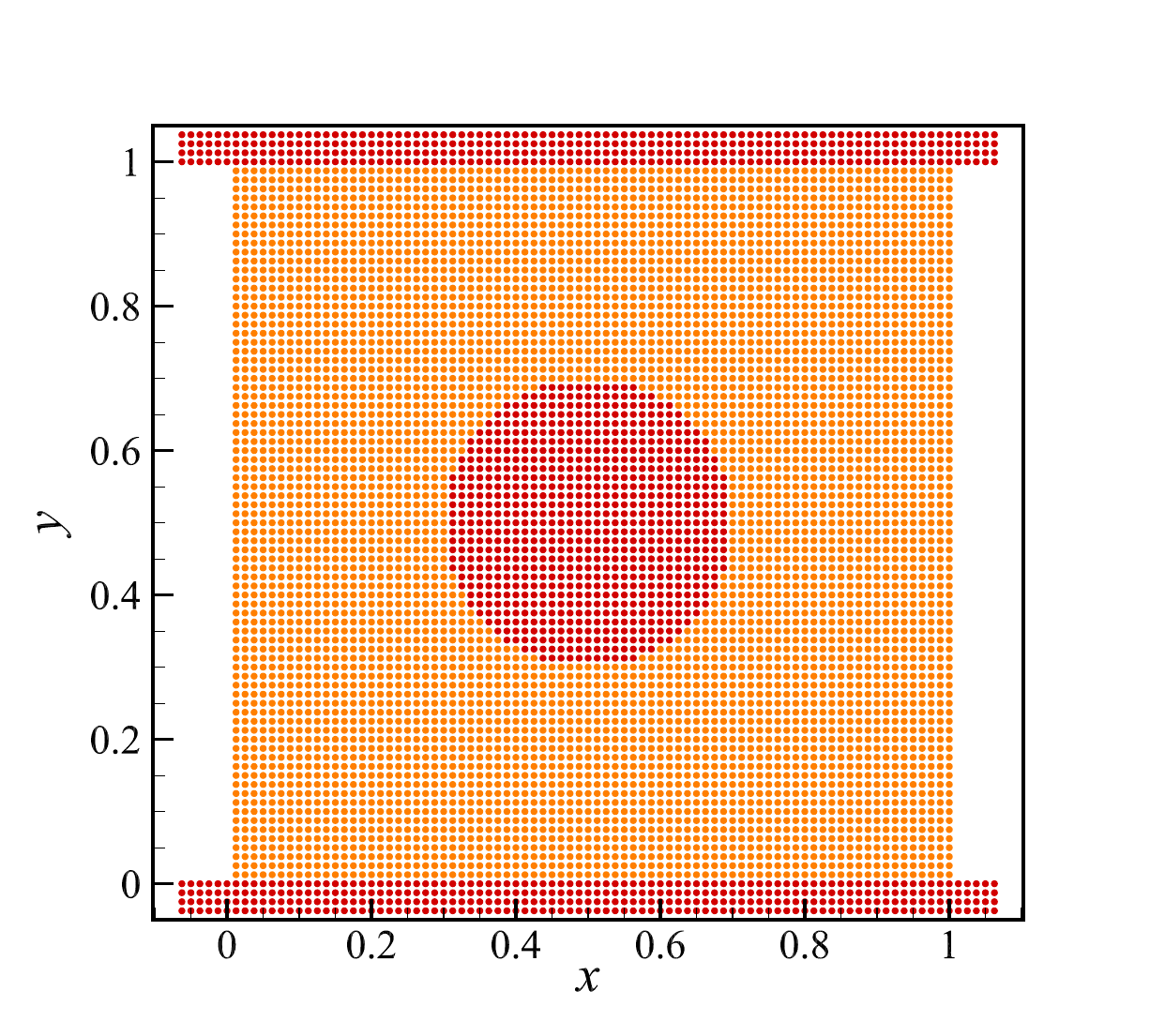}
			\caption{}
			\label{fig:cy_initial_b}
		\end{subfigure}%
	\end{minipage}
	\caption{The initial setup of the flow around a periodic array of cylinders: (a) the computational model and (b) the particle distribution at the initial instant.}
	\label{fig:cy_initial}
\end{figure}

For further validation of the effectiveness of the ${\rm{G^2ALSPH}}$ method in simulating complex flows, the flow around a periodic array of cylinders is chosen as the additional test~\cite{li_Extension_2024,vazquez-quesada_SPH_2012,ellero_SPH_2011}. The initial setup is illustrated by~\cref{fig:cy_initial}. The physical parameters of the computational model are set as follows: $L_{\rm{H}}=0.5$, $L_{\rm{C}}=1.0$, and $r_{\rm{C}}=0.2$. The width of the domain to be studied is set to be $1.0$. The artificial sound speed is set as $c_{\rm{s}}=5.0$. The Reynolds number $Re=1.0$ and the Weissenberg number $Wi=0.1$ are set. Three probe points, E1, E2, and E3, are set in the channel, whose distances to the lower boundary are $L_{\rm{H}}/2$, $L_{\rm{H}}$, and $3L_{\rm{H}}/2$, respectively. As shown in~\cref{fig:cy_initial_b}, the initial particle spacing is set as $d_0=0.0125$, corresponding to $79$ fluid particles along the y-direction. The total numbers of the fluid and wall particles are $5527$ and $1521$, respectively. The smoothing length is set as $h=1.5d_0$. The time step is set as $\Delta t=5\times {10}^{-5}$. The total simulation time is set to be $1.2$, which allows the flow to reach a steady state. The other settings are the same as those in the Poiseuille flow in~\cref{sec4.1}.
\subsubsection{The accuracy and convergence of the proposed SPH method}\label{sec4.3.1}
\begin{figure}[H]
	\centering
	\begin{subfigure}[b]{0.45\linewidth}
		\includegraphics[width=\linewidth]{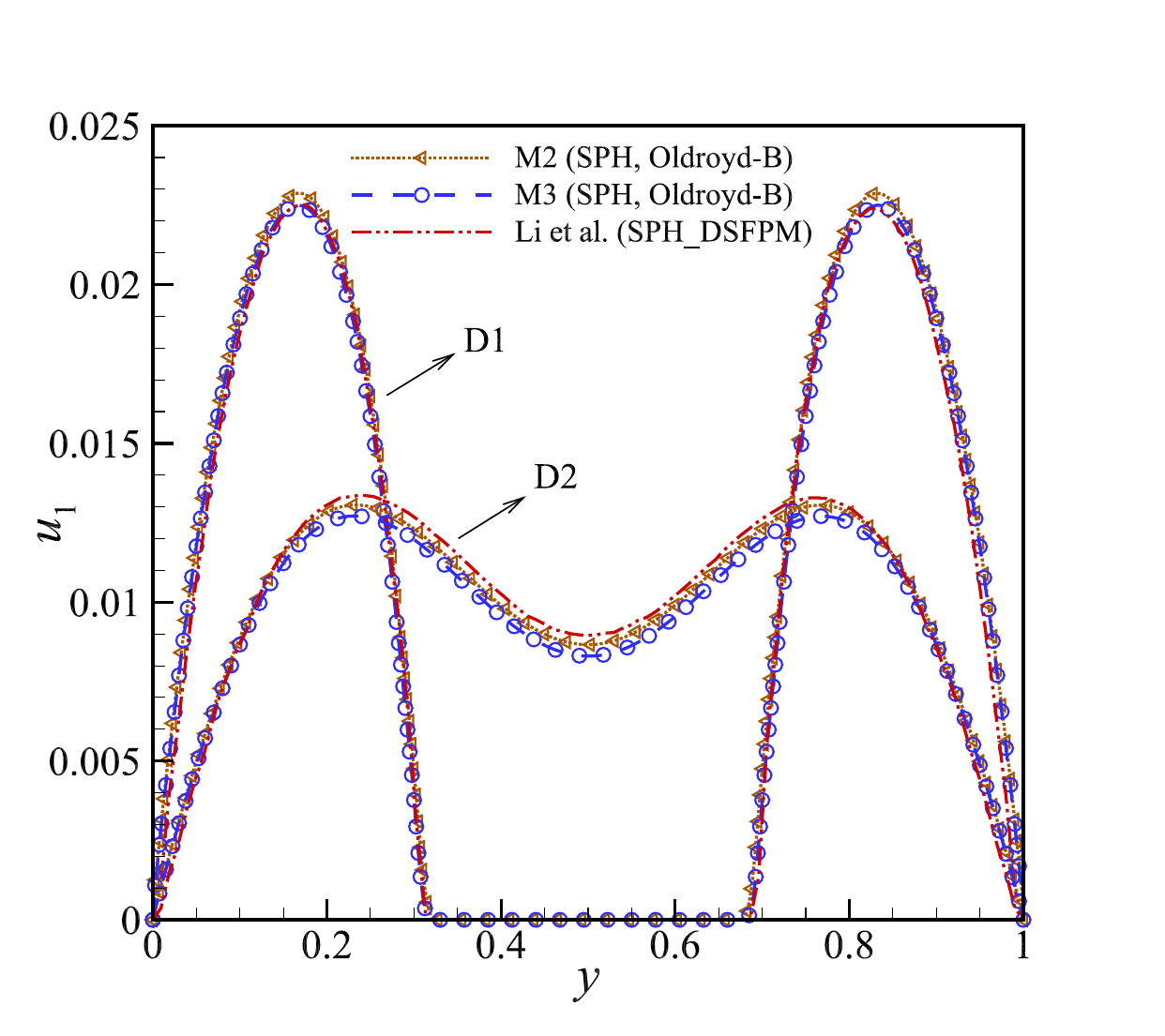}
		\caption{}
		\label{fig:cy_convergence_a}
	\end{subfigure}%
	\begin{subfigure}[b]{0.45\linewidth}
		\includegraphics[width=\linewidth]{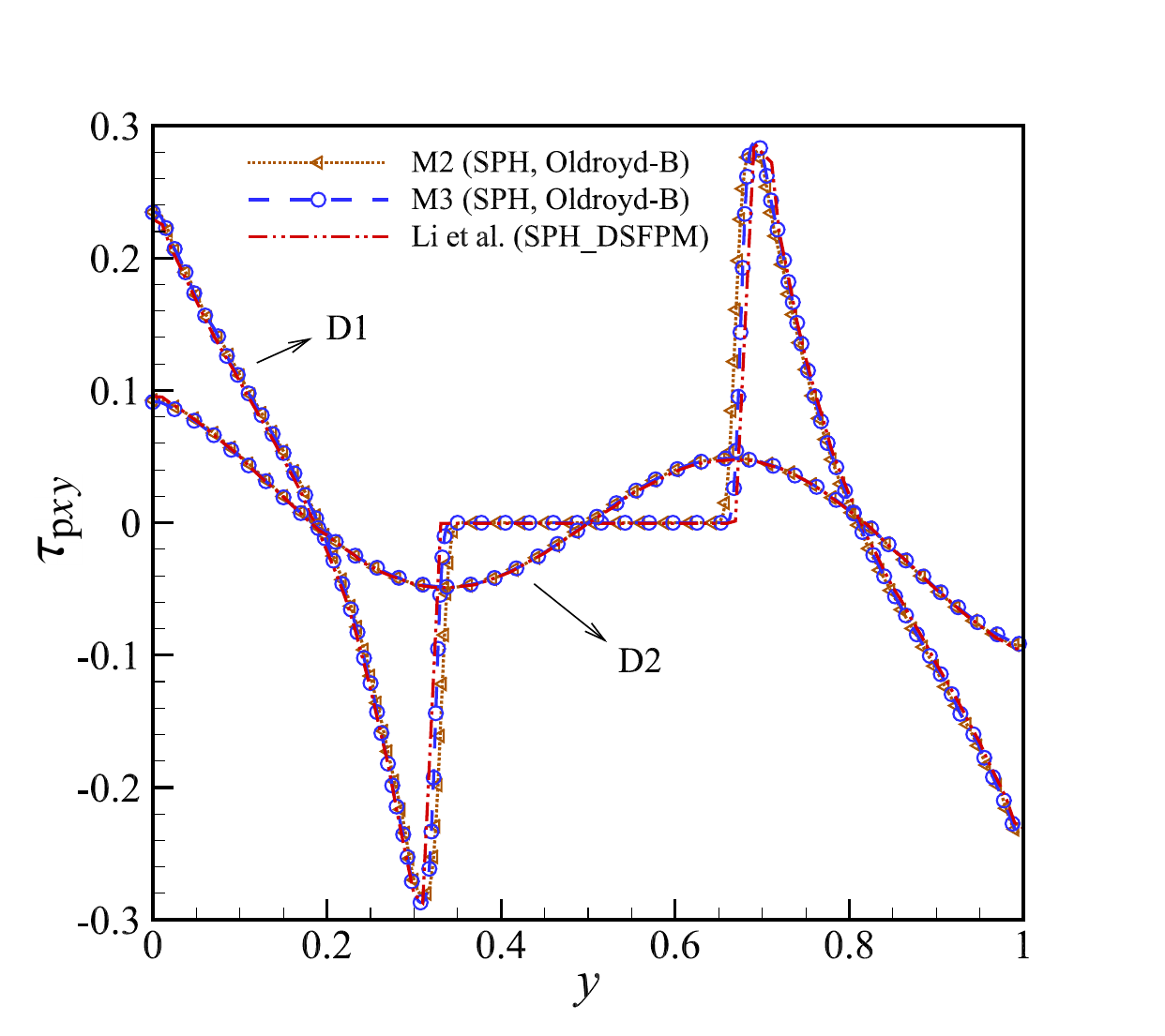}
		\caption{}
		\label{fig:cy_convergence_b}
	\end{subfigure}%
	\caption{Comparison among the results at the steady state in the probe lines D1 and D2 obtained by the SPH method with the Oldroyd-B model and the results presented in the literature~\cite{li_Extension_2024}: (a) the distributions of the velocity $u_1$ and (b) the distributions of the shear stress $\tau_{{\rm{p}}xy}$.}
	\label{fig:cy_convergence}
\end{figure}

The effectiveness of the SPH method in simulating complex viscoelastic flows is assessed in this subsection. Two resolutions, M2 and M3, defined in~\cref{sec4.1.1} are employed, which correspond to initial particle spacing settings of $0.0125$ and $0.01$, respectively. In~\cref{fig:cy_convergence}, the distributions of the velocity $u_1$ and the shear stress $\tau_{{\rm{p}}xy}$ in the probe lines D1 and D2 at the steady state are presented, where the results obtained by the SPH method with the Oldroyd-B model match well those obtained by an extended numerical scheme of smoothed particle hydrodynamics and finite particle method with density smoothing (SPH$\_$DSFPM) in the literature~\cite{li_Extension_2024}. These results prove that the proposed SPH method is accurate in simulating the flow around a periodic array of cylinders and can provide reference analytical results for the ${\rm{G^2ALSPH}}$ method. 
\subsubsection{The flows around a periodic array of cylinders with three different tolerances ($\mathit{Wi=0.1}$)}\label{sec4.3.2}
Compared to the cases in~\cref{sec4.1.3,sec4.2.1}, the flow around a periodic array of cylinders with $Wi=0.1$ is a more complex numerical case with extensive variation in eigenvalues $c_1$ and $c_2$. Three tolerances, $\delta_{\rm{tol}}=0.05$, $\delta_{\rm{tol}}=0.01$, and $\delta_{\rm{tol}}=0.005$, are chosen in this subsection. The initial local learned constitutive relations are kept the same as R1. The initial particle spacing is set as $d_0=0.0125$. The other physical parameters and computational model settings are kept unchanged.
\par
\begin{figure}[H]
	\centering
	\begin{subfigure}[b]{0.333\linewidth}
		\includegraphics[width=\linewidth]{figures/GPRTWOIN530_238.pdf}
		\caption{$\delta_{\rm{tol}}=0.05$ with R1}
		\label{fig:cy_al-results2_a}
	\end{subfigure}%
	\begin{subfigure}[b]{0.333\linewidth}
		\includegraphics[width=\linewidth]{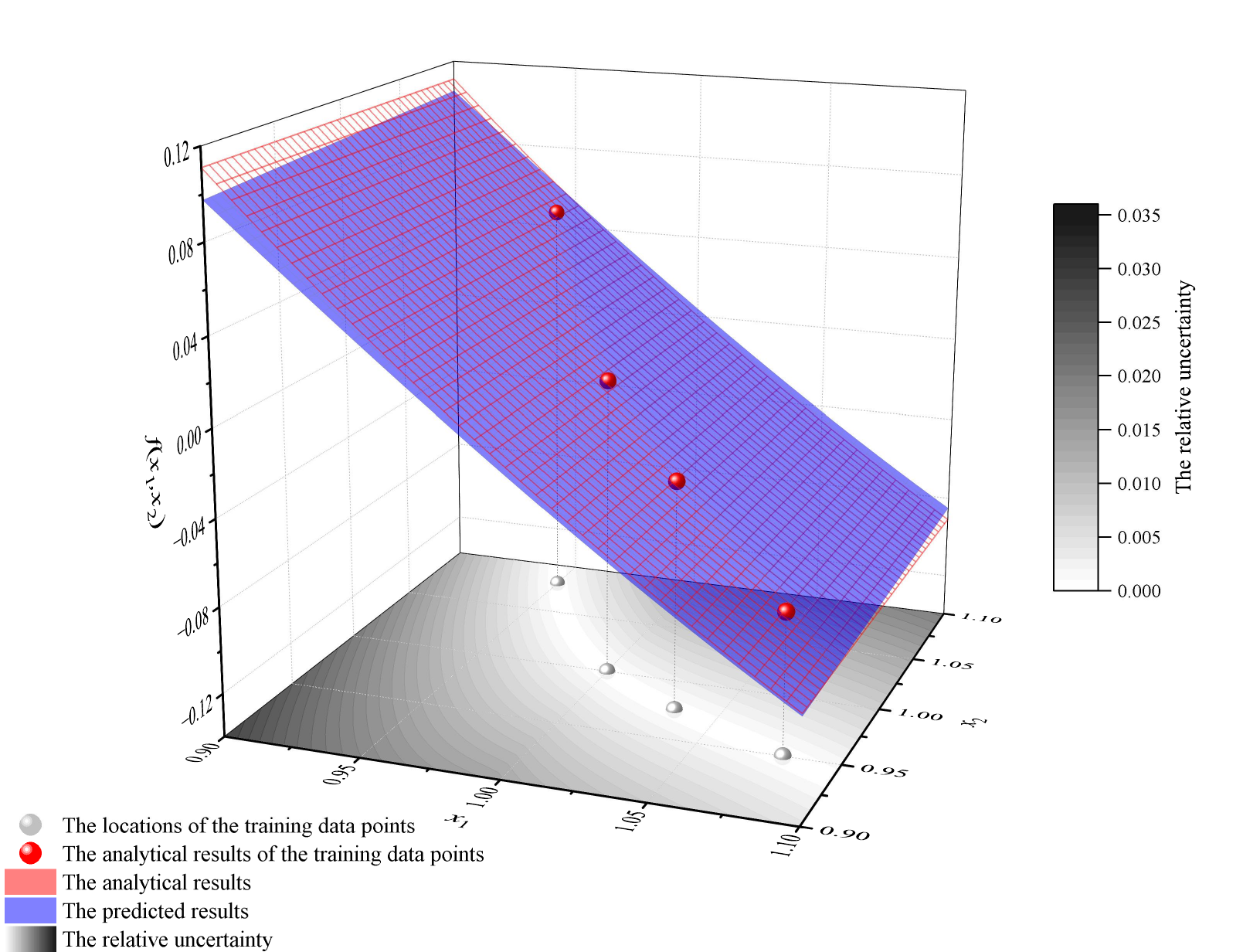}
		\caption{$\delta_{\rm{tol}}=0.01$ with R1}
		\label{fig:cy_al-results2_b}
	\end{subfigure}%
	\begin{subfigure}[b]{0.333\linewidth}
		\includegraphics[width=\linewidth]{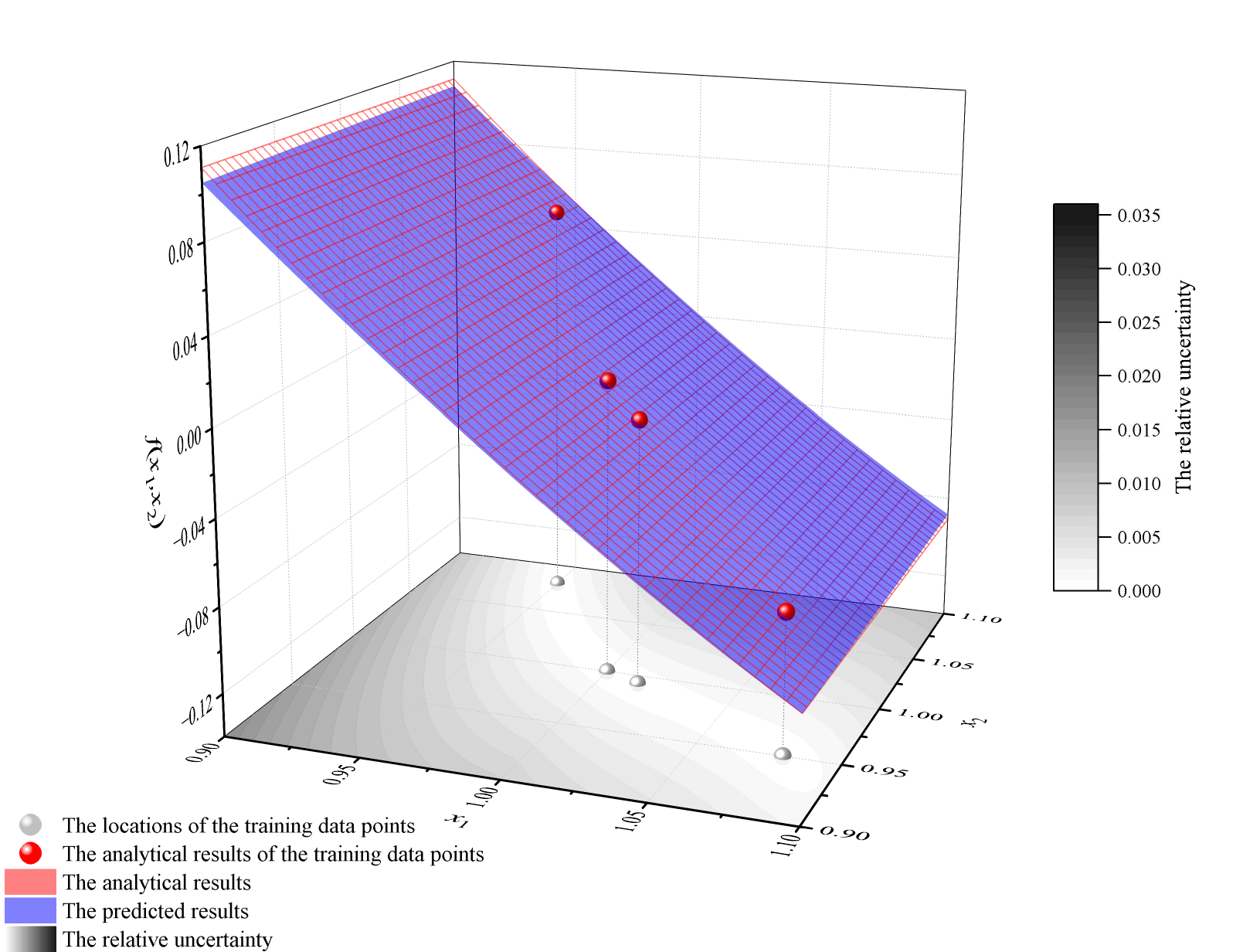}
		\caption{$\delta_{\rm{tol}}=0.005$ with R1}
		\label{fig:cy_al-results2_c}
	\end{subfigure}%
	\caption{The final learned constitutive relations obtained by the ${\rm{G^2ALSPH}}$ method with three different tolerance settings for the flows around a periodic array of cylinders ($Wi=0.1$): (a) $\delta_{\rm{tol}}=0.05$, (b) $\delta_{\rm{tol}}=0.01$, and (c) $\delta_{\rm{tol}}=0.005$, where three initial local constitutive relations are the same as R1.}
	\label{fig:cy_al-results2}
\end{figure}

In~\cref{fig:cy_al-results2}, three final learned constitutive relations are presented. The same findings as in~\cref{sec4.1.3,sec4.2.1} can be obtained. As the tolerance decreases, more training data points are required, and adding more data improves the relative uncertainty results in some regions.
\par
\begin{figure}[H]
	\centering
	\begin{subfigure}[b]{0.5\linewidth}
		\includegraphics[width=\linewidth]{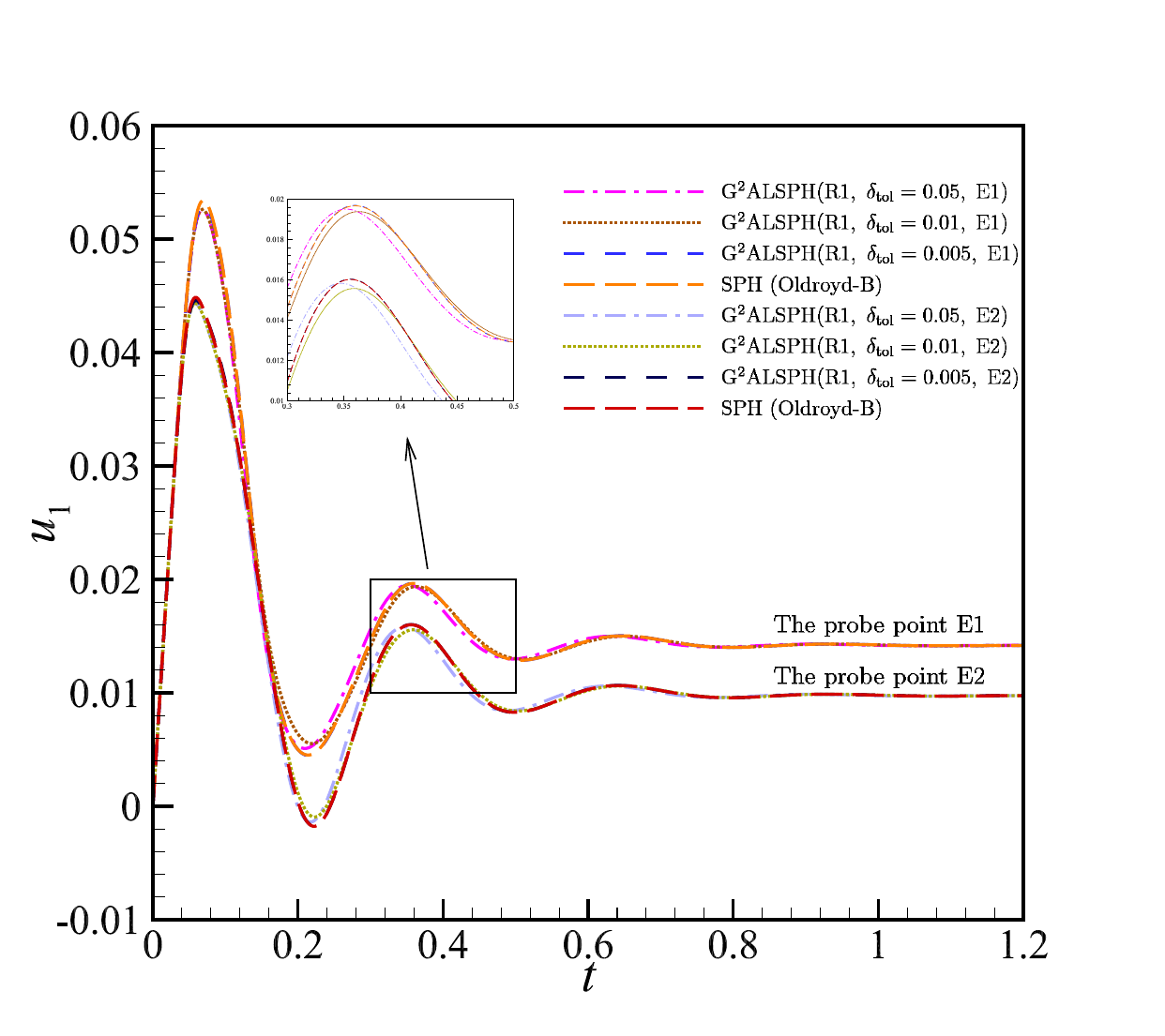}
		\caption{}
		\label{fig:cy_result1_a}
	\end{subfigure}%
	\begin{subfigure}[b]{0.5\linewidth}
		\includegraphics[width=\linewidth]{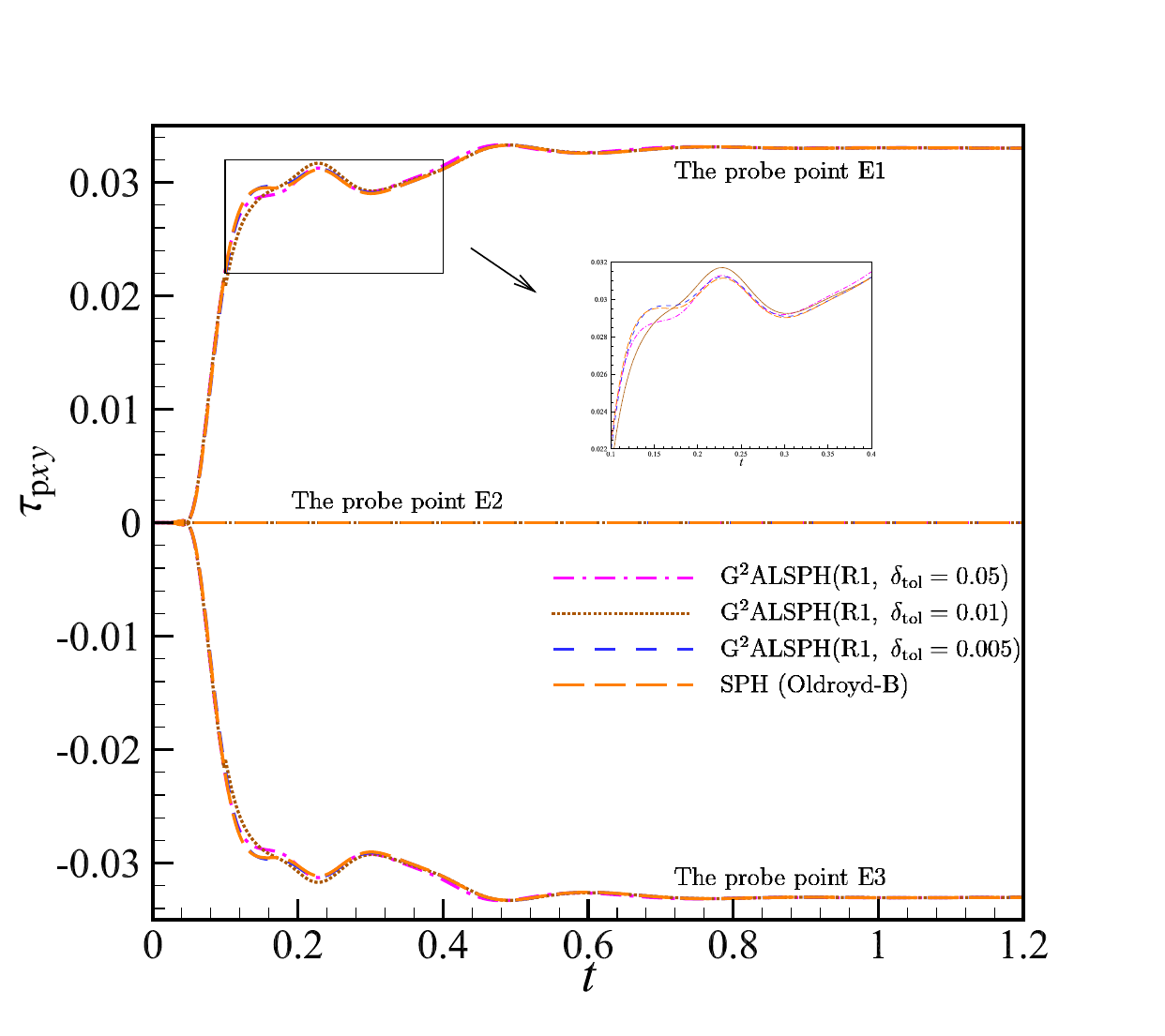}
		\caption{}
		\label{fig:cy_result1_b}
	\end{subfigure}%
	\caption{Comparison among the results in the probe points E1, E2, and E3 obtained by the ${\rm{G^2ALSPH}}$ method with three different tolerance settings and the SPH method with the Oldroyd-B model: (a) the variations of the velocity $u_1$ over time and (b) the variations of the shear stress $\tau_{{\rm{p}}xy}$ over time.}
	\label{fig:cy_result1}
\end{figure}

Regarding the results during the simulation process, the variations of the velocity $u_1$ and the shear stress $\tau_{{\rm{p}}xy}$ over time at probe points E1, E2, and E3 are depicted in~\cref{fig:cy_result1}. In~\cref{fig:cy_result1_a}, as the tolerance decreases, the results of the variations of the velocity $u_1$ over time obtained by the ${\rm{G^2ALSPH}}$ method converge to the reference analytical results. In~\cref{fig:cy_result1_b}, as shown in the enlarged zone, the results obtained by $\delta_{\rm{tol}}=0.05$ are more accurate than those obtained at $\delta_{\rm{tol}}=0.01$ for $0.1<t<0.35$, but less accurate than those obtained at $\delta_{\rm{tol}}=0.01$ for $0.35<t<0.4$. The results fluctuate within a certain accuracy range, and the simulations can run smoothly until the steady state. The presence of fluctuations is due to the fact that the constitutive relations are updated at different instants for different tolerance settings. Despite them, the results satisfy the respective accuracy limits. When the tolerance is lowered to $0.005$, the results obtained by the ${\rm{G^2ALSPH}}$ method match well the results obtained by the SPH method with the Oldroyd-B model, which proves the final accuracy of the ${\rm{G^2ALSPH}}$ method.
\par
\begin{figure}[H]
	\centering
	\raisebox{0.9\height}{\rotatebox{90}{\scriptsize{The eigenvalue $c_1$}}}
	\begin{minipage}[t]{0.9\textwidth}
		\centering
		\begin{subfigure}[b]{0.33\textwidth}
			\raisebox{1.5\height}{\rotatebox{0}{\scriptsize{${\rm{G^2ALSPH}}$}}}
			\centering
			\includegraphics[width=\textwidth]{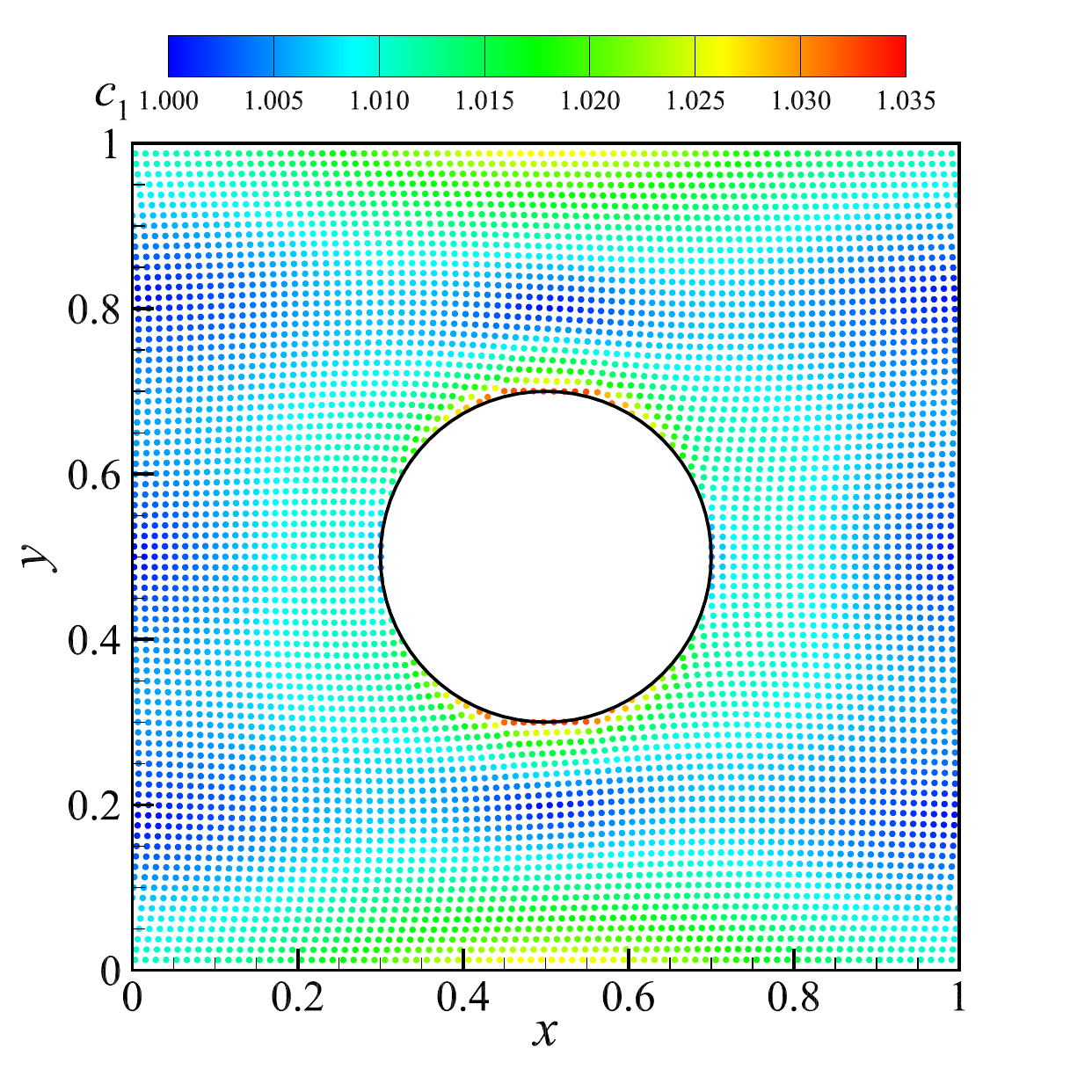}
			\label{fig:cy_distribution_a1}
		\end{subfigure}%
		\begin{subfigure}[b]{0.33\textwidth}
			\raisebox{1.5\height}{\rotatebox{0}{\scriptsize{SPH (Oldroyd-B)}}}
			\centering
			\includegraphics[width=\textwidth]{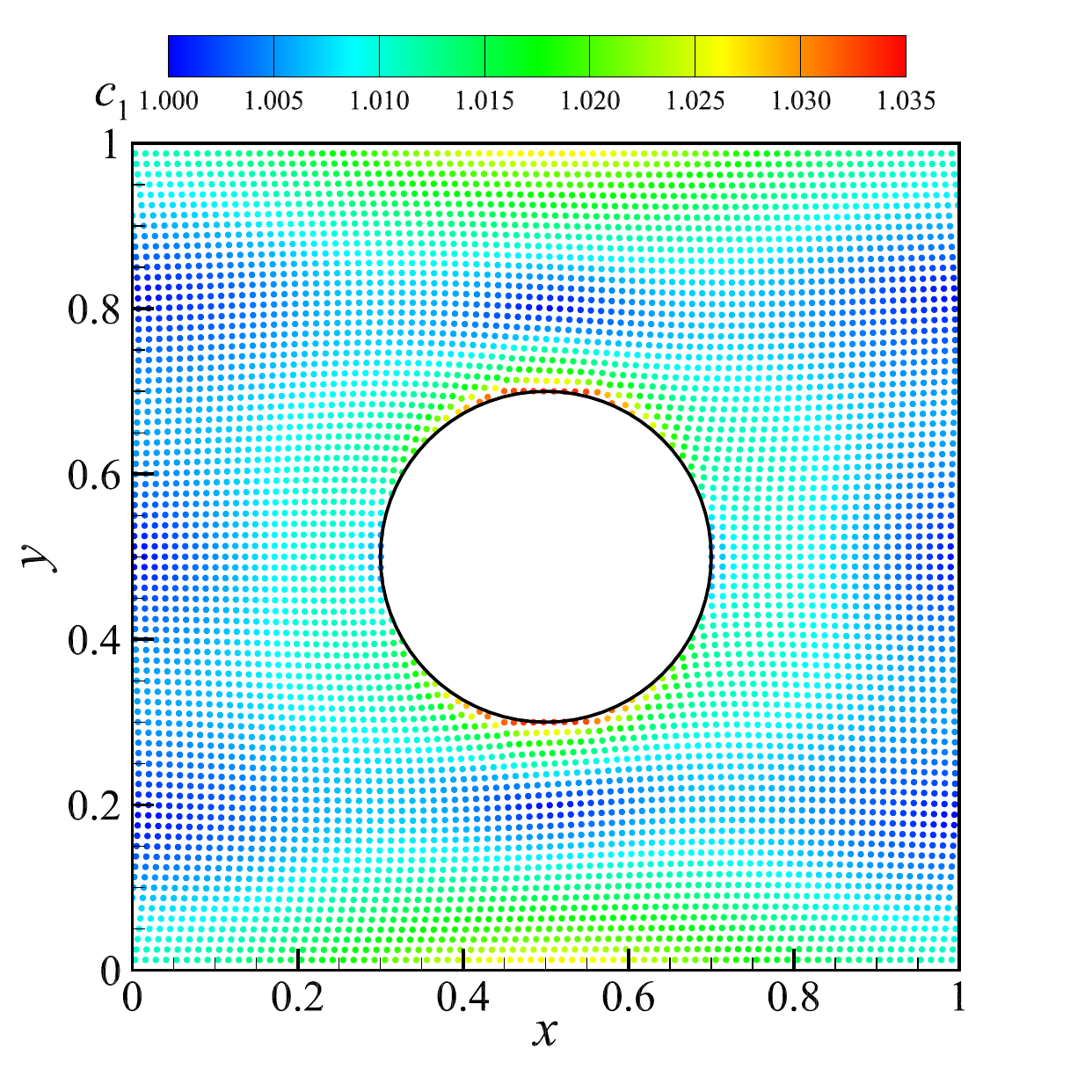}
			\label{fig:cy_distribution_b1}
		\end{subfigure}%
		\begin{subfigure}[b]{0.33\textwidth}
			\raisebox{1.5\height}{\rotatebox{0}{\scriptsize{The relative errors}}}
			\centering
			\includegraphics[width=\textwidth]{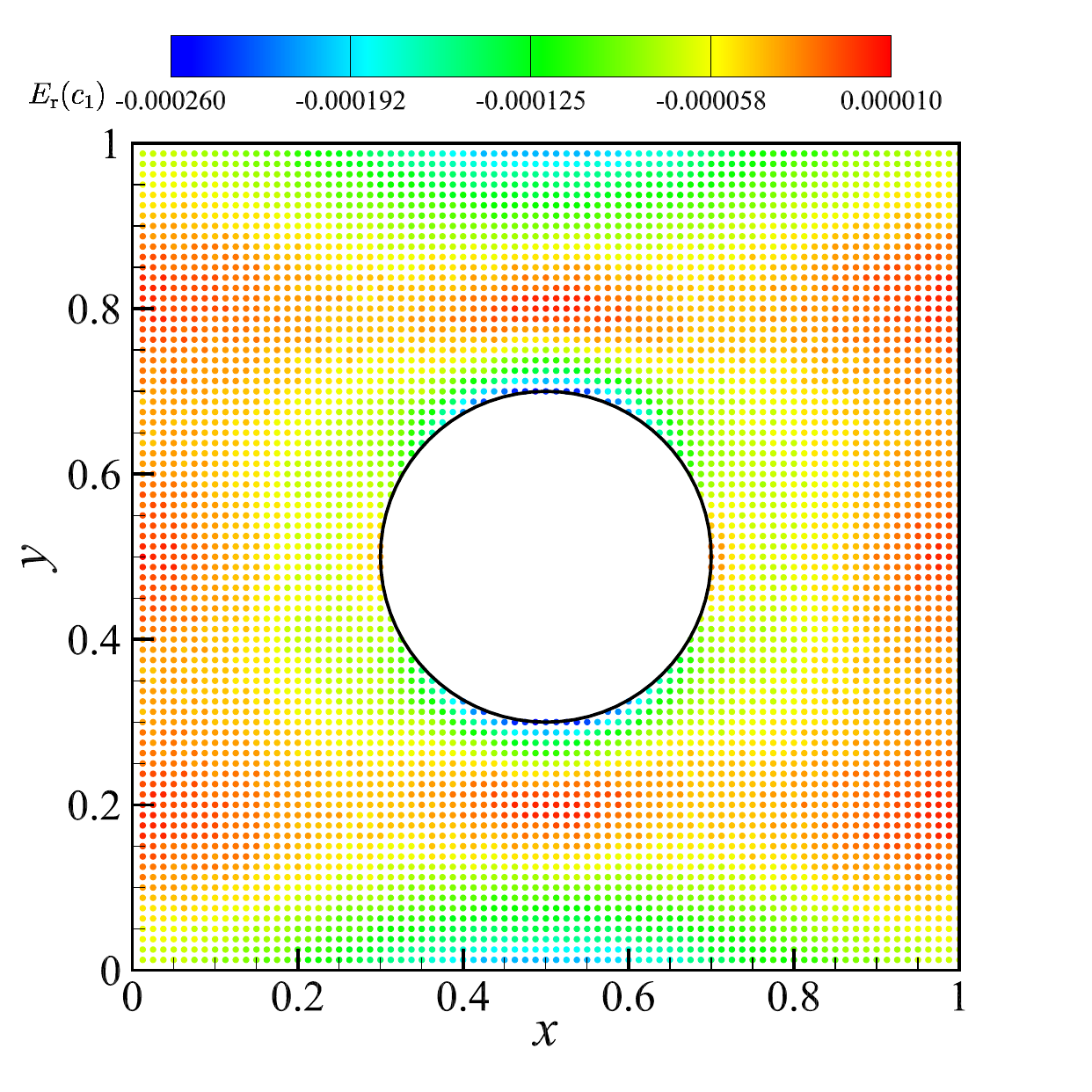}
			\label{fig:cy_distribution_c1}
		\end{subfigure}%
	\end{minipage}
	\vspace{-6mm}
	\vfill
	\centering
	\raisebox{0.9\height}{\rotatebox{90}{\scriptsize{The eigenvalue $c_2$}}}
	\begin{minipage}[t]{0.9\textwidth}
		\centering
		\begin{subfigure}[b]{0.33\textwidth}
			\centering
			\includegraphics[width=\textwidth]{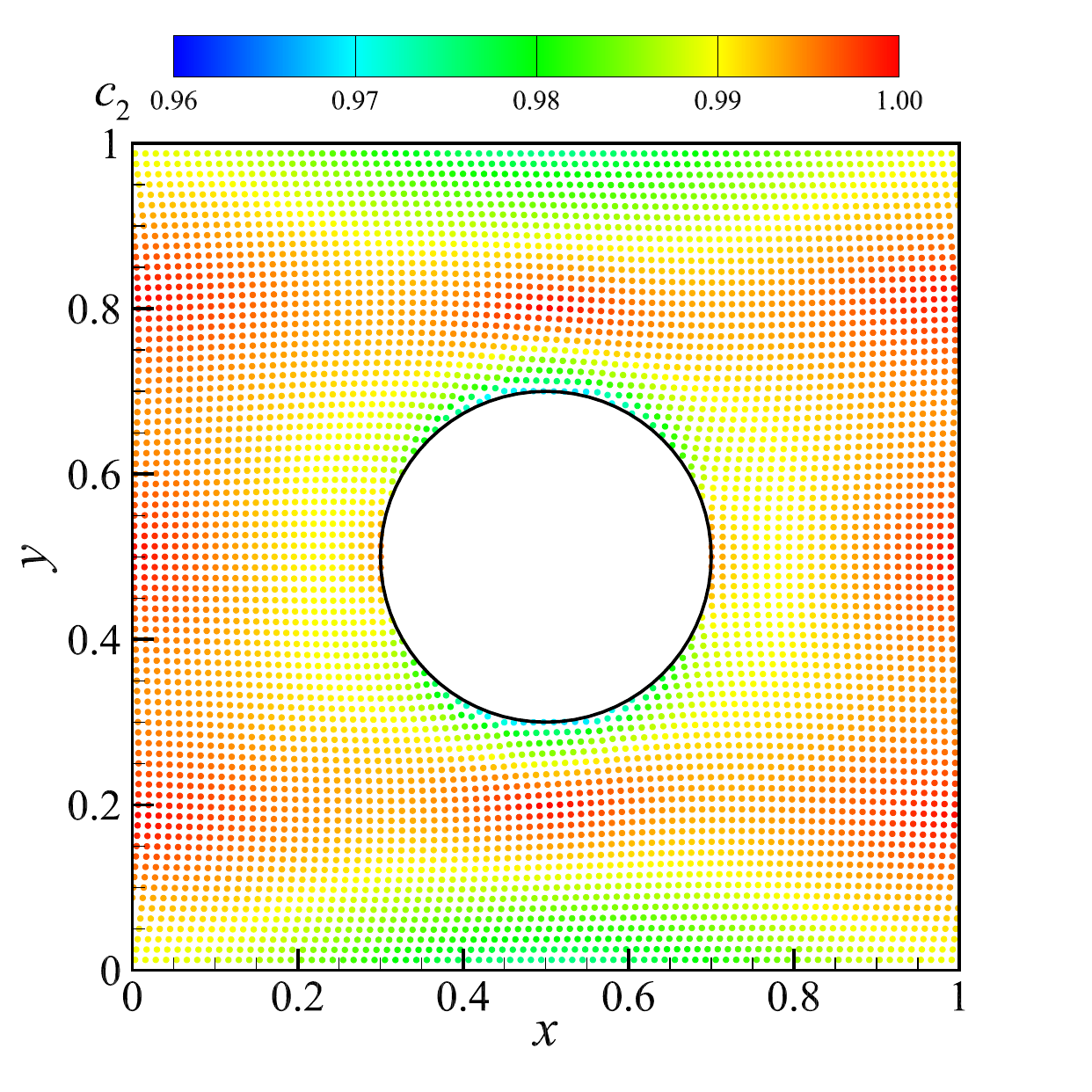}
			\label{fig:cy_distribution_a2}
		\end{subfigure}%
		\begin{subfigure}[b]{0.33\textwidth}
			\centering
			\includegraphics[width=\textwidth]{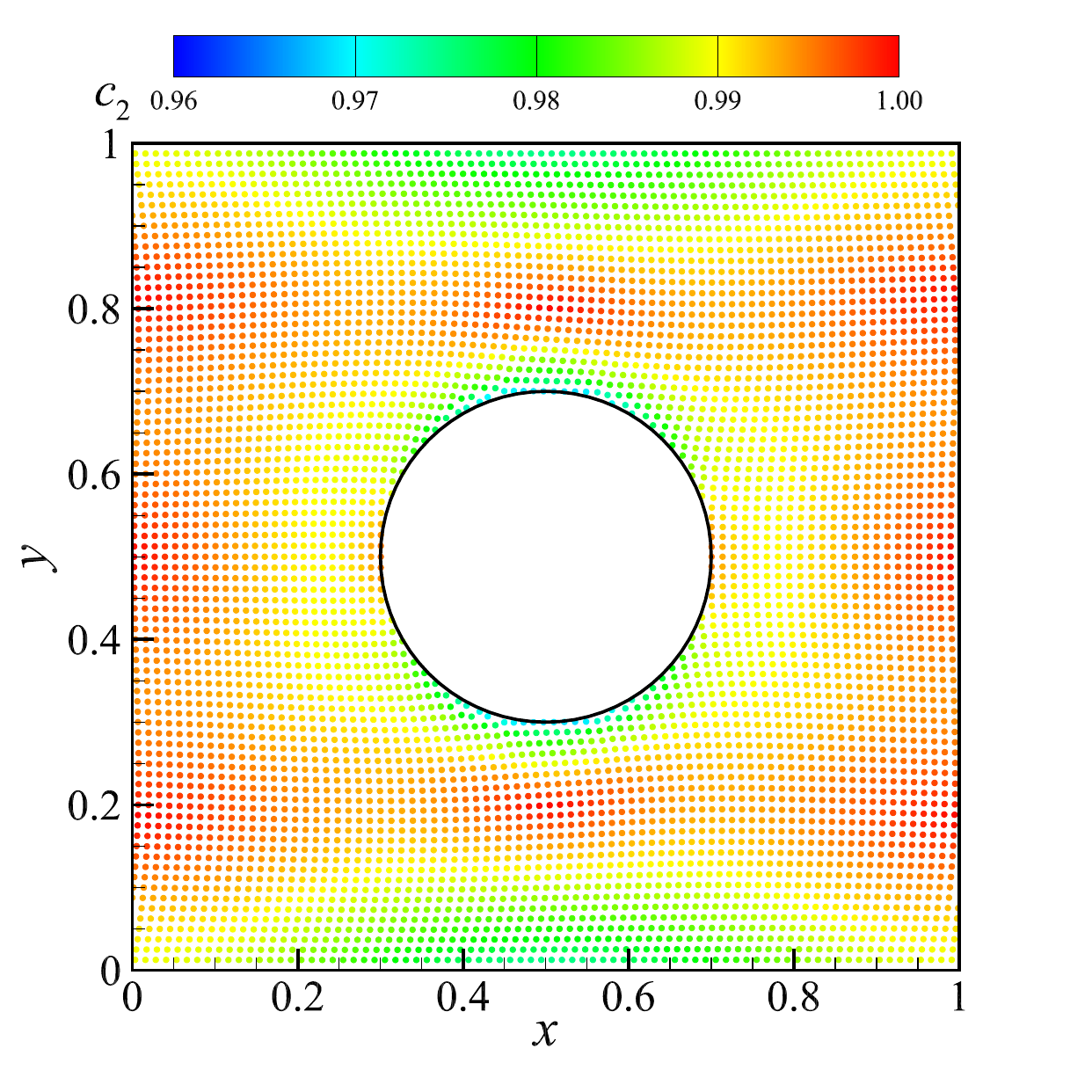}
			\label{fig:cy_distribution_b2}
		\end{subfigure}%
		\begin{subfigure}[b]{0.33\textwidth}
			\centering
			\includegraphics[width=\textwidth]{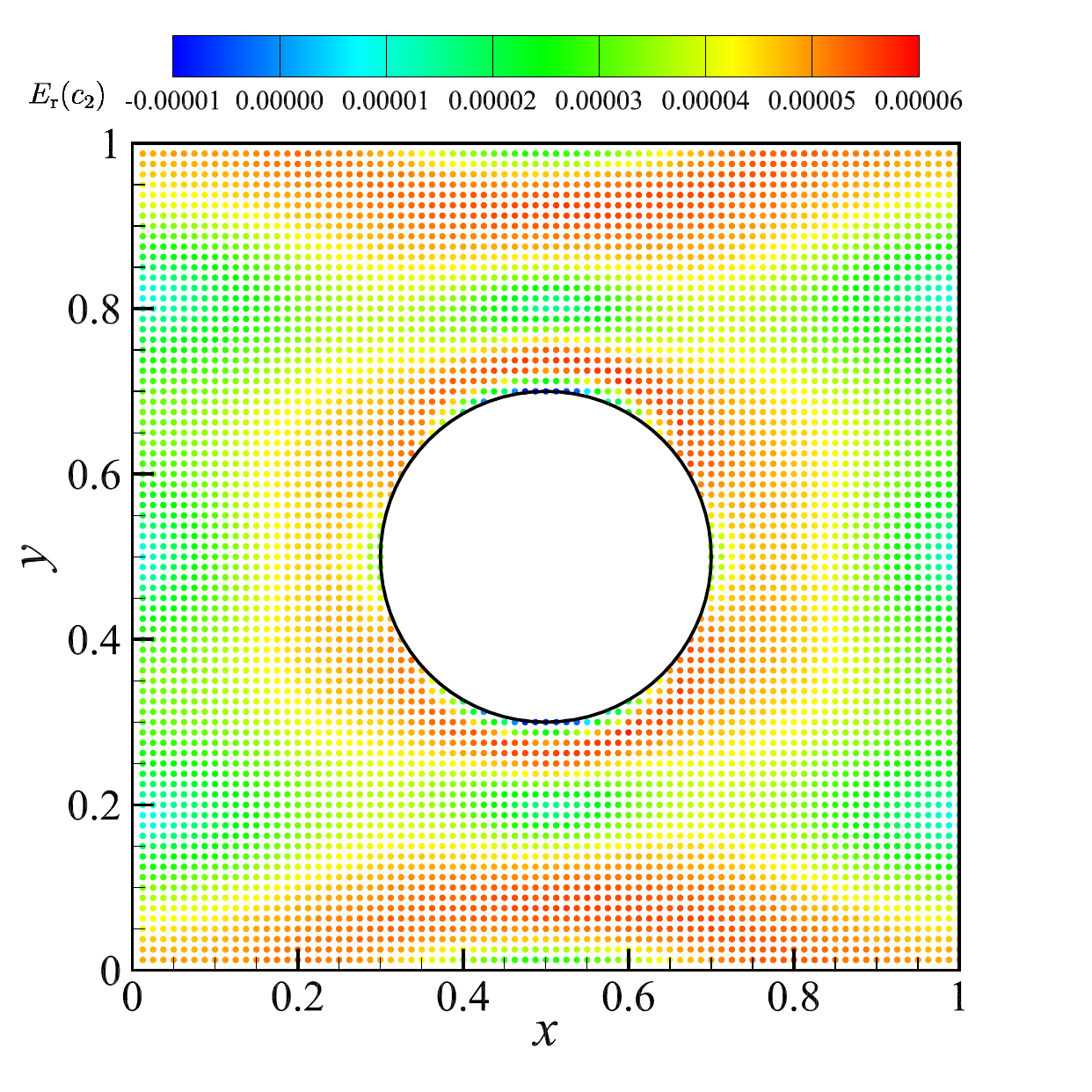}
			\label{fig:cy_distribution_c2}
		\end{subfigure}%
	\end{minipage}
	\vspace{-6mm}
	\vfill
	\centering
	\raisebox{1.2\height}{\rotatebox{90}{\scriptsize{The velocity $u_1$}}}
	\begin{minipage}[t]{0.9\textwidth}
		\centering
		\begin{subfigure}[b]{0.33\textwidth}
			\centering
			\includegraphics[width=\textwidth]{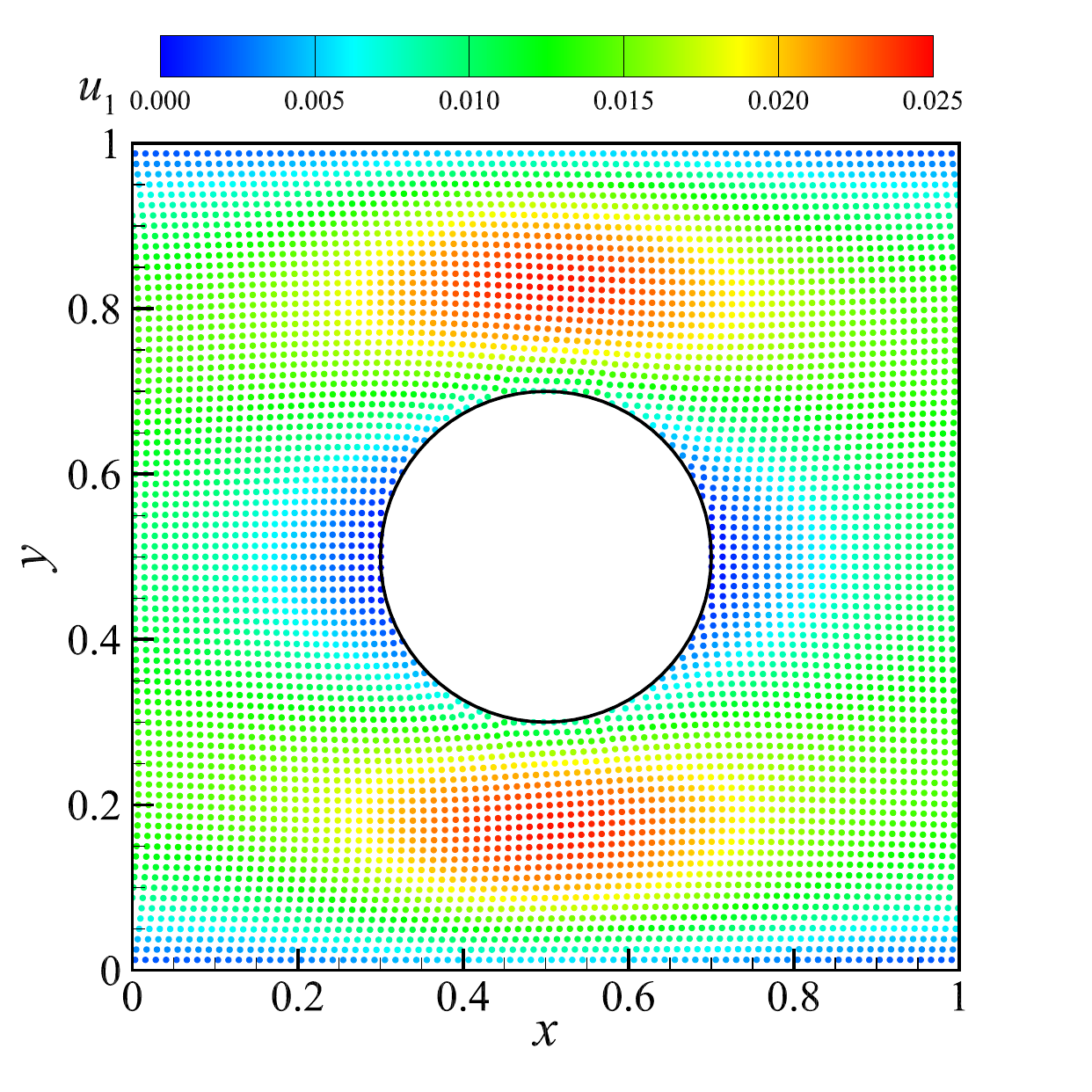}
			\label{fig:cy_distribution_a3}
		\end{subfigure}%
		\begin{subfigure}[b]{0.33\textwidth}
			\centering
			\includegraphics[width=\textwidth]{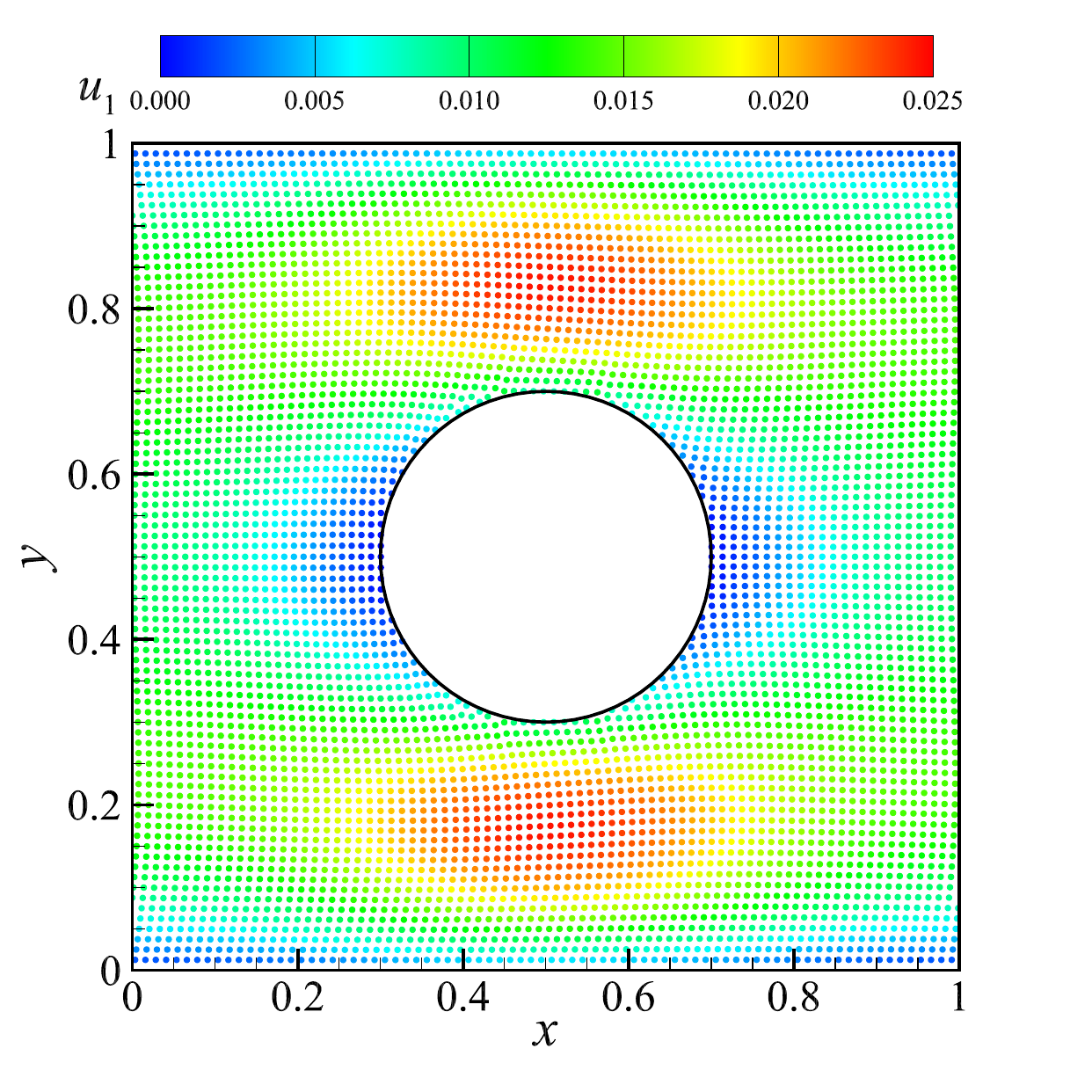}
			\label{fig:cy_distribution_b3}
		\end{subfigure}%
		\begin{subfigure}[b]{0.33\textwidth}
			\centering
			\includegraphics[width=\textwidth]{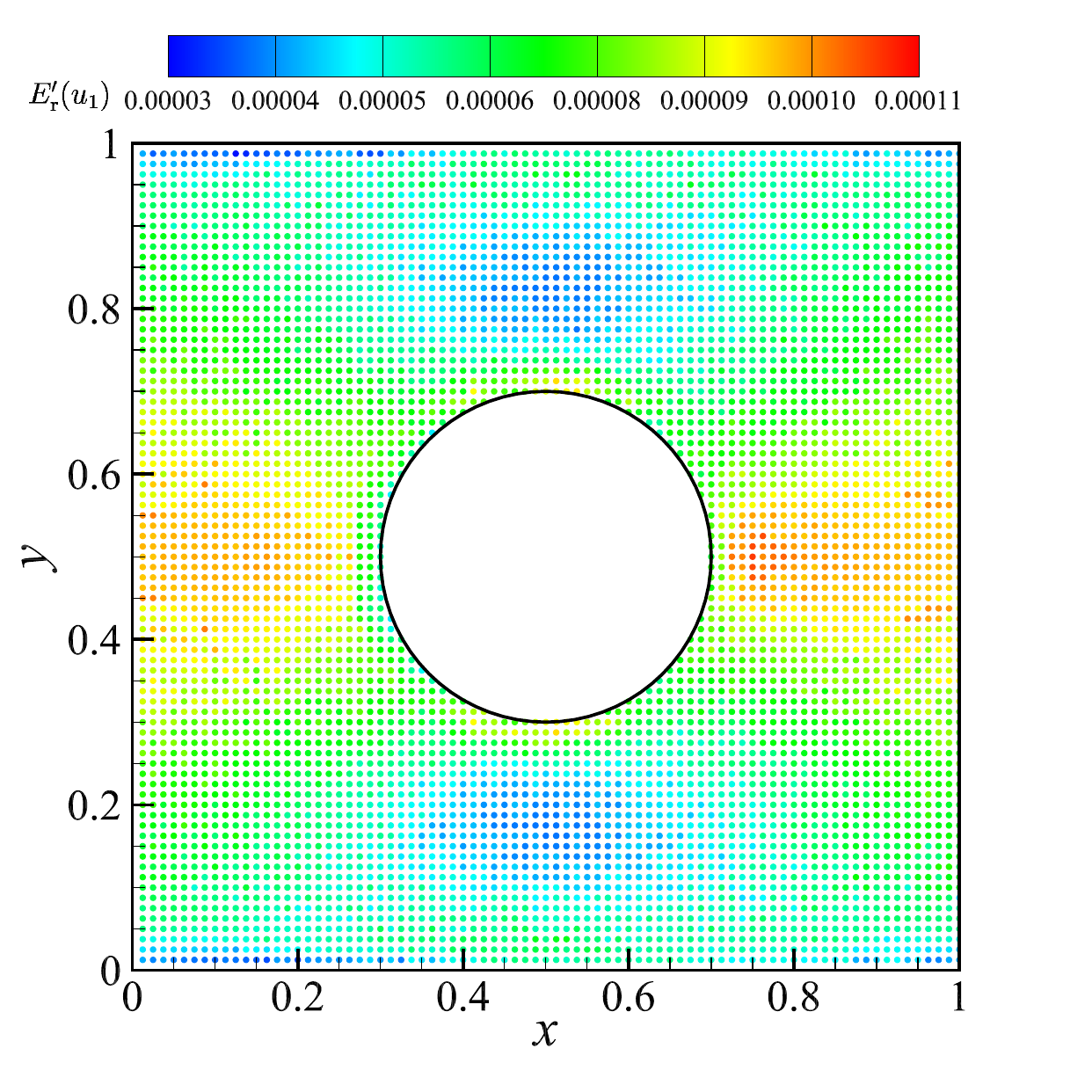}
			\label{fig:cy_distribution_c3}
		\end{subfigure}%
	\end{minipage}
	\vspace{-6mm}
	\vfill
	\centering
	\raisebox{0.2\height}{\rotatebox{90}{\scriptsize{The first normal stress difference $N_1$}}}
	\begin{minipage}[t]{0.9\textwidth}
		\centering
		\begin{subfigure}[b]{0.33\textwidth}
			\centering
			\includegraphics[width=\linewidth]{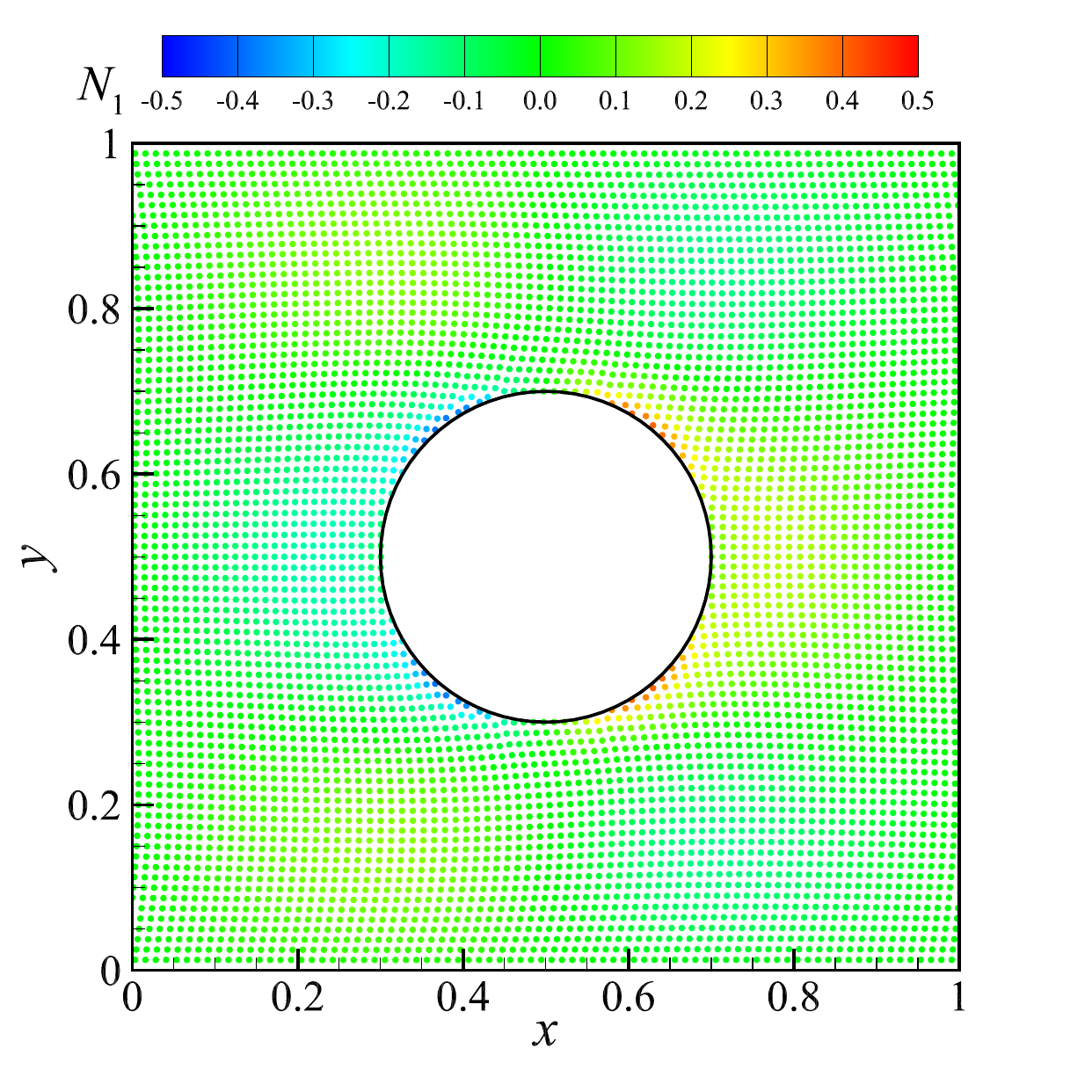}
			\caption{}
			\label{fig:cy_distribution_a4}
		\end{subfigure}%
		\begin{subfigure}[b]{0.33\textwidth}
			\centering
			\includegraphics[width=\textwidth]{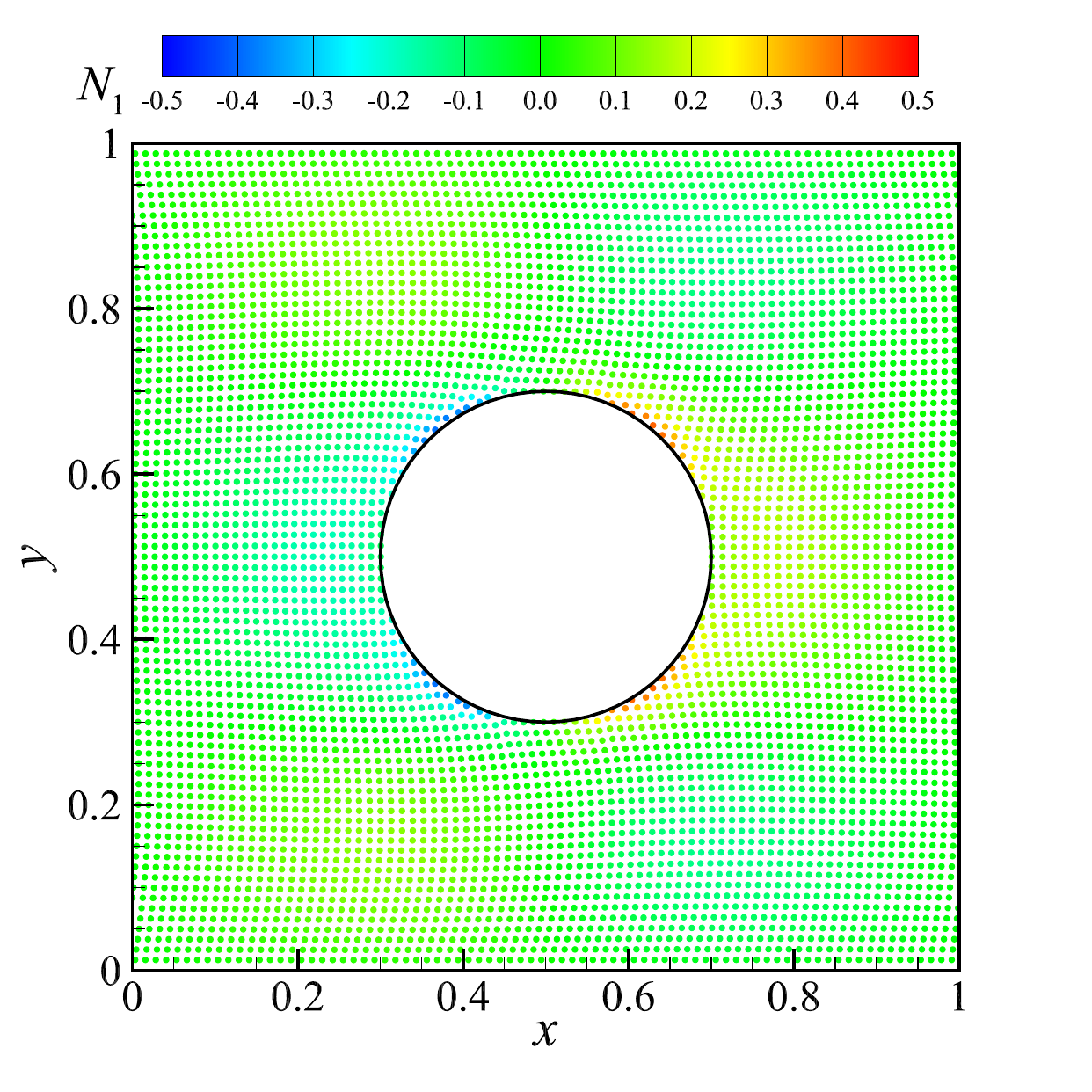}
			\caption{}
			\label{fig:cy_distribution_b4}
		\end{subfigure}%
			\begin{subfigure}[b]{0.33\textwidth}
			\centering
			\includegraphics[width=\textwidth]{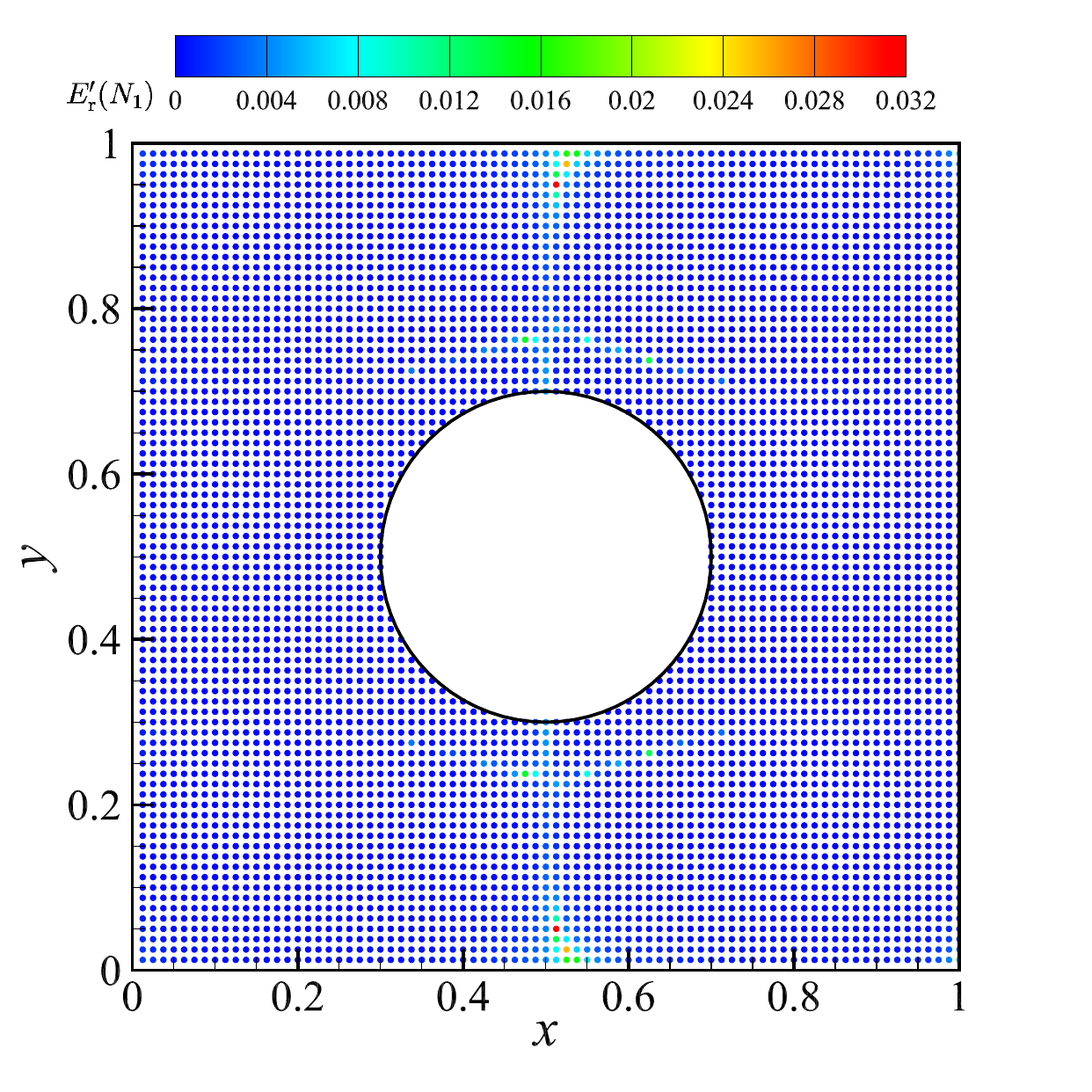}
			\caption{}
			\label{fig:cy_distribution_c4}
		\end{subfigure}%
	\end{minipage}
	\vspace{-1mm}
	\caption{The distributions of the eigenvalues $c_1$ and $c_2$, the velocity $u_1$, and the first normal stress difference $N_1$ at the steady state for the flows around a periodic array of cylinders ($Wi=0.1$): (a) the results obtained by the ${\rm{G^2ALSPH}}$ method ($\delta_{\rm{tol}}=0.005$, with R1 serving as the initial local constitutive relation), (b) the results obtained by the SPH method with the Oldroyd-B model, and (c) the relative errors.}
	\label{fig:cy_distribution}
\end{figure}

As to the results at the steady state, the distributions of the eigenvalues $c_1$ and $c_2$, the velocity $u_1$, and the first normal stress difference $N_1$ at the steady state for the flows around a periodic array of cylinders at $Wi=0.1$ are given in~\cref{fig:cy_distribution}. The results of the first normal stress difference are calculated in the same way as in~\cref{sec4.1.3}. As shown in~\cref{fig:cy_distribution_c4}, for the eigenvalues $c_1$ and $c_2$, all the absolute values of the relative errors on fluid particles are below $0.026\%$. For the velocity $u_1$, all the values of the relative errors on fluid particles are below $0.011\%$, whereas, for the first normal stress difference $N_1$, they are less than $0.4\%$ for most particles. The distributions of every physical quantity at the final instant obtained by the ${\rm{G^2ALSPH}}$ method at $\delta_{\rm{tol}}=0.005$ and the SPH method with the Oldroyd-B model are in very good agreement.
\subsubsection{Analysis of the space explored by the eigenvalues $c_1$ and $c_2$ in the flow around a periodic array of cylinders}\label{sec4.3.3}
\begin{figure}[H]
	\centering		
	\includegraphics[width=9.0cm]{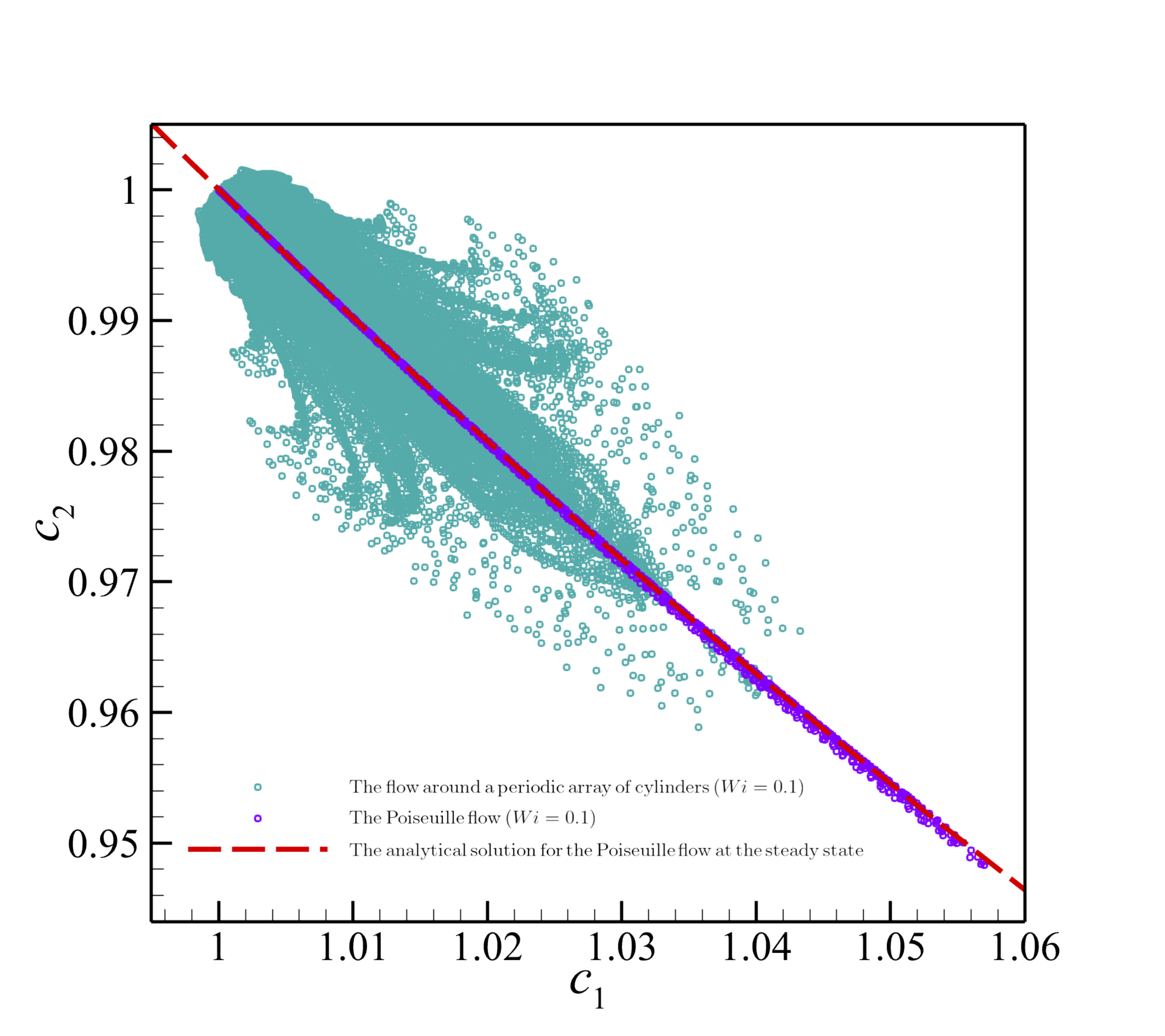}
	\caption{The locations of the eigenvalues $c_1$ and $c_2$ outputted every $500$ time iterations in the flow around a periodic array of cylinders ($Wi=0.1$) and in the Poiseuille flow ($Wi=0.1$), and the analytical solution for the Poiseuille flow at the steady state.}
	\label{fig:cy_c1c2}
\end{figure}

To further study the application process of the learned constitutive relation, the locations of the eigenvalues $c_1$ and $c_2$ in the transient Poiseuille flow at $Wi=0.1$ and the flow around a periodic array of cylinders at $Wi=0.1$ are outputted every $500$ time iterations, as depicted in~\cref{fig:cy_c1c2}. In GPR, the closer the region to the training data points, the higher the predictive accuracy. As shown in~\cref{fig:cy_c1c2}, the accuracy of the learned constitutive relation R1 can be well ensured in the regions that correspond to the region of purple cycle symbols and its neighborhood with $\delta_{\rm{tol}}=0.05$. Meanwhile, in~\cref{fig:cy_c1c2}, the region of cyan cycle symbols is exactly in that neighborhood. This is the reason why the active learning procedure is not initiated throughout the numerical simulation with $\delta_{\rm{tol}}=0.05$ in the flow around a periodic array of cylinders in~\cref{sec4.3.2}. In this case, the learned constitutive relation established in a small region performs well in a larger region, which demonstrates the generalization ability and applicability of the learned constitutive relation. The learned constitutive relation can therefore be used as a well-predicted initial local constitutive relation in other flow simulations for the same viscoelastic fluid. For the cases with $\delta_{\rm{tol}}=0.01$ and $\delta_{\rm{tol}}=0.005$ in~\cref{sec4.3.2}, only one additional data point is sampled during the respective simulations. The proposed relative uncertainty strategy greatly extends the applicability of the learned constitutive relation.
\vspace{1cm}
\subsection{The flows around a periodic array of cylinders at larger Weissenberg numbers}\label{sec4.4}
The locations of the eigenvalues $c_1$ and $c_2$ vary more complexly in the flows around a periodic array of cylinders than in the Poiseuille flows, which makes it a good way to further study the relative and direct uncertainty strategies. In this subsection, the flows around a periodic array of cylinders at three Weissenberg numbers, $Wi=0.1$, $Wi=0.5$, and $Wi=1.0$, are simulated. The corresponding values of the total simulation time are set to be $1.2$, $2.4$, and $2.4$, respectively. The smoothing lengths are all set as $h=1.5d_0$. The other settings are the same as those in~\cref{sec4.3}.
\subsubsection{The superiority of the relative uncertainty strategy over the direct uncertainty strategy}\label{sec4.4.1}
The flows around a periodic array of cylinders at $Wi=0.1$, $Wi=0.5$, and $Wi=1.0$ are simulated by the ${\rm{G^2ALSPH}}$ method with the relative uncertainty strategy and the direct uncertainty strategy, separately. The initial local constitutive relations are kept the same as R1. Except for the different uncertainty strategies and the corresponding tolerance settings, the other physical parameters and model settings are the same. The accuracy evaluation tools and the settings of the pre-set tolerance are the same as in~\cref{sec4.2.3}. 
\par
\begin{figure}[H]
	\centering
	\begin{minipage}[t]{0.9\linewidth}
		\centering
		\raisebox{2.5\height}{\rotatebox{90}{\scriptsize{$Wi=0.1$}}}
		\begin{subfigure}[b]{0.49\linewidth}
			\raisebox{1.5\height}{\rotatebox{0}{\scriptsize{The relative uncertainty strategy}}}
			\centering
			\includegraphics[width=\linewidth]{figures/GPRTWOIN513_403.pdf}
			\label{fig:cy_consitutive_a1}
		\end{subfigure}%
		\hfill
		\begin{subfigure}[b]{0.49\linewidth}
			\raisebox{1.5\height}{\rotatebox{0}{\scriptsize{The direct uncertainty strategy}}}
			\centering
			\includegraphics[width=\linewidth]{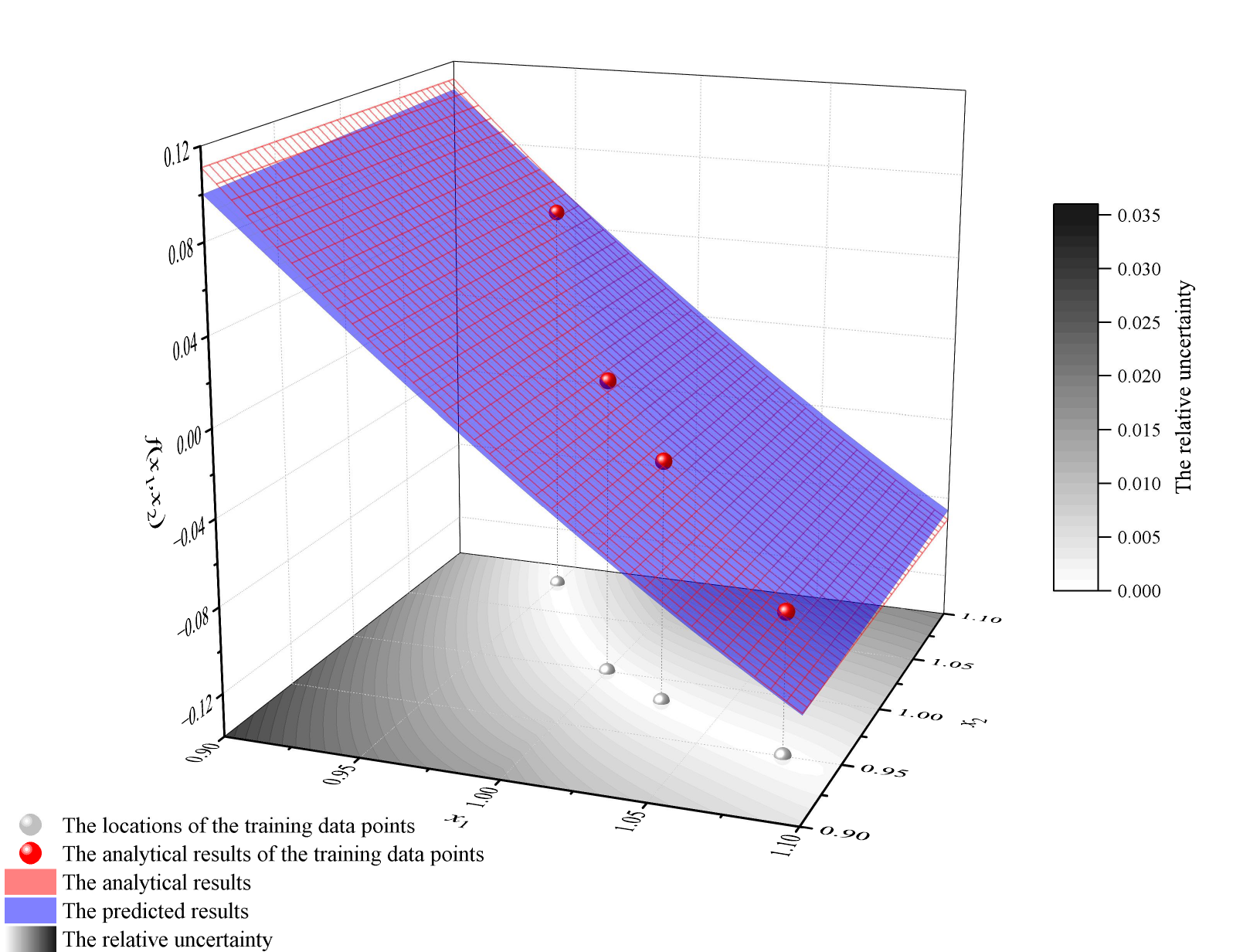}
			\label{fig:cy_consitutive_b1}
		\end{subfigure}%
	\end{minipage}
	\vspace{-6mm}
	\vfill
	\centering
	\raisebox{2.5\height}{\rotatebox{90}{\scriptsize{$Wi=0.5$}}}
	\begin{minipage}[t]{0.9\linewidth}
		\centering
		\begin{subfigure}[b]{0.49\linewidth}
			\centering
			\includegraphics[width=\linewidth]{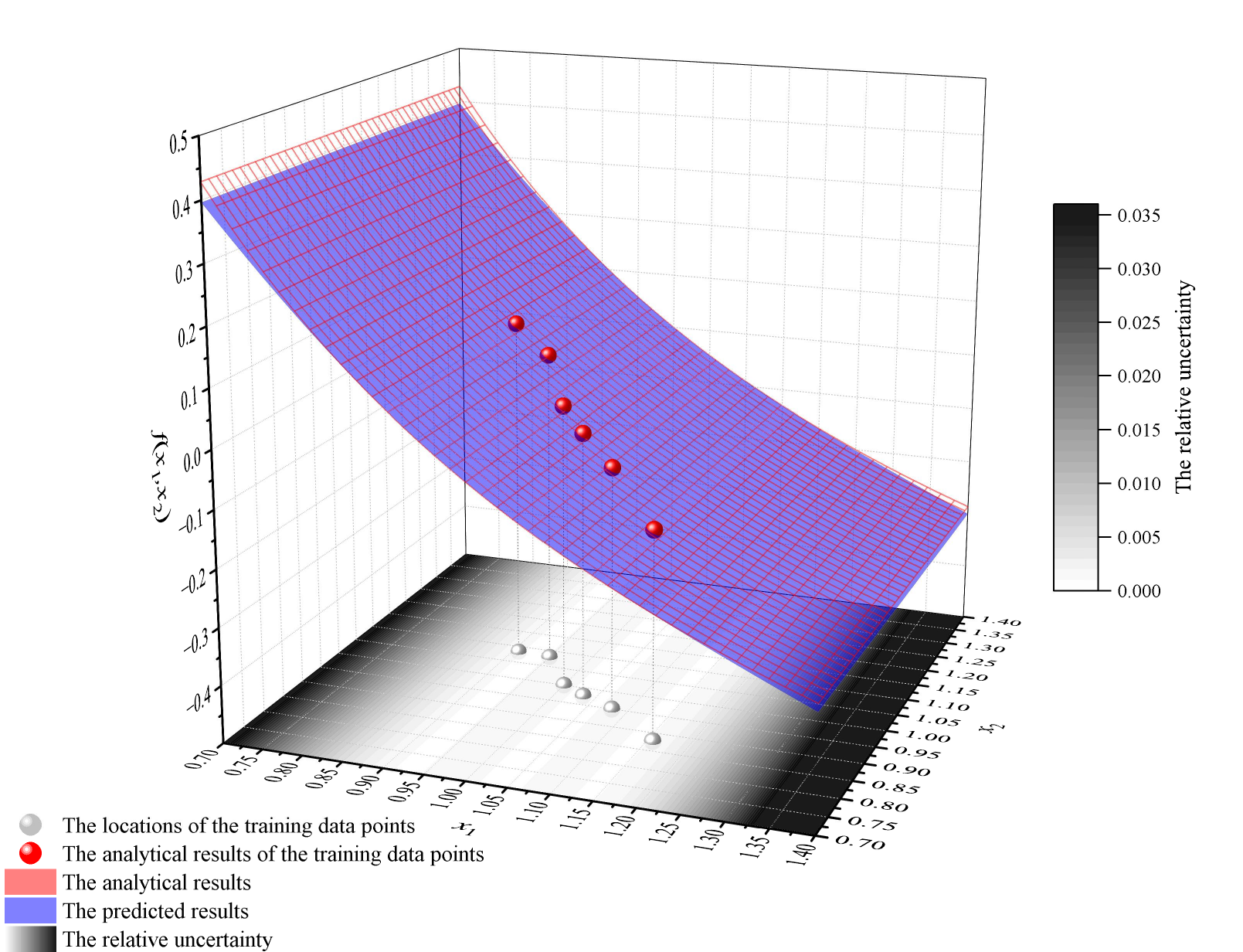}
			\label{fig:cy_consitutive_a2}
		\end{subfigure}%
		\hfill
		\begin{subfigure}[b]{0.49\linewidth}
			\centering
			\includegraphics[width=\linewidth]{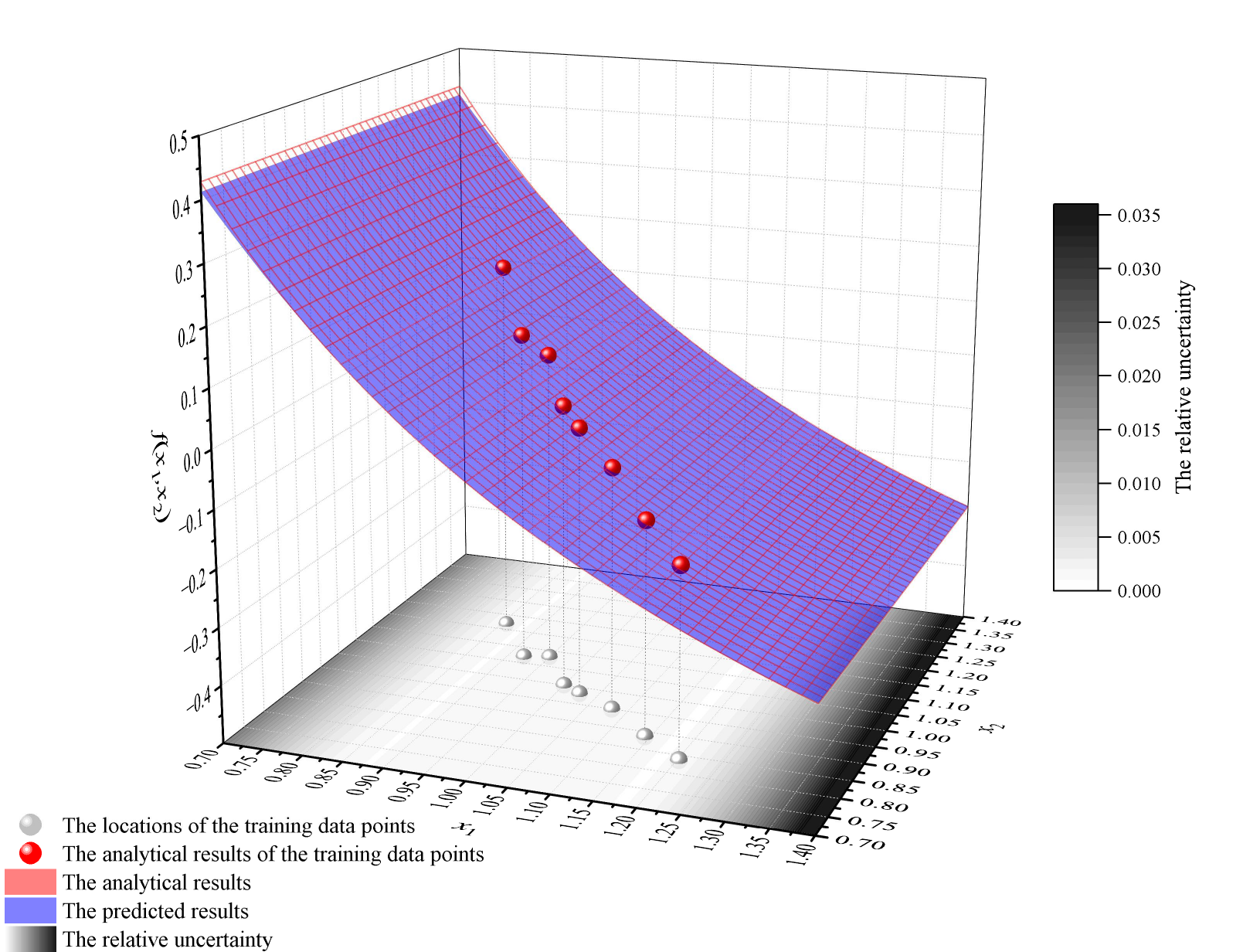}
			\label{fig:cy_consitutive_b2}
		\end{subfigure}%
	\end{minipage}
	\vspace{-6mm}
	\vfill
	\centering
	\raisebox{2.5\height}{\rotatebox{90}{\scriptsize{$Wi=1.0$}}}
	\begin{minipage}[t]{0.9\linewidth}
		\centering
		\begin{subfigure}[b]{0.49\linewidth}
			\centering
			\includegraphics[width=\linewidth]{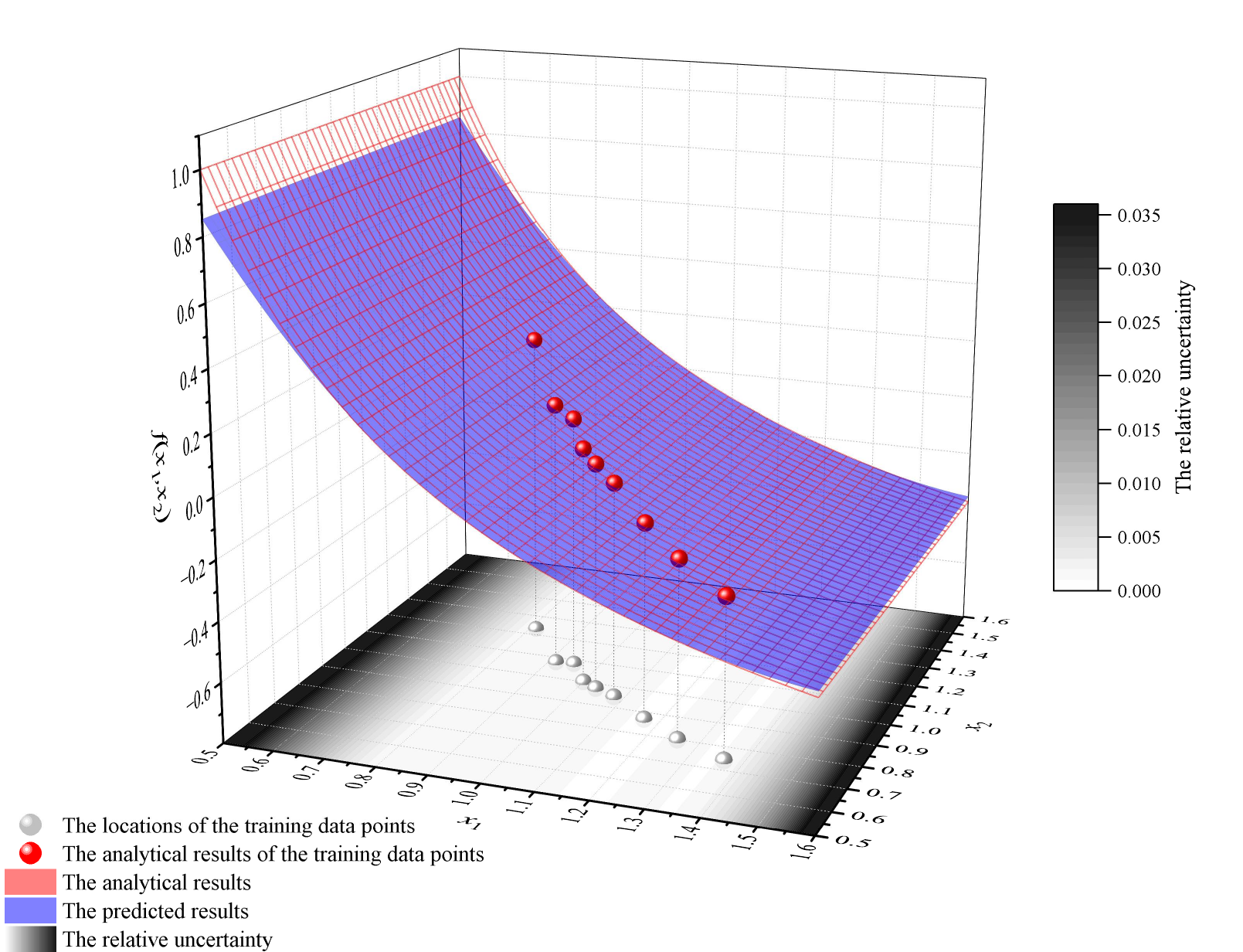}
			\caption{$\delta_{\rm{tol}}=0.01$ with R1}
			\label{fig:cy_consitutive_a3}
		\end{subfigure}%
		\hfill
		\begin{subfigure}[b]{0.49\linewidth}
			\centering
			\includegraphics[width=\linewidth]{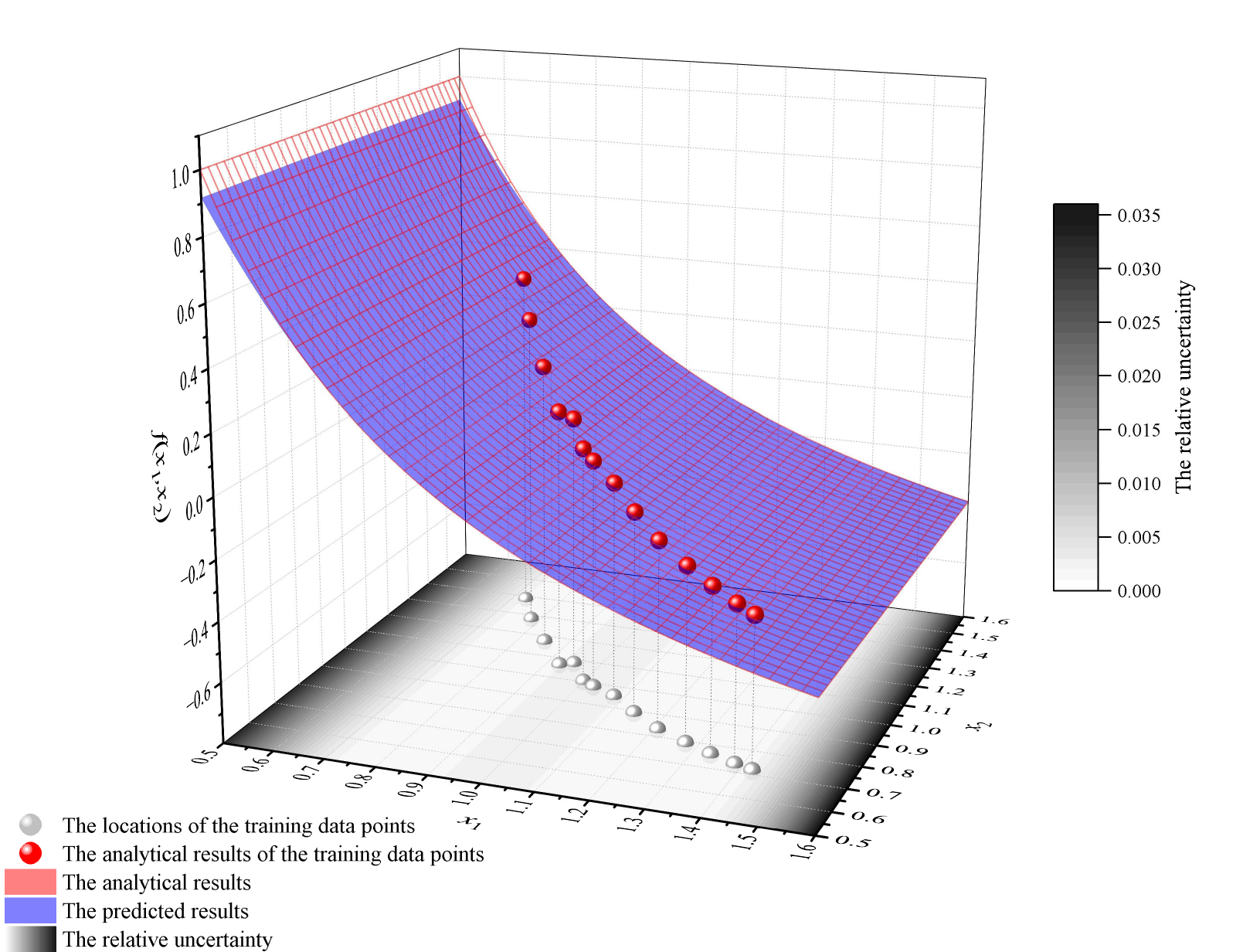}
			\caption{$\delta^{\prime}_{\rm{tol}}=0.002$ with R1}
			\label{fig:cy_consitutive_b3}
		\end{subfigure}%
	\end{minipage}
	\vspace{-1mm}
	\captionsetup{justification=justified}
	\caption{The final learned constitutive relations obtained by the ${\rm{G^2ALSPH}}$ method for the flows around a periodic array of cylinders ($Wi=0.1$, $Wi=0.5$, and $Wi=1.0$): (a) the relative uncertainty strategy ($\delta_{\rm{tol}}=0.01$ with R1 serving as the initial local constitutive relation) and (b) the direct uncertainty strategy ($\delta^{\prime}_{\rm{tol}}=0.002$ with R1 serving as the initial local constitutive relation), where $\delta_{\rm{tol}}=0.01$ plays the equivalent role in the relative uncertainty strategy as $\delta^{\prime}_{\rm{tol}}=0.002$ does in the direct uncertainty strategy.}
	\label{fig:cy_consitutive}
\end{figure}
\begin{figure}[H]
	\centering
	\begin{minipage}[t]{0.9\linewidth}
		\centering
		\raisebox{2.9\height}{\rotatebox{90}{\scriptsize{$Wi=0.1$}}}
		\begin{subfigure}[b]{0.49\linewidth}
			\raisebox{0.1\height}{\rotatebox{0}{\scriptsize{The velocity variations}}}
			\centering
			\includegraphics[width=\linewidth]{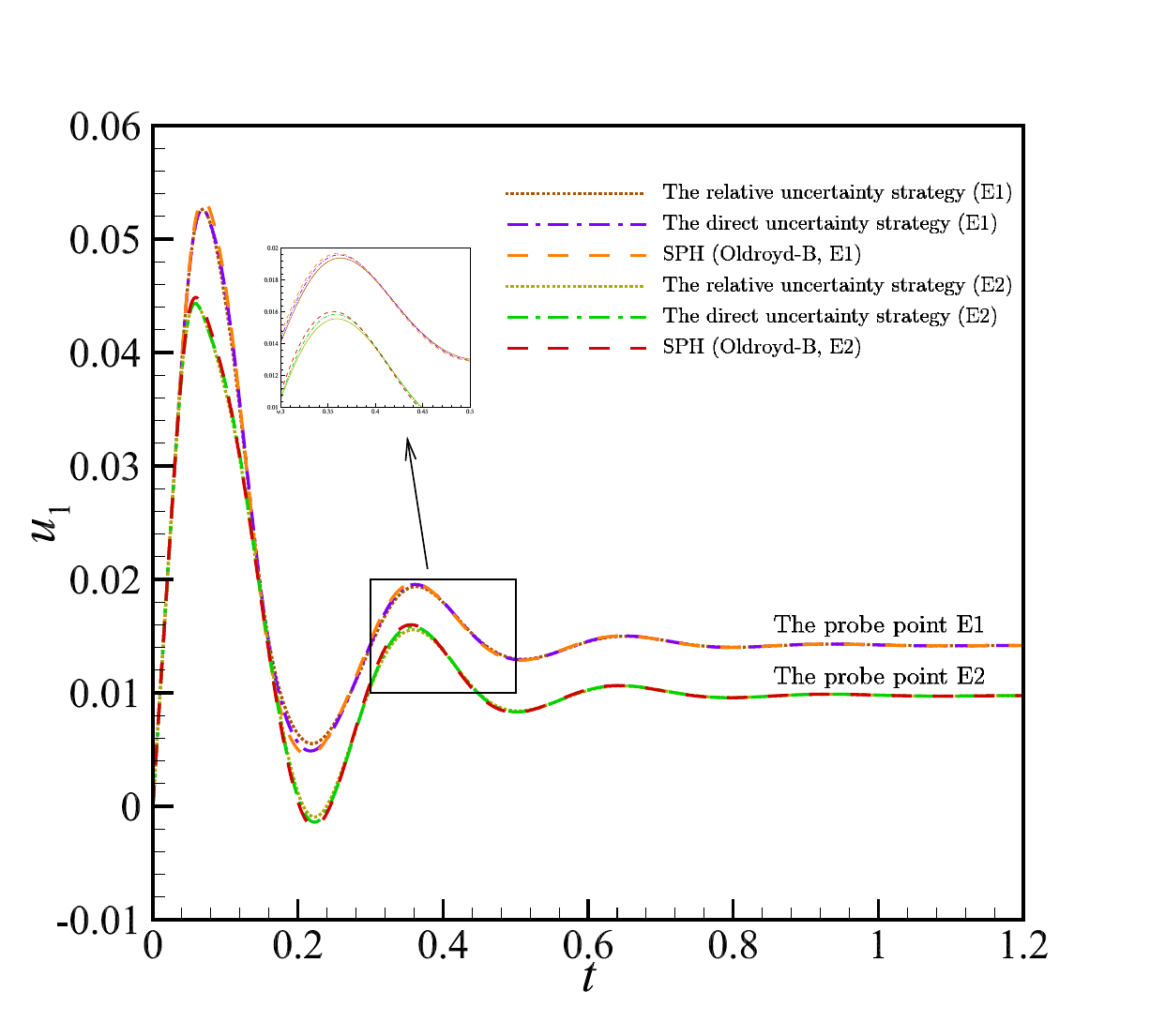}
			\label{fig:cy_results_a1}
		\end{subfigure}%
		\hfill
		\begin{subfigure}[b]{0.49\linewidth}
			\raisebox{0.1\height}{\rotatebox{0}{\scriptsize{The shear stress variations}}}
			\centering
			\includegraphics[width=\linewidth]{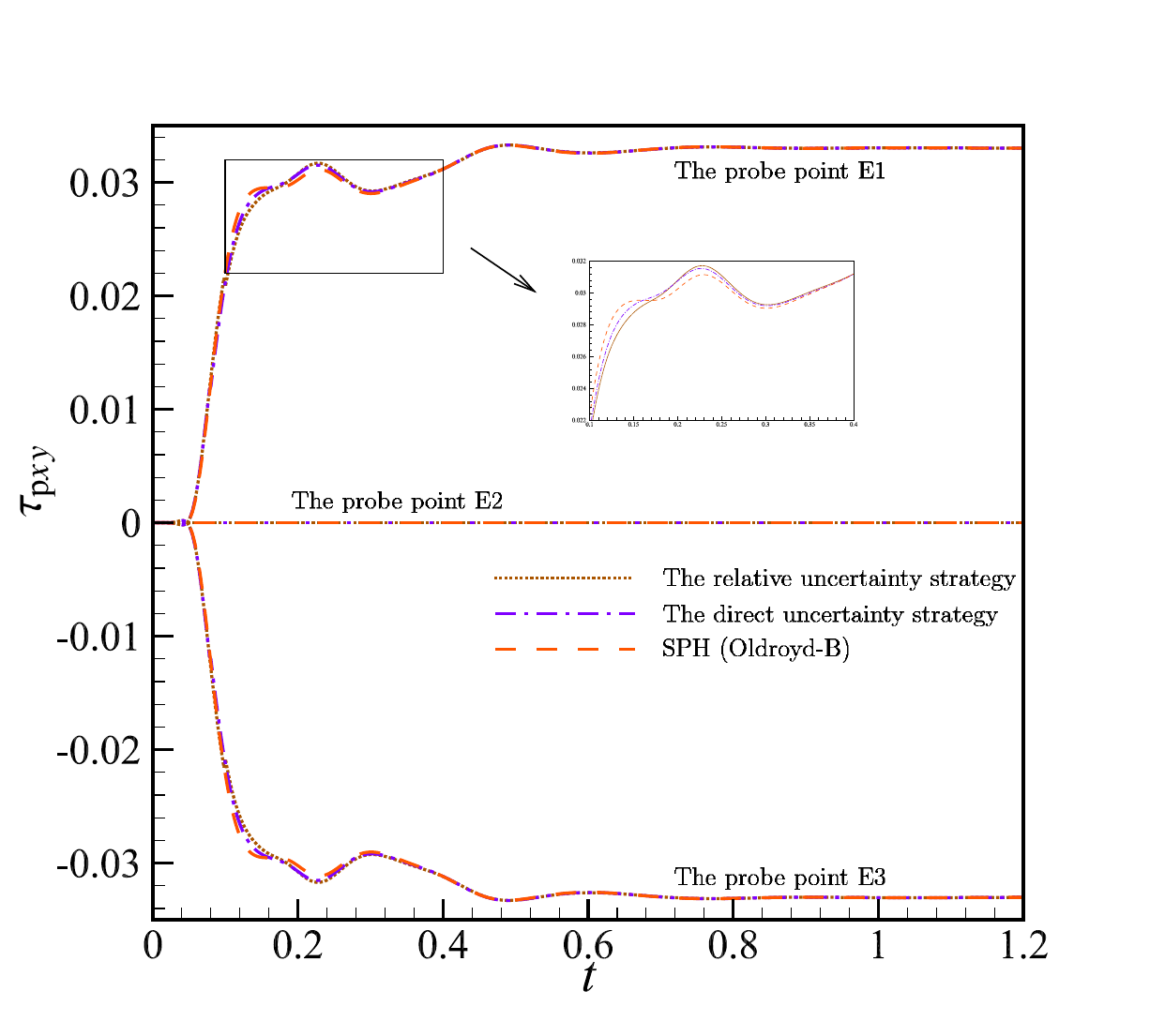}
			\label{fig:cy_results_b1}
		\end{subfigure}%
	\end{minipage}
	\vspace{-6mm}
	\vfill
	\centering
	\raisebox{2.9\height}{\rotatebox{90}{\scriptsize{$Wi=0.5$}}}
	\begin{minipage}[t]{0.9\linewidth}
		\centering
		\begin{subfigure}[b]{0.49\linewidth}
			\centering
			\includegraphics[width=\linewidth]{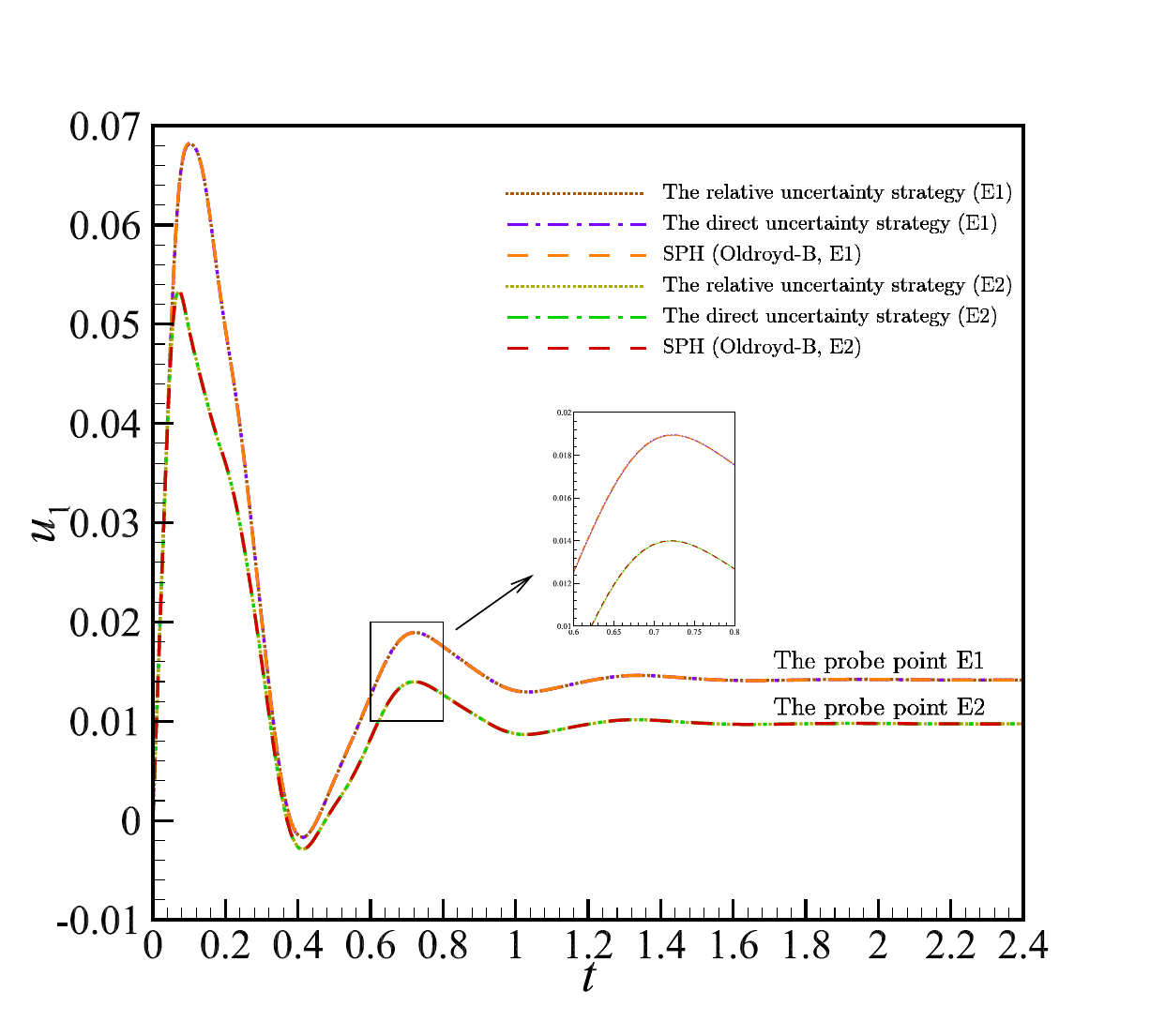}
			\label{fig:cy_results_a2}
		\end{subfigure}%
		\hfill
		\begin{subfigure}[b]{0.49\linewidth}
			\centering
			\includegraphics[width=\linewidth]{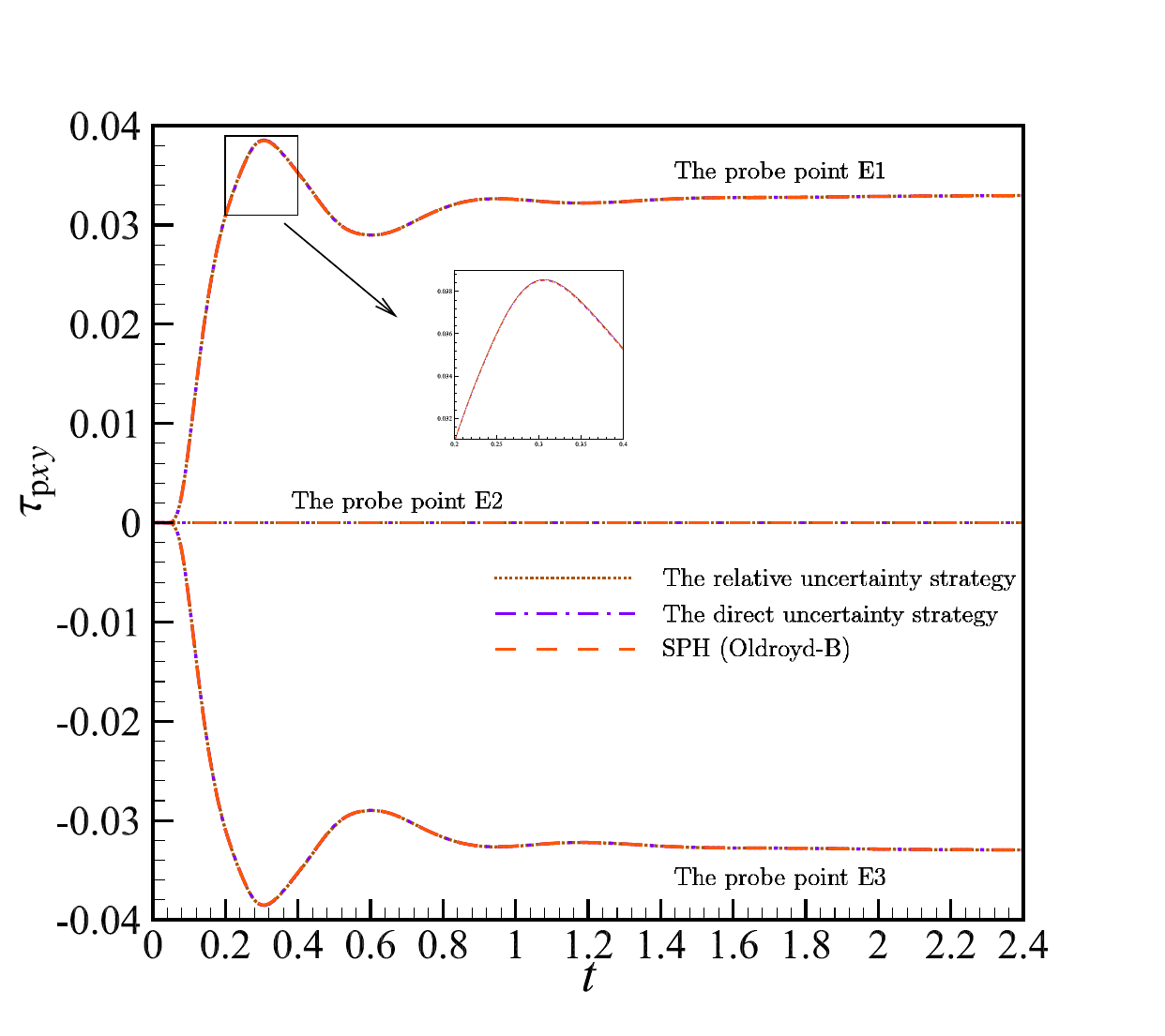}
			\label{fig:cy_results_b2}
		\end{subfigure}%
	\end{minipage}
	\vspace{-6mm}
	\vfill
	\centering
	\raisebox{2.9\height}{\rotatebox{90}{\scriptsize{$Wi=1.0$}}}
	\begin{minipage}[t]{0.9\linewidth}
		\centering
		\begin{subfigure}[b]{0.49\linewidth}
			\centering
			\includegraphics[width=\linewidth]{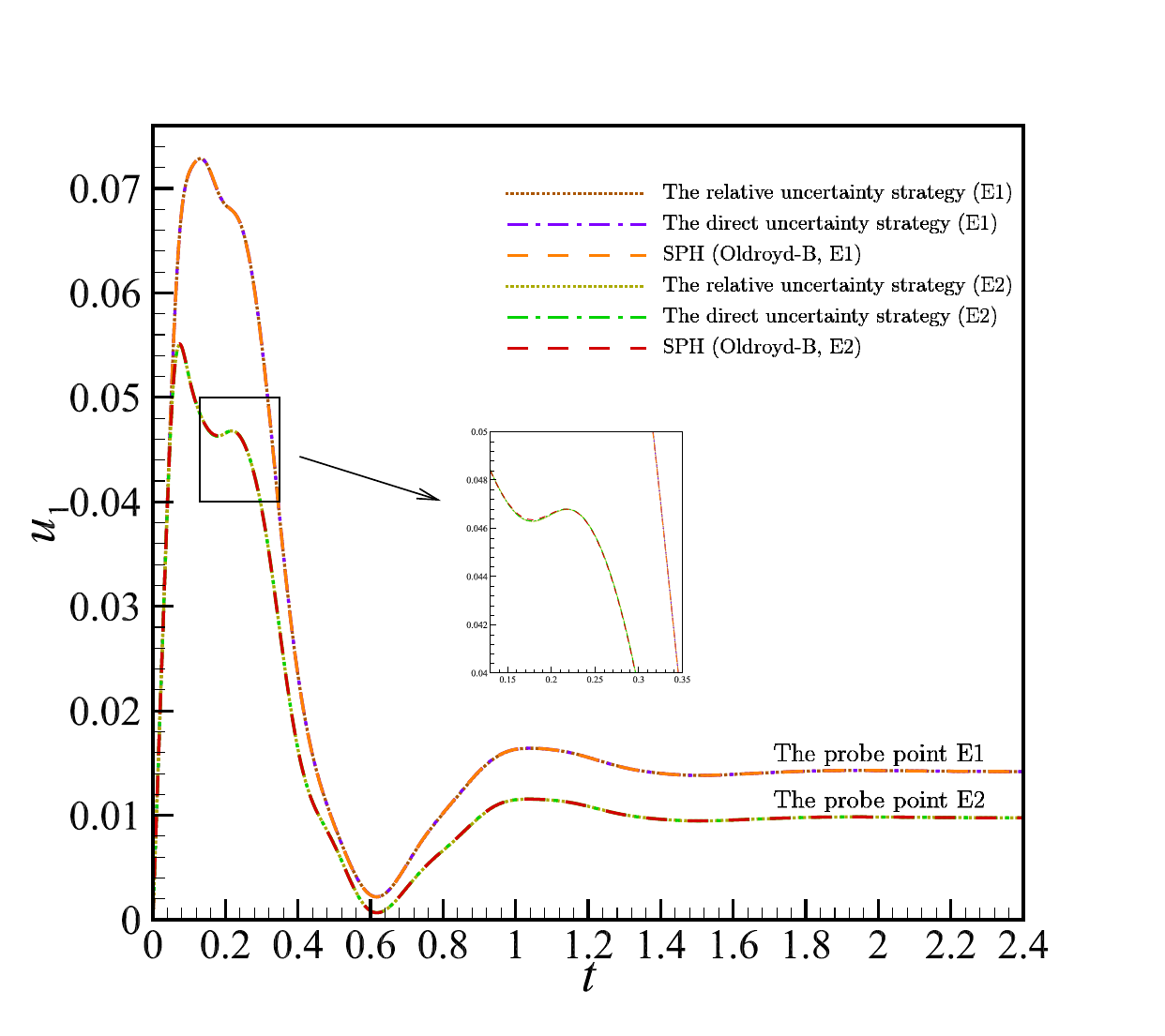}
			\caption{}
			\label{fig:cy_results_a3}
		\end{subfigure}%
		\hfill
		\begin{subfigure}[b]{0.49\linewidth}
			\centering
			\includegraphics[width=\linewidth]{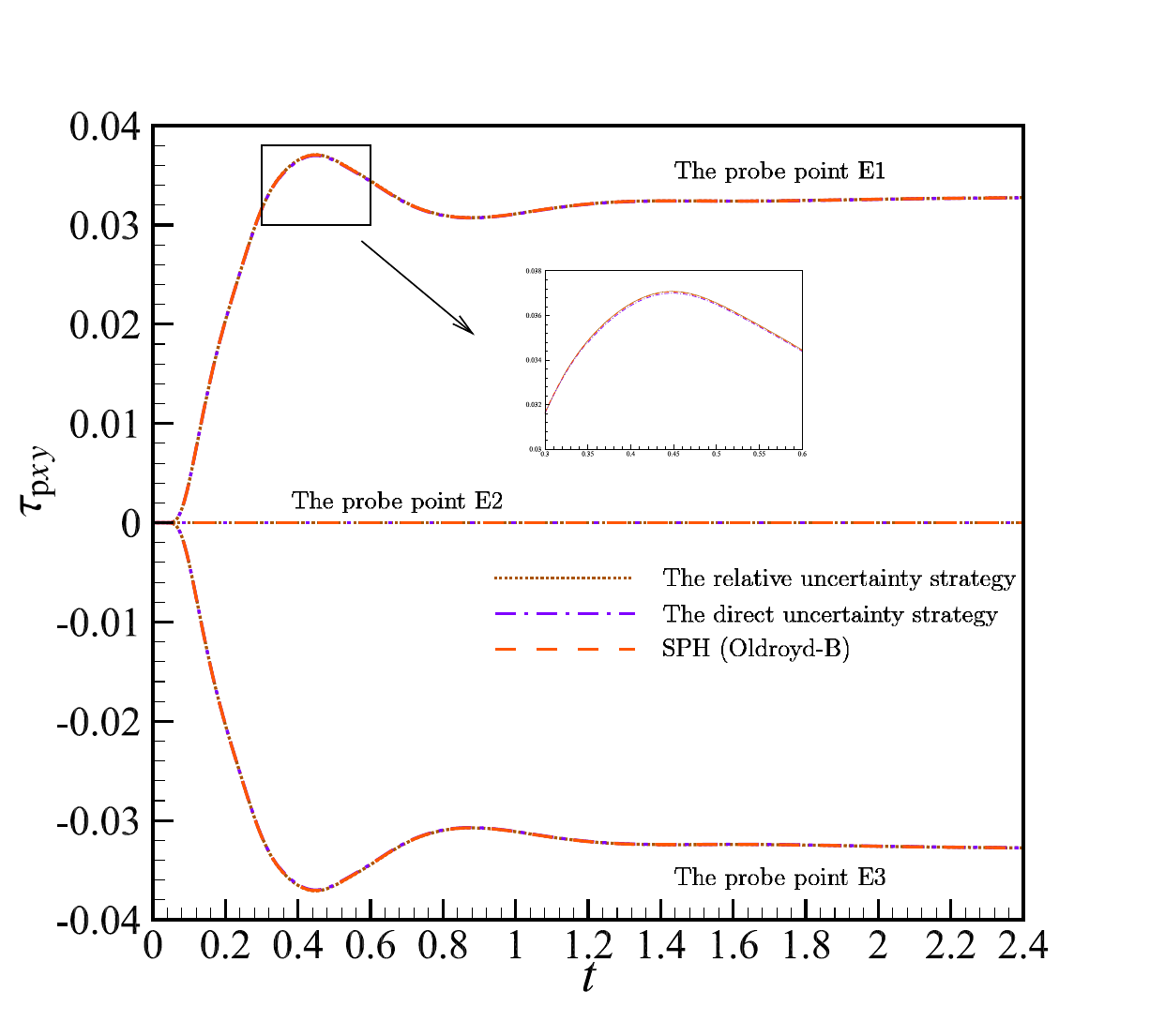}
			\caption{}
			\label{fig:cy_results_b3}
		\end{subfigure}%
	\end{minipage}
	\vspace{-1mm}
	\captionsetup{justification=justified}
	\caption{Comparison among the results in the probe points E1, E2, and E3 in the flows around a periodic array of cylinders ($Wi=0.1$, $Wi=0.5$, and $Wi=1.0$), which are obtained by the ${\rm{G^2ALSPH}}$ method with relative and direct uncertainty strategies and the SPH method with the Oldroyd-B model: (a) the variations of the velocity $u_1$ over time and (b) the variations of the shear stress $\tau_{{\rm{p}}xy}$ over time.}
	\label{fig:cy_results}
\end{figure}
\begin{figure}[H]
	\centering		
	\includegraphics[width=9cm]{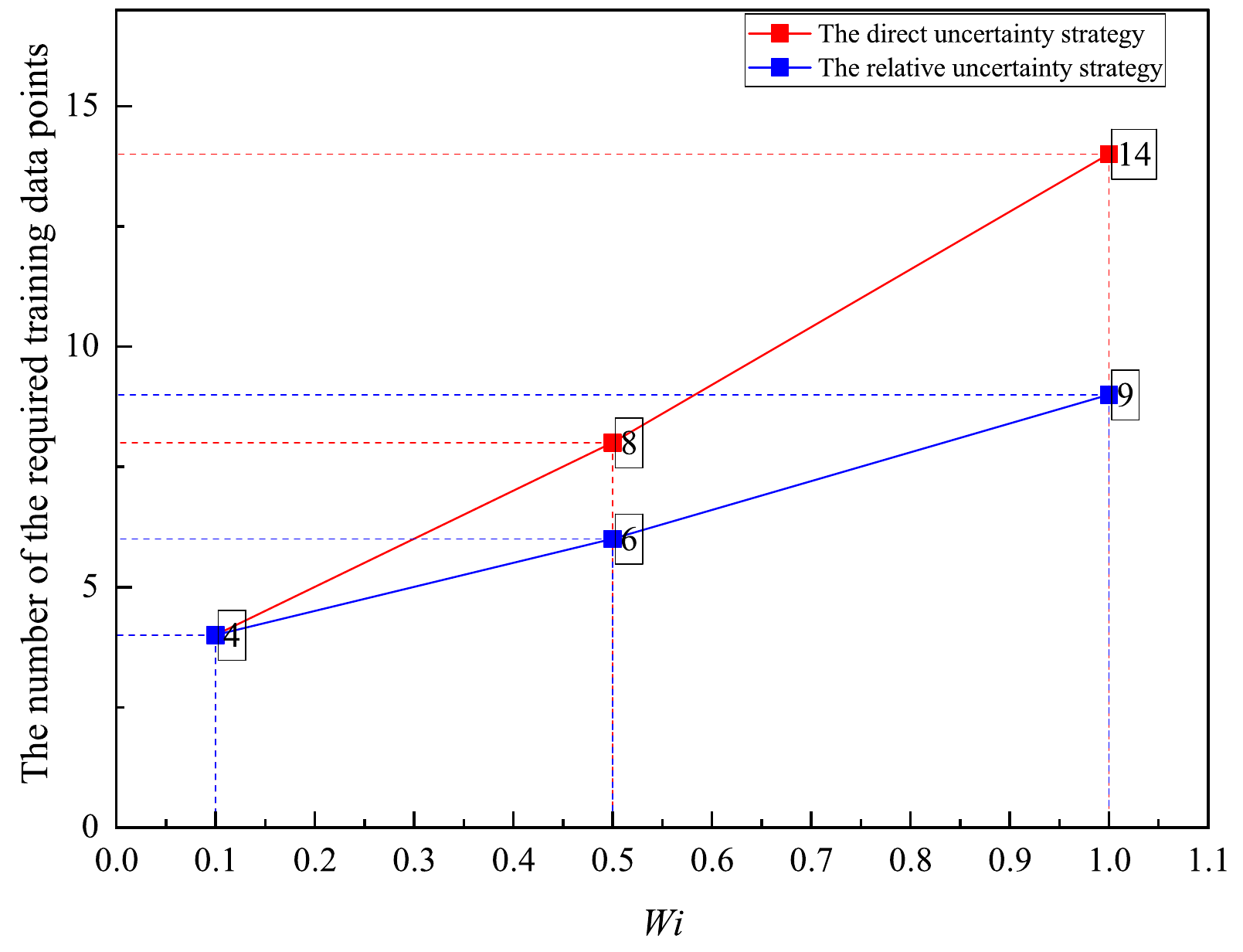}
	\caption{Comparison of the numbers of the final required training data points obtained by the relative and the direct uncertainty strategies in the flows around a periodic array of cylinders ($Wi=0.1$, $Wi=0.5$, and $Wi=1.0$).}
	\label{fig:number2}
\end{figure}

The same findings as in~\cref{sec4.2.3} can be obtained here. In~\cref{fig:cy_consitutive}, the final learned constitutive relations are presented. The direct uncertainty strategy yields more accurate constitutive relations in larger regions due to its stricter tolerance, along with more required training data points. In~\cref{fig:cy_results}, the results of the variations of the velocity $u_1$ and the shear stress $\tau_{{\rm{p}}xy}$ over time in the probe points E1, E2, and E3 are given. The direct uncertainty strategy with $\delta^{\prime}_{\rm{tol}}=0.002$ yields results slightly closer to those reference analytical results than the relative uncertainty strategy with $\delta_{\rm{tol}}=0.01$, due to its stricter limit. However, both strategies achieve high accuracy and closely agree. \cref{fig:number2} shows the comparison of the numbers of the training data points in the final learned constitutive relations. As the Weissenberg number increases, the difference in the number of data points required for the two strategies to complete the simulation becomes larger. Overall, the relative uncertainty strategy reduces required training data while maintaining accuracy compared to the direct uncertainty strategy. The superiority of the relative uncertainty strategy over the direct uncertainty strategy is fully validated.
\subsubsection{Analysis of the space explored by the eigenvalues $c_1$ and $c_2$ in the flows around a periodic array of cylinders}\label{sec4.4.2}
\begin{figure}[H]
	\centering		
	\includegraphics[width=9.0cm]{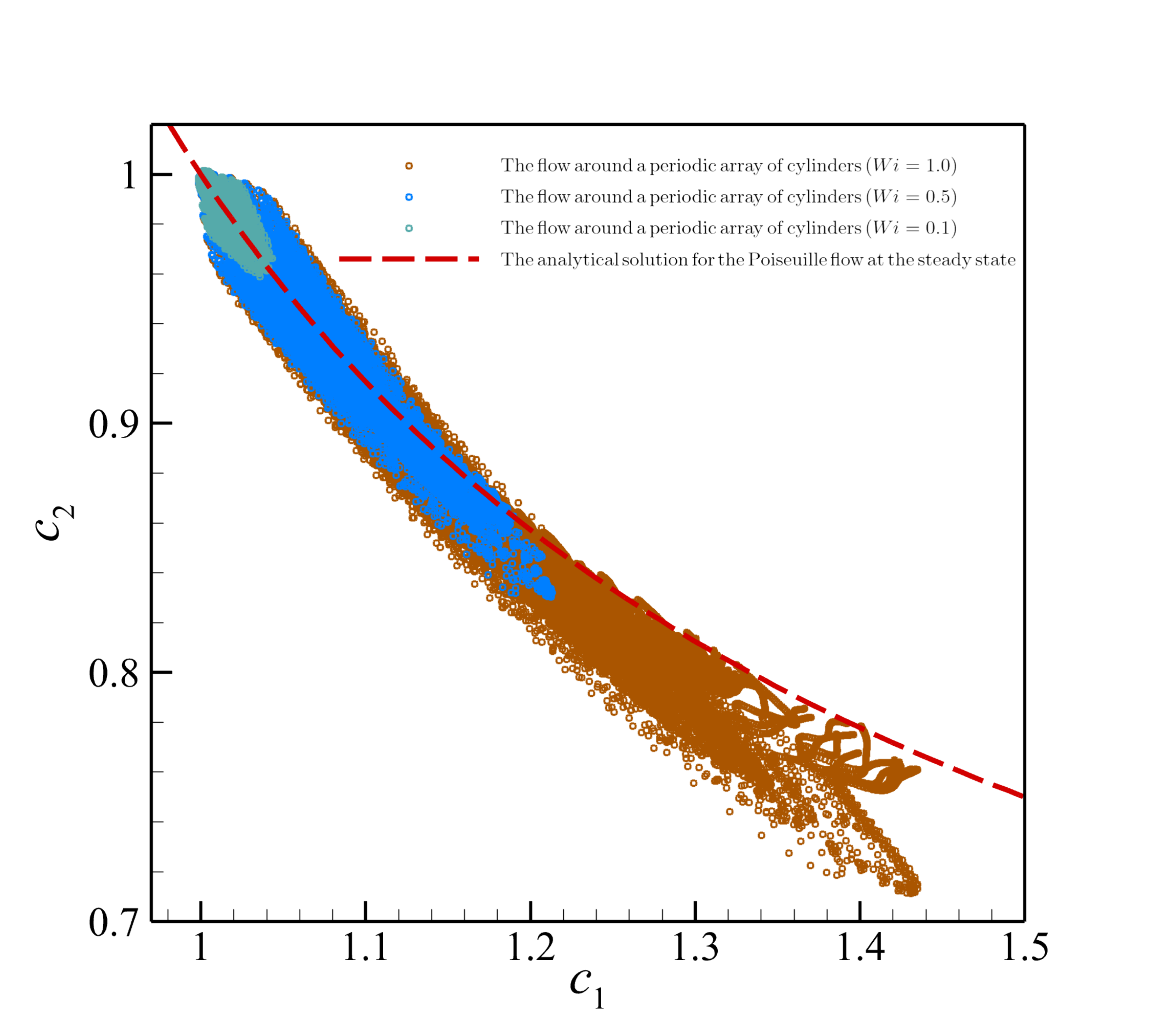}
	\caption{The locations of the eigenvalues $c_1$ and $c_2$ outputted every $500$ time iterations in the flows around a periodic array of cylinders ($Wi=0.1$, $Wi=0.5$, and $Wi=1.0$) and the analytical solution for the Poiseuille flow at the steady state.}
	\label{fig:c1c2_3}
\end{figure}

As shown in~\cref{fig:c1c2_3}, the locations of the eigenvalues $c_1$ and $c_2$ in the transient flows around a periodic array of cylinders at $Wi=0.1$, $Wi=0.5$, and $Wi=1.0$ are outputted every $500$ time iterations, which correspond to the cyan, blue, and brown circle symbols, respectively. The same conclusion as in~\cref{sec4.2.4} can be drawn here. As the Weissenberg number increases, the elasticity of viscoelastic fluids increases, and the corresponding regions covered by the eigenvalues $c_1$ and $c_2$ during the numerical simulation expand. The learned constitutive relation R1 is applied as the initial local constitutive relation in the numerical cases in~\cref{sec4.4.1}. The learned constitutive relation R1 can satisfy a given pre-set tolerance limit in a small region, which is presented in~\cref{sec4.2.4}. Thus, for the flows around a periodic array of cylinders at $Wi=0.1$, $Wi=0.5$, and $Wi=1.0$ in~\cref{sec4.4.1}, more data points need to be dynamically sampled to guarantee that the updated learned constitutive relations satisfy the pre-set tolerance limit. In this way, the accuracy of the entire simulation process can be ensured at every instant.
\section{Conclusions}\label{sec5}
In this paper, a novel data-driven numerical method, i.e., the ${\rm{G^2ALSPH}}$ method, is presented for modeling viscoelastic fluids guided by the GENERIC framework. Moreover, a new GPR active learning strategy based on the relative uncertainty is proposed. The GENERIC framework simplifies the target constitutive relation to a function with two inputs and one output for two-dimensional flow problems, incorporating the flow-history-dependent memory effect without specifying any predescribed model. Guided by the GENERIC framework, the ${\rm{G^2ALSPH}}$ method can be generalized to any dilute polymer solution. And the target function can be accurately reconstructed by the new active learning strategy. Compared to the existing machine learning methods that are applied to learn the viscoelastic constitutive relation, the ${\rm{G^2ALSPH}}$ method greatly simplifies the complexity of the target constitutive relation, and the generalization ability can be ensured by actively updating the learned constitutive relation. The Poiseuille flows and the flows around a periodic array of cylinders at different Weissenberg numbers are rigorously investigated. The good agreement with the reference analytical results and the even particle distribution prove that the ${\rm{G^2ALSPH}}$ method is an effective tool to simulate viscoelastic flows with the constitutive relation learned from data.
\par
Regarding the new active learning strategy, its accuracy evaluation tool for GPR prediction results consists of the novel relative uncertainty. The simulation results show that the new active learning strategy reduces the required data while maintaining accuracy compared to the choice of the direct uncertainty. Fewer data result in fewer matrix calculations and fewer mesoscopic simulations when the ${\rm{G^2ALSPH}}$ method is possibly applied to multi-scale simulations. Furthermore, the training data set, the optimized hyperparameters, and the selected covariance function precisely and uniquely establish an accurate learned constitutive relation. As presented in the simulation results, the learned constitutive relation can be used as a well-predicted initial local constitutive relation in other flow simulations for the same viscoelastic fluid, which significantly widens the ${\rm{G^2ALSPH}}$ method's applicability.
\par
The initial local constitutive relation and the pre-set tolerance are two necessary initial settings for the ${\rm{G^2ALSPH}}$ method. The simulation results demonstrate their accuracy is hardly affected by the choice of the initial local constitutive relation but is determined by the setting of the pre-set tolerance. A proper tolerance setting can ensure the accuracy of the learned constitutive relation at every time instant and hence the accuracy of the entire simulation process. Moreover, by decreasing the pre-set tolerance, the predictive accuracy can be improved while more training data points need to be sampled. In the future, the required data are expected to be provided by concurrent mesoscopic simulations, where the method proposed here represents an accurate and robust data-driven framework.
\section*{Declaration of competing interest}
The authors declare that they have no known competing financial interests or personal relationships that could have appeared to influence the work reported in this paper.
\section*{Data availability}
Data will be made available on request.
\section*{Acknowledgments}
This research is supported by the China Scholarship Council (No. 202206290140), by the National Natural Science Foundation of China (Grant Nos. 12471389 and 11971387), and by the Basque Government through the BERC 2022-2025 program, the ELKARTEK 2024 program (ELASTBAT: KK-2024/00091), and the IKUR 2025 program. The research is also partially funded by the Spanish State Research Agency through BCAM Severo Ochoa excellence accreditation CEX2021-001142-S/MICIN/AEI/10.13039/501100011033 and through the project PID2020-117080RB-C55 (\verb+"+Microscopic foundations of soft-matter experiments: computational nano-hydrodynamics\verb+"+ - acronym \verb+"+Compu-Nano-Hydro\verb+"+). The authors would like to thank Dr. Adolfo Vázquez-Quesada and Dr. Elnaz Zohravi for their helpful discussions.
\appendix
\section{The non-dimensionalization of the governing equations}
\label{app:non-dimensionalization}
The non-dimensionalization can decrease the number of free parameters and facilitate the study of the problem. In this paper, the non-dimensional parameters are defined as follows:
\begin{equation}
	\begin{aligned}
		& \bm{r}^*=\frac{{\bm{r}}}{L}, \quad \bm{u}^*=\frac{{\bm{u}}}{V}, \quad \rho^*=\frac{\rho}{\rho_0}, \quad \nabla^*=L \nabla,\\
		& p^*=\frac{p L}{\eta V}, \quad \tau^*=\frac{\tau L}{\eta V}, \quad t^*=\frac{V}{L} t,
		\label{eq:non-dimensional}
	\end{aligned}
\end{equation}
where the variables $L$, $V$, and $\eta$ are length scale, velocity scale, and viscosity scale, respectively. The non-dimensional Reynolds number is defined as $Re=\rho_0 V L/\eta$. The non-dimensional Weissenberg number is defined as $Wi=\lambda_{\rm{p}} V/L$, where $\lambda_{\rm{p}}$ is the polymeric relaxation time. The solvent viscosity ratio is defined as $\beta=\eta_{\rm{s}}/\eta$. The total viscosity $\eta$ is defined by the expression $\eta=\eta_{\rm{s}}+\eta_{\rm{p}}$, where $\eta_{\rm{s}}$ and $\eta_{\rm{p}}$ are the solvent and polymer contributions to the total solution viscosity, respectively. By employing the~\cref{eq:non-dimensional} into ~\cref{eq:mass,eq:momentum2,eq:evaluation_eigenvalue_new,eq:target_function_eigen}, and omitting the asterisk \verb+"+$\ast$\verb+"+, the non-dimensional mass and momentum conservation equations and the non-dimensional constitutive equation can be given as
\begin{equation}
	\dot{\rho} = -\rho\nabla\cdot\bm{u},
	\label{eq:mass-non-dimensional}
\end{equation}
\begin{equation}
	\bm{\dot{u}} = -\frac{1}{Re}\nabla p + \frac{\beta}{Re}{\nabla ^2}{\bm{u}} + \frac{1}{Re}\nabla\cdot\bm{\tau}_{\rm{p}} + Ga\frac{\bm{F}}{g},
	\label{eq:momentum-non-dimensional}
\end{equation}
\begin{equation}
	\dot{c}_\alpha=2 c_\alpha \kappa_{\alpha \alpha}+\frac{1}{Wi} c_\alpha {\tilde{\sigma}}_\alpha,
	\label{eq:evaluation_eigenvalue_new-non-dimensional}
\end{equation}
\begin{equation}
	\bm{\tau}_{\rm{p}}=-\frac{1-\beta}{Wi}\sum_\alpha c_\alpha \tilde{\sigma}_\alpha \bm{v}_\alpha \bm{v}_\alpha^{\mathrm{T}},
	\label{eq:target_function_eigen-non-dimensional}
\end{equation}
where the parameter $Ga=\rho g L^2 / (\eta V)$ is defined and $g$ is the value of the gravity acceleration~\cite{ellero_SPH_2005,simavilla_Nonaffine_2023,otto_Machine_2024}.
\section{The analytical solutions of the Poiseuille flow with the Oldroyd-B model}
\label{app:Poiseuille_analytical}
For a viscoelastic flow with the Hookean dumbbell model or the Oldroyd-B model, the analytical solution for the transient velocity in the transverse flow direction of the Poiseuille flow in a two-dimensional channel with respect to the position $r_y$ is given by~\cite{xue_Numerical_2004,ellero_SPH_2005}
\begin{equation}
	u_{1}(r_y, t)=-\frac{Re}{2}r_y(r_y-1.0) F_0-\sum_{n=0}^{\infty} \frac{4 Re F_0}{N^{3}} \sin (N r_y) \exp \left(-\frac{\alpha_{n}}{2 W i} t\right) G(t),
	\label{eq:function_u}
\end{equation}
where
\begin{equation}
	\alpha_{n}=1+N^{2} \beta \frac{W i}{Re},\quad N=(2 n+1) \pi,
	\label{eq:alpha_N_N}
\end{equation}
\begin{equation}
	G(t)= \begin{cases}\cosh \left(\frac{\beta_{n} t}{2 W i}\right)+\frac{\gamma_{n}}{\beta_{n}} \sinh \left(\frac{\beta_{n} t}{2 W i}\right), & \alpha_{n} \geq 2 N \sqrt{W i / Re}, \\ \cos \left(\frac{\beta_{n} t}{2 W i}\right)+\frac{\gamma_{n}}{\beta_{n}} \sin \left(\frac{\beta_{n} t}{2 W i}\right), & \alpha_{n}<2 N \sqrt{W i / Re},\end{cases}
	\label{function_G}
\end{equation}
and
\begin{equation}
	\beta_{n}=\sqrt{\left|\alpha_{n}^{2}-4 N^{2} \frac{W i}{Re}\right|}, \quad \gamma_{n}=1-N^{2}(2-\beta) \frac{W i}{Re},
	\label{eq:theta_gamm}
\end{equation}
and the variables $Re$, $Wi$, $F_0$, and $\beta$ are the Reynolds number, the Weissenberg number, the external force acceleration in the $y$ direction, and the solvent viscosity ratio, respectively. The solvent viscosity ratio $\beta$ is defined as $\beta=\eta_{\rm{s}}/\eta$. The total viscosity $\eta$ is defined by the expression $\eta=\eta_{\rm{s}}+\eta_{\rm{p}}$, where $\eta_{\rm{s}}$ and $\eta_{\rm{p}}$ are the solvent and polymer contributions to the total solution viscosity, respectively.
\par
Based on the above variable definitions, the analytical solution for the transient shear stress in the transverse flow direction of the Poiseuille flow is given by~\cite{xue_Numerical_2004,ellero_SPH_2005}
\begin{equation}
	\tau_{{\rm{p}}xy}(r_y, t)=- F_0 (1-\beta) (r_y-0.5) +\sum_{n=0}^{\infty} \frac{4 Re F_0}{N} \cos (N r_y) \exp \left(-\frac{\alpha_n}{2 Wi} t\right) Q(t),
\end{equation}
where
\begin{equation}
	Q(t)=\left(\frac{\beta}{Re N}-\frac{\alpha_n}{2 Wi N^3}\right) G(t)+\frac{1}{N^3} \frac{\partial G(t)}{\partial t}.
\end{equation}
\par
When the viscoelastic Poiseuille flow with the Hookean dumbbell model or the Oldroyd-B model reaches the steady state, the analytical solutions for the velocity, the shear stress, and the first normal stress difference are as follows~\cite{xue_Numerical_2004,ellero_SPH_2005}: 
\begin{equation}
	u_{x}(r_y)=-\frac{Re F_0}{2}r_y\left(r_y-1\right),\quad
	\tau_{{\rm{p}}xy}(r_y)=\frac{1-\beta}{Re}\left(\frac{\partial u_{x}}{\partial r_y}\right),\label{analytical1}
\end{equation}
\begin{equation}
	N_{1}(r_y)=\tau_{{\rm{p}}xx}(r_y)-\tau_{{\rm{p}}yy}(r_y)=\frac{2Wi(1-\beta)}{Re}\left(\frac{\partial u_{x}}{\partial r_y}\right)^{2}.\label{analytical2}
\end{equation}
\bibliographystyle{elsarticle-num} 
\bibliography{references}
\end{document}